\documentclass[3p,times,sort&compress,number]{elsarticle}
\usepackage{H1}
\usepackage[pdftex,colorlinks=true]{hyperref}
\usepackage{bbold}
\usepackage{natbib}

\usepackage{graphicx}

\usepackage{array}

\newcommand{\Sztot}{S^z_{\rm tot}}

\hypersetup{
citecolor = blue
}

\newcommand{\Heff}{H_{\rm eff}}

\newcommand{\subsubsubsection}[1]{\emph{\textbf{#1}---} }

 \newcommand{\mcite}[1]{\cite{#1}}
 \newcommand{\mlabel}[1]{\label{#1}}

\begin{document}
\begin{frontmatter}

\title{A Brief History of Time Crystals}

\author[Har,Sta]{Vedika Khemani\corref{cor1}}
\cortext[cor1]{Corresponding Author}
\ead{vkhemani@stanford.edu}

\author[MPI]{Roderich Moessner}
\author[PU]{S.~L.~Sondhi}

\address[Har]{Department of Physics, Harvard University, Cambridge, Massachusetts 02138, USA}
\address[Sta]{Department of Physics, Stanford University, Stanford, California 94305, USA}
\address[MPI]{Max-Planck-Institut f\"{u}r Physik komplexer Systeme, 01187 Dresden, Germany}
\address[PU]{Department of Physics, Princeton University, Princeton, New Jersey 08544, USA}

\begin{abstract}
The idea of breaking time-translation symmetry has fascinated humanity at least since ancient proposals of the perpetuum mobile. Unlike the breaking of other symmetries, such as spatial translation in a crystal or spin rotation in a magnet, time translation symmetry breaking (TTSB) has been tantalisingly elusive. 
We review this history up to recent developments which have shown that 
{\it discrete} TTSB does takes place in periodically driven (Floquet) systems in the presence of many-body localization (MBL). Such Floquet time-crystals represent a new paradigm in quantum statistical mechanics --- that of an intrinsically \emph{out-of-equilibrium} many-body phase of matter with no equilibrium counterpart. 

We include a compendium of the necessary background on the statistical mechanics of 
phase structure in many-body systems, before specializing to a detailed discussion of the nature, and diagnostics, of
TTSB. In particular, we provide precise definitions that formalize the notion of a time-crystal as a stable, \emph{macroscopic, conservative clock} --- explaining both the need for a many-body system in the infinite volume limit, and for a lack of net energy absorption or dissipation. 
Our discussion emphasizes that TTSB in a time-crystal is accompanied by the breaking of a spatial symmetry --- so that time-crystals exhibit a novel form of \emph{spatiotemporal} order. 

We also cover a range of related phenomena, including various types of long (but not infinitely long)-lived prethermal 
time-crystals, and expose the roles played by symmetries -- exact and (emergent) approximate -- and their breaking. We clarify 
the distinctions between macroscopic many-body time-crystals and other ostensibly similar dynamical phenomena dating as far back as the works of Faraday and Mathieu. \emph{En route}, we encounter Wilczek's suggestion from a few years ago that macroscopic systems should exhibit time translation symmetry breaking in their ground states, together with a theorem that ruled this out.
We also analyse pioneering recent experimental work detecting signatures of time crystallinity in a variety of different platforms,
providing a detailed theoretical explanation of the physics in each case. In all existing experiments, the system does \emph{not} realize a `true' time-crystal phase in an asymptotic sense, and our analysis helps identify necessary ingredients for improvements in future experiments.

\end{abstract}

\end{frontmatter}

\tableofcontents

\section{Introduction}
\mlabel{sec:introduction}
\subsection{Another symmetry, another symmetry breaking?}
One of the great themes of modern physics is the role of symmetries in understanding the behavior
of systems with large numbers of degrees of freedom, be they the quantum field theories of particle physics or the assemblies of atoms that constitute the subject of condensed matter physics. A central piece of the importance of symmetries is the possibility that they are spontaneously broken, \emph{i.e.} that while the dynamics are exactly symmetric, the actual state of the system is not. Broadly, the subject of time crystals has to do with whether time translation invariance or time translational symmetry (TTS) can be spontaneously broken.  Specifically, what it is about has to do with whether the TTS broken state is periodic in time, whence ``time \emph{crystal}'' (TC). This represents the spontaneous emergence of a time-dependent ``clock" within a time-invariant system. The intellectual origins of this question arguably hark all the way back to early musings about perpetuum mobiles. In 2012 Wilczek raised this possibility for the ground states of macroscopic quantum systems~\mcite{Wilczek12} (as well as classical systems~\mcite{WilczekClassical}) and our subject was reborn in its most recent incarnation.

This review is primarily about the work that followed and hence about \emph{conservative} systems---those which do not absorb net energy from their environment. Clocks which run on a continuing energy input are, of course. legion---we all rely on them to get out of bed in the morning. They are particularly interesting in the context of chemical and biological systems where their systematic study can be traced at least to Prigogine \mcite{prigogine}, who coined the term ``time order'' which then inspired work on synchronization in oscillators such as in the celebrated Kuramoto model \mcite{Kuramoto1975, Kuramoto1987, KuramotoRMP}. While such open systems are interesting in their own right,  they do not present the question of principle that conservative clocks do. We will for the most part stay away from them here --- until the very end of this review where we discuss a large compendium of related dynamical phenomena.

This formulation of the TTSB question is deceptively simple: another symmetry, another symmetry breaking. Why would that not be possible? If space crystals, why not time crystals? But this turns out to be not so simple. Consider the following three correctives to the naive view. First, even in Lorentz invariant systems, which one might consider the best bet for swapping space for time, the two are not interchangeable---causality distinguishes between spacelike and timelike separations\footnote{More formally, generators of space translations and generators of time translations are very different---the latter is the Hamiltonian which is exquisitely system dependent, as it defines the system, while the former is the momentum which is much less system dependent.}. Second, the standard account of symmetry breaking is a set of statements about systems that reach thermal equilibrium. By their very nature, the late time equilibrium states of these systems cannot break TTS in any useful sense: the operational definition of reaching equilibrium is precisely that one can no longer detect the passage of time by making local measurements on the system! 
Third, finite-sized (few-body) systems routinely exhibit oscillations or recurrences, which naively implies that TTS is almost always broken in finite systems; to wit, consider a single simple harmonic oscillator or pendulum. This is in sharp contrast to traditional symmetries which are generally {\it not} broken in finite systems. 

What these considerations suggest is that any phenomena we wish to identify with TTSB are going to be outside the traditional equilibrium framework---hence intrinsically non-equilibrium phenomena (or possibly strictly ground state phenomena). And that they are going to be meaningful \emph{only} in the many-body infinite system limit. Once we abandon equilibrium though, the reader may worry that TTSB will be ubiquitous. After all one of the explanations for why there {\it is} time (the thermodynamic arrow of time) is that the universe is not in equilibrium. 

The solution to bounding the space of phenomena will be to a) restrict to local Hamiltonian systems with locally bounded Hilbert spaces, b) require at least some stability to choice of interactions as well as initial conditions so as to define a \emph{phase} of matter with TTSB and c) to require non-trivial time dependence at asymptotically long times so as to rule out transient phenomena (or adjacent phenomena such as phase ordering~\mcite{Bray}). 

With these constraints we will find that there is a relatively small set of macroscopic and conservative systems that need to be discussed. Among Hamiltonian systems,  only many body localized~\mcite{Anderson58, Basko06, Gornyi, PalHuse, OganesyanHuse, Prelovsek, Imbrie2016, NandkishoreMBLReview, EhudMBLRMP}, periodically driven (Floquet)~\cite{Shirley:1965cy,ZelDovich:1967va,Sambe:1973hi} systems appear to give rise to time crystals---stable. macroscopic, conservative clocks---in a fully sharp sense~\mcite{KhemaniPRL, ElsePRL, CVS}. 

Floquet systems are not time-invariant and only exhibit a \emph{discrete} time-translation symmetry (dTTS), corresponding to time-translations by integer multiples of the driving period: $H(t+nT)=H(t), \;n\in \mathbb{Z}$.  The dTTS is spontaneously broken in a Floquet time-crystal, in which observables show a ``subharmonic" response and oscillate with a period that is a multiple of the driving period\footnote{A spatial crystal with atoms arranged in a lattice has only a discrete spatial translation symmetry corresponding to translations by integer multiples of the lattice spacing, analogous to a Floquet system with dTTS. A lattice antiferromagnet or charge density wave can further break the discrete spatial translation symmetry by displaying a periodicity that is a multiple of the lattice spacing, in analogy to a discrete time-crystal with broken dTTS.}. In addition, these phases also exhibit long-range correlations in space, altogether displaying a novel form of \emph{spatiotemporal} order.  
Interestingly, the discovery of these discrete time crystals came just as various no-go theorems~\mcite{bruno, Nozieres, Oshikawa15} had established that the early proposals, by 
Wilczek and others~\mcite{Wilczek12, LiTCProposal}, for realizing time crystals would not work.

One the most exciting upshots of the discovery of Floquet time-crystals is that they represent new -- intrinsically non-equilibrium -- entries into the catalog of possible phases of matter~\cite{KhemaniPRL, MoessnerSondhiReview}. 
At its broadest level, the subject of time crystals has to do with the late-time states of isolated many-body quantum systems, and notions of order therein. Within traditional equilibrium thermodynamics, it is assumed that systems equilibrate to a thermal state at late times, and these can display patterns of order characteristic of different phases (and sharp changes at phase transitions as the system parameters are varied). 
In contrast, as we will discuss in the course of this review, Floquet time-crystals reflect a new paradigm in quantum statistical mechanics --- that of an \emph{out-of-equilibrium} many-body phase of matter. 

We note here that if we relax our temporal requirements from infinitely long lived phenomena to merely exponentially long lived phenomena, then the class expands to what are known as ``prethermal'' time crystals~\mcite{ElsePrethermal, LuitzInfT}. And if we allow for TTSB without strict periodicity in time, then it is possible to realize time-\emph{glasses}, a phenomenon we will touch on briefly. 

\subsection{Aims and outline}

We have two aims in this review. Our primary aim is to recount recent progress in ascribing a sharp meaning to the intuitive idea of time crystals, which give mathematical form to our comments leading to this point. In particular, this allows us to describe a series of recent advances that have culminated in a concrete demonstration of TTSB in periodically driven systems---where, far from {Prigogine's} and Wilczek's context, time crystals find their most solid home. A secondary aim is to survey some old and new phenomena which are clarified by these conceptual advances. Our main focus is on closed quantum systems where most of the recent action has taken place. But, in the interest of completeness, we also briefly survey classical systems as well as both classical and quantum open systems--- focusing in particular on a set of ostensibly related  phenomena that often come up in discussions of time crystals, such as Faraday waves~\mcite{faradaywaves} or period-doubling bifurcations {\it en route} to chaos~\mcite{Feigenbaum1978}. By the end, we hope that our discussion, particularly the statistical mechanical lens for examining phase structure that we adopt throughout, will convince the reader of what is special about macroscopic, conservative quantum time crystals.

This review is structured as follows. In Sec.~\ref{sec:history}, we first provide an overview of precursors of time crystals, starting with the venerable perpetuum mobile, and ending with the idea of a symmetry-protected time crystal. This journey will, in particular, help us sharpen the minimal ingredients that are needed for realizing a time-crystal, and the questions that must be answered in the remaining sections in our quest for a time-crystal. 

In order to keep the article self-contained, also in view of a certain lack of agreed-upon terminology in the field, we provide copious background material on quantum many-body systems. Sec.~\ref{sec:thermalisation} is a compendium of material on four central actors of our plot, namely: thermalisation, symmetry-breaking and order, many-body localisation, and Floquet systems. It provides an account of  equilibration and thermalization (or the absence thereof) in closed quantum systems, both with static and periodic Hamiltonians. It is written to provide a brief overview of those concepts, rather than a detailed exposition. The expert reader can safely skip this section but we note that an appreciation of  many-body localisation to help evade thermalization, and the notion of eigenstate order for defining phases and transitions between them out-of-equilibrium, are crucial  building blocks on which the following discussions draw extensively.  

Sec.~\ref{sec:DefineTTSB} then addresses the notion of time translation symmetry breaking in quantum systems. This builds upon, and is juxtaposed 
to the case of standard symmetry breaking. This is inspired by the discussion of Watanabe and Oshikawa in Ref.~\cite{Oshikawa15}, but substantively extends their treatment, including a number of hitherto unpublished considerations. We believe this section presents one the most comprehensive accounts of the issue of defining TTSB, to date. This also includes a discussion of the central no-go theorems of  Watanabe and Oshikawa, and a correction of one of their results, for finite temperatures. 

With all of these in hand, Sec.~\ref{sec:FMBLTC} is devoted to the presentation and analysis of the Floquet many-body localized time crystal, also known as a Floquet/discrete time crystal (DTC). This section starts with a presentation of a periodically driven MBL Ising chain, which realizes four different many-body phases~\cite{KhemaniPRL}. One of these, the so-called ``$\pi$ spin-glass" phase, spontaneously breaks both dTTS and spatial Ising symmetry --- demonstrating a novel form of non-equilibrium spatiotemporal order and furnishing the first concrete realization of a TC. We discuss the properties of this phase at length, focusing in particular on its stability, dynamics, and on various concrete diagnostics of TTSB.  

Next, we  relax some of the most stringent conditions in the definition of time-crystals, considering an expanded class of systems in Sec.~\ref{sec:prethermalTC}. This includes (i) Floquet \emph{prethermal} time-crystals that are not infinitely long-lived~\cite{ElsePrethermal, LuitzInfT}, and (ii) \emph{symmetry-protected} time-crystals in time-independent Hamiltonian systems, which are exceptions to the previously discussed no-go theorems for TCs in equilibrium. 

Sec.~\ref{sec:expt_real} is devoted to the various experimental realisations of time crystal physics that have appeared over the last couple of years, as well as the theoretical concepts invoked in their discussions~\mcite{MishaTCExp, MonroeTCExp, RovnyPRL, RovnyPRB, SreejithStars, CriticalTCPRL,LuitzInfT}. This section again contains some new analyses which has not yet appeared in print elsewhere, especially in relation to the experiment studying TCs in a trapped ion setup~\mcite{MonroeTCExp}. We perform an in-depth cross-platform analysis of the different experiments, which leads us to identify and sharpen the desired extensions of such experiments towards an unambiguous future realisation of a time-crystal.

Sec.~\ref{sec:zoology} presents a broader zoo of ``time crystals", 
extending our discussion beyond the setting of closed quantum systems. It presents phenomena of a broader provenance, and in particular  considers classical analogs, as well as (effectively) few-body versions of TCs. As a main conceptual enlargement of the idea,  it also addresses the vast field of open systems. This is concerned with nature and role of a ``bath" or reservoir as a matter of principle, and addresses issues such as stability of the TC phenomena, or their relation to the physics of clocks. 

This material will hopefully also allow the reader to get a sense of the wide variety of dynamical phenomena that are ostensibly similar to time-crystals, dating as far back as observations of period doubling in the works of Faraday, and ranging from chaos in mathematical biology to Rabi oscillations to complex many-body engineering.  These examples explicitly demonstrate how the omission of genuine many-body effects, or the permission of dissipation, or other departures from the Hamiltonian nature of the dynamics, enlarge the
class of phenomena to include many rather familiar settings which one may, or perhaps rather may not as we do not, choose to attach the label ``time crystal" to.

We conclude with closing remarks about broader lessons learnt in the search for time crystals, and an outlook for future research in Sec.~\ref{sec:outlook}.

\section{Time crystals: The first 864 years} 
\mlabel{sec:history}

\begin{figure}
    \centering
     \includegraphics[width=\columnwidth]{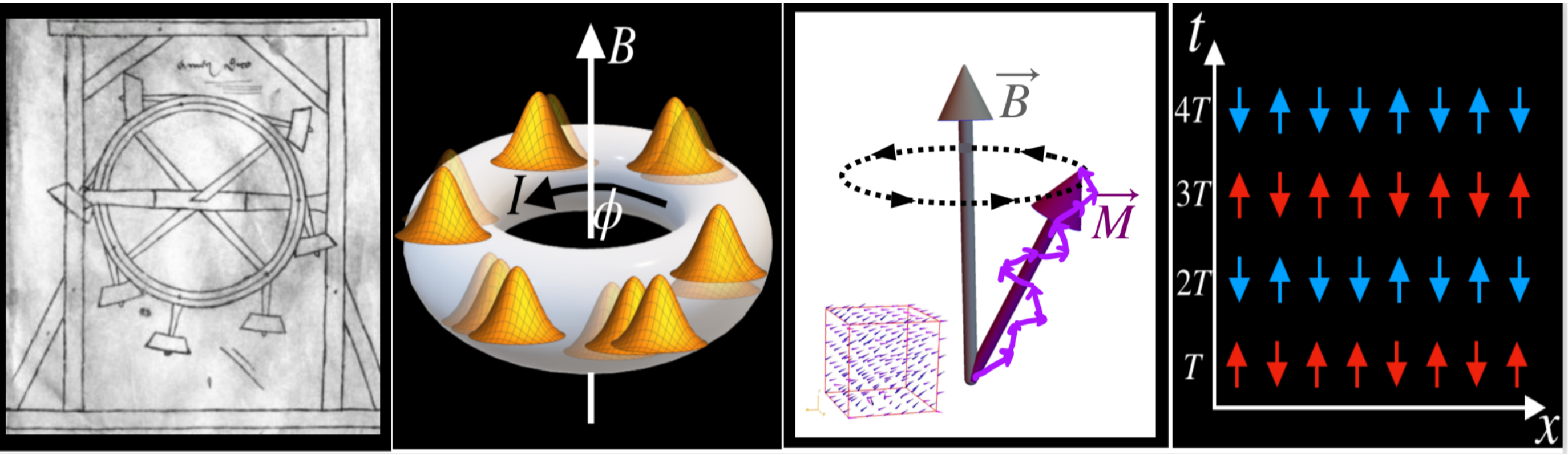}
     \caption{From the perpetuum mobile to the discrete time crystal. From left to right: (a)  A 13th century sketch of a perpetual motion machine~\cite{perpetualWiki}. (b) A schematic of Wilczek's proposal: a perenially rotating charge-density wave on a superconducting ring threaded by a fractional flux. (c) symmetry-protected time crystal: the conserved component of the total moment in the XY plane precesses around the applied field. (d) Discrete Floquet time crystal: stroboscobically observed spin state exhibits spatiotemporal order, glassy in the horizontal (space) direction, and period-doubled in the vertical (time) direction.}
    \mlabel{fig:perpetuum_mobile}
\end{figure}

In this section we discuss some highlights of the history leading up to the discovery of Floquet phases and the DTC.

\subsection{What happened to the perpetuum mobile?}

The idea of evading \emph{thermalization}, \emph{i.e.} evading the emergence of a time-independent
thermal steady state, has exerted a fascination across the ages. In order to appreciate how tall an order the realization of a time crystal is, let us start by constructing one from the ancient idea of a perpetuum mobile, and assemble necessary ingredients in the process. 

A perpetuum mobile is an isolated machine supposed to keep in motion indefinitely, on its own accord.  Typically, the idea was to implement a set of mechanical constraints which would conspire by acting asymmetrically over a motion cycle in a manner designed to keep a wheel turning forever, see Fig.~\ref{fig:perpetuum_mobile}(a). Perhaps the first person to propose one was the Indian Mathematician/Astronomer Bhaskaracharya in 1150---this is the device known as Bhaskara's Wheel or ``overbalanced wheel".
For the next 625 years, the possibility of such a device remained an open question. But, by 1775, frontier scientific opinion had converged on the idea that such a machine could not be constructed, and in that year the 
the French Academy of Sciences noted that ``the construction of perpetual motion is absolutely impossible" and stopped examining such ideas~\cite{PerpetuumHistory}. We note that this was even before the laws of thermodynamics that oppose various aspects of such perpetual motion were formulated. 
But with the first and second laws in hand, the case against perpetual motion came to rest on both.

The first part of the case is that perpetual motion needs a source of energy to oppose dissipation, which is inevitably present. While this is a matter of practical interest\footnote{If our readers wish to buy a perpetuum mobile on the internet, they should check whether a battery is included or needs to be purchased separately!}, for the questions of principle central to this review, we will sidestep this issue by permitting the study of an idealized model where friction between the physical system and its environment -- for example, in the form of friction with air molecules or in the bearings supporting the turning wheel -- is neglected. For one thing, one can typically imagine tuning such dissipation over large dynamic ranges and extrapolating the results to the limit of zero dissipation. For another, even with dissipation removed, there is still a fundamental obstacle to perpetual motion which has a genuinely macroscopic origin. We are referring to the second law of thermodynamics, which requires any macroscopic system to head towards an entropy maximizing equilibrium state.

To see how the second law enters, let us turn to a set of familiar devices almost as old as time 
itself --- clocks. These are, by construction, non-equilibrium in nature, and they depend on predictably repetitive motion: of the sun in the sky, of the vibrations of a quartz crystal in a watch, or -- perhaps most easily visualised -- of a pendulum in a grandfather clock. Indeed, an ideal frictionless pendulum \emph{will} exhibit periodic motion for any 
amplitude out to infinite time.

Now, if the pendulum keeps oscillating away to infinite time, surely it does 
{\it not} reach equilibrium? Indeed this would be true if the pendulum 
was described by exactly {\it one} degree of freedom (Fig.~\ref{fig:pendulum}(a)). But a \emph{physical} pendulum 
is made up of a macroscopic number of degrees of freedom coupled non-linearly, such as the atoms of the material of which the clock's pendulum is made (Fig.~\ref{fig:pendulum}(b)). 
During its oscillatory motion, subject to acceleration, the energy initially contained 
in the rigid center-of-mass motion will ``leak'' into the other degrees of freedom. As 
energy is the only conserved quantity,  
the late time state should reach thermal equilibrium with a ``hotter" pendulum 
hanging more or less vertically, up to small fluctuations: the kinetic energy of the rotational motion of the pendulum on its bearing has been distributed between  its -- much more numerous -- internal modes~(Fig.~\ref{fig:pendulum}(c)). 

Of course, if the pendulum is very rigid, this transfer 
could take a very long time; or if the full set of modes are well described by an
integrable system, one might again expect anomalous behavior---we will return to these
issues in a Sec.~\ref{sec:prethermalTC} on prethermal time crystals.
However, the essential point is that there is no principled reason to expect anything except 
the vanishing of oscillations even in an isolated frictionless pendulum, provided it is treated as a macroscopic object with many constituent degrees of freedom.  

This discussion focuses attention on twin aspects of a {macroscopic} clock. 
First, as is well known from general discussions of ordering, it is necessary to think of a thermodynamic limit of infinite system size to define a sharp phase which breaks TTS. While this is an idealized setting, the examples of clocks given above approximate this concept as well as any of the systems which have been the subject of investigations in thermodynamics and statistical
mechanics over the years. However, secondly and importantly, the large system size must go along with a large number 
of \emph{interacting} degrees of freedom constitutive of \textit{many-body} physics. 

\begin{figure}
    \centering
     \includegraphics[width=\columnwidth]{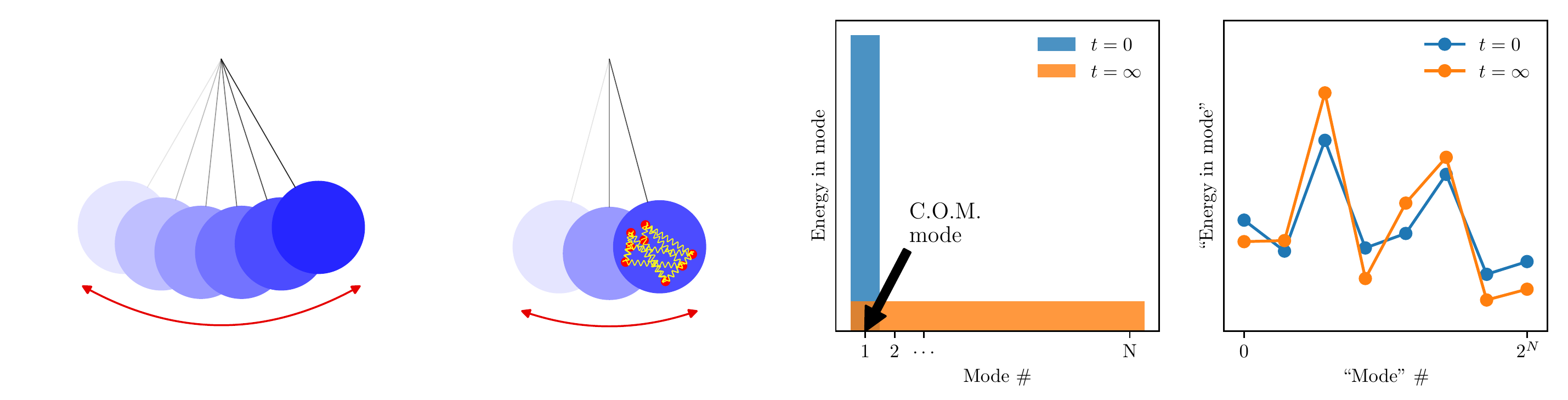}
     \caption{The many-body  pendulum in motion. From left to right: (a) (Initial) motion of a pendulum with the energy concentrated in its centre-of-mass oscillation mode as kinetic energy. (b) (Very) late time motion, where this energy has partially leaked into heating the internal modes of the pendulum. (c) Sketch of the distribution of the energy between modes. At $t=0$, energy is concentrated in one mode. At late times, a high-entropy state is reached with energy equipartitioned between all modes. (d) By contrast, in a many-body localized system, such energy equipartition does not take place, and  excited  ``modes" retain most of their energy forever.}
    \mlabel{fig:pendulum}
\end{figure}

This recapitulation helps us abstract the necessary ingredients, and brings us to the challenges faced by a putative time crystal. We are looking for an interacting many-body system which exhibits oscillations forever, without requiring a constant source of energy. In other words, as we noted in the introduction, a time crystal is best defined as a 
\begin{center}
    {\it stable, conservative, macroscopic clock.} 
\end{center}
Stability with respect to microscopic details embeds a given construction in a \emph{phase}, energy conservation rules out everyday oscillators powered by external power~\footnote{While we will consider periodically driven Floquet systems which do not strictly conserve energy, we will discuss below how the relevant cases nevertheless reach a steady state with no net energy pumped from the drive.}, and the requirement that the system be macroscopic selects for a genuinely collective effect while also screening out the ubiquity of TTSB in few-body systems which generically exhibit oscillations or recurrences (see discussion below Eq.~\eqref{eq:Ot-diag} for a more quantitative discussion of this). 
Such a time crystal will essentially require a mechanism for contending with the second law of thermodynamics, which threatens to drive our macroscopic clock towards a maximum entropy equilibrium state. We now consider various escape routes from this fate.

\subsection{Symmetry-protected time crystals}

We begin with a well-known setting where it is, in fact, quite simple to exhibit time crystalline behavior fully consistent with the second law. The key to achieving this is to consider thermal equilibrium in the presence of additional conserved quantities, and the breaking of the associated symmetries. This adds additional structure to the late-time states and can, in turn, enable oscillations. While opinions may differ on whether time crystals realized in this manner are ``sufficiently interesting'', for our purposes they will do what time crystals should do.\footnote{We thank David Huse for focusing our attention on this class of ``symmetry-protected time crystals".} 

Stripped to its basics, let us consider a simple macroscopic spatial crystal which spontaneously breaks (spatial) translational symmetry. To set it in uniform motion, one can endow it with a finite (linear) momentum density. This momentum density \textit{cannot} be degraded as it is a constant of motion--the motion goes on forever. The spontaneously broken translation symmetry in the crystal, combined with the finite momentum density, means a stationary observer sees a time-varying profile forever. Note that the spontaneous breaking of translational symmetry is only sharply defined in the infinite size limit and, strictly speaking, the motion lasts forever only in this limit. This example can equivalently be understood by transforming to a moving frame in which a stationary crystal appears to be in motion: the conservation of linear momentum means that this is a valid picture for all time.  

A situation where a particular type of behaviour is enabled by the presence
of a symmetry is known as symmetry-protected, hence the name symmetry-protected time crystal (SPTC). This can also be arranged by using internal symmetries. For example, consider a lattice easy-plane Heisenberg magnet (XXZ magnet) with $U(1)$ symmetry in three spatial dimensions. This system can spontaneously break $U(1)$ spin rotational symmetry by choosing a direction for its macroscopic magnetisation vector in the XY plane. A Larmor rotation of this magnetisation vector, say due to the addition of an external magnetic field in the $Z$ direction, corresponds to a finite angular momentum density as a constant of motion (Fig.~\ref{fig:perpetuum_mobile}(c)).

Before concluding this section, three comments are in order. First we note that not all symmetries will do the trick. One which may not is rotational symmetry. For example consider a rigid dumbbell rotating in free space. By angular momentum conservation, it would appear that it will rotate at a fixed frequency forever. But once again, upon taking the internal structure of the dumbbell into account, we find that the motion is not stable;  we should expect the dumbbell to respond to the internal stresses placed upon it by the motion, for example the centrifugal force in the non-inertial frame rotating with the dumbbell.\footnote{Indeed, such stresses from the accelerated motion of the pendulum in the last section are also key to the transfer of energy to the other modes of the system. In our SPTC examples in this section, such stresses are absent.}  Indeed, in the limit of an infinitely large dumbbell, the centrifugal force will scale up with the distance from the center and break the system apart, illustrating an important difference between internal and spatial angular momentum conservation. Of course, the time scale over which the system deforms under these stresses could be quite long. The planet we live on ourselves is sufficiently rigid and exhibits a periodic rotation to pretty good accuracy.\footnote{At this point it is worth noting that our interest in this review is primarily on systems with short ranged interactions of the kind usually discussed in statistical mechanics. Strictly speaking gravity is outside our purview. In giving the example of the earth we wish to assume its rigidity and not a specific theory of it.} 

Second, it is interesting to consider the fate of symmetry protected time crystals to weak symmetry breaking perturbations. It turns out that some symmetries (like $U(1)$ conservation of discrete charge) allow for long-lived prethermal SPTCs even when the symmetry is weakly broken~\cite{ElsePrethermal}, while other symmetries may not afford such stability. We will discuss this in Section~\ref{sec:prethermalTC}.

Finally, note that SPTCs exhibit \emph{spatiotemporal} order \emph{i.e.} they exhibit spontaneous symmetry breaking in both time \emph{and} space. The long-range spatial order comes from the spontaneous breaking of the additional protecting symmetry. This gives a non-zero order parameter, which then exhibits motion under the action of the symmetry.  
We will see later that Floquet MBL time-crystals, the protagonists of our story, do not require a protecting symmetry. However,  in these cases, the system nevertheless has an additional \emph{emergent}  symmetry that is spontaneously broken in the time-crystal phase~\mcite{CVS}, see Sec.~\ref{s:ases}. Hence, 
as emphasized in Refs~\mcite{KhemaniPRL,CVS, Oshikawa15}, all known time crystals are best described as spatiotemporally  ordered phases of matter. 
Such a description also naturally emphasizes the necessity of an infinite-size limit, a nuance that is often lost while distinguishing  time-crystals from other phenomena, such as that of period doubling in few-body systems, see Sec.~\ref{sec:zoology}. 

\subsection{Are there ground state time crystals?}
Next, we note that even for systems without additional symmetries, the \emph{ground} state is not strictly subject to the strictures of thermodynamics and could possibly exhibit motion. 
In 2012 Wilczek, and Wilczek and Shapere, raised the question of whether ground states---quantum or classical---could exhibit perpetual motion for macroscopic systems~\cite{Wilczek12, WilczekClassical}. Both papers refer to their proposals being ``perilously close" to being perpetual motion machines. However as they dealt with ground states they were not required to address thermodynamic constraints.

Aside from raising the general question in the modern language of symmetry breaking, Wilczek proposed a concrete construction of a time crystal which contained two ingredients: flux induced supercurrents and translational symmetry breaking. He first observed that a superconducting ring threaded by a fraction of a flux quantum is known to exhibit a persistent current in its ground state.\footnote{Even absent a flux, an experiment aimed at measuring the time constant of the supercurrent decay in a non-equilibrium state found a value in excess of $10^5$ years~\mcite{supercurrentLifetime}. It would seem that by comparison to waiting for a supercurrent to decay, watching grass grow is downright exciting.}
If one now managed to superpose a charge-density wave (which breaks spatial translation symmetry) on this supercurrent, an observer would start seeing an AC component on top of the DC supercurrent, as a result of having maxima and minima of the charge-density wave pass by in alternation. While Wilczek considered a model with infinite range attractive interactions to induce a CDW, he suggested that crystallization with short ranged repulsive interactions could perhaps produce the same result. 
While there is clearly a family resemblance between Wilczek's proposal and the SPTCs considered above, it is important to note that his proposal did {\it not} postulate a current carrying ensemble; we will discuss the key technical difference in Section \ref{sec:nogo} below.

Wilczek's paper attracted immediate interest, with discouraging results. It was shown by Bruno~\cite{bruno} and clarified by Nozieres~\cite{Nozieres} that Wilczek's specific proposal would not work---and that the actual ground state of his toy model for example did not exhibit time crystallinity. Then Watanabe and Oshikawa formulated the problem in some generality for many body systems---discussed below in Sec.~\ref{sec:DefineTTSB}---and showed in 2014 that ground time crystals could not exist for static Hamiltonian systems with short range interactions, by means of a correlation function diagnostic~\cite{Oshikawa15}. More precisely, they ruled out spatio-temporal order. Thus 864 years after Bhaskara's idea the only known time crystals were the SPTCs fully consistent with the laws of thermodynamics---equilibrium states with symmetry protected motion. The window opened by Wilczek had closed.

However {\it mirabile dictu} at almost the same time an entirely different set of developments in quantum statistical mechanics provided an entirely new angle on the search for a time crystal. These developments were centrally concerned with the discovery of unitary ``many body localized'' quantum dynamics that does not lead to thermalization~\cite{NandkishoreMBLReview, EhudMBLRMP}. Colloquially, the initial energy in different localized ``modes" remains stuck forever (Fig.~\ref{fig:pendulum}(d)). It was shown that one can still define a notion of phase structure in these systems via the notion of ``eigenstate order''~\cite{lpqo, PekkerHilbertGlass}. One of the forms of order identified in this fashion for a periodically driven system ---originally termed ``$\pi$-spin glass'' order~\cite{KhemaniPRL}---could be identified with the breaking of a discrete time translation symmetry of the dynamics, in additional to breaking spatial Ising symmetry (Fig.~\ref{fig:pendulum}(e)). While this phase was entirely outside Wilczek's framework---there are no ground states in these systems as they have time dependent Hamiltonians---their identification with time crystals built on his framing the question in terms of symmetry breaking.

In the next section we present the background material needed to understand these developments (Sec.~\ref{sec:thermalisation}), while the one after is devoted to a careful disussion of how to define TTSB (Sec.~\ref{sec:DefineTTSB}). Then, Sec.~\ref{sec:FMBLTC} lays out the physics of the DTC.  This is followed by discussions of prethermal TCs (Sec.~\ref{sec:prethermalTC}), experiments (Sec.~\ref{sec:expt_real}) and a mini-survey of phenomena ostensibly similar to TCs (Sec.~\ref{sec:zoology}), as outlined in detail at the end of the previous section.

\section{Thermalization, localization and order in closed quantum systems}
\mlabel{sec:thermalisation}

\begin{figure} [t]
    \centering
     \includegraphics[width=0.29\textwidth]{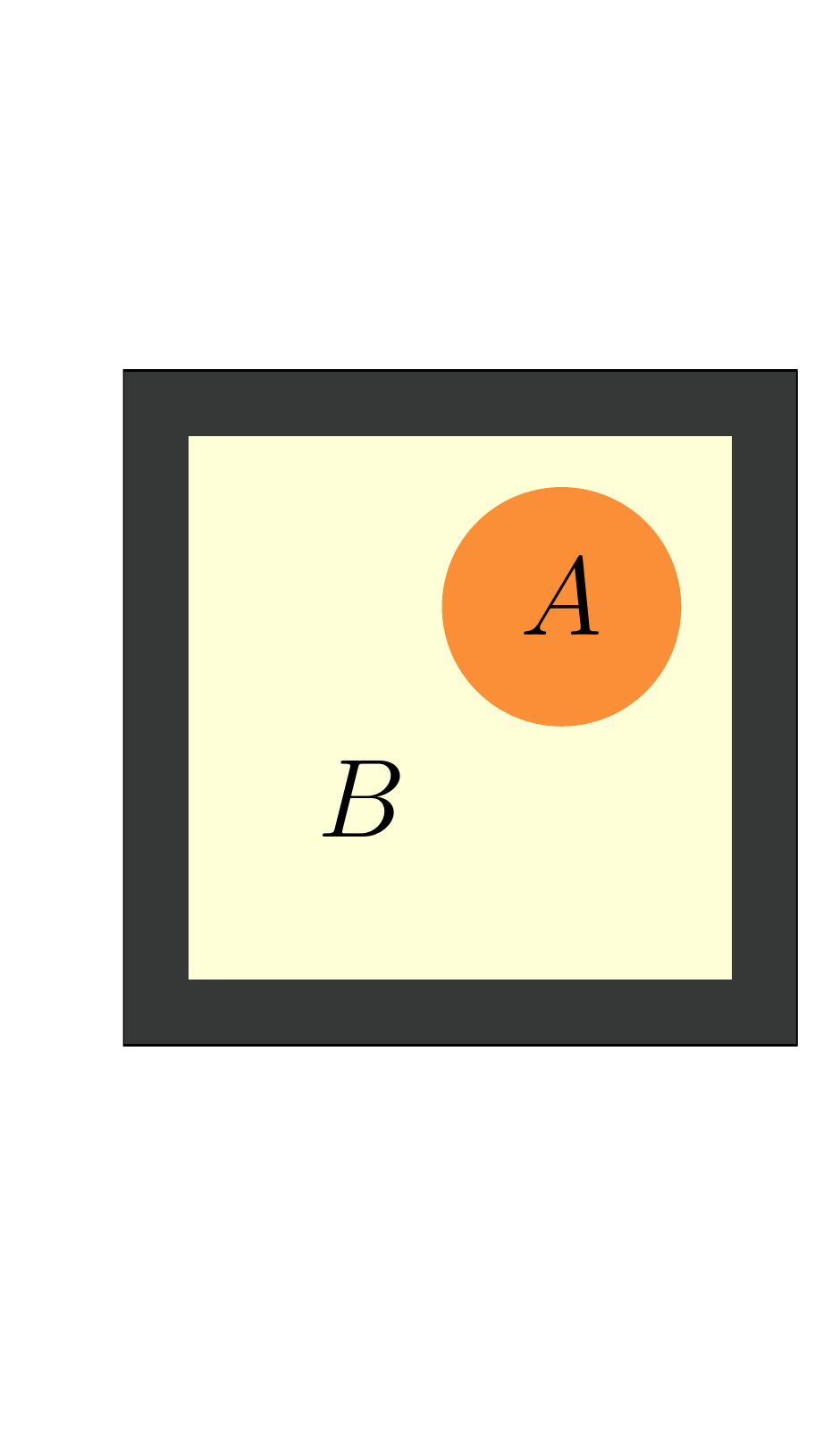}
     \includegraphics[width=0.35\textwidth]{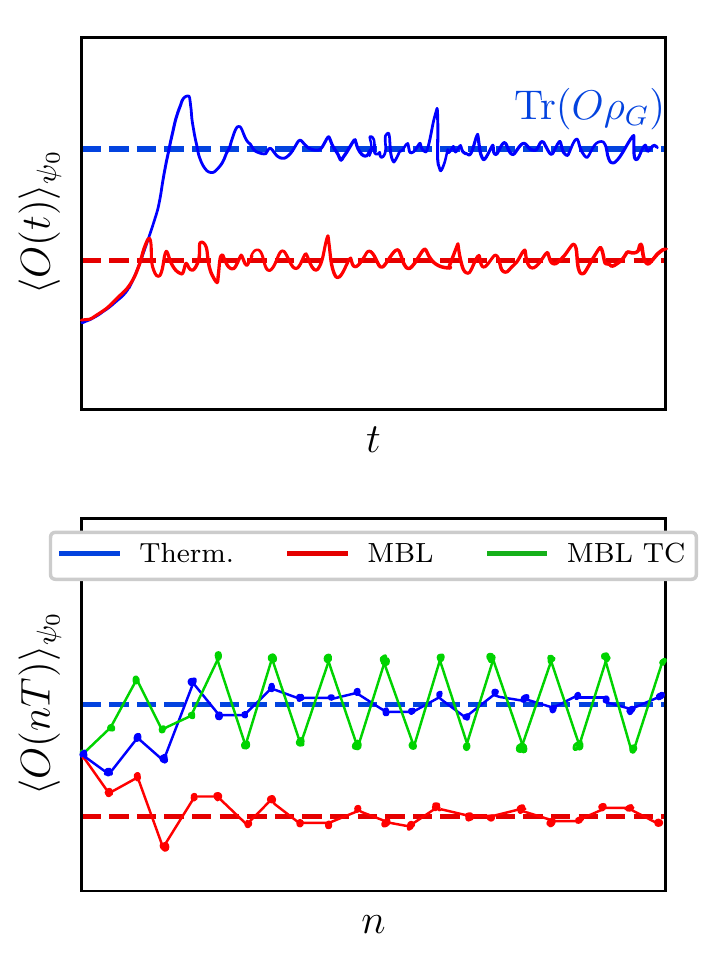}
     \includegraphics[width=0.35\textwidth]{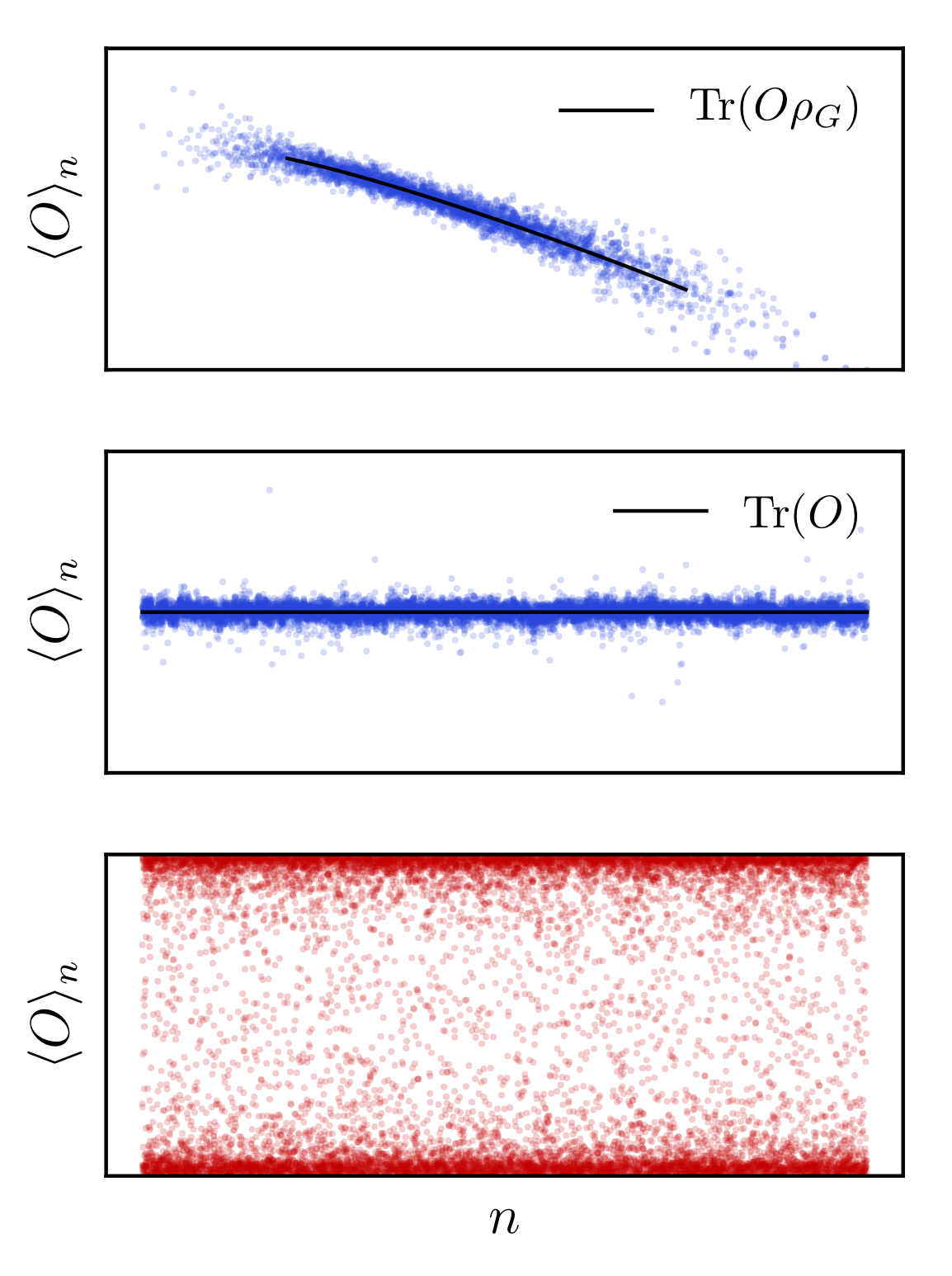}
     \caption{Thermalization, and its absence, in a closed quantum system. Left: even in the absence of an external bath, a subsystem A can exhibit  effectively thermal behaviour, with the remainder of the system, B, effectively acting as a bath. In this case, the reduced density matrix of A, upon tracing out region B, describes a Gibbs distribution $\rho_G = \frac{1}{Z}e^{-\beta H}$. Middle: Schematic sketches of the approach to late-time behaviour of the expectation value of a local observable for a static Hamiltonian (top) and Floquet (bottom) system. For a thermal system, a steady state described by a Gibbs distribution (trivial `infinite-temperature' in the Floquet case) is reached. By contrast, the MBL system reaches a non-thermal steady state determined by the value of the conserved l-bits. Observables are probed continuously in time for the static case and stroboscopically every period for the Floquet case, with the approach to a steady state corresponding to synchronization in the latter.  
     Crucially, the stroboscopic late-time state in the MBL time crystal is not synchronised: it exhibits period-doubling with respect to the drive, and hence discrete TTSB. Right: eigenstate thermalisation in thermal static Hamiltonian (top) and  Floquet (middle) systems: the expectation value of a local observable $\langle O\rangle$ plotted versus eigenstate number, arranged according to growing (quasi-)energy is a smooth function, with fluctuations small in system size. The functional form is given by the Gibbs distribution at inverse temperature $\beta$ set by the energy density of the eigenstate, with $\beta=0$ for the Floquet case. For a (Floquet-)MBL system (bottom), adjacent states exhibit `eigenstate chaos', i.e.\ strong variations in observable expectation values between eigenstates adjacent in (quasi-)energy.  }
    \mlabel{fig:thermalization}
\end{figure}

In this section we  review the building blocks needed for the construction of the Floquet DTC. These are  (i) the basics of thermalization in isolated quantum systems, (ii) the diagonal ensemble and its relationship to late time states, (iii) the basics of many body localization, (iv) the notion of eigenstate/eigensystem order in static systems, (v) the basic formalism of Floquet systems and (vi) the existence of Floquet many-body localized systems. We address these in turn. Readers familiar with individual elements are encouraged to skip ahead---the truly new material in this review, including some original content, starts in the next section. 

\subsection{Thermalization in closed systems}

Some central questions in the area of many-body quantum dynamics have to do with understanding whether the unitary dynamics of an isolated or closed quantum system is able to establish thermal equilibrium at late times~\cite{PolkovnikovRMP, DAlessio:2016aa, NandkishoreMBLReview}; if so, \emph{how} does the approach to equilibrium proceed? If not, when and how can thermalization be impeded?  While these are very fundamental and general issues in quantum statistical mechanics, they have received renewed focus in recent years --- motivated, in large part, by tremendous advances in techniques for engineering and controlling well-isolated many-body experimental systems~\cite{BlochRMP, GirvinQED, Kinoshita:2006ve, Langen:2013aa, Schreiber:2015aa, Choi2016, Neill2016, Kaufman:2016aa, Misha51atom, LevPrethermal, GreinerMBLEntanglement2018}. 

In the standard lore, most textbook treatments of quantum statistical mechanics assume that the system of interest is coupled to an external ``bath" which acts as a reservoir of conserved charges, like energy $E$ or particles $N$. Over time, the system ergodically explores all available microstates and reaches a thermal equilibrium state, such as the Gibbs state, as a result of exchanging charges with the bath (or, in more modern parlance, as a result of getting entangled with the bath). The late-time equilibrium state is characterized by only a few thermodynamic parameters such as temperature or pressure, and hence only a few macroscopic properties of the initial state survive to late times. 

However, there is no external bath in a closed quantum system, and unitary time evolution formally preserves all information about the initial state for infinitely late times. Thus, the notion of thermalization as an \emph{erasure} of memory about initial conditions as the system approaches an equilibrium state needs refining. Instead, in a closed system, thermalization is better viewed as the ``scrambling" or spreading of quantum information over highly non-local (experimentally inaccessible) degrees of freedom, so that \emph{local} memory about the initial state can get {hidden} in non-local variables, and local observables can still reach thermal equilibrium at late times. In essence, the rest of the system is able to act as an internal bath for local subregions and bring them to thermal equilibrium~\cite{PolkovnikovRMP,DAlessio:2016aa, NandkishoreMBLReview} (Fig.~\ref{fig:thermalization}(a)).  Formally, 
\begin{equation}
    \lim_{t \rightarrow \infty} \lim_{V\rightarrow \infty} \mbox{Tr}\left[O \; \rho(t)\right] = \lim_{V\rightarrow \infty} \mbox{Tr} \left[O  \; \rho_{\rm eq}(\beta, \mu \cdots) \right].
\mlabel{eq:thermalize}
\end{equation}
where $\rho_{\rm eq}$ is the density matrix for the the equilibrium ensemble, such as the canonical Gibbs ensemble, $\rho_G = \frac{1}{Z}e^{-\beta H}$, and a limit of large system volumes $V$ and times $t$ is taken\footnote{Traditionally, temperature is denoted by $T$, but we reserve the symbol $T$ for the period of our Floquet drive. We hence use the inverse temperature, $\beta$, here and in the following.}. The thermodynamic parameters ($\beta,\mu \cdots$)  in $\rho_{\rm eq}$ are set by the expectation values of the conserved densities in the initial state; for example, the relation $\mbox{Tr}[H \rho(0)] = \frac{1}{Z}\mbox{Tr}[e^{- \beta H} H] $ is used to determine the inverse temperature $\beta$ in a system defined by a time-independent Hamiltonian $H$ which only conserves energy\footnote{This requires the initial state to have sub-extensive uncertainty in all extensive conserved quantities so that thermodynamic parameters like temperature are well defined (for example, $\langle \Delta E\rangle_0 \sim V^\alpha$, with $\alpha < 1$. 
All short range correlated initial states, like product states, satisfy this property. However, superpositions of states with extensively different expectation values of conserved quantities (``cat states") do not.}.
Thermalization is the expected outcome in generic many-body systems with only one or a few local conservation laws. By ``generic" we mean strongly-interacting, non-integrable Hamiltonians that are not fine-tuned in any way. This in contrast to integrable systems that represent special points in the parameter space of all Hamiltonians, and display a constrained dynamics due to the presence of extensively many conservation laws~\cite{Calabrese_2016}.

\subsection{Late time states and the diagonal ensemble}
\mlabel{sec:diagonalensemble}

We will now see in detail how thermalization takes place as a consequence of real time dynamics.  We start with a general (out-of-equilibrium) initial state $|\psi_0\rangle$ and ask what happens to operator expectation values at late times. We will focus on two ideas: (i) equilibration -- the tendency of observables to settle down to late-time steady states in macroscopic systems, and (ii) thermalization -- that the steady state value agrees with the appropriate thermodynamic expectation value, in accordance with Eq.~\ref{eq:thermalize}.
Again, we assume that the operators of interest are few-body and the states of interest have well defined expectation values for conserved quantities. 

Now, any state $|\psi_0\rangle$ can be expanded in the eigenbasis of the Hamiltonian.  In an interacting many-body (MB) system, there are $O(\rm exp(V))$ eigenstates labeled $|\nu \rangle$ with energy eigenvalues $E_{\nu}$:
\begin{equation}
    |\psi_0(t)\rangle = \sum_{\nu } c_\nu e^{-i E_\nu t} |\nu\rangle.
\end{equation}
A generic initial state at finite temperature has large overlap on exponentially many eigenstates that have the same energy density as the initial state, so that $c_\nu(E_\nu)$ is peaked about the initial energy density, with a width set by the energy fluctuations in $|\psi_0\rangle$. 

The expectation value of a local operator $\langle O(t) \rangle$  has the form: 
\begin{align}
    \langle \psi_0(t)|O|\psi_0(t)\rangle &= \sum_{\nu \mu } c_\mu^* c_\nu e^{-i (E_\nu-E_\mu) t} \langle \mu|O|\nu\rangle \nonumber, \\
    &= \sum_{\nu \mu: E_\nu = E_\mu } c_\mu^* c_\nu \langle \mu|O|\nu\rangle + \sum_{\nu \mu: E_\nu \neq E_\mu }c_\mu^* c_\nu e^{-i (E_\nu-E_\mu) t} \langle \mu|O|\nu\rangle.
\mlabel{eq:Ot-diag}
\end{align}
The second term represents a sum over exponentially many time-varying phase factors. For a sufficiently generic energy spectrum, the energy differences $E_\mu - E_\nu$ are incommensurately distributed. In this case, in the limit of large $V$, we expect that the incommensurate phase factors in the second term above will dephase at late times, and sum to an exponentially small in $V$ value. This leads to \emph{equilibration} to a time-independent late-time steady state value for $O(t)$ determined by the \emph{block diagonal ensemble} (first term in Eq.~\eqref{eq:Ot-diag}), along with exponentially small in $V$ fluctuations about this value (from the second term in Eq.~\eqref{eq:Ot-diag}). In contrast, for a small and finite $V$, the second sum is over finitely many energy differences, and hence will exhibit quasiperiodic revivals in time. This formalizes our prior statements that finite-sized systems generically show time-dependence in local observables and break TTSB. 

The requisite energy spectrum for a large system to \emph{equilibrate} to the block-diagonal ensemble (upto exponentially small in $V$ fluctuations) does not need a particularly narrow form, e.g. it does \emph{not} have to be of the Wigner-Dyson form charactersistic of chaotic systems. If the spectrum is non-degenerate, the late time ensemble is simply diagonal $O_{\rm DE} = 
 \sum_{\nu} |c_\nu|^2 \langle \nu|O|\nu\rangle$. But, in the presence of symmetry dictated spectral degeneracies (or rational commensurations in energy spacings) we will need to be more careful, with much profit by the time we get to the DTC. 
 
Indeed this discussion already illustrates that in order to obtain non-trivial time dependence, such as in a TC, something special must happen to prevent this dephasing induced equilibration - which is the generic outcome for a typical finite-temperature initial state evolving under a many-body Hamiltonian. 

Thus far we have not made use of the thermalization properties of the system, only that energy level spacings in the MB spectra are generic enough (i.e. typically incommensurate) for the system to dephase and reach equilibrium. To connect the equilibrium block-diagonal ensemble with the correct thermodynamic ensemble as in Eq.\ref{eq:thermalize}, we need to appeal to the notion of the celebrated \emph{eigenstate thermalization hypothesis} (ETH)~\cite{Deutsch, Srednicki, Rigol}. The ETH implies that local observables evaluated in energy eigenstates vary smoothly as a function of energy density (and other conserved densities), so that eigenstates nearby in energy have identical local properties (again, upto exponentially small in $V$ fluctuations). Further, eigenstate expectation values (EEV) agree with expected thermodynamic ensemble averages at the temperature associated with the energy density of the eigenstate. This is shown in Fig.~\ref{fig:thermalization}(d).  
In other words, in an ETH obeying system, each individual eigenstate is separately a fine microcanonical ensemble for the system!\footnote{Note that if Eq.~\eqref{eq:thermalize} holds for \emph{all} initial states with a well defined energy density, then it must also hold for individual many-body eigenstates, which implies that each eigenstate separately looks thermal.} 

When the ETH holds, we can further simplify the diagonal ensemble to: 
\begin{equation}
O_{\rm DE} =\sum_{\nu} |c_\nu|^2 \langle \nu|O|\nu\rangle = \langle \nu|O|\nu\rangle \sum_{\nu} |c_\nu|^2 = \langle \nu|O|\nu\rangle, 
\label{eq:diag_ensemble}
\end{equation}
where we have used the fact that the initial state has a well-defined energy density and hence substantial weight only over eigenstates at the same energy density with the same EEV. Thus, the late-time diagonal ensemble answer agrees with the expectation value in any eigenstate with the same energy density, which in turn agrees with the statistical mechanical thermal average in the infinite volume limit (Fig.~\ref{fig:thermalization}(b))\footnote{We should note that ETH is not all or nothing. For example, certain special Hamiltonians can display a lack of thermalization and long-lived oscillations in local observables starting from particular initial states, but not others. This may stem from the presence of certain ETH violating ``scarred" eigenstates in the spectrum~\cite{ShiraishiMori, ck, ChoiAbanin, TDVPScars, lerose, ichinose, konik, Iadecola, IadecolaScarsTC, kn2019, Munich}, interspersed between thermal eigenstates, and possibly with commensurate energy spacings that prevent dephasing. 
However, these are fine tuned Hamiltonians~\cite{ck}, and only very special initial states with large overlap on the scars show oscillations. Then there are also fine tuned integrable Hamiltonians for which the long time states depend on an extensive number of parameters and which thermalize instead to ``generalized Gibbs ensembles"~\cite{Calabrese_2016}. We do not discuss them here.}.

The ETH is by no means universally obeyed, and we next turn to a setting where the smooth variation of observables as a function of eigenstate energy is  replaced by discontinuous jumps, also known as eigenstate chaos. 

\subsection{Many-body Localization}
\mlabel{sec:mbl}
Many-body localization plays a central role in defining Floquet phases by providing a mechanism for limiting energy absorption typically expected in driven many-body systems. While a complete discussion of MBL is beyond the scope of this review --- we direct the reader to several excellent recent treatments of this subject~\cite{NandkishoreMBLReview, EhudMBLRMP, SarangSid} --- below we highlight some features of MBL most salient to our discussion of Floquet heating and ordering.

In a nutshell, the phenomenon of many-body localization refers to the generalization of Anderson localization~\cite{Anderson58} for disordered, non-interacting systems to the interacting setting~\cite{Basko06, Gornyi, PalHuse, OganesyanHuse, Prelovsek, Imbrie2016}. MBL systems are unable to act as a bath for themselves and \emph{never} reach local thermal equilibrium. Specifically, while local operators still equilibrate to a steady-state given by the block diagonal ensemble, the equilibrium value does not agree with the expected thermal answer, so that some memory of local initial conditions is retained forever (Fig.~\ref{fig:thermalization}(b))\footnote{There is subtlety worth noting. For ETH obeying systems, both indices in the second term in Eq.~\ref{eq:Ot-diag} run over $O(e^V)$ values. For MBL systems, for each $\mu$ there are only $O(1)$ sizeable matrix elements to states $\nu$. So while there are enough terms to justify dephasing, they are fewer in number. Readers can use the l-bit formalism we describe below to understand how this distinction comes about.}. Likewise, these systems do not obey the ETH so that local properties fluctuate strongly from state to state at the same energy density and there is no well defined notion of ``temperature" (Fig.~\ref{fig:thermalization}(f)).

\begin{figure}[t]
    \centering
     \includegraphics[width=\columnwidth]{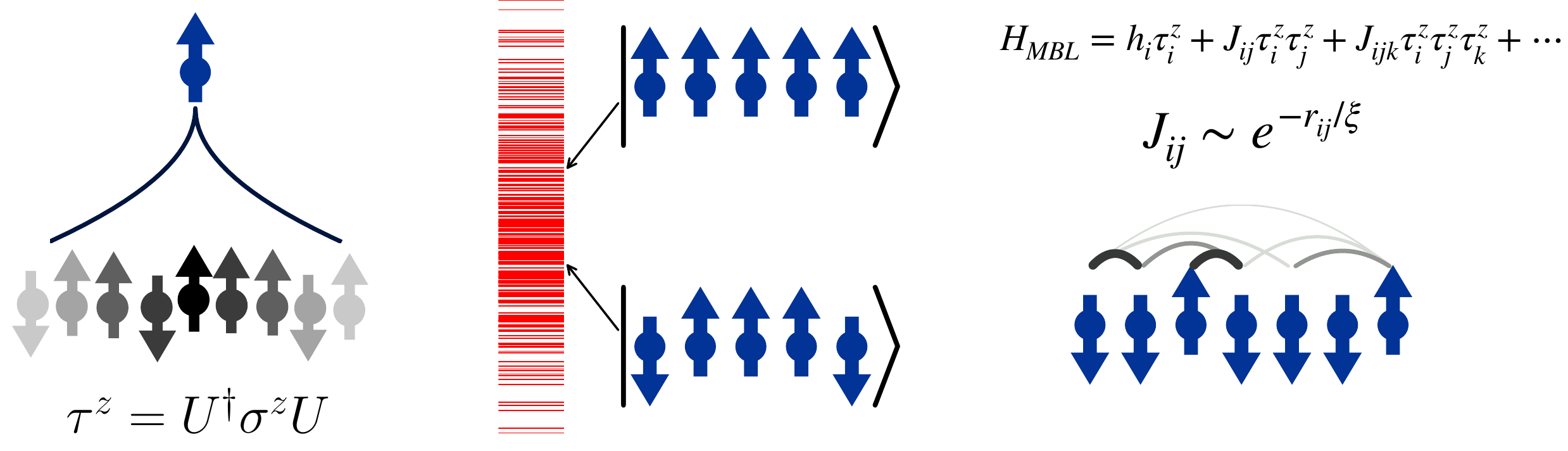}
     \caption{MBL and local integrals of motion (l-bits). Left: Each l-bit $\tau^z_i$ (in blue) is only comprised of  physical spins (in black) from an exponentially decaying envelope around its location $i$. Middle: eigenstates throughout the bulk of the spectrum can simply be specified by their `classical' list of l-bits $\{\tau_i^z\}$. This is because the Hamiltonian can be expressed in terms of these emergent integrals of motion like a classical Ising magnet (right), with coupling strenghts decaying exponentially in l-bit separation $r_{ij}$.}
    \mlabel{fig:lbits}
\end{figure}

\subsubsection{Local integrals of motion, emergent integrability, and stability of the MBL phase}
The best understood localized systems are ``fully'' MBL wherein the lack of thermalization in MBL systems is credited to an \emph{emergent} integrability: the existence of an extensive set of local integrals of motion--the so-called ``l-bits"---that commute with the Hamiltonian and with each other~\cite{Huse14, Serbyn13cons}. These generalize the localized single-particle orbitals of an Anderson insulator, the occupation number of each of which is separately conserved.  

Concretely, consider a paradigmatic example of a localized spin-chain in one dimension~\cite{Prelovsek, PalHuse}: 
\begin{align}
    H = \sum_{i=1}^L J \left(\sigma_i^x \sigma_{i+1}^x + \sigma_i^y \sigma_{i+1}^y\right) + J_z \sigma_i^z \sigma_{i+1}^z + h_i \sigma_i^z,
\mlabel{eq:HXXZ}
\end{align}
where $L$ is the system size, $\sigma_i^{x/y/z}$ are Pauli spin 1/2 operators on site $i$, and the fields $h_i$ are drawn randomly and independently from a disorder distribution of width $W$. This can be mapped to a model of spinless fermions via a Jordan Wigner transformation, where $J$ sets the nearest-neighbor hopping, $J_z$ is the strength of nearest-neighbor density-density interactions, and $h_i$ is the magnitude of the on-site potential. When $J_z = 0$, this is an Anderson localized model of non-interacting fermions in a disordered potential. MBL refers to the persistence of localization for large enough disorder $W$, even in the presence of weak interactions  $J_z \neq 0$. 

In the localized regime there exists a finite-depth local unitary transformation, $\mathcal{V}$, relating the physical ``p-bits", $\sigma$, to new variables, the localized ``l-bits", $\tau$, such that $\tau_i = \mathcal{V}^{\dagger} \sigma_i \mathcal{V}$ and 
\begin{align}
    H = \sum_i \tilde{h}_i \tau_i^z + \sum_{ij} J_{ij} \tau_i^z \tau_j^z+ \sum_{ijk} J_{ijk} \tau_i^z \tau_j^z \tau_k^z + \cdots
\mlabel{eq:lbit}
\end{align}
is purely diagonal in the new l-bit basis~\cite{Bauer13, Huse14, Serbyn13cons, Imbrie2016}.\footnote{A local (or low depth) unitary is a unitary which can be written as $\mathcal{V}=\mathcal{T} e^{-i \int^t_0 ds K(s)}$ for some local bounded Hamiltonian $K(t)$, with $t$ finite in the thermodynamic limit. Note that the locality of such a unitary is a subtle business outside of the very strongly localized region due to proliferating resonances and Griffiths effects\mcite{Bauer13,Gopalakrishnan15, GopalakrishnanGriff, deroeckAvalanche, Imbrie2016}.} In this basis, $\tilde{h}_i \sim O(W)$, the interactions $J_{ij}$ fall off exponentially with distance $|i-j|$, and the $\cdots$ denote exponentially decaying higher order terms. The emergent integrability is now manifest as $[H, \tau_i^z] = 0 \; \forall \; i$. The l-bits $\tau_i$ are ``dressed" versions of the physical ``p-bits" $\sigma_i$, with support that is typically exponentially localized near site $i$ deep in the MBL phase (Fig.~\ref{fig:lbits}). All local observables generically have finite overlap with the conserved l-bits, and hence the system retains some memory about local initial conditions for infinitely late times, never reaching local thermal equilibrium. In other words, the system approaches a non-thermal steady state determinined by the value of all the lbits in the initial state (Fig.~\ref{fig:thermalization}(b)).

In an Anderson insulator,  the interaction terms $J_{ij}$ are zero so that $H = \sum_i \tilde{h}_i \tau_i^z$, and the $\tau^z$s represent the localized single-particle fermionic orbitals (Jordan-Wigner transformed). Thus, in the single-particle limit, the transformation $\mathcal{V}$ is simply the linear unitary transformation relating the ``physical'' fermions in the real-space basis to the ``localized" eigenbasis. With interactions, $\mathcal{V}$ is generically non-linear, but still local\footnote{
While localization continues to be perturbatively stable to interactions in all dimensions\cite{Basko06}, no rigorous proof exists for dimensions $>1$ or systems with longer ranged interactions. In fact, in these settings, there are arguments suggesting that MBL may be asymptotically unstable due to non-perturbative ``avalanche" instabilities arising from the effect of rare disorder fluctuations that embed locally thermal regions in large localized samples\cite{deroeckAvalanche}. Nevertheless, such instabilities also predict asymptotically long thermalization times, making this a mostly academic issue ---  and one that can potentially be avoided altogether by using other, deterministic, onsite potentials (such as quasiperiodic modulations) which still show localization but are less susceptible to dangerous large deviation effects\cite{IyerQP, KhemaniCPQP}}.     

Finally, note that by ``emergent" integrability, we mean that the conserved l-bits depend on the details of the Hamiltonian --- weakly perturbing the Hamiltonian will slightly modify the spatial structure of all the l-bits $\tau$ but preserve integrability, and hence localization. This is why MBL defines an extended dynamical \emph{phase} of matter, as opposed to being a fine-tuned property of special points in parameter space. As various parameters like disorder or interaction strength are tuned, the system can transition to a thermalizing one. This novel phase transition is completely invisible to thermodynamics, but is instead a \emph{dynamical} phase transition having to do with whether or not the system is able to thermalize locally under its own dynamics. The nature of this transition remains an active area of inquiry~\cite{PalHuse, Luitz15, ClarkBimodal, KhemaniTransition, KhemaniCPQP, VHA, PVP, DVP, MuellerAvalanche, ZhangToyRG, SerbynKTPRL, KTPRB, MorningstarKT}

\subsubsection{Dynamical and eigenstate properties of MBL systems}
\label{sec:mbldynamics}
The l-bit picture provides a phenomenological explanation for many of the characteristic features of MBL systems. From the l-bit Hamiltonian~\eqref{eq:lbit}, it is clear that all eigenstates are product states of the dressed $\tau^z$ operators: $|\nu\rangle~=~|\uparrow \downarrow \downarrow\uparrow\downarrow~\cdots~\uparrow\rangle$ (Fig.~\ref{fig:lbits}). Thus, the bipartite entanglement entropy of eigenstates will typically exhibit an ``area law", stemming primarily from the l-bits that straddle the entanglement cut\cite{PalHuse, Bauer13}. Thus, in this sense, even highly excited MBL eigenstates resemble gapped ground states of quantum systems\footnote{The term highly excited refers to eigenstates with a finite energy density above the ground state.}. 
In contrast, the ETH implies that the von Neumann entanglement entropy of a finite temperature eigenstate in a thermalizing system must agree with the thermal entropy (which is extensive) and hence thermalizing systems have ``volume-law entanglement". The MBL phase transition is a novel \emph{eigenstate phase transition} involving a singular change in the entanglement properties of \emph{individual}, highly-excited many-body eigenstates---even those at energy densities corresponding to infinite temperature---which change from thermal, “volume-law” entangled states to non-thermal, ``area-law" entangled eigenstates in the MBL phase.

We expect the statistics of the eigenenergy spacings to obey a Poisson distribution due to a lack of level repulsion stemming of the presence of extensively many local conservation laws~\cite{PalHuse, OganesyanHuse}. Further we expect that states that are nearly adjacent in energy have vastly different local patters of $\tau_i^z$ --- in violation of the ETH\footnote{Starting from a reference product eigenstate, flipping a single spin costs an $O(1)$ energy $\sim W$ which is much larger than the average level spacing; by contrast, an exponentially small energy difference is typically only attained by a cancellation process involving $O(L)$ spin flips with positive/negative energy differences relative to the reference state. More technically, this implies that MBL systems have a local spectral gap, even in the absence of an actual gap.}. 

We close this whirlwind recap of MBL by briefly mentioning that the l-bit model \eqref{eq:lbit} and the structure of the eigenstates has distinctive, measurable consequences for the dynamics of the system. The most striking of these has to do with the growth of bipartite entanglement entropy with time in quenches starting from a generic unentangled product state. The exponentially weak interactions between l-bits $J_{ij}\sim e^{-|i-j|/\xi}$ leads to a slow dephasing dynamics which causes to the entanglement to only grow logarithmically in time~\cite{Prelovsek, Bardarson_loggrowth, Serbyn13Entanglement} (as opposed to thermalizing systems, where this growth is ballistic in time). A similar dephasing mechanism leads to a slow power law approach of the expectation values of local observables to their late-time, non-thermal, saturation values~\cite{SerbynQuenches}. 

Since we will frequently refer to the dynamics of local observables in MBL systems, we briefly review here the argument for the power law approach to equilibrium. The basic idea is the $\tau_j^{x/y}$ effectively look like Rabi oscillators that feel an ``effective field" which depends on $\tilde{h}_j$ and the state of all other l-bits (Eq.~\ref{eq:lbit}). These operators precess in this effective field, as is apparent from their Heisenberg equations of motion $O(t) = e^{iHt}Oe^{-iHt}$:
\begin{align}
    \tau^z_j(t) &= \tau^z_j\nonumber\\
    \tau^x_j(t) &= \tau_j^x \cos(2H_jt) - \tau_j^y \sin(2H_jt)\nonumber\\
    \tau^y_j(t) &= \tau_j^x \sin(2H_jt) + \tau_j^y \cos(2H_jt)
\end{align}
where $H_j$ is a functional of the l-bits, with longer and longer range terms coming in with weaker and weaker strength: 
$H_j = \tilde{h}_j + \sum_{r} J_{j,j-r} \tau^z_{j-r} + \sum_{r} J_{j,j+r} \tau^z_{j+r}  + \cdots$~\cite{SerbynQuenches}.

For an Anderson insulator, all the higher-order terms are absent and these operators just oscillate at the frequency $\tilde{h}_j$. Even with interactions, if we evaluate two-point auto-correlators such as $\langle \nu|\tau_j^x \tau_j^x(t)|\nu\rangle$ in the l-bit \emph{eigenstates}, then these oscillate forever with a frequency set by an effective field that is determined by the configuration of l-bits in the particular state $|\nu\rangle$ --- a point we will return to below in a discussion on time-glasses. 

In contrast, if we start with a generic initial state that is a superposition of eigenstates with different l-bit configurations, then the effective precession rate will vary incommensurately across these eigenstates. The superposition over all these incommensurate frequencies leads to dephasing and decay towards zero. At a given time $t$, only terms in the expansion of $H_j$ that are comparable to $1/t$ will dephase; the hierarchical exponential decay of couplings then straightforwardly gives a power law approach of l-bit expectation values towards zero, when evaluated starting from generic initial states~\cite{SerbynQuenches}.

Any local physical operator can be expanded in the l-bit basis. For such operators, the part that overlaps with $\tau^z$ will be conserved forever, while the parts that overlap with the off-diagonal operators decay as power-laws.

\subsection{Floquet Basics}

As advertised already, a principal focus of this review is on periodically time-dependent 
``Floquet" systems for which $H(t)= H(t+T)$, with time-period $T$~\cite{Shirley:1965cy,ZelDovich:1967va,Sambe:1973hi}. Such systems have \emph{discrete} time-translation symmetry (dTTS) because the system only returns to itself after translations of time by integer multiples of the period, $t \rightarrow t + n T, n \in \mathbb{Z}$. This is in contrast to time-independent Hamiltonians which have \emph{continuous} time-translation symmetry (cTTS) since the system is invariant under arbitrary translations of time, $t \rightarrow t + \delta t, \delta t \in \mathbb{R}.$ Despite the time-dependence in the Hamiltonian parameters in a Floquet system, the system is still treated as ``closed" in that it undergoes unitary time evolution with a time-dependent $H(t)$; in this sense, we define closed as ``unentangled with the environment".

Energy is no longer conserved in a Floquet system because of the time dependence in $H(t)$. Thus, in the absence of any other local conservation laws, the late time equilibrium state that maximizes entropy is simply the ``infinite temperature" ensemble $\rho_{\infty} \propto \mathbb{1}$\footnote{We are being casual in our use of infinite ``temperature" here, since energy is no longer a good quantum number. By the infinite temperature ensemble, we simply mean an equal probability density matrix proportional to the identify.}. More colloquially, standard linear response theory predicts that a generic MB system will absorb energy from the periodic drive, ultimately heating to infinite temperature. Thus, by Eq.~\eqref{eq:thermalize}, thermalization in MB Floquet systems means that local observables look featureless at late times with no non-trivial temporal correlations, regardless of the initial state \emph{i.e.} there is a single trivial Floquet ergodic phase~\cite{D_Alessio_2014,Lazarides_2014,PonteChandran_2015}.

Just as the dynamics of a static time-independent Hamiltonian $H_0$ can be understood via the eigensystem properties of $H_0$, the dynamics of Floquet systems are intrinsically governed by the eigenvalues and eigenvectors of the so-called Floquet unitary $U_F \equiv U(T)$, which is the time evolution operator over one period: 
\begin{align}
   U_F= U(T) = \mathcal{T} e^{-i \int_{t_0=0}^{t_0+ T} dt' H(t')}.
   \mlabel{eq:Uf}
\end{align}
These are denoted $\{ |\nu \rangle, e^{-i E_\nu T} \}$ so that $U_F |\nu \rangle = e^{-iE_\nu T} |\nu\rangle$. The difference between static systems with cTTS and Floquet systems with dTTS is analogous to the difference between continuous and discrete spatial translation symmetry. Thus, the \emph{quasi}-energies $E_\nu$ are only defined modulo $\omega = 2\pi/T$, analogous to the restriction of the quasimomentum to the first Brillouin zone in a periodic spatial crystal. One can formally define a Floquet Hamiltonian, $H_F$, via $U_F = e^{-iH_FT}$, although this is generally a highly non-local operator and one that is not uniquely defined due to the non-uniqueness of the $E_\nu$~\cite{PolkovnikovFloquetReview}. 

The argument for the reduction to the late time diagonal ensemble in Eq.~\eqref{eq:Ot-diag} can be run using Floquet eigenstates and eigenvalues. The key is to examine the time-dependence at a sequence of \emph{stroboscopic} times, $t = nT$ with $n \in \mathbb{Z}$. Once again, any initial state can be expanded in the eigenstates of $U_F$ which form a complete basis: $|\psi_0\rangle = \sum_{\nu} c_{\nu}|\nu\rangle$, so that: 
\begin{align}
    \langle \psi_0(nT)|O|\psi_0(nT)\rangle 
    = \sum_{\nu \mu: E_\nu = E_\mu} c_\mu^* c_\nu \langle \mu |O|\nu  \rangle + \sum_{\nu \mu: E_\nu \neq E_\mu} c_\mu^* c_\nu \langle \mu |O|\nu\rangle e^{-i (E_\nu-E_\mu) n T}  
\mlabel{eq:Ot_diag_Floquet}
\end{align}
Now we can make use of the same dephasing arguments as before (see below Eq.~$\eqref{eq:Ot-diag}$) to argue that $O(nT)$ becomes independent of $n$ at late times in a generic system, upto exponentially small in $V$ fluctuations. In other words, observables look \emph{synchronized} with the periodic drive, which is the driven analog of equilibration to a steady state in the undriven problem\footnote{We note that, in principle, the late-time steady-states in periodically driven systems could show a non-trivial periodic time-dependence in local observables with period $T$, when probed as a function of continuous time. In defining $U_F$ in Eq.~\eqref{eq:Uf}, we have implicitly picked a ``starting point" for a time period, $t_0=0$. In principle, we can define Floquet unitaries $U_F(t_0)$ for different starting times $0 \leq t_0 <T $, which would be relevant for describing the system at times $t= t_0 + nT$. The eigenstates of these different unitaries are related by a gauge choice~\cite{PolkovnikovFloquetReview}. Thus, while each stroboscopic time-series for a given $t_0$ becomes time-independent at large $n$, the system could still show time-dependence as a function of $t_0$, corresponding to synchronization with non-trivial time dependence \emph{within} a period. We note that this does \emph{not} happen for Floquet ergodic systems which thermalize to infinite temperature regardless of $t_0$ and are hence trivially synchronized.
A non-trivial intra-period variation is obtained, however, in non-interacting periodically driven integrable systems which do not thermalize to $\rho_{\infty}$ but instead to a periodic Gibbs ensemble~\cite{AchilleasPGE}.}. Having discussed synchronization, we turn now to thermalization. In an ergodic Floquet system, each Floquet eigenstate must reproduce $\beta=0$ expectation values in the Floquet version of ETH, so that the $|\nu\rangle$ all look locally featureless and maximally entangled.  Fig.~\ref{fig:thermalization}(c) shows a schematic sketch of a stroboscopically probed observable in a Floquet system relaxing to the infinite temperature equilibrium value, while Fig.~\ref{fig:thermalization}(e) shows Floquet ETH in that all eigenstates reproduce infinite temperature expectation values. 

The question of MBL in Floquet systems is whether the trivial late time states of the ergodic limit can be evaded by localization, and the question of finding time crystals is, heuristically, whether such late time states are temporally perdiodic without being synchronized. For example, if observables show a time-dependence with periodicity $kT$, with $k > 1 \in \mathbb{Z}$, then we expect that the discrete TTS of the drive is spontaneously broken from the group $\mathbb{Z}$ (representating translations by integer multiples of $T$) to the subgroup $k\mathbb{Z}$ (representating translations by integer multiples of $kT$). Figure.~\ref{fig:thermalization}(c) shows schematically depicts non-synchronous evolution in a DTC. 

We re-emphasize here that a lack of synchronization alone is not sufficient for realizing a time crystal, cf.\ Sec.~\ref{sec:zoology}, which highlights similarities and differences between time crystals and various more familiar examples of period doubling.  Nevertheless, our central point here is that lack of synchronization in a closed and driven many-body quantum system necessarily requires a mechanism for preventing thermalization to the trivial infinite temperature state.

\subsection{Avoiding Heating: Floquet MBL}
\label{sec:floqMBL}
\begin{figure}[t]
    \centering
     \includegraphics[width=0.6\columnwidth]{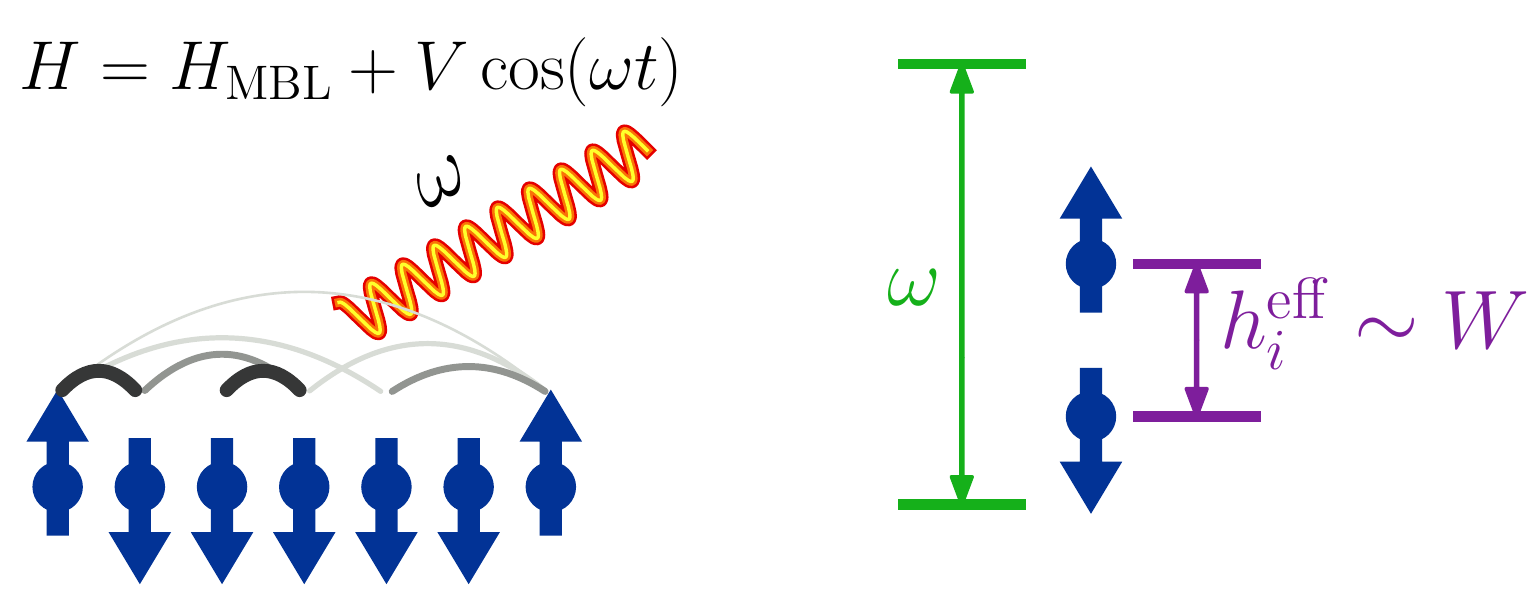}
     \caption{Floquet MBL: persistence of MBL upon adding an external drive to the l-bit Hamiltonian (cf.\ Fig.~\ref{fig:lbits}). If the driving frequency $\omega$ is much larger than the typical local bandwidth $W\sim h_i^{\mathrm{eff}}$, the localised system cannot absorb an energy quantum of the drive field, and hence fails to heat up.}
    \label{fig:floquet_lbits}
\end{figure}

We are now ready to understand how interacting MBL systems can evade heating even in the presence of a time-dependent drive. It is well understood how this happens in a \emph{non-interacting} Floquet system with single-particle (SP) energy eigenmodes that lie in some finite $O(1)$ bandwidth set by the local couplings in the system $\sim J$. Many-body states are constructed by simply populating these SP modes. For such systems, if one drives at a frequency much bigger than the SP bandwidth,  $\omega \gg J$, then the drive is ineffectual at exciting the SP modes, and energy absorption can be limited. More formally, one can show that there exists a convergent Floquet-Magnus expansion in powers of $1/\omega$ which can be used to construct a local, non-interacting $H_F$~\cite{PolkovnikovFloquetReview, EckardtMagnus}. By contrast, such a high-frequency expansion obviously breaks down for any \emph{finite} $\omega$ in a generic, interacting system with a dense and \emph{extensive} MB spectrum as the system size is taken to infinity --- yet another restatement of the heating result. 

Remarkably, it has been shown that localization can prevent such heating, even in many-body interacting systems, via a picture that shares qualitative similarities with the SP case discussed above\cite{Lazarides_2015,PonteHuveneers_2015,abanin2014theory}. The l-bit picture provides an intuitive understanding for why a localized system can remain localized even when subjected to a weak periodic drive: $H(t) + H_{\rm MBL} + V \cos(\omega t)$. The basic idea is that a local drive $V$ can only locally flip groups of l-bits, which leads to an $O(1)$ energy difference set by the local disorder scale $\sim W$, as is apparent from Eq.~\eqref{eq:lbit}. For weak enough amplitude and  $\omega >> W$, the drive is again ineffectual at generating resonances to heat the system. The picture is similar to the SP case:  an MBL system (roughly) looks like a set of weakly coupled local Rabi oscillators, the l-bits $\tau$, with finite $O(1)$ energy scales set by $W$, and driving at frequencies larger than this $O(1)$ scale can avoid indefinite heating (Fig.~\ref{fig:floquet_lbits}). 

In the Floquet MBL regime, one can construct a local time-independent Floquet Hamiltonian $H_F$ which is itself many-body localized\footnote{Although one might need to move to a suitable rotating frame to construct a local Floquet Hamiltonian.}, so that many of the stroboscopic properties of the driven system can still be understood within the l-bit framework. 
Equivalently, the system still has l-bits which commute with the Floquet unitary: $[U_F, \tau_i^z]=0$. 
In this case, the system reaches a late-time steady state with no net energy absorption over a drive period; this is the sense in which Floquet systems without energy conservation can still be ``conservative" in the parlance of our prior discussion.

As the frequency is lowered so that $\omega \ll W$, the system is able to absorb multiple energy quanta of size $\omega$ to locally straddle the energy difference $W$ and heat up. While a rigorous proof is still lacking, there is by now substantial analytical and numerical evidence for the existence of a Floquet MBL phase which can transition to a delocalized one as the drive parameters are changed. 

Finally we note that, as before, synchronization follows quite generally from dephasing in a generic MB system, and MBL alone is not enough to prevent this. However, it does prevent thermalization to an infinite temperature state, so that the late-time stead state is non-thermal and set by the l-bit expectation values (Fig.~\ref{fig:thermalization}(c)). Likewise, Floquet ETH is not obeyed, so that EEVs of local observables fluctuate strongly and are non-thermal (Fig.~\ref{fig:thermalization}(f)).

\subsection{Eigensystem Order and Transitions}
\mlabel{sec:MBLorder}

Now that we have discussed the basics of thermalization -- or the lack thereof -- in closed quantum systems, we turn next to phase structure. 
Just as phases of matter are defined for thermalizing systems, one can also ask whether it is possible to define a further notion of phase structure \emph{within} the class of out-of-equilibrium MBL systems. One might think that the absence of equilibrium would render this impossible. Remarkably, that is not so and there is a rich notion of phase structure even within the class of localized systems~\cite{lpqo, PekkerHilbertGlass, Bauer13, Chandran14, Bahri15, Kjall14}. What's more, the lack of equilibrium constraints can permit order in settings where it it may otherwise be disallowed, a phenomenon termed ``localization protected quantum order"~\cite{lpqo} --- with the DTC being perhaps one of the most striking examples.

\begin{figure}[t]
    \centering
    \includegraphics[width=\columnwidth]{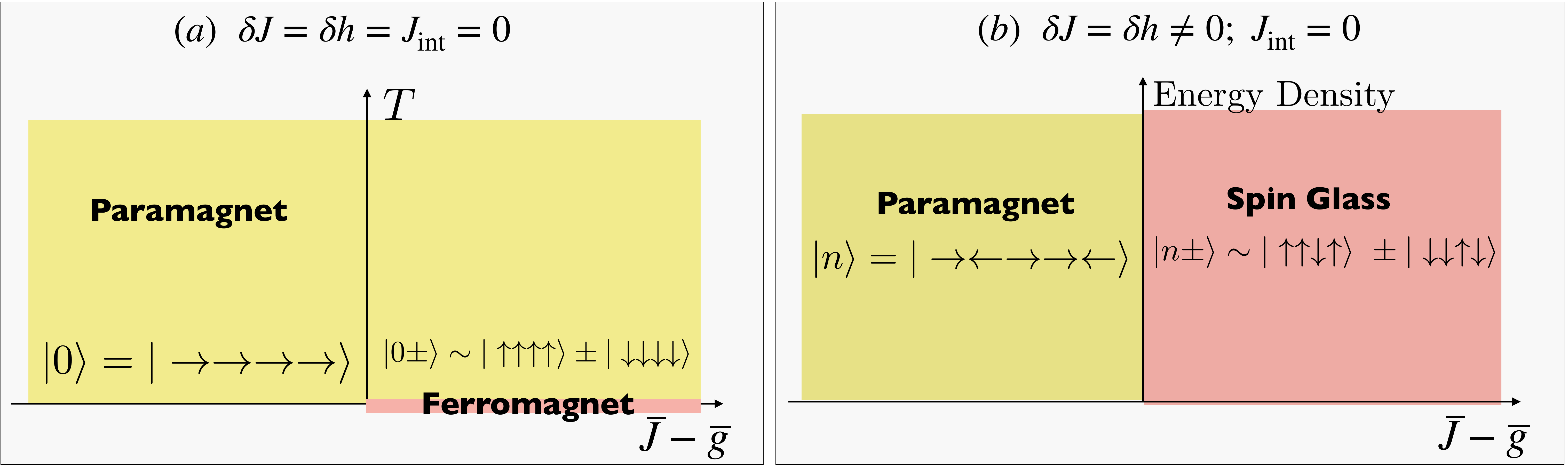}
    \caption{Phase structure of non-interacting Ising chain. Left: for clean system, symmetry-breaking ordering only occurs at zero temperature. Here, the finite-size ground state doublet is made up of a pair of Schr\"odinger cat states. This order disappears upon switching on nonzero temperature, or for sufficiently strong transverse fields $g$ larger than exchange $J$. In the resulting paramagnet, the ground state is a unique state polarised in the field direction. Right: In the strongly disordered localised case, symmetry broken eigensystem order exists at all energy densities in the spin glass state, with quasi-degenerate states in pairs of Schr\"odinger cat states throughout the spectrum. In the paramagnet, eigenstates have no such pairing.}
    \label{fig:ising}
\end{figure}

Traditionally, phase structure is studied within the framework of quantum statistical mechanics, so that phases are classified by order parameters evaluated in equilibrium Gibbs ensembles, and phase transitions are detected by non-analyticites in the partition function.
Since such ensembles are ill defined out-of-equilibrium, we will need a broader framework for defining phases and transitions in MBL systems. 
This can be done via the notion of \emph{eigensystem order}~\cite{lpqo, PekkerHilbertGlass}---which reduces to the usual notion of phases and transitions in equilibrium ensembles for ergodic systems that obey ETH. For MBL systems, however, this will yield phases and transitions which will be \emph{invisible} to the standard ensembles of statistical mechanics, consistent with their irrelevance in this setting.

Eigensystem order subsumes three complementary measures of order in general
quantum Hamiltonian systems:
\begin{enumerate}
    \item The first is the notion of \emph{eigenstate order}, wherein one measures physically meaningful quantities: order parameters, correlation functions, entanglement entropies and
    such in \emph{individual} many-body eigenstates of an MBL system instead of averaging over the entire spectrum as in an equilibrium Gibbs state. The suppression of thermal fluctuations enables {individual} highly-excited eigenstates to display long-range order (LRO), both symmetry breaking and topological, even in dimensions and at energy densities where order is forbidden in equilibrium, say by the Peierls-Mermin-Wagner theorems. Hence MBL eigenstates can show non-trivial signatures of order, even while thermal averages look trivial. In an ergodic system, the two will agree via ETH. 

    \vspace{1pt}
    \item The second, related, measure has to do with the properties of the \emph{eigenspectrum i.e} features such as degeneracies in the distributions of the energy eigenvalues of the many-body system or the statistics of the level spacing distribution. 
    
    \vspace{1pt}
    \item The third diagnostic probes \emph{measurable dynamical} signatures of different forms of order. Indeed, while the eigensystem diagnostics referenced above are valuable for formally defining order out-of-equilibrium, experiments do not have access to individual MB eigenstates. Thus, the translation of these eigensystem properties to measurable dynamical quantities is essential from a practical point of view. Indeed, eigenstate phases and phase transitions are equally well thought of as dynamical phases and phase transitions. 
\end{enumerate}

The ordering itself follows our general understanding of order. It can involve breaking a symmetry, or be topological in some appropriate sense. The transitions can involve singular changes in patterns of order in eigenstates as a function of system parameters or conserved densities, e.g. between a symmetry broken and a symmetric phase. The MBL-to-thermal transition is itself an example of an eigenstate phase transition.
The disjunction between individual eigenstates and ensemble averages arises as the ETH breaks down and strong fluctuations between proximate eigenstates ensue. Finally, dynamical signatures arise because the late time diagonal ensembles average over eigenstates in a manner that allow the absence/presence of ordering to be detected by starting with suitably chosen \emph{classes} of initial states.

We note that the notion of eigenstate order applies to both static Hamiltonian systems and periodically driven Floquet systems, where the relevant eigenstates for the latter are those of the Floquet unitary $U_F$. We focus on the static case here, deferring a discussion of the driven case to Section~\ref{sec:FMBLTC}. 
While a completely general exposition of phase structure and order is beyond the scope of this review, it may help the reader to have in mind the $d=1$ transverse field Ising model, with uniform and random couplings~\cite{lpqo, PekkerHilbertGlass, Kjall14}. We 
summarise the salient features of this model to help uncompress the above statements. The reader can consume this now or, alternatively, simply wait until we discuss the DTC in Section~\ref{sec:FMBLTC} to directly see how eigensystem order works in the somewhat more elaborate context of a Floquet Ising model, referring back to this subsection as needed.
\begin{figure}
    \centering
    \includegraphics[width=0.35\columnwidth]{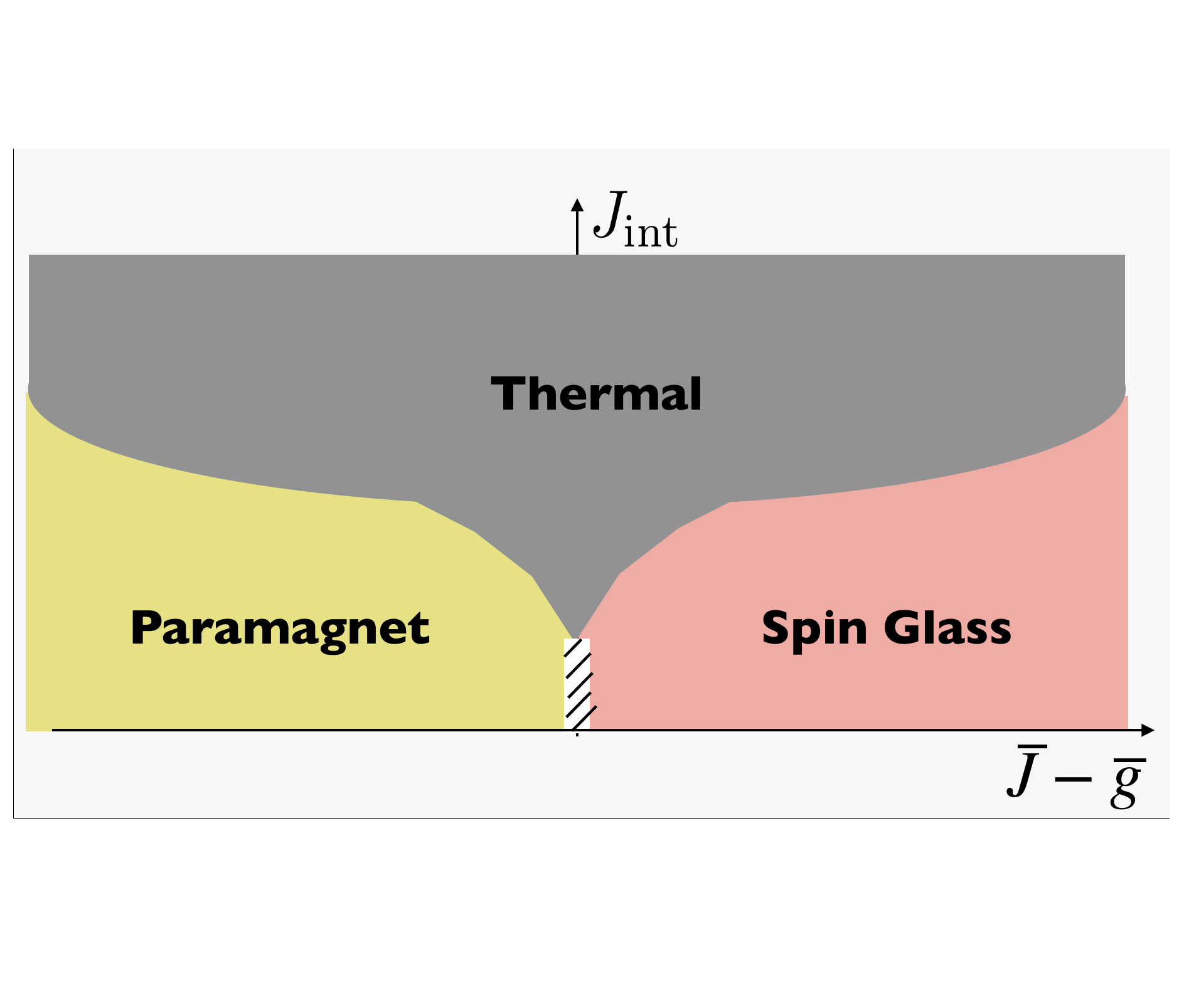}
    \includegraphics[width=0.64\columnwidth]{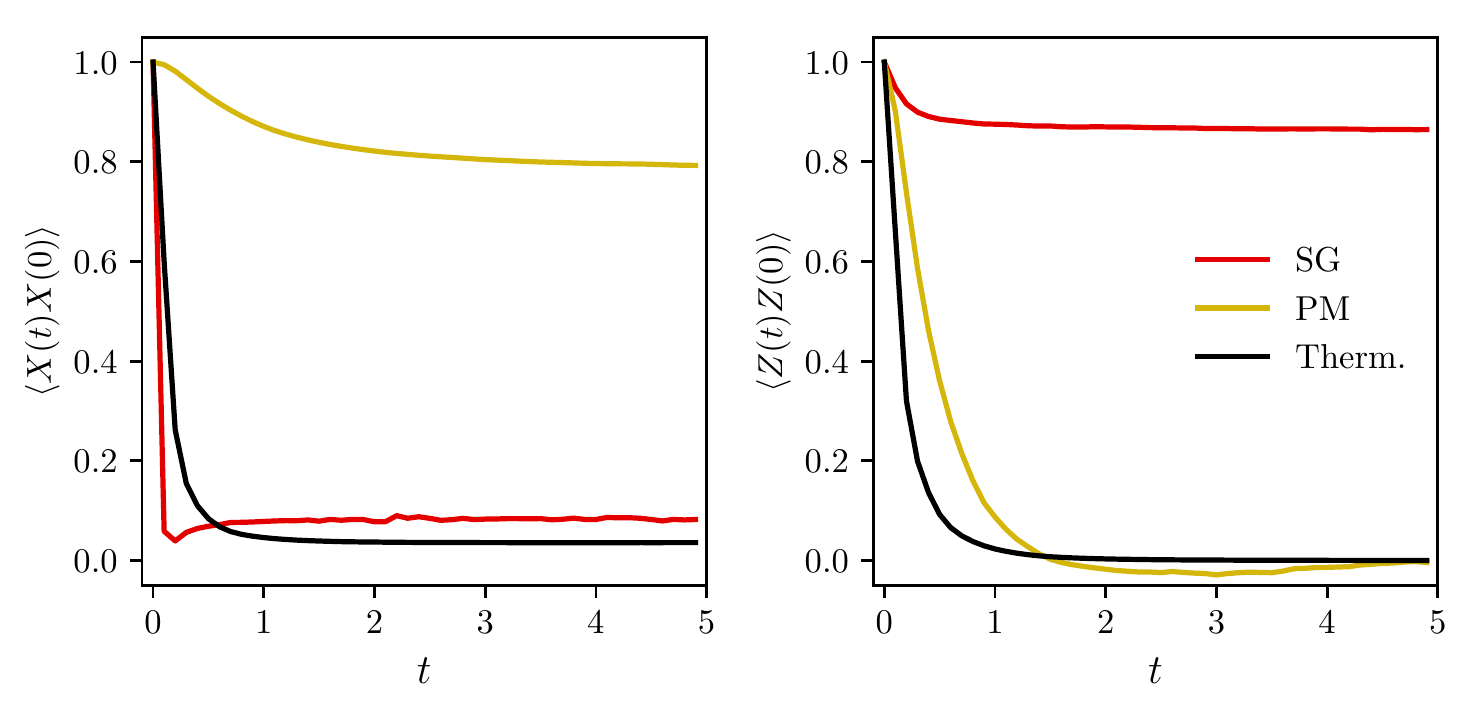}
    \caption{Interacting Ising chain. Left: Schematic (self-dual) phase diagram. Adding interactions of strength $J_{int}$ to  the non-interacting chain (cf.\ Fig.~\ref{fig:ising}) eventually leads to delocalisation, replacing both spin glass and paramagnet with a thermal phase. The nature of the transition between the interacting paramagnet and spin glass is at present unclear. Right: Dynamical diagnostics for distinguishing between the thermal, paramagnetic and spin glass phases by examining the persistence of $\sigma^x$ and $\sigma^z$ correlators at long times. The thermal phase remembers neither, the spin-glass phase remembers only $\sigma^z$, while the paramagnet strongly remembers $\sigma^x$ and weakly remembers $\sigma^z$ -- related to the dominant terms in the Hamiltonian and the overlap of the physical p-bits with the intrinsic lbits. The autocorrelators are averaged over all states, so $\langle \rangle$ reflects an infinite temperature average, which can be non-zero for \emph{dynamical} correlators as in the Edwards Anderson order parameter. The data in (b,c) reflect simulations of Eq.~\eqref{eq:ising} with random fields $h_i$ and $J_i$ drawn from log-normal distributions, with equal variances $\delta \ln(h_i) = \delta \ln(J_i)=1$ and means $\overline{\ln(h_i)}=0$ and $\overline{\ln(J_i)} \equiv \ln \overline{J}$ respectively. 
    The SG and PM curves have $\ln \overline{J} = \pm 2$, respectively, and $J_{\rm int} = 0.1\overline{J}$. The thermalising curve has $\overline{J} = 1$ and $J_{\rm int}=3$.
    }
    \label{fig:ising_interacting}
\end{figure}

Consider a one-dimensional transverse field magnet in a system of length $L$:
\begin{equation}
    H = -\sum_i J_{i} \sigma_i^z \sigma_{i+1}^z + h_i \sigma_i^x + J_{\rm int} (\sigma_i^z \sigma_{i+2}^z +  \sigma_i^x\sigma_{i+1}^x).
\mlabel{eq:ising}
\end{equation}
This model has a discrete $\mathbb{Z}_2$ Ising symmetry $P_x = \prod_i \sigma_i^x$. The local magnetization on a site is taken to be `up' or `down' corresponding to $\sigma_i^z = \pm 1$, and the symmetry operator $P_x$ flips the magnetization of all spins. 
The phase where the fields $h_i$ dominate is a paramagnet (PM), while the one with dominant exchanges $J_i$ is an ordered ferromagnet or, in the disordered case, a \emph{spin glass} (SG). When $J_{\rm int} = 0$, the model is mappable to a system of non-interacting Majorana fermions (the Kitaev chain/p-wave superconductor) via a Jordan Wigner mapping. We allow for the possibility of disorder in the fields and couplings $h_i$ and $J_i$ respectively, which may be drawn from distributions with means $\overline{J}, \overline{h}$ and widths $\delta J, \delta h$. 
While one can generically pick any interaction terms as long as they are Ising symmetric, the particular choice in Eq.~\eqref{eq:ising} statistically respects the self-dual nature of the non-interacting Ising model\footnote{The Ising duality interchanges the exchange terms $\sigma_i^z \sigma_{i+1}^z$ and the transverse field $\sigma_i^x$.}.

Consider first the non-interacting clean model with no disorder so that $J_i =J$, $h_i = h$ and $J_{\rm int}=0$. Deep in the paramagnet, the ground state is gapped, paramagnetic and nondegenerate, having all spins (nearly) aligned with the transverse field in the $x$ direction, $|\rm PM\rangle_0 \simeq |\leftarrow \leftarrow \cdots \leftarrow \rangle$. This phase has no long-range order, so that connected correlation functions of the order parameter  $C(i,j) = \langle \sigma_i^z \sigma_j^z \rangle - \langle \sigma_i^z \rangle \langle \sigma_j^z \rangle $ are exponentially decaying with $|i-j|$ in every eigenstate

By contrast, the system spontaneously breaks Ising symmetry deep in the ferromagnet wherein the two symmetry broken ground states with spins nearly aligned/anti-aligned along the $z$ axis are $|\uparrow\rangle \simeq |\uparrow \uparrow \cdots \uparrow \rangle$ and $|\downarrow\rangle \simeq |\downarrow \downarrow\cdots \downarrow\rangle$. At any finite system size, the system can tunnel between these and the two lowest lying eigenstates are $\mathbb{Z}_2$ symmetric ``Schrodinger cat" states that are eigenstates of the Ising symmetry $P_x$ with eigenvalues $\pm1$:  $|0_\pm\rangle=\frac{1}{\sqrt{2}}\left[|\uparrow\rangle\pm|\downarrow\rangle\right]$. These states display long-range order (LRO), so that $\lim_{|i-j|\rightarrow \infty} \lim_{L\rightarrow \infty} C(i,j) \neq 0$. 
The sharply different behavior of this correlator in the ground states of the two phases thus serves as \emph{eigenstate} diagnostic of quantum order  (\emph{cf.} Item 1 in the list of measures of LPQO above)
Relatedly, because the two symmetry broken states can only be connected by sweeping a domain wall across the entire system (which requires $O(L)$ steps visiting intermediate states at energy cost $\sim J$) the resulting energy splitting between the cats is exponentially small, $\sim(h/J)^L$.  Thus, the quasidegeneracy of the ground state of a finite-size system is also sharply different between the PM and FM phases, and this feature of the \emph{eigenspectrum} can serve as a complementary diagnostic for discriminating between the two phases (\emph{cf.} Item 2 in the list of measures of LPQO above).   

Now consider finite temperatures. It is well known from the Landau-Peierls theorem that one dimensional systems in equilibrium cannot show spontaneous symmetry breaking at any finite temperature. At a finite temperature, there is a finite density of excitations above the ground state, which look like domain walls (DWs) in the FM phase. In a clean system, the energy of a domain wall is, to a first approximation, independent of its location so that these can move freely across the sample. The finite temperature eigenstates look like superpositions of states with finite densities of DWs in differing spatial locations. 
The presence of these delocalized fluctuating domain walls destroys the long-range correlation between spins, leading to $\lim_{|i-j|\rightarrow \infty} C(i,j) \rightarrow 0$ at any finite temperature. The phase diagram of the clean model is shown in Fig.~\ref{fig:ising}(a). 

Things change fundamentally when disorder is added. Consider again the non-intracting regime with weak transverse fields $h_i$. If the sign of $J_i$ is random, the ground state -- along with \emph{all} excited states -- acquires a glassy Edwards-Anderson order parameter~\cite{lpqo, PekkerHilbertGlass, Kjall14}; this is the eigenstate order. Absent interactions, the domain walls in the free system get Anderson localized and become ``pinned" with the disorder. The symmetry broken phase with strong exchange fields is now a  \emph{spin-glass}. When $h=J_{\rm int} = 0$, each eigenstate looks like a random glassy pattern of locally symmetry broken ``frozen" magnetizations, which is strictly degenerate in energy with its Ising reversed partner:
\begin{align}
    |\nu_z \rangle &=  |\{s_i^z\}_\nu \rangle =  |\uparrow \downarrow \downarrow \cdots \uparrow\rangle \nonumber \\
    |\overline{\nu}_z \rangle &=  |\{-s_i^z\}_\nu\rangle =  P_x|\nu^z\rangle = |\downarrow \uparrow \uparrow \cdots \downarrow\rangle
\label{eq:glassystates}
\end{align}
where $\nu$ refers to the eigenstate number, $s_i^z = \pm1$, and $\sigma^z_i|\{s_i^z\}_\nu\rangle=s_i^z$. Upon adding weak transverse fields, the system can tunnel between these so that the finite-size $\mathbb{Z}_2$ eigenstates look like Ising even/odd pairs of ``Schrodinger cat states". At the exactly solvable point, the cats are strictly degenerate and look like: 
 \begin{equation}
|\nu_\pm\rangle =  \frac{1}{\sqrt{2}}\left(|\{s_i^z\}_\nu \rangle \pm |\{-s^z_i\}_\nu \rangle \right) = \frac{1}{\sqrt{2}} \left( |\uparrow \downarrow\downarrow \cdots \uparrow\rangle \pm |\downarrow \uparrow\uparrow \cdots \downarrow\rangle\right) \equiv |\{d_i\}_\nu, p_\nu=\pm1\rangle.
\mlabel{eq:catstates}
\end{equation}
These are labeled by the expectation values of the Ising symmetry, $P_x$, and the $(L-1)$ Ising symmetric domain wall operators $D_i \equiv \sigma_i^z \sigma_{i+1}^z$.  These are the l-bits in this exactly solvable limit and completely specify each state.

This phase spontaneously breaks Ising symmetry and shows LRO, as diagnosed by the non-vanishing two point correlators of the Ising order parameter in each Schrodinder cat eigenstate: 
\begin{equation}
    C_{ij} = \langle \{d\}_\nu, p_\nu| \sigma^z_i  \sigma^z_j|\{d\}_\nu, p_\nu \rangle_c \ne 0 \mbox{ for arbitrarily large } |i-j|.
\mlabel{eq:sgcorr}
\end{equation}
Note that this correlator is non-zero for arbitrarily distant $|i-j|$, but the sign of $C_{ij}$ fluctuates depending on whether the state $\nu$ has an even/odd number of pinned domain walls between sites $i$ and $j$. 
In other words, all eigenstates break the global Ising symmetry, but differently in each state--- whence the appellation spin-\emph{glass}. If we average the correlator over all eigenstates such as in a thermal Gibbs state (say for a fixed $i,j$), then the fluctuating signs cause the average to vanish at any finite temperature consistent with the Peierls theorem. We reiterate:  properties visible in \emph{individual} highly-excited eigenstates of localized systems may be invisible to thermodynamic ensemble averages. 

Turning to eigenspectrum properties, we see that the \emph{entire} eigenspectrum is paired so that the 
even and odd parity cat states are strictly degenerate when $h_i=J_{\rm int}=0$~\footnote{Both $|\nu_+\rangle$ and $|\nu_-\rangle$ have eigenvalues $E_\nu(d,p) = -\sum_i J_i d_i$ independent of $p$. In other words, the eigenenergies \emph{within} any given parity sector are randomly distributed (with no level repulsion), but the entire spectrum is \emph{paired} reflecting the degeneracy between the Ising even and odd cat states, $E(d,+1) = E(d,-1)$. In the dual free-fermionic description, the SG phase is a topological phase with a Majorana zero mode which is responsible for the degeneracy between the Ising even/odd states in the many-body spin language.} .
On perturbing with weak transverse fields, it can be shown that the fundamental properties of this phase---the breaking of the $\mathbb{Z}_2$ symmetry and the associated doublet structure of the eigenspectrum---continue to hold as long as the average couplings obey $\overline{J}>\overline{h}$ (Fig.~\ref{fig:ising}(b)). However, in the perturbed model, the the paired cat states now have an energy splitting that is exponentially small in $L$ rather than being strictly degenerate\footnote{Note that the intra-energy splitting between the oppposite-parity members of a doublet and the inter-energy splitting between different eigenstates of the same parity are now \emph{both} exponentially small in $L$, but with different decay constants. This leads to the possibility of an additional ``pairing" transition within the 0SG phase, depending on the relative magnitudes of the inter- and intra- energy gaps\cite{lpqo}}. The physical $D_i$ operators get weakly dressed to form new l-bits, $\tilde{D}_i = \tau_i^z \tau_{i+1}^{z}$ with exponential tails. In the infinite size limit, the paired states become exactly degenerate, and the system can spontaneously break Ising symmetry to select a member of the pair with spins aligned randomly in the $\tau^z$ basis.

In the opposite limit of strong transverse fields, $h \gg J$, the system is still localized for strong enough disorder, but it is paramagnetic with no LRO or SSB. For $J= J_{\rm int} =0$, the eigenstates look like random patterns of spins pointing along the $\sigma^x$ direction: $|\nu\rangle = | \{ s_i^x\}_\nu \rangle = |\leftarrow\leftarrow\rightarrow\leftarrow \cdots \rightarrow\rangle$. Upon introducing weak exchange terms, these continue to weakly dressed versions labeled by their expectation value under the l-bit operators $\tau_i^x$ . These states show no LRO so that connected correlation functions $C_{ij} = 0 $ in all eigenstates for large $|i-j|$. And the eigenspectrum has no special pairing properties. 

Thus, the disordered Ising model Eq.~\eqref{eq:ising} shows two distinct localized phases: the SG with spontaneous symmetry breaking and long-range glassy order, and the disordered paramagnet. The two are distinguished by sharp signatures in the eigensystem properties, specifically, connected correlation functions measuring LRO in individual eigenstates, and spectral pairing in the eigenvalues. The symmetry broken SG phase is \emph{more} stable than its equilibrium FM counterpart, and can persist to all energy densities because the thermal fluctuations that destroy LRO are suppressed in the localized phase. Without interactions, the entire spectrum is localized for any $W>0$, and the SG phase transitions to the PM at $\overline{J} = \overline{h}$ via an infinite randomness phase transition that involves a singular change in the properties of every eigenstate~\cite{Fisher95, lpqo, PekkerHilbertGlass}~(Fig.~\ref{fig:ising}).  

Upon turning on weak interactions, the system remains localized deep in the MBL PM and SG phases. However, it can transition to a thermalizing phase for strong enough interactions. The fate of the SG-to-PM transition in the presence of interactions is presently unsettled, and it is likely that the transition proceeds via an intervening sliver of thermal phase between the two localized phases. A sketch of a putative phase diagram for the self-dual Ising model is shown in Fig.~\ref{fig:ising_interacting}(a).

Finally, before concluding this section, we emphasize that the eigensystem diagnostics distinguishing between the localized PM, SG and thermalizing phases have \emph{measurable} dynamical signatures, and the transitions between these phases are hence also dynamical phase transitions.
We will consider the interacting system for which the dynamics are more generic. 
Specifically for the Ising case, one can measure the expectation value of a local Ising odd operator in time, say $M_i = \sigma_i^z$ with $P_x M_i P_x = -M_i$, together with local Ising even operators such as $\sigma_i^x$, upon starting from experimentally accessible initial states. The behavior of $\langle \psi_0| O(t)| \psi_0 \rangle$ with $O = \sigma_i^{x/z}$ can be used to distinguish between the three phases, and is easily understood within the l-bit picture: 
\begin{itemize}
    \item In the PM, the l-bits are Ising even $\tau^x_i$ operators which are exactly conserved. Thus, physical $\sigma^x_i$ operators have large overlap on these and saturate to a non-zero constant in time, say upon starting from product states in the $\sigma^x$ basis. 
    In contrast,  Ising odd $\sigma^z_i$ operators are orthogonal to the Ising even l-bits and decay to zero as a power-law in time for all initial states, as discussed in Sec.~\ref{sec:mbldynamics}. Note that local Ising odd operators are order parameters for Ising symmetry breaking. Thus, the decay of these operators is behooving of a \emph{paramagnet} for which the late-time expectation values of Ising order parameters should be zero.
    
    \item In contrast, in the SG phase, we can choose the l-bits to be either Ising even operators $D_i =  \tau_i^z\tau_i^z$ or the Ising odd operators $\tau_i^z$ (upto exponentially small in $L$ errors). Thus, a physical Ising odd $\sigma_i^z$ operator has large overlap on $\tau_i^z$ and saturates to a non-zero constant, provided one starts with a symmetry broken initial state with a non-zero initial expectation value of $\tau_i^z$. On the other hand, an Ising even operator like $\sigma_i^x$  has non-zero (but small) overlap on Ising even l-bits $D_i$ in the SG phase, and hence saturates to a small non-zero value from Ising even initial states.
    
    \item Finally, in the thermal phase, it follows from the ETH that both $\sigma_i^{x/z}$ saturate to zero if we start from random initial product at infinite temperature. 
\end{itemize}  
To summarize, we can distinguish between the SG, PM and thermal phases by measuring the dynamics of \emph{physical} operators $\sigma_i^z$ and $\sigma_i^x$ starting from experimentally accessible initial states. Simulations of these, averaged over all initial states, are shown in Fig.~\ref{fig:ising_interacting}, clearly distinguishing the three phases.

For completeness, we note that the dynamics of these operators from specific initial states can also be understood by expanding the initial states in terms of the eigenstates, as in Eq.~\eqref{eq:Ot-diag}. We note that a symmetry broken initial state such as a product state in the $\sigma_i^z$ basis has non-zero overlap on both Ising symmetric and antisymmetric members of the nearly degenerate `cat' doublets. If we measure Ising odd operators like $M_i$, it can only connect between (finite-size) eigenstates with opposite eigenvalues under the Ising symmetry. Thus, in this phase, a non-zero late-time expectation value for $M_i$ relies on using the \emph{block}-diagonal ensemble in \eqref{eq:Ot-diag} to connect between the degenerate Ising symmetric cat states: $\lim_{t\rightarrow\infty}\lim_{L\rightarrow\infty} \langle \psi_0|\sigma_i^z(t)|\psi_0\rangle = \sum_\nu c_{\nu_+}^* c_{\nu_-} \;\langle \nu_+|\sigma_i^z|\nu_-\rangle + h.c.$\footnote{In contrast, in the PM phase, the eigenstates $ |\nu\rangle$ are product states in the $\tau^x$ basis. The $\sigma_i^z$ operator locally flips an odd number of spins in the $\tau^x$ basis, connecting to states that are typically $O(1)$ away in energy. These contribute only to the off-diagonal non-degenerate (second) term in Eq.~\eqref{eq:Ot-diag} which is exponentially small at late times. Moreoever, there are generically no degeneracies in the MB spectrum in the PM phase, so the first term in Eq.~\eqref{eq:Ot-diag} only receives a contribution from diagonal eigenstate expectation values $\langle \alpha|O|\alpha\rangle$, which are necessarily zero because the eigenstates have a definite value under Ising symmetry and $M_i$ is Ising odd. Thus, in the paramagnetic phase,  $\langle \sigma_i^z(t)\rangle$ decays to zero for all short range correlated initial states.}

In sum, the example of the static, disordered Ising chain illustrates how localization enables long-range order to persist to high energy densities where it would otherwise be disallowed in equilibrium, and how one can define concrete dynamical and eigensystem diagnostics for measuring this order. We reemphasize that in thermalizing systems, the notion of eigenstate order is congruent with the usual definition of phases. This is because the ETH implies that local observables (such as order parameters) computed in individual eigenstates agree with those computed in the Gibbs state at a temperature appropriate to the energy density of the eigenstate. On the other hand, MBL systems do not obey the ETH and nearby many-body eigenstates have very different local properties. This is what enables individual MBL eigenstates to display order even if thermodynamic averages are forbidden from doing so.

\section{Defining Time Translation Symmetry Breaking in Quantum Systems}
\mlabel{sec:DefineTTSB}

At this point we have recapitulated several classic arguments against energy conserving macroscopic time crystals, as well as given some special examples of such time crystals protected by additional symmetries. 
We would now like to specialize to quantum systems, and to sharpen and explicate criteria for identifying time crystals. 

The task of defining a sharp diagnostic for TTSB in analogy with other, more conventional, forms of  symmetry breaking was first systematically addressed by Watanabe and Oshikawa (WO) in Ref.~\mcite{Oshikawa15}. In the present section we will present their diagnostic, and also present two additional, complementary, diagnostics of TTSB, of which one is experimentally the most useful one. We will note that all three diagnostics agree that thermalizing systems do not exhibit TTSB. In contrast, they allow static MBL systems to exhibit temporal glassiness but not time crystallinity. We will then review WO's twin theorems ruling out ground state and excited state time crystals in systems governed by static Hamiltonians. We will show that a) their excited state proof has a subtle error which renders it invalid, even though the conclusion is still likely to be correct and b) that their results leave various loopholes, which allow for TTSB (or TTSB-like) behavior in several settings;  of these, we have already discussed one (SPTCs) and we will discuss two more in this section. Finally their results did not address Floquet systems at all, and we will discuss TTSB in those in
Sec.~\ref{sec:FMBLTC}.

\subsection{Diagnostics and standard lore on spontaneous symmetry breaking}
\mlabel{sec:standardSSB}

We start with a recapitulation of standard ideas on spontaneous symmetry breaking, before generalizing these to the case of time translation symmetry breaking. 

In standard discussions of broken symmetries, one typically considers two kinds of symmetries: (i) internal symmetries, for example spin rotational or Ising symmetry in a model with spin degrees of freedom, and (ii) spatial symmetries, like translation or rotation.
In both cases, while the Hamiltonian $H$ respects the symmetry, the state of the system in a symmetry broken phase does not. For example, a solid in a crystalline phase spontaneously breaks the continuous translation symmetry of the (real) space it inhabits by exhibiting a periodically modulated spatial structure. Likewise, spins in a ferromagnetic phase become aligned, giving rise to a net magnetization which points along some direction and breaks spin rotational symmetry. This is in contrast to a paramagnetic phase with uncorrelated and fluctuating spins, which has no net magnetization and does not exhibit SSB.

In other words, \emph{in a symmetry broken phase, the macroscopic equilibrium state of the system is non-invariant under the symmetry transformation.} And the ``spontaneous" refers to the fact that the Hamiltonian or dynamics of the system respects the symmetry. 

The above discussion implies two intrinsic features of SSB: (i) a degeneracy, or non-uniqueness, of the system's equilibrium states and (ii) the need to take the thermodynamic limit to meaningfully define SSB. In general, a symmetry of the dynamics must map equilibrium states onto other equilibrium states. Consequently, if an equilibrium state is non-invariant under an exact symmetry of the dynamics, it must be one of a set of equilibrium states related by the action of the symmetry. Additionally, it turns out that non-invariant states are strictly in equilibrium only in the infinite size limit. At any finite size, there is a finite tunneling probability between different non-invariant states --- so that the equilibrium state of a finite system is always unique and symmetric. Thus, SSB is a property that can only be exhibited in the thermodynamic limit. 

These statements are best illustrated with an example. Consider the lattice quantum Ising model discussed earlier, $H = -J \sum_{\langle ij \rangle} \sigma_i^z \sigma_j^z - h \sum_i \sigma_i^x$ with internal discrete Ising symmetry $P_x = \prod_i \sigma_i^x$. The two degenerate symmetry-broken ground states in the ferromagnetic phase (in the infinite size limit),  $|\uparrow\rangle \simeq |\uparrow\uparrow \cdots \uparrow\rangle$ and $|\downarrow\rangle \simeq |\downarrow\downarrow \cdots \downarrow\rangle$, are non-invariant under the spin-flip symmetry of the Hamiltonian, and are mapped to one another by the action of the symmetry: $P_x|\uparrow\rangle = |\downarrow\rangle$. However, at any finite size, the system can tunnel between the $|\uparrow\rangle, |\downarrow\rangle$ states, and the ground states of $H$ look like ``Schr\"odinger cat'' or superposition states which are also eigenstates of $P_x$ with eigenvalue $\pm 1$, $|\pm \rangle \sim|\uparrow\rangle \pm |\downarrow\rangle$. The energy splitting between the $|\pm\rangle$ states is exponentially small in the volume of the system, and these become strictly degenerate in the infinite size limit enabling SSB. More generally, a ferromagnet that spontaneously breaks a \emph{continuous} spin rotational symmetry has an infinite set of equilibrium states related by rotations.

The considerations above indicate that one must take care in defining suitable diagnostics for SSB. Naively, one might think to diagnose ferromagnetism by measuring an order parameter such as the global magnetization, $M~=~\sum_i \sigma_i^z$, in an equilibrium state such as the Gibbs state at some temperature. However, $\langle M \rangle$ is zero in any finite sized system because the equilibrium state is an eigenstate of the Ising symmetry, while the order parameter is Ising odd: $P_x M P_x = -M$\footnote{More formally, an order parameter for diagnosing the spontaneous breaking of a symmetry $G$ must transform as a \emph{non-trivial} irreducible representation (irrep) of $G$. This implies that one-point expectation values of order parameters necessarily vanish in eigenstates or Gibbs states of $H$, which are symmetric with respect to $G$ at any finite size.}. Moreover, although a symmetry broken equilibrium state  can have a non-zero $\langle M \rangle$ (in the infinite size limit), this state is not-unique; hence the infinite size limit must be approached with care. 

We discuss how these issues are resolved by presenting three convenient measures that are used, in practice, for diagnosing ``ordinary" SSB of spatial or internal symmetries in time-independent Hamiltonians. We recapitulate them here, and discuss how they apply to TTSB in the next subsection:

\begin{enumerate}
\item \mlabel{it:SSBdiagcorr}\underline{Long-range order in correlation functions:} While one-point expectation values of the order parameter are zero in equilibrium states (at finite size), two-point correlation functions need not be. Indeed, a symmetry broken phase is associated with \emph{long-range order}. For example, the magnetization in a ferromagnet gets ``locked" into a global pattern, with long-range correlations between the local magnetization of distant spins. As a result, the presence of long-range correlations of the order parameter in equilibrium states of the system is a standard diagnostic for SSB:
\begin{align}
\lim_{|{\bf r - r'}|\rightarrow\infty}\lim_{V\rightarrow\infty} \langle C({\bf r} ,{\bf r'})\rangle \equiv \lim_{|{\bf r - r'}|\rightarrow\infty}\lim_{V\rightarrow\infty}\langle O({\bf r })O({\bf r'})\rangle_c 
\equiv \lim_{|{\bf r - r'}|\rightarrow\infty}\lim_{V\rightarrow\infty} \langle O({\bf r })  O({\bf r'})\rangle - \langle O({\bf r }) \rangle \langle  O({\bf r'})\rangle
\neq 0 .
\mlabel{eq:SSBcorr}
\end{align}
Here $O({\bf r})$ is a local order parameter, such as the local magnetization $\sigma^z_{i}$ for the case of an Ising magnet or the local density $\varrho(\bf r)$ for a spatial crystal, and $\langle \rangle_c$ denotes the connected correlation function. We note that short-range correlated states in which the connected correlator between distant operators tends to zero are said to \emph{cluster}, while states with long-range correlations (such as Schrodinger cats) are non-clustering~\footnote{An operator independent diagnostic of clustering is furnished by examining the mutual information between distant subregions. This is exponentially small in the distance between subregions in clustering states, and serves as an upper bound on the connected correlator between any operators in the subregions.}. 

\hspace{6pt}We also recall that in some cases, particularly when spatial symmetries are involved, it is more natural to consider correlation functions of a spatially integrated order parameter. For example, the global magnetization $M = \sum_{i} \sigma^z_{i}$ for an Ising magnet, or a density modulation, $\varrho({\bf G}) = \int d^dx \; \varrho({\bf x}) e^{-i {\bf G\cdot x}}$ at wavevector $\mathbf{G}$ for a spatial crystal. The LRO in a spatial crystal is then diagnosed by non-trivial connected correlations in: $\lim_{V \rightarrow \infty} \langle \varrho({\bf -G}) \varrho({\bf G}) \rangle_c/V^2 \neq 0$. However, even in this case, one can still work with local operators as the information in the global correlator can equally be reconstructed by Fourier transforming the correlators $\langle O({\bf r}) O({\bf r+ r'}) \rangle$.

\vspace{6pt}
\item \mlabel{it:SSBdiagsusc}\underline{Susceptibility to a symmetry-breaking field:} Alternatively, one can diagnose SSB by adding a small \emph{explicit} global symmetry breaking field to $H$ and computing the response of the system. The response is measured via the expectation value of the (global) order parameter, $\langle O \rangle_h$, evaluated in equilibrium states \emph{in the presence of the field}, taken to be of strength $h$. Because the symmetry is explicitly broken, the equilibrium states can have a non-zero expectation value $\langle O \rangle_h$. 
The symmetry broken phase is diagnosed by a \emph{divergent} susceptibility: \emph{i.e.} a finite, non-zero expectation value for $\frac{1}{V}\langle O \rangle_h$ even as the magnitude of the symmetry-breaking field is taken to zero, provided the system size is taken to infinity \emph{before} $h$ is taken to zero: 
\begin{align}
\lim_{h\rightarrow 0} \lim_{V \rightarrow \infty} \frac{1}{V}\langle O \rangle_h \neq 0.
\mlabel{eq:ssbfield}
\end{align}
For example, in the case of an Ising magnet, one adds a small global field $h \sum_{i} \sigma_{i}^z$ to the Hamiltonian and measures the global magnetization $M$ in the presence of this field. A non-zero value in the order of limits discussed above defines a phase with spontaneous Ising symmetry breaking.  Similarly, to diagnose a spatial crystal, one could apply a periodic potential $V({\bf x}) = h \cos({\bf G \cdot x})$, which probes the susceptibility of the system to a periodic modulation with wavevector ${\bf G}$. 

\hspace{6pt} A diverging susceptibility of the order parameter directly follows from the fact that symmetric eigenstates in different symmetry sectors become degenerate in the infinite size limit in the symmetry broken phase. Thus, a vanishing symmetry-breaking field is sufficient to mix between these and produce symmetry broken eigenstates with non-zero expectation values of the order parameter.
The opposite order of limits, $\lim_{V \rightarrow \infty} \lim_{h\rightarrow 0} \langle O \rangle_h$, does not give a non-zero order parameter, consistent with the fact that finite-sized eigenstates are symmetric when $h \rightarrow 0$: the vanishing field cannot surmount their finite-size splitting. Loosely speaking, the field does not ``cause" the magnetization, but simply orients it. 

\hspace{6pt} Finally, we remind the reader that, more broadly, the fundamental object being probed here is the finite volume real time susceptibility
\begin{equation}
    \langle \chi(t;V) \rangle = \theta(t) \sum_{i j}\langle  [\sigma_i^z(t), \sigma_j^z(0)] \rangle \ ,
\end{equation}
whose Fourier transform is the complex susceptibility $\langle \chi(\omega;V)\rangle $. This describes the response of the system, as measured by the change in the expectation value of $\langle M(t) \rangle$ to switching on a perturbation $H_p = M(t') h(t')$ via the linear response expression
$$
\delta \langle M(t) \rangle = \int_{-\infty}^t \langle \chi(t-t';V)\rangle \, h(t') \ .
$$
The static susceptibility discussed above is obtained by taking the low-frequency limit: $$\left.\chi_0 = \frac{d \langle M \rangle }{d h}\right|_{\omega =0} = \lim_{\omega \rightarrow 0}\langle \chi(\omega; V)\rangle,$$ and SSB is defined by a divergent response in $\chi_0/V^2$ in the infinite $V$ limit. Note the factor of $1/V^2$, which accounts for the fact that we are probing the susceptibility of the \emph{global} magnetization $M$ to turning on a global field, and this should receive a contribution from every pair of distant spins if there is long-range order in the system.  

\vspace{6pt}
\item \mlabel{it:SSBdiagbound}\underline{Response to a boundary field:}
In a similar vein as the diganostic above, one can apply a small symmetry breaking field at the boundaries of the system and examine its impact on order parameter expectation values deep in the bulk. In the SB phase, these expectation values have nonzero limiting values even as the system size is taken to infinity, for the same reasons as discussed above. This criterion was first discussed by Peierls in the context of classical Ising models, and the choice of boundary conditions (`up' or `down') was used to select one of multiple degenerate non-invariant equilibrium states in the infinite size limit. We note here that the best choice of boundary conditions may not always be obvious. For example, spin-glasses are characterized by a multitude of different broken symmetry equilibrium states, with different ``frozen" patterns of local magnetization. In such systems, the equilibrium state selected in the bulk could be quite sensitive to the choice of boundary conditions.

\end{enumerate}

In sum, we have explicated three complementary ``standard" diagnostics of SSB in systems with spatial or internal symmetries. In equilibrium, we expect that all three criteria will consistently diagnose a SB phase.

\begin{figure}

\makebox[\textwidth][c]{\includegraphics[width=1.1\textwidth]{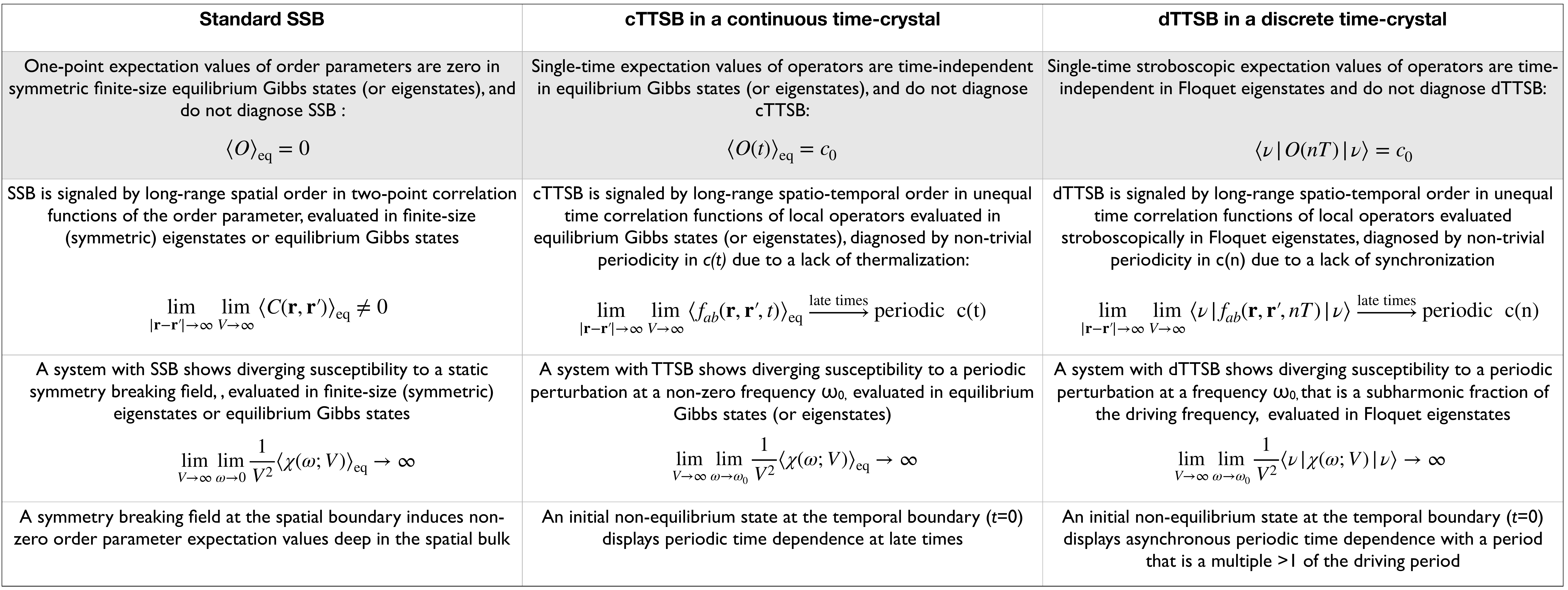}}
     \caption{Table comparing diagnostics for standard symmetry breaking with analogous quantities for the case of cTTSB and dTTSB.}
    \mlabel{fig:diagnostics}
\end{figure}

\subsection{Diagnostics for time translation symmetry breaking}
\mlabel{sec:TTSB}

We now try to carry over this set of ideas to the case of time translation symmetry to define analogous diagnostics for TTSB.  In a system with continuous time-translation symmetry, the Hamiltonian is time-independent and energy is conserved. If cTTS is spontaneously broken, then states of the system should display time dependence. 

Thus, naively, one might expect to diagnose a time-crystal by simply looking for periodic time-dependence in observables, $\langle O(t)\rangle = \langle e^{iHt} O e^{-iHt}\rangle$, where the expectation value is taken in an appropriate equilibrium state. However, by definition, observables are time-independent in equilibrium states so this measure will never show TTSB. (Each eigenstate of $H$, $|\nu\rangle$ with eigenvalue $E_\nu$, simply picks up a phase factor $e^{-iE_\nu t}$ under time evolution.  This phase cancels against its conjugate, leaving no time-dependence in expectation values $\langle O(t)\rangle$ evaluated in any eigenstate or Gibbs average over eigenstates.) However, this alone should not be sufficient grounds for rejecting the possibility of TTSB. The time independence of $\langle O(t) \rangle$ is exactly analogous to the case of spatial or internal symmetries, where one-point expectation values of order parameters are zero in equilibrium states at any finite size, as discussed above.  Instead, one should consider alternate diagnostics for TTSB, analogous to the three defined in the prior subsection.

To do this, we begin by considering the simplest possible quantum clock---the harmonic oscillator, and then ask how the ideas can be generalized to many body systems in search of a \emph{macroscopic} quantum clock.
The harmonic oscillator is governed by the Hamiltonian
\begin{equation}
    H = \frac{p^2}{2m} + \frac{1}{2} k x^2 \ ,
\end{equation}
with a natural frequency $\omega_0 = \sqrt{k/m}$. 
The oscillator exhibits time translation symmetry since $H$ is time independent, and thus the dynamics is time independent. As before, the expectation value of an operator $O(t)$ in any eigenstate is $t-$independent. 
Translating our statements on order parameters in the previous subsection yields:
\begin{enumerate}[(I)]
    \item Two time expectation values exhibit oscillations, e.g. $\langle x(t) x(0) \rangle \propto \cos{\omega_0 t}$, just as two-point spatial correlations functions are non-zero in a system with ordinary SSB (Eq.~\eqref{eq:SSBcorr})
    \item The oscillator responds to a drive at a frequency $\omega$, $H_p = \lambda x \cos{\omega t}$ with a susceptibility $\chi(\omega) = {1 \over m (-\omega^2 + \omega_0^2 + i \epsilon \omega)}$ which diverges at the natural frequency $\omega = \omega_0$. This analogous to a divergent \emph{static} susceptibility in systems with ordinary SSB (Eq.~\eqref{eq:ssbfield})
    \item If we start the oscillator in a linear combination of eigenstates then all expectation values are periodic for all times with $T=2 \pi /\omega_0$. If we start the system in a coherent state the motion is, in addition, monochromatic \emph{i.e.} only the fundamental frequency enters the motion. The choice of a non-equilibrium initial state is akin to selecting a symmetry-broken ``boundary condition" \emph{in time} at $t=0$. The persistence of time dependence at infinitely late times is analogous to spontaneous symmetry breaking infinitely deep in the bulk of a system seeded by an appropriate boundary choice, as in Peierls construction. 
\end{enumerate}
Thus, we see that the oscillator exhibits time-translation symmetry in its eigenstates, but its ``hidden'' periodicity can be uncovered by examining two-point temporal correlations, by studying the response of the system to time-dependent fields that break TTS, or by looking at dynamics starting from general initial states -- in direct analogy to the three diagnostics presented for SSB of spatial and internal symmetries.  The next task is to generalize this set of observations to many body systems. In addition, as we discuss below, measures are also readily generalized to the Floquet case to diagnose dTTSB. 
The three diagnostics for standard symmetry breaking, together with their counterparts for cTTSB and dTTSB are summarized in Figure \ref{fig:diagnostics}.

\subsubsection{Temporal and spatio-temporal order}
\mlabel{sec:TTSBdef_spatiotemporal}
As mentioned above, while equal time expectation values are time-independent in equilibrium states, \emph{unequal} time correlators can still show non-trivial temporal dependence and serve as a diagnostic for TTSB. Thus, analogous to the first diagnostic for SSB in Eq.~\eqref{eq:SSBcorr}, we examine the correlator: 
\begin{align}
    \langle f_{ab}({\bf r},{\bf r'},t) \rangle &=   \langle a({\bf r}, t) b({\bf r'},0) \rangle - \langle a({\bf r}, t)\rangle \langle b({\bf r'},0) \rangle , \nonumber \\  
    &=   \langle a({\bf r}, t) b({\bf r'},0) \rangle - \langle a({\bf r},0)\rangle \langle b({\bf r'},0) \rangle , 
\mlabel{eq:fabdef1}
\end{align}
where $a({\bf r})$ and $b({\bf r})$ are local operators near position ${\bf r}$, and the second line follows because the expectation value is taken in a finite volume equilibrium state with time-independent one-point expectation values. Indeed, this correlator was first examined as a diagnostic for TTSB by WO~\cite{Oshikawa15}, who defined a time crystal as a system in which 
\begin{align}
    \lim_{|{\bf r}-{\bf r'}|\rightarrow \infty} \lim_{ V\rightarrow \infty}  \langle f_{ab}({\bf r},{\bf r'},t)\rangle  \xrightarrow[]{\text{large times}} c(t),
    \mlabel{eq:f-spatiotemp}
\end{align}
where $c(t)$ shows periodic time-dependence for arbitrarily large times\footnote{Note that the limits in Eq.~\eqref{eq:f-spatiotemp} may not actually exist. Ideally we hope to obtain enough information to examine the behavior of $f_{ab}({\bf r},{\bf r'}, t;V)$ for all $|{\bf r}-{\bf r'}|<L$ and all $t$ for a given system size and given eigenstate. We are being this careful here in order to allow consideration of disordered systems which can exhibit ``chaos'' in the variation of their correlations with system size as well as with eigenstate choice when localized, see
Sec.~\ref{sec:TTSBdef_initstate}). We will address these issues when we discuss Floquet MBL time crystals.}. Since we are interested in non-trivial time-dependence as a diagnostic of TTSB, we will drop the ``connected" part which is the second term on the RHS of Eq.~\ref{eq:fabdef1} which only gives a static contribution.

Note that Eq.~\eqref{eq:f-spatiotemp} is a true \emph{many-body} diagnostic, probing correlations between distant constituents in a large system.  This is a stringent definition for time-crystals entailing long-range correlations in both space \emph{and} time. In the parlance of Ref.~ \mcite{CVS}, non-trivial time dependence in this limit defines a system with long-range \emph{spatio-temporal} order. Intuitively, a macroscopic many-body clock, should have all its constituents moving periodically in lockstep, satisfying Eq.~\eqref{eq:f-spatiotemp}. A related measure was discussed in the language of C* algebras in Ref.~\cite{ElsePrethermal}.

We note that while WO studied $f_{ab}$ in ground states and in equilibrium Gibbs states for time independent Hamiltonians, these can also be examined in \emph{individual} many-body highly excited eigenstates, which will be the more appropriate choice for MBL systems.  In thermalizing systems, we expect the two measures to show similar behavior\footnote{This is not guaranteed. Generalizing the ETH to unequal time spatio-temporal correlation functions is subtle in that  that it requires grappling with various unsolved issues pertaining to off-diagonal ETH~\cite{DAlessio:2016aa, ChalkerButterfly}.}. Finally, in the periodically driven setting, the appropriate eigenstates are those of the Floquet unitary $U_F$ defined in Eq.~\eqref{eq:Uf}, and the time should be probed stroboscopically $t=nT$. To obtain dTTSB in the Floquet case, one would need $c(nT)$ to display time dependence with a period that is a multiple of the driving period, corresponding to a lack of synchronization with the drive. 

A few matters deserve further exposition. 
\begin{itemize}
 \item First, we do not expect generic thermalizing many-body systems to display spatiotemporal order on quite general grounds, due to dephasing arguments similar to those considered in the discussion of the diagonal ensemble below Eq.~\ref{eq:Ot-diag}. 
 
 In detail, consider $f({\bf r},{\bf r'},t)$ evaluated in an eigenstate $|\nu \rangle$ of a time-independent Hamiltonian $H$ (the Gibbs measure works analogously): 
\begin{align}
    \langle \nu | f_{ab}({\bf r},{\bf r'},t)|\nu\rangle &= \langle \nu | a({\bf r},t) b({\bf r'},0) | \nu\rangle \nonumber  = \sum_\mu \langle \nu | e^{iHt} a({\bf r}) e^{-iHt}|\mu\rangle \langle \mu| b({\bf r'}) | \nu\rangle \nonumber \\
 &= \sum_{\mu: E_\mu =  E_\nu} \langle \nu |  a({\bf r}) |\mu\rangle \langle \mu| b({\bf r'}) | \nu\rangle + \sum_{\mu: E_\mu \neq E_\nu} e^{-i (E_\nu - E_\mu)t}\langle \nu |  a({\bf r}) |\mu\rangle \langle \mu| b({\bf r'}) | \nu\rangle,
    \mlabel{eq:fab-parsed}
\end{align}
where we have inserted a complete set of many-body eigenstates $\{|\mu\rangle\}$ with energy eigenvalues $\{E_\mu\}$.
For local operators $a({\mathbf r})$ and $b({\mathbf r'})$, we see that the sum over  eigenstates in the second term combines $O(\exp(V))$ phase factors set by the energy differences $\Delta_{\nu\mu}= E_\mu-E_\nu$. Our analysis now parallels our earlier discussion of the diagonal ensemble. The energy differences $\Delta_{nm}$ are generically incommensurate and continuously distributed (except for special degeneracies between eigenstates in different symmetry sectors in SSB phases, in the infinite size limit). For a small and finite $V$, the sum in Eq.~\eqref{eq:fab-parsed} is over finitely many frequencies, and hence $f_{ab}(t)$ will exhibit quasiperiodic revivals in time ---  again illustrating that any finite-size system generically exhibits TTSB. Thus, a non-trivial diagnostic of TTSB necessarily requires us to look for non-trivial temporal correlations \emph{after} the infinite volume limit has been taken first, as in Eq.~\eqref{eq:f-spatiotemp}. In this limit, the $\exp(V)$ incommensurate phase factors in the second term in Eq.~\eqref{eq:fab-parsed} will generically dephase and combine to give a late-time contribution which vanishes exponentially as $V \rightarrow \infty$. Thus, at late times in a large system, $\langle f_{ab}(t)\rangle $ is almost time-independent and dominated by the first term in Eq.~\eqref{eq:fab-parsed}, the block-diagonal ensemble. The same statements apply to Floquet systems upon considering Floquet eigenstates and stroboscopic times, as in Eq.~\ref{eq:Ot_diag_Floquet}.  

\item Second, we note that if we do not require locality for the operators $a$ and $b$, then it is very easy to get periodic time-dependence in $\langle \nu| f_{ab}(t)| \nu \rangle $. Specifically, if we pick $a = b^\dagger = |\nu\rangle \langle \mu_0|$ for some eigenstate $\mu_0$, then $\langle f_{ab}(t) \rangle$ displays periodic oscillations with frequency $\Delta_{\nu\mu_0} = (E_\nu-E_{\mu_0})$. Of course operators of the form $|\nu\rangle \langle \mu|$ are usually highly non-local in a MB system. But they direct our attention to the question of whether it might be possible to find {\it local} operators (or sums of local operators) that might---for time crystals---exhibit strictly periodic oscillations. The issue of how best to define a suitable order parameter for TTSB is a subtle and interesting question in its own right which was examined in Ref.~\mcite{TTSBRep} and furnishes yet a fourth diagnostic of time crystals. 

\item Third, the discussion above emphasizes that any system that displays TTSB according to Eq.~\eqref{eq:f-spatiotemp} must have special correlations in its eigenspectrum to prevent dephasing. We will see later that this is indeed the case in the driven MBL $\pi$-SG phase, where the eigenstates come in pairs with a quasienergy difference that is a rational fraction of the driving period, and local operators can efficiently connect different members of a pair. Even within thermalizing systems, there has recently been some interest in systems with ``many-body scars", which are special ETH violating eigenstates that are embedded within an otherwise thermalizing spectrum. These can have commensurate energy spacings relative to each other and display non-trivial temporal correlations. However, this behavior is not expected to be robust to general perturbations. Indeed, while the full range of dynamical possibilities is still a major open question, it is currently believed that the only \emph{robust} outcomes for closed MB systems are either many-body localization or full thermalization with no special structure in the eigenspectrum.

\item Fourth, we note that one can also naturally consider unequal time correlators involving global spatially averaged observables $A = \sum_{{\bf r}} a({\bf r})$, and $B = \sum_{{\bf r}} b({\bf r})$: 
\begin{align}
\langle f_{AB}({t}) \rangle &= \frac{1}{V^2} \langle  A(t)  B(0)\rangle = \frac{1}{V^2}\sum_{\bf r}\sum_{\bf r'} \langle  a({\bf r}, t)  b({\bf r'},0)\rangle. 
\mlabel{eq:fabdef2}
\end{align}
In this case, we average all terms in a finite volume to obtain a function of $t$ alone. Then, a TC is defined by the presence of late-time oscillations in the averaged correlator in the limit of infinite volume: $$\lim_{V \rightarrow \infty} \langle f_{AB}({t}) \rangle = c(t),$$ again with the same generalizations as before for the Floquet case. Now, it is possible that the spatial averaging in $f_{AB}$  introduces cancellations between different terms and thus hides interesting time dependence. 
However, ignoring this for the moment, it is intuitively clear that the spatially averaged correlator will generally be dominated by unequal time correlators between local observables at large separations, and hence show non-trivial time-depence when the diagnostic $f_{ab}(t)$ shows spatiotemporal order.

\item Finally, it is interesting to consider the behavior of $f_{ab}({\bf r}, {\bf r'}, t)$ along large temporal but fixed spatial separations, but still in the limit of infinite $V$. If $\lim_{V \rightarrow \infty} \langle f_{ab}({\bf r},{\bf r'},t)\rangle$ only shows non-trivial $t-$dependence for fixed $|{\bf r}-{\bf r'}|$ but arbitrarily late times, we will refer to this as strictly \emph{temporal} order. We note that WO do not consider the case of strictly temporal order on the grounds that it would classify ``trivial" systems such as a single oscillator (or a chain or independent, decoupled oscillators) as a time crystal. However, while it is true that finite-sized (or decoupled) systems will exhibit TTSB in this sense, thermalizing macroscopic systems with many-body interactions will {\it not},  due to the dephasing arguments outlined below Eq.~\eqref{eq:fab-parsed}. Thus, the definition for purely temporal order is still informative in the setting of infinitely large MB systems. Again, this can be considered for the Floquet case by evaluating the correlator in Floquet eigenstates and probing time stroboscopically.

\end{itemize}

\subsubsection{Susceptibility for TTSB}
\mlabel{sec:TTSBdef_susceptibility}
Next, we generalize the suceptibility diagnostic Eq.~\eqref{eq:ssbfield} discussed for ordinary SB to the case of TTSB in a many-body system. Once again, the object of interest is the finite-volume real time susceptibility
\begin{equation}
    \langle \chi_{ab}({\bf r},{\bf r;}t; V)\rangle  = \theta(t) \langle [a({\bf r},t), b({\bf r'},0)] \rangle \ ,
\end{equation}
whose Fourier transform is the complex susceptibility $\chi_{ab}({\bf r},{\bf r'},\omega; V)$. This describes the response as measured by the change in the expectation value $\langle a({\bf r},t) \rangle$ to switching on a perturbation $H_p = b({\bf r'},t') h({\bf r'},t')$.
As before, one can also consider a global susceptibility:
\begin{equation}
   \langle  \chi_{AB}(t; V) \rangle = \theta(t) \langle [A(t), B(0)] \rangle \ ,
\end{equation}
and in both cases expectation values are measured in either a Gibbs state, or an eigenstate of $H$ or $U_F$. 

The question of interest for a time-crystal is whether the response diverges for a \emph{periodic} perturbation $h(t)$ with frequency $\omega_0 >0$, again in the infinite size and infinite spatial separation limit. This corresponds to a divergence in $\chi_{\omega_0} \equiv\lim_{V\rightarrow \infty} \lim_{\omega \rightarrow \omega_0} \frac{1}{V^2} \langle \chi_{AB}(\omega;V)\rangle $, which measures the tendency of a static system to break into spontaneous oscillation at frequency $\omega_0$. For a Floquet system, the response would need to diverge at a frequency that is different from (and generally a fraction of) the driving frequency to obtain dTTSB: $\omega_0 \neq \omega$.  

For a thermalizing system, this will not be the case. We can work up an argument to this effect by starting with our considerations in Sec.~\ref{sec:TTSBdef_spatiotemporal} and noting that if the system is prepared in a Gibbs state, the fluctuation dissipation theorem relates the functions $\langle f_{ab}\rangle $ and $\langle \chi_{ab}\rangle $ via their Fourier transforms. By ETH we can also prepare such systems in an eigenstate and reach the same conclusion.

For potential time crystals we see that the spectrum will again need special features. For a time crystal to exhibit a singular response at $\omega_0$, eigenstates separated by  $\hbar \omega_0$ must exhibit some form of spectral pairing. We will see later in our discussion of the DTC how such a feature arises in the Floquet setting.

\subsubsection{Non-trivial late time evolution}
\mlabel{sec:TTSBdef_initstate}
Finally, we consider our last diagnostic for TTSB, which asks whether there is non-trivial temporal dependence in the dynamics of local operators at arbitrarily late times,  starting from out-of-equilibrium initial states $|\psi_0\rangle$:
\begin{align}
 \lim_{V\rightarrow \infty} \langle \psi_0| O(t) |\psi_0\rangle \xrightarrow[]{\text{late times}} c(t). 
\mlabel{eq:ttsb-latetime}
\end{align}
The limit of large system sizes is taken as before and by non-trivial we mean that $\lim_{t \rightarrow \infty} c(t)$ fails to exist. Because we are starting from a non-equilibrium initial state, even one-point expectation values can show time-dependence. This measure is analogous to the Peierls diagnostic for SSB presented earlier, wherein a symmetry breaking boundary field is sufficient to give a non-zero expectation value for an order parameter deep in the bulk of an infinitely large system. For the case of TTSB, the appropriate ``boundary" is the boundary of time at $t=0$, and a symmetry broken boundary condition corresponds to a non-equilibrium initial state. The question of interest for TTSB is whether the system can remain out-of-equilibrium even at arbitrarily late times; and in particular, for time-crystals, whether the out-of-equilibrium state at late times displays periodic oscillations. For experiments, this is clearly the relevant diagnostic. Again, for periodic systems, we would need $c(nT)$ to show oscillations in a DTC instead of being time-independent (synchronized).

There is however one immediate challenge in using this diagnostic---what class of initial states should we use? As ``Hilbert space is a big place''~\cite{Caves} we cannot speak with confidence to completely general choices. There are issues of preparability from the viewpoint of theory and from the viewpoint of experiment, and then there is the issue of correctly ``seeding'' the desired order. We will not attempt a careful discussion of the first question here, except to note that historically it has been useful to restrict attention to states that exhibit clustering of expectation values of local operators on the grounds that superpositions of macroscopically distinct states are unstable (Schrodinger cats are not commonly observed \footnote{We note that clustering does not imply low entanglement. Finite energy density eigenstates of local quantum systems simultaneously exhibit clustering and volume law entanglement.}). 
On the second question we note that, in the case of usual SB, one has a set of degenerate symmetry broken equilibrium states in the thermodynamic limit, and the goal is to select one of these states by appropriately choosing boundary conditions. One reasonable choice for the boundary condition is to match the desired expectation value of a local order parameter deep in the bulk of the non-invariant equilibrium state. In a putative time crystal, we would be trying to ``seed'' a periodic steady state which looks like a ``limit cycle". How might we do this by a choice of initial state?

Before commenting further on this let us consider the case of a thermalizing system. 
We return to the earlier discussion based on Eqn.~(\ref{eq:Ot-diag}). 
For a generic thermalizing system, the late time expectation values are governed by the time-independent block diagonal ensemble and the function $c(t)$ in Eq.~\eqref{eq:ttsb-latetime} is time independent. In other words, systems that thermalize, by definition, reach time independent late time steady states\footnote{It is possible to pick states where this approach holds for spatially local operators but not for general few body operators. The classic example is an initial state that leads to coarsening or phase ordering dynamics.}. 

We also note that the diagnostic above is readily generalized to the Floquet case, where observables are measured stroboscopically, again starting from an out-of-equilibrium initial state. This stroboscopic time series reaches a steady state value if the system synchronizes with the drive, which is the generic outcome expected in a thermalizing Floquet system -- while a TC stems from a lack of synchronization.  We will see in subsequent sections how Floquet MBL systems in a TC phase display special eigenspectrum correlations that can prevent dephasing and enable arbitrarily long-lived oscillations at multiples of the driving period.

Returning to TTSB, in all known examples of time-crystals, symmetry protected or otherwise, TTSB is accompanied by the breaking of an additional symmetry, either internal or spatial.  In all such cases, the late-time oscillations in the ``limit cycle" correspond to the system cycling between the different symmetry-broken equilibrium states of the additional symmetry. The states that appear in the cycle are related by the action of the symmetry. For example, in an XY magnet with U(1) symmetry, the precessing system cycles through the symmetry-broken states related by the action of rotation. 
Moreover, as we discuss later, in Floquet MBL TCs the additional symmetry is \emph{emergent}, the best studied example being that of an emergent $\mathbb{Z}_2$ Ising symmetry in a perturbed $\pi$-spin glass model. This model of a DTC shows period doubled dynamics, as the system cycles between a particular Ising symmetry broken configuration  
and its Ising reversed partner (Fig.~\ref{fig:perpetuum_mobile}(d)). Thus, in these known cases we need the initial states to break the additional/emergent symmetries in the system.  Since the MBL TC is a glass, in principle there are exponentially many non-invariant broken symmetry states corresponding (roughly) to random `up'-'down' orientations of the local magnetization on each site.

In any case, while such considerations are useful to bear in mind, we do not wish to be overly restrictive in specifying the ``right" choice of initial states, to leave room for other mechanisms for TTSB that may be discovered in the future. 

\subsection{TTS and MBL}

In the above discussion we have introduced three diagnostics for TTSB and commented on how they operate in thermalizing systems. We will now briefly review their application to static MBL systems. The relevant results are as follows:
\begin{enumerate}
    \item Correlations: There are now two interesting choices of time independent states in which we can evaluate the correlation function $f_{ab}({\bf r},{\bf r'},t)$ in Eq.~\ref{eq:fabdef1}. 
    
    First, we can work in individual many body eigenstates. As these eigenstates have a product description in the l-bit basis, local operators like $a({\bf r})$ and $b({\bf r'})$ primarily connect a given eigenstate to other eigenstates in which l-bits within a localization length of their location are disturbed. Thus a) their connected correlator decays exponentially with separation and b) if we keep the two locations spatially nearby, the temporal evolution exhibits response at a discrete set of generically incommensurate frequencies corresponding to the local fields and couplings in the l-bit Hamiltonian~\eqref{eq:lbit}. The weight of the Fourier response on these incommensurate frequencies dies off rapidly as the l-bits involved become distant from ${\bf r}$ and ${\bf r'}$. This was explicitly discussed in terms of l-bit operators in Section~\ref{sec:mbldynamics}. In this sense, all static MBL systems are \emph{time-glasses}, exhibiting purely temporal TTSB with several incommensurate local frequencies, but \emph{not} long-range spatiotemporal order.  In such systems, global correlators like Eq.~\ref{eq:fabdef2} are  suppressed by a factor of $V$ due to the lack of long-range spatial order.\footnote{More generally, following the work on velocity dependent Lyapunov exponents in \mcite{VDLE}, it would be interesting to examine the correlator along ``rays" defined by $|{\bf r}-{\bf r'}|=v t$ for a fixed $v$. We will not discuss matters in this generality here.} 
    
    Second, we can work in the Gibbs state. Now we end up averaging the answer in the earlier discussion over exponentially many nearby eigenstates and the result is time independent at long times, and no longer exhibits any glassiness.
  
    \item Susceptibility: If we begin with an eigenstate, the local susceptibility $\chi(\omega)$ is a sum of delta functions at the incommensurate frequencies discussed above. Essentially we are probing something close to the response of independent oscillators. In the Gibbs state we have an additional average over eigenstates and upon some minimum smoothing we lose the singular response.
  
    \item Late time dynamics: If the initial state is spread over a sufficiently large number of exact eigenstates, the late time state is captured by the diagonal ensemble as discussed in Sec.~(\ref{sec:mbl}), and we do not observe the glassiness. This will true for a generic short-range correlated initial state that is easily preparable in experiment. 
    In contrast, a non-interacting Anderson insulator \emph{does} show glassiness even in dynamics from general initial states, as discussed in Section~\ref{sec:mbldynamics}. 
\end{enumerate}

To summarize: MBL systems are temporal glasses in principle, but as a practical matter observing this requires preparing something close to an exact eigenstate. 

\subsection{No-go theorems and exceptions}
\mlabel{sec:nogo}
We now turn to Watanabe and Oshiwaka's no-go theorems that were argued to rule out time-crystals~\cite{Oshikawa15}. Of the three separate diagnostics for TCs we have presented, OW only examine the first, namely the possibility of long-range spatiotemporal order in unequal space-time correlation functions $f_{ab}({\bf r},{\bf r'},t)$, as defined in \eqref{eq:f-spatiotemp}. Moreover, in their analysis, these correlators are always evaluated in ground states or thermal Gibbs states.

WO's first theorem bounded the expectation value of $f_{AB}(t)$~ \eqref{eq:fabdef2}
\begin{equation}
    {1 \over V^2} | \langle 0| A(t) B(0) |0 \rangle - \langle 0| A(0) B(0)|0 \rangle | \leq C {t \over V}
    \mlabel{eq:OWgs}
\end{equation}
in any exact ground state.  Assuming the LHS has a limit as $V \rightarrow \infty$, it must vanish. From this, WO conclude that time-independent Hamiltonian systems do not exhibit spatio-temporal order in their ground states in the infinite volume limit, thus excluding Wilczek's original scenario. This result also suggests that we can ``unbundle'' \eqref{eq:OWgs} to conclude that $f_{ab}({\bf r},{\bf r'},t)$ factors to $\langle a({\bf r},0) \rangle \langle b({\bf r'},0) \rangle$ as $|{\bf r}-{\bf r'}| \rightarrow \infty$. However OW's technique itself cannot be directly applied to get this result.

WO's second theorem attempted to generalize their result to finite temperature Gibbs states to show that $\langle f_{AB}(t)\rangle $ at late times is a constant that does not depend on $t$. However, their proof relies on a different technique at finite temperature. At finite temperature, their proof uses only two ingredients: (i) the fluctuation dissipation theorem, which relates $\chi_{AB}(\omega)$ to $f_{AB}(\omega)$ evaluated in Gibbs states, and (ii) locality and the Lieb-Robinson bound, which ensures that $\lim_{V\rightarrow \infty} \chi_{AB}(t;V) = 0$ as long as $t=o(V^{1/d})$. Note that, surprisingly, they do \emph{not} use any aspect of the thermalization properties (or lack thereof) of the system. 
However, as we discuss below, their proof has a subtle error so that a finite temperature time crystal is not actually ruled out by their arguments. 

More generally, there are various interesting exceptions to WO's results. These do \emph{not} rule out:

\begin{itemize}
    \item Strictly temporal (as opposed to spatiotemporal) order as we noted in our discussion of many body localized systems above. As discussed previously, temporal order refers to non-trivial time dependence in $f_{ab}({\bf r},{\bf r'},t)$ in the limit $V\rightarrow \infty$, but at finite $|{\bf r}-{\bf r'}|$. While we do not find a time crystal in this setting, we do find time glasses.
    
    \item Symmetry protected time-crystals. As discussed previously, one can get macroscopic long-range spatiotemporal order in systems endowed with an additional symmetry, when the additional symmetry is spontaneously broken. 
    For example, consider a system endowed with an additional $U(1)$ symmetry corresponding to particle number conservation $N$. We consider equilibrium states in the grand canonical ensemble at a particular temperature $T$ and chemical potential $\mu$: $\rho_{GC} \propto e^{-\beta (H-\mu N)}$. Operators which change the particle number $N$ (such as the raising operator $a^\dagger$) will ``trivially" display oscillations at a frequency set by $\mu$. A condensate which spontaneously breaks $U(1)$  conservation of $N$ below a critical temperature $T_c$ will have a non-zero expectation value for $\langle a^\dagger \rangle$, and hence states below $T_c$ will show oscillations in $\langle a^\dagger(t) \rangle$ with frequency $\mu$. OW do consider this example, but exclude it on the grounds that such a time dependence can never actually be measured. In essence, this is akin to measuring the phase of a condensate with broken $U(1)$ symmetry (which is known to have an oscillatory time dependence), but this cannot be measured~\cite{Volovik}. On the other hand, the condensate can be coupled it to a different one and the relative phase difference can be measured, $\mu_1 -\mu_2$ --- this is simply the AC Josephson effect. Closely related ideas have been considered in proposals for realizing TCs in superfluid systems~\cite{WilczekSF, Volovik, Volovik_timeqc, Svistunov_tc, vanderstraten}. 
    
    We note that the mathematics of the single condensate problem looks identical if one instead considers the spontaneous breaking of an internal $U(1)$ spin symmetry, such as in an $XY$ magnet. And here the oscillations can be measured! For concreteness, consider a model of spins interacting with nearest neighbor exchange interactions in three dimensions:  $$H_{XY} = \sum_{\langle ij \rangle} J ( S_i^x S_j^x + S_i^y S_j^y + \Delta S_i^z S_j^z).$$ This model conserves $S^z_{tot}$, and has a symmetry broken phase below a critical temperature $T_c$. In the SB phase, the magnetization vector points along a fixed direction in the $XY$ plane, and an order parameter for the SSB is provided by $S_{tot}^{+} = S^x_{tot} + i S^y_{tot}$. This order parameter exhibits Larmor precession in the presence of an external magnetic field in the $Z$ direction: $e^{i ht S_{tot}^z } S_{tot}^+ e^{- i ht S_{tot}^z} = e^{i h t} S_{tot}^+$. 
    Thus, if we prepare a symmetry-broken initial state $|\psi_0\rangle$ at a temperature below $T_c$ and consider the time-evolution of $S_{tot}^{+}(t)$ in the presence of an external magnetic field,  we get:
    \begin{align}
     \langle \psi_0| e^{i t (H_{XY} + h S_{tot}^z)} S_{tot}^+ e^{i t (H_{XY} + h S_{tot}^z)} |\psi_0\rangle 
     = e^{i h t} \langle\psi_0| e^{i t H_{XY}} S_{tot}^+ e^{i t H_{XY}} |\psi_0\rangle \xrightarrow{\text{late times}} c_0 e^{iht}.
     \nonumber \\
    \end{align}
    In the first step, we have used the the Larmor precession of $S_{tot}^+$ and the fact that $[S_{tot}^z, H_{XY}]=0$, while the second step reflects the saturation of the order parameter to a non-zero value $c_0$ for a symmetry broken initial state prepared below $T_c$ in a thermalizing system. Thus, the field $h$ plays the role of the chemical potential $\mu$ in the condensate example, but the oscillations here are directly measurable unlike the phase of the condensate. In fact, such oscillations were reported in NMR experiments several decades ago~\cite{urbina}.
    
    While the calculation above considers the dynamics of a symmetry broken initial state, we can equally well consider two point expectation values $\langle S_{tot}^+(t) S_{tot}^-(0)\rangle$ in symmetric equilibrium states below $T_c$, and these again show oscillations in the presence of a magnetic field $h$. 
    
    Finally, as emphasized several times, a symmetry broken state is strictly only in equilibrium in the limit of infinite size. Any finite size system will eventually tunnel between different symmetry broken states destroying the oscillations discussed above -- but this will take a time parametrically long in the system size,  which could be much longer than experimentally accessible coherence times in realistic experiments. Likewise, one might argue that the preparation of a symmetry broken state requires the addition of weak \emph{explicit} symmetry-breaking terms in the Hamiltonian, which destroys the symmetry protection that we rely on above. However,
    as discussed in Sec.~\ref{sec:prethermalTC}, such systems still display an approximate ``prethermal" $U(1)$ symmetry for exponentially long times in the applied field $h$, and hence may again display oscillations for lifetimes exceeding experimental coherence times~\cite{ElsePrethermal}.

    \item Analogs of coherent states with an energy strictly above that of the ground state can exhibit periodic expectation values. Indeed, we have already exhibited such states in our discussion of SPTCs above. They are not ruled out by the WO theorem. There is also the recently discovered phenomenon of ``quantum scars'' which are set of highly excited states in special fine-tuned Hamiltonians. Initial states with large overlap on the scar states can display macroscopic oscillations. While these are even more non-generic, they have the virtue of apparently not requiring a protecting symmetry. These cases display time-crystal-like behavior according to our third definition of TCs, which examines dynamics from out of equilibrium initial states (Sec.~\ref{sec:TTSBdef_initstate}) --- but only for a special class of atypical initial states and for special fine-tuned Hamiltonians. 
    
    \item Finally, WO's arguments do not rule out spatio-temporal order in Gibbs states. While this forms the crux of WO's results, the reason it is not excluded is because there is a subtle error in the $T>0$ proof in their paper. WO use the fluctuation dissipation theorem to relate $\chi_{AB}(\omega)$ and $f_{AB}(\omega)$ evaluated in Gibbs states, and try to argue that $f_{AB}(\omega)$ has no response at any non-zero frequency in the infinite volume limit. Their argument uses the Lieb-Robinson theorem, which guarantees that $\chi_{AB}(t)$ decays to zero in the limit of large $V$ as long as $t = o(V^{1/d})$. 
    
    However, in passing from $\chi_{AB}(t)$ to $\chi_{AB}(\omega)$, they assume that they can interchange the time integration needed to obtain the Fourier transform with the large spatial separation limit. \emph{However, the interchange of limits they perform is only valid if one assumes a lack of temporal order i.e.} they implicitly \emph{assume} that the susceptibility $\chi_{ab}(t)$ decays sufficiently rapidly at late times for operators at finite spatial separations (see Appendix~\ref{sec:OW} for more details). Their assumption is intimately related (though not identical)  to the result they're trying to prove in the first place, namely the lack of spatiotemporal order.   Indeed, our discussion below Eq.~\ref{eq:fab-parsed} discusses how dephasing in a many-body system generically destroys \emph{both} temporal and spatiotemporal order.    
    
    Since the input into WO's argument is \emph{only} the Lieb-Robinson bound on spatially well separated commutators, they do not have enough information on the time dynamics or thermalization properties of the system to justify this step. 
    
    Relatedly, while the passage from $\chi(t)$ to $f(t)$ is problematic in WO's paper, one might wonder whether their results are sufficient to directly give a bound on $\chi_{AB}(t)$ which, as discussed in Sec.~\ref{sec:TTSBdef_susceptibility}, is a fine diagnostic for TTSB in its own right. However, the Lieb-Robinson theorem they rely on only bounds $\chi_{AB}(t)$ for times that scale polynomially with system size, $t = o(V^{1/d})$. On the other hand, obtaining the possibility of a diverging susceptibility in $\chi(\omega)$ requires one to consider all times $t>0$. Indeed, even in the case of standard breaking of an Ising symmetry, the diverging susceptibility at $\omega \rightarrow 0$ follows from the behavior of $\chi(t)$ at times $O(\exp(V))$. One needs to wait until such times to probe the tunneling between different symmetry broken states, the mechanism that is ultimately responsible for the divergent susceptibility.
    
    Of course, none of this is to say that the considerations from the Lieb Robinson theorem in WO's paper are wrong --- simply that such considerations are not sufficient to make statements about the susceptibility of the system towards spontaneous oscillation. In retrospect, this is perhaps not surprising since the Lieb-Robinson theorem only assumes locality but no other information about the system at hand. Naively, one would expect to need at least some information about the thermalization properties of the system to rule out TCs.     
    
    That said, WO's conclusion of the absence of TCs at finite temperature is likely correct anyway, as ergodicity/thermalization is a much stronger constraint on the search for time crystals. This was apparent, for example, in the presentation of the dephasing arguments in the prior subsections. 
    
    Note also that if we wish to bound the time-dependence of the correlator $f_{ab}(x,y,t) = \langle a(x,t) b(y,0)\rangle $ in the limit where the spatial separation $|x-y|$ is taken to infinity first, followed by the late time limit, as in Eq.~\eqref{eq:f-spatiotemp}, then one only needs to assume that correlations cluster in the Gibbs state \emph{i.e.} $\lim_{|x-y|\rightarrow\infty} \langle a(x) b(y)\rangle \rightarrow \langle a(x) \rangle \langle b(y)\rangle$. This is because $a(x,t)$ viewed in the Heisenberg picture has support in some finite ball of radius $v_{LR} t$ about position $x$, where $v_{LR}$ is the Lieb-Robinson or butterfly speed. If we take the $|x-y| \rightarrow \infty$ limit for finite $t$, then the support of $a(x,t)$ is roughly disjoint from that of $b(y,0)$. In that case, $\langle a(x,t) b(y,0)\rangle \rightarrow \langle a(x,t)\rangle \langle b(y,0)\rangle = \langle a(x,0)\rangle \langle b(y,0)\rangle$. The first step follows from clustering, while the second relies on the time-independence of equal time correlators in Gibbs states. Indeed, there are various results proving the clustering of correlations in Gibbs states at high enough temperatures, away from phase transitions and/or symmetry breaking. Thus, at the least, the absence of spatiotemporal order in Gibbs states in all such cases follows quite directly.

        \item And, of course, anything to do with Floquet MBL systems, which are entirely outside the purview of their proof.
    
\end{itemize}

To summarize the state of the play at the end of this section: static Hamiltonian systems without additional symmetries do not appear to allow for time crystal; at best they allow for time glasses. This follows from combining the second law for finite energy densities above the ground state, the WO theorem for ground states, and the explicit example of MBL systems. All of this requires us to look even further afield---to Floquet systems---and that is what we do next.


 \section{Floquet Many-Body Localized Time Crystals} 
 \mlabel{sec:FMBLTC}
 
We have now assembled all the ingredients needed to present the central actors in our story---the new phases of Floquet MBL systems that exhibit TTSB. As mentioned earlier, these phases reflect a new paradigm in quantum statistical mechanics, that of a many-body Floquet phase.  These phases are most crisply defined via the notion of eigenstate order --- now generalized to the eigenstates of the Floquet unitary ---  which creates sharp distinctions between different parameter regimes. However, there is sufficient universality in late time states so that different phases can be distinguished by ``seeding'' the time evolution with selected families of experimentally preparable initial states. This work builds on the discovery of topologically non-trivial Floquet single 
particle systems and recent advances in their classification \cite{Thouless:1983hb,Oka:2009kc,Jiang:2011cw,Kitagawa:2011fj,Lindner:2011ip,kitagawa2010topological,rudner2013anomalous, titum2015disorder, titum2015anomalous, Nathan:2015tk}. However, analogous to the clean, non-interacting Ising chain of Section~\ref{sec:MBLorder}, the system can display trivial many-body dynamics even for non-trivial single particle drives\footnote{For example, the non-trivial `topological' phase of a clean, non-interacting Kitaev chain/p-wave superconductor is related by duality to the ferromagnetic  phase of a clean transverse field Ising model. Nevertheless, the many-body spin states at any finite temperature are delocalized and do not show long-range order.}. Thus, the full framework of 
disorder and interactions is required for the many-body problem.

In the following we  discuss these ideas in the context of the simplest setting---that of a driven/Floquet Ising chain
---which will lead us to the DTC whose properties we  then discuss at some length. Readers interested in the more general universe of Floquet phases may wish to consult a recent review~\cite{SondhiFloqReview} in addition.

Since we will discuss a number of developments as our presentation of the DTC phase unfolds, it is useful to present a concrete and generic model for a DTC here, along with a road map for how we will arrive at this model in the course of this section. 
Consider a chain of spin 1/2's, with the Floquet drive from Ref.~\cite{CVS}: 
\begin{align}
    U(T) &= \exp\left[{-i g \sum_j \sigma_j^x}\right]\exp\left[-iT H_{\rm int}\right]\nonumber\\
    H_{\rm int}& = \sum_j J_{j} \sigma_j^z \sigma_{j+1}^z + h_j^x \sigma_j^x + h_j^y \sigma_j^y + h_j^z \sigma_j^z,
\mlabel{eq:TCgeneral}
\end{align}
where an ``interaction" Hamiltonian $H_{\rm int}$ acts for a time $T$, followed by an instantaneous rotation of all spins about the $x$ axis by an angle $2g$. Consider various limits: 
\begin{itemize}
    \item When $J_j=0$, the system represents a chain of decoupled spins acted upon by local fields. While individual spins can show Rabi oscillations at frequencies set by the local fields, such a system lacks the long-range spatiotemporal order that is required of a TC (\emph{cf.} Eq.~\eqref{eq:f-spatiotemp}). This is consistent with our prior statements: obtaining a TC requires us to consider a genuinely interacting many-body system. 
    
    \item When $h_i^x=h_i^y=0$ and $g=\pi/2$, a product state prepared in the $z$ basis picks up a phase under evolution by $H_{\rm int}$, but then gets exactly flipped ($\sigma^z \rightarrow -\sigma^z$) under the action of the $\pi$ rotation about the $x$ axis. Thus, the system trivially shows period doubling for this choice of parameters since the state returns to its itself (upto a phase) every two periods. 
    
    Additionally, it is easy to check (and explained below) that the eigenstates of the Floquet unitary for this set of parameters look like global superposition or ``Schrodinger cat" states between a given $z$ product state and its flipped partner. These display long-range order in unequal spacetime correlation functions, as desired in a TC phase (Eq.~\eqref{eq:f-spatiotemp}). 
    
    \item When $h_i^y = h_i^z = 0$, the system has $\mathbb{Z}_2$ Ising symmetry, $P_x = \prod_i \sigma_i^x.$ and can be mapped to a model of non-interacting fermions
\end{itemize}
We show below that the spatiotemporal order apparent for the parameters in the second bullet above survives weak perturbations, so that there is an extended and generic TC \emph{phase} in parameter space that displays both long range order in space and symmetry breaking in time. We further show that this phase is an extension of the so-called ``$\pi$ spin-glass" ($\pi$SG) phase of a periodically driven Ising model, discussed in Ref.~\cite{KhemaniPRL} and below. The spatiotemporal order in the $\pi$SG comes from a combination of (i) long-range spatial order due to Ising symmetry breaking and (ii) period doubling dynamics of the Ising order parameter in time. Remarkably, the $\pi$SG survives \emph{all} weak perturbations of the drive~\cite{ElsePRL, CVS} including, for example, those that explicitly break the Ising symmetry, as in Eq.~\eqref{eq:TCgeneral}, or those that enact 
imperfect rotations by angles that are not exactly $\pi$\footnote{The perturbations must, however, respect the periodicity of the drive}. The perturbed phase nevertheless displays LRO and oscillations that are locked to twice the period. 
The underlying reason is similar to why MBL is stable as a phenomenon: the perturbed system displays an \emph{emergent} Ising symmetry, which is then broken in the TC phase to give both long-range order and period doubling dynamics for the emergent order parameter. In other words, the phase continues to be described as a $\pi$SG with respect to the emergent Ising symmetry. 

In the subsections below, we start by discussing the phase diagram of the simplest and most natural driven Ising model which shows three other phases in addition to the $\pi$SG. We then explain the stability of the $\pi$SG phase, emphasizing the role of emergent symmetries. This, in turn, allows us to demonstrate how the perturbed $\pi$SG  explicitly displays all the diagnostics of a TC phase discussed in Sec.~\ref{sec:DefineTTSB}. We will end with a set of generalizations of the $\pi$SG.

\begin{figure}[h]
\centering
\includegraphics[width=0.40\columnwidth]{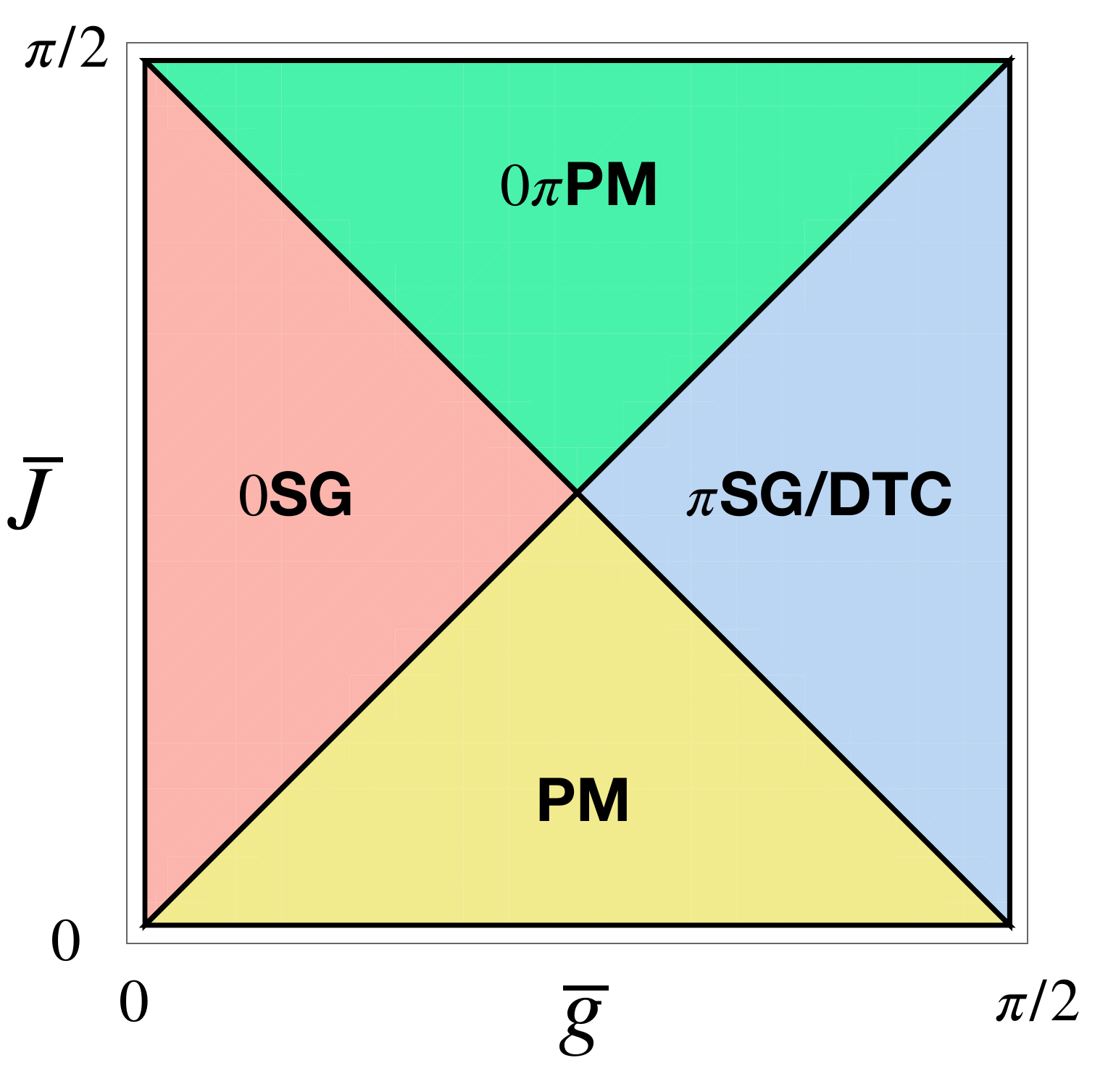}
\caption{Phase diagram for the non-interacting version of the binary drive of Eq.~\eqref{eq:drivenIsing}, with transverse field $\overline{g}$ and exchange $\overline{J}$ as tuning parameters. The phases are the paramagnet (PM), the spin glass (0SG), the symmetry-protected topological paramagnet ($0\pi$PM) and the $\pi$SG/discrete time crystal. Both SG phases break Ising symmetry and show LRO, while the $\pi$SG additionally also shows period doubling and furnishes a realization of a spatiotemporally ordered time-crystal.  
Transverse field $\overline{g}=\pi/2$ corresponds to an Ising spin flip. } 
\mlabel{fig:DrivenIsing}
\end{figure}   
 
\subsection{Phases in a driven Ising chain}

We begin with a one-dimensional driven Ising model~\mcite{KhemaniPRL,vonKeyserlingk2016a} with a binary drive that alternates between the Hamiltonians $H_x$ and $H_z$ every period, and $H_{x,z}$ are picked analogous to the static Ising model in Eq.~\eqref{eq:ising}:
\begin{align}
    U_F &= e^{-iH_x} e^{-iH_z} \nonumber \\
    H_x &= \sum_i g_i \sigma_i^x + \sum_i J^x_{\rm int}\sigma_i^x\sigma_{i+1}^x \nonumber \equiv H_0^x + H_{\rm int}^x \\
    H_z &= \sum_i J_i \sigma_i^z\sigma_{i+1}^z + \sum_iJ^z_{\rm int} \sigma_i^z \sigma_{i+2}^z \equiv H_0^z + H_{\rm int}^z.
    \mlabel{eq:drivenIsing}
\end{align}
Both $H_x$ and $H_z$ commute with a $\mathbb{Z}_2$ global Ising symmetry generator $P_x=\prod_i \sigma_i^x$, and are related to each other by Ising duality when $J^x_{\rm int}=J^z_{\rm int}$. We have separated the Hamiltonians $H_{x/z}$ into non-interacting ($H^{x/z}_0$) and interacting pieces ($H^{x/z}_{\rm int}$) respectively.  The couplings $g_i$, $J_i$ are chosen randomly to produce localization, and have means $\overline{g}$ and $\overline{J}$ respectively. This model was studied in Ref.~\mcite{KhemaniPRL} and shown to have four dynamical phases. Two, the paramagnet (PM) and 0-spin-glass (0SG), are analogs of the undriven MBL Ising phases found earlier in Sec.~\ref{sec:MBLorder} and are denoted `Type I' phases. And two new `Type II' phases, the $0\pi$PM and the $\pi$SG/DTC, are novel to the driven setting. Both spin-glass phases spontaneously break Ising symmetry and show long-range order, while the $\pi$SG additionally also exhibits TTSB. The $0\pi$ PM does not exhibit LRO or spatial symmetry breaking, but exhibits TTSB on the \emph{boundaries} of the system. Our choice of nomenclature will become apparent in the discussion of these phases below.  
Our treatment will focus on the three features of eigensystem order discussed in Sec.~\ref{sec:MBLorder}, namely (i) eigenstate properties, (ii) eigenspectrum properties and (iii) measurable dynamical properties.

\subsubsection{Non-interacting limit}
When $J^x_{\rm int} = J^z_{\rm int} = 0$, the drive alternates between the two terms of the canonical transverse field Ising chain. Much as the latter is solvable by reduction to free fermions, so is this drive~\cite{Bastidas2012, Thakurathi:2013dt}. Especially simple phase boundaries are obtained in the free limit~\cite{KhemaniPRL}, in which the system is Anderson localized for arbitrarily weak disorder in the couplings. These are depicted in the non-interacting phase diagram in Fig.~\ref{fig:DrivenIsing}a), from~\cite{KhemaniPRL}.  We begin by taking an initial tour of this phase diagram along its outer boundary, which allows us to describe the phases that populate it in particularly transparent limits. The subsequent subsections perturb away from these limits along various axes, for example by adding interactions and/or Ising symmetry breaking couplings. 
We will not explicitly use the mapping to free fermions, focusing instead on the generic features of this model which survive integrability breaking.
Indeed, while the eigensystem properties look qualitatively similar between the interacting and non-interacting cases, the lack of dephasing in the free case makes the dynamics quite different. Hence, we begin by discussing the eigensystem properties, deferring a discussion of dynamics to the interacting case.

 \subsubsubsection{Paramagnet} The first phase we consider is a driven analog of the localized paramagnet discussed in Sec.~\ref{sec:MBLorder}. This phase has no spontaneous symmetry breaking, and there is no long-range order in any Floquet eigenstate.

We start along the lower horizontal boundary in Fig.~\ref{fig:DrivenIsing}, setting $J_i = 0$. 
The Floquet unitary is 
\begin{equation}
    U_0^{PM} \equiv U_F  = e^{-iH_0^x},
    \mlabel{eq:U0pm}
\end{equation} 
and the eigenstates of $U_F$ are simply the eigenstates of the local Hamiltonian $H_0^x$, which is the Floquet Hamiltonian for this drive (\emph{i.e.} the logarithm of $U_F$). The properties of $H_0^x$ were discussed at length in Section~\ref{sec:MBLorder}. The eigenstates look like product states in the $\sigma^x$ basis, $|\{ s_i^x\}_\nu\rangle $ , with no long-range order, and are hence deserving of the appellation paramagnetic. There are no special pairings in the energy spectrum. The quasienergies are $E_\nu = (\sum_i g_i s_i^x) \; \rm{mod}\; 2\pi$, which are arranged randomly on a circle of radius $2\pi$, with generically no degeneracies. The quasienergies also do not exhibit any level repulsion --- as is appropriate for a localized system with $L$ conserved operators, the l-bits $\{\sigma_i^x\}$.

On perturbing away from the $J_i=0$ line,
the fundamental property of this phase---the lack of $\mathbb{Z}_2$ eigenstate order---extends to the entire region marked paramagnetic in Fig.~\ref{fig:DrivenIsing} in the free model. Away from the $J_i=0$ line, the l-bits get weakly dressed and pick up exponentially decaying tails as discussed in Sec.~\ref{sec:mbl}:  $\tau_i^x = \mathcal{V}^\dagger \sigma_i^x\mathcal{V}$ for a local unitary $\mathcal{V}$. 

\subsubsubsection{0-Spin Glass:} Next, we consider a driven analog of the localized spin-glass phase from Sec.~\ref{sec:MBLorder}, which exhibits spontaneous Ising symmetry breaking and bulk long-range order. The entire Floquet eigenspectrum comprises exponentially degenerate pairs of (cat) states that are even/odd under the Ising symmetry. These look like superpositions of classical low entanglement states with spin glass order, and their Ising reversed counterparts.

Along the left vertical boundary, we set $g_i=0$ and the single period unitary is 
\begin{equation}
  U_0^{0SG}\equiv U_F = e^{-iH_0^z},  
  \mlabel{eq:U0sg}
\end{equation}
corresponding to the Floquet Hamiltonian $H_0^z$, discussed in Sec.~\ref{sec:MBLorder}. The $\mathbb{Z}_2$ symmetric eigenstates of $U_F$ (or equivalently, $H^z_0$) come in degenerate pairs, $|\nu_\pm\rangle = |\{d_i\}_\nu, p_\nu=\pm1\rangle$, as in Eq.~\eqref{eq:catstates}, labeled by the eigenvalues of $(L-1)$ domain-wall operators, $D_i = \sigma_i^z\sigma_{i+1}^z$, with eigenvalues $\{d_{i}\}_\nu$, and the parity $P_x$ with eigenvalue $p_\nu$. The phase spontaneously breaks Ising symmetry and shows glassy LRO in two point correlators as in Eq.~\ref{eq:sgcorr}. 

Turning to the eigenspectrum properties, we see the even and odd parity cat states are strictly degenerate along the line $g_i=0$, corresponding to quasienergies $E_\nu^\pm = (\sum_i J_i d_i) \mbox{ mod } 2\pi.$ In other words, the quasienergies \emph{within} any given parity sector are randomly distributed on a circle of radius $2\pi$ (with no level repulsion), but the entire spectrum is \emph{paired}. In the dual free-fermionic description, the $0$SG phase is a topological phase with a Majorana zero mode which is responsible for the degeneracy between the Ising even/odd states in the many-body spin language. 
The degeneracy between the cats means that we can equally well pick symmetry broken eigenstates that are the $z$ product states, $|\{s_i^z\}_\nu\rangle$, with non-zero expectation values for local Ising order parameters $\sigma_i^z$.

Putting all of this together, the term $0$SG thus describes the combined eigensystem properties of this phase:  symmetry breaking \emph{glassy} order in the eigenstates, and spectral pairing with \emph{zero} quasienergy difference between the paired cat states.

On perturbing away from the $g_i=0$ line by adding transverse fields, the fundamental properties of this phase---the breaking of the $\mathbb{Z}_2$ symmetry and the associated doublet structure of the eigenspectrum---continue to hold throughout the region labeled $0$SG 
in Fig.~\ref{fig:DrivenIsing} in the free model. However, in the perturbed model, the paired cat states  have an energy splitting that is exponentially small in $L$ rather than strictly vanishing.
As before, the l-bits (domain wall operators) get dressed to pick up exponential tails:  $\tilde{D}_i^x = \mathcal{V}^\dagger D_i^x\mathcal{V}$.

To summarize, thus far we have discussed two phases, the PM and the $0$SG, which can be sharply distinguished by the absence or presence of $\mathbb{Z}_2$ eigenstate order. These `Type I' phases are both continuations of phases already present in static systems---what is new is that Floquet localization allows the eigenstate order/disorder distinction to persist in the driven non-equilibrium setting. Now we turn to the two remaining phases which are genuinely new to the Floquet setting with no static analog---denoted as `Type II Floquet phases'.

\subsubsubsection{ $\pi$ Spin Glass/Time Crystal:} Next, we turn to the first novel driven phase -- the protagonist of our story -- the spatiotemporally ordered $\pi$SG phase, which spontaneously breaks both the Ising symmetry and TTS. This phase is closely related to the $0$SG, but with one crucial difference. 

Along the right vertical boundary we set $g_i= \pi/2$. Then, using the fact that $e^{-i\frac{\pi}{2}\sigma_j^x}= -i \sigma_j^x$, we see that 
$$
\exp \left[-i \frac{\pi}{2} \sum_j\sigma_j^x\right] = (-i)^L \prod_j \sigma_j^x \propto P_x. 
$$
Thus, up to a factor of $i$ for odd length chains, the Floquet unitary is  
\begin{equation}U_0^{\pi SG} \equiv U_F = P_x e^{-iH_z}.
\mlabel{eq:U0sg2}
\end{equation} 
Since $[P_x, U_F]=0$, the eigenstates of this unitary are the \emph{same} as the $\mathbb{Z}_2$ symmetric eigenstates of the $0$SG. These were 
 worked out by noting that all the domain wall operators $D_i \equiv \sigma^z_i \sigma^z_{i+1}$ commute with $P_x$,  $U_{F}$ and with one another and form the ``l-bits". Thus, the eigenstates look like the even/odd Schrodinger cat states in Eq.~\ref{eq:catstates}, $|\{d_i\}_\nu, p_\nu\rangle $. The non-clustering Floquet eigenstates exhibit long range connected correlations and symmetry breaking spin glass (SG) order in $\sigma^z_{i}$ via the non-vanishing two point functions with random (glassy) sign in each eigenstate,  Eq.~\eqref{eq:sgcorr}.

The eigenstates $|\{d_i\}_\nu, p_\nu\rangle $ have corresponding unitary eigenvalues $u(d, p) =  p e^{-i\sum_{r=1}^{L-1}J_{r}d_{r}}$.  Note that the presence of the $P_x$ term in the unitary means that the opposite parity cat-state partners now have unitary eigenvalues differing by a minus sign $u(d,-1)=-u(d,-1)$ and hence quasienergies differing by $\pi$.  We refer to this phenomenon as a $\pi$ spectral pairing of cat states. 
Due to this splitting, one is {\it not} free to pick 
linear combinations of the cats that break the $\mathbb{Z}_2$ symmetry explicitly. In the dual free-fermionic description, the $\pi$SG phase is a topological phase with a $\pi$ Majorana zero mode responsible for the $\pi$ spectral pairing between the Ising even/odd states in the many-body spin language~\footnote{A Majorana mode at quasienergy $\epsilon$ must satisfy $\epsilon$ = -$\epsilon$. Because of the $2\pi$ periodicity of the quasienergy spectrum, this permits Majorana modes at both zero and $\pi$ quasienergy~\cite{Jiang:2011cw}.}.

Along this line in parameter space, it is particularly transparent to see why the $\pi$SG furnishes a realization of a discrete TC. If one starts with a product state in the $\sigma^z$ basis, $|\psi_0\rangle = |\{s_i^z\}\rangle$, it picks up a phase along the part of the drive that evolves with $H_z$, but is exactly flipped to its Ising reversed partner $|\{-s_i^z\}\rangle$ under the action of $P$, thereby showing oscillations with a period $2T$ (Fig.~\ref{fig:perpetuum_mobile}(d)).  Thus, while $\langle \sigma^z_i (nT) \rangle =0$ in the Floquet eigenstates, the observable shows a periodic time dependence with period $2T$ starting short-range correlated product states of the form $|\psi_0\rangle =  |\{s_i^z\} \rangle \sim |\{d_i\},+ \rangle + | \{d_i\},- \rangle$. The $\pi$ quasienergy splitting between the even/odd parity cat states is essential for obtaining period doubled dynamics and preventing dephasing of the off-diagonal matrix elements between non-degenerate eigenstates in the expansion of $\langle\psi_0|\sigma_i^z(nT)|\psi_0(nT)\rangle$ in Eq.~\ref{eq:Ot_diag_Floquet}.

The period doubling can also be seen directly by noting that the Ising order parameter in $\pi$SG oscillates with frequency $\pi$ or period $2T$, as indicated by the stroboscopic equation of motion $\sigma^z_r(nT)=(-1)^n \sigma^z_r$\mcite{KhemaniPRL,CVS}. This follows from the fact that $\sigma^z_r$ anticommutes with $U_{F}$.

The $\pi$SG is then characterized by this combination of state dependent long range $\mathbb{Z}_2$ order, and the presence of $\pi$ doublets in the spectrum. Indeed, the moniker $\pi$SG is meant to highlight the complementary spatiotemporal order inherant in this phase: the SG order in space, and period doubling in time resulting from the $\pi$ spectral pairing.  

While the period doubling and time-crystalline nature of the $\pi$SG phase is particularly transparent along the $g=\pi/2$ line, the reader might worry that this model is too trivial a rendition of a TC -- after all, we're explicitly and precisely flipping all spins ever period, so there is no surprise there is period doubling! \emph{However, the striking feature is that the fundamental properties of this phase --- the $\mathbb{Z}_2$ symmetry breaking, and $\pi$ spectral pairing --- persist throughout the entire region labeled $\pi$SG in Fig.~\ref{fig:DrivenIsing}.} The perturbed system is still characterized by l-bits that are domain wall operators, but with exponential tails, as for the $0$SG.

\begin{figure}
    \centering
    \includegraphics[width=0.9\columnwidth]{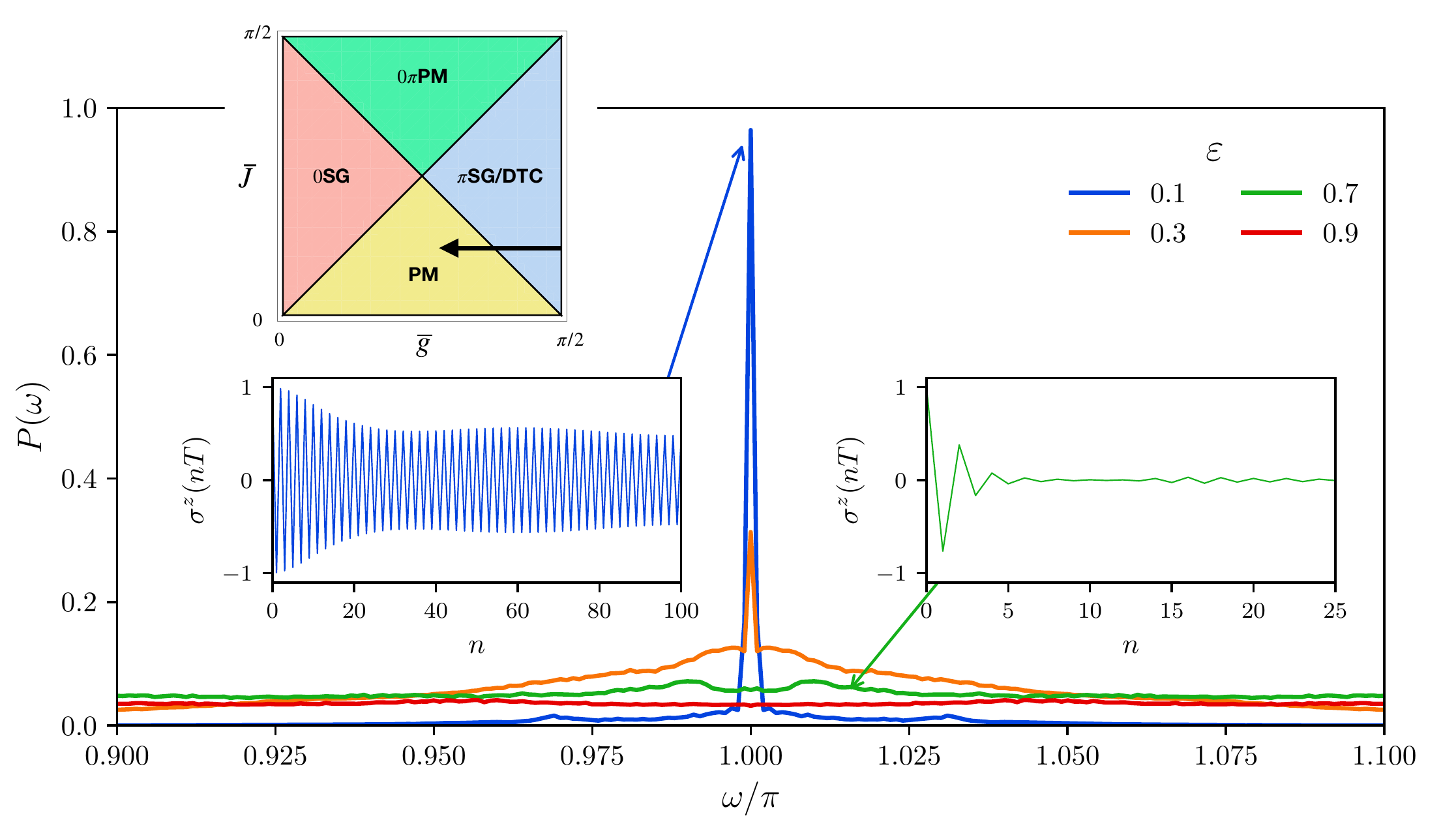}
    \caption{$\pi$SG in the presence of imperfect Ising flip, parametrised by  $\overline{g}=(\pi-\epsilon)/2$ with  $\epsilon>0$, along the trajectory in the phase diagram denoted by an arrow in the inset. Data reflect simulations of Eq.~\ref{eq:imperfect} with $\overline{J} = \pi/8$ with uniform $J_i$ disorder of full width $\delta J = \pi/16$. The real-time data is averaged over all initial states for a given sample, so that $\langle \rangle$ represents an infinite temperature autocorrelator, and also averaged across 2000 independent disorder samples. The power spectra are obtained from a discrete Fourier transform of the averaged time-traces.
    For small $\epsilon=0.1$ (blue traces), the subharmonic response at period doubling persists, shown in real time (inset) and in the Fourier spectrum (main panel). By contrast, a larger $\epsilon=0.7$ (green traces) suppresses the long-time subharmonic signal completely.}
    \label{fig:perturbedPiSG}
\end{figure}

To elaborate on the stability in a physically illuminating way, it is instructive to consider a specific cut through the phase diagram in Fig.~\ref{fig:DrivenIsing} at fixed $\overline{J}$, but $g = \pi/2 - \epsilon/2$. 
Thus, instead of enacting a perfect spin flip (\emph{i.e.} a rotation about the $x$ axis by $\pi$), we enact an \emph{imperfect} rotation by an angle $\pi-\epsilon$: 
\begin{equation}
    U_F = \exp\left[{-i \frac{(\pi-\epsilon)}{2}\sum_j \sigma_j^x}\right]\exp\left[{-i \sum_j J_j \sigma_j^z\sigma_{j+1}^z}\right].
    \label{eq:imperfect}
\end{equation}
Within a purely semiclassical picture, the spins would get rotated by $(\pi - \epsilon)$ so that each successive period would pick up an ``error" in the rotation of the $z$ spins, giving beats in the time dynamics with frequencies $\pi \pm \epsilon$. Instead, due to the long-range Ising order, the spins remain robustly locked to a period $2T$ for a finite range in the deviation $\epsilon$, before transitioning to the PM phase with no LRO and no period doubling. We will explain this further within the l-bit formalism below. Fig.~\ref{fig:perturbedPiSG} shows the dynamics of $\sigma^z(nT)$ averaged over starting product $z$ states, for different deviations $\epsilon$. The real-time signal shows period doubling for a range of $\epsilon$'s, with a peak in the Fourier transformed signal locked at a frequency $\omega = \pi$, while this is washed out in the PM phase for large enough deviations $\epsilon$.

\subsubsubsection{ $0 \pi$ Paramagnet:} 
Finally, the other novel Floquet phase is a bulk paramagnet which exhibits TTSB on the \emph{boundaries} of the system, forming a symmetry protected topological (SPT) Floquet phase~\cite{SenthilSPTRev}. We discuss the properties of this phase below, but this is a digression from our main story. Readers can safely skip ahead if they so desire.

Along the upper horizontal boundary of the free phase diagram, we set $J_i= \pi/2$, while the $g_i$ are random as before. Then, using the fact that $\exp[{-i\frac{\pi}{2} \sigma_j^z\sigma_{j+1}^z}] = -i\sigma_j^z\sigma_{j+1}^z$, we see that 
$$e^{-iH_z} = \prod_j (-i)\sigma_j^z \sigma_{j+1}^z = 
\begin{cases}
(-i)^L \;\;\;\;\;\; \; &\text{for periodic boundary conditions}\\
(-i)^{L-1} \sigma^z_1 \sigma^z_L\;\;\;\; &\text{for open boundary conditions}
\end{cases}$$
Thus, for closed chains or periodic boundary conditions, the Floquet unitary is $U_F =  e^{-iH_x}$ (up to a factor of $i$ for odd length chains), and thus the analysis proceeds exactly as for the paramagnetic phase. 

However, the situation is more interesting for open chains and thus in the presence of a boundary. Now the unitary receives a boundary contribution and takes the form 
\begin{equation}
    U_0^{0\pi PM} \equiv U_F=\sigma^z_1 \sigma^z_L e^{-i\sum_j g_j \sigma_j^x}.
    \mlabel{eq:U0pm2}
\end{equation}
Away from the edges, the eigenstates look like product states in the $x$ basis, just as in the paramagnetic phase. However, the action of the unitary flips the state of the boundary spins (in the $x$ basis), so that the eigenstates look like even/odd superpositions (or cat states) of $x$ spins on the edges. Thus, for a given configuration of bulk spins, the eigenstates come in \emph{quartets}, labeled by (i) the product of the two edge spins $s_1^x$ and  $s_L^x$ (for example, aligned  $|\rightarrow \rightarrow\rangle$, or antialigned $|\rightarrow \leftarrow\rangle$) and (ii) a parity $p$ indicating whether the edge cat states are even/odd superpositions of the edge spins and their reverses:
$$
|\nu_{\pm}\rangle_{s^x_1s^x_L} =   \frac{1}{\sqrt{2}}\left(e^{i (s_1^x h_1 + s_L^x  h_L)}
|s_1^x, \{s_2^x \cdots s_{L-1}^x\}_\nu, s_{L}^x\rangle \pm 
e^{-i (s_1^x h_1 + s_L^x  h_L)}
|-s_1^x, \{s_2^x \cdots s_{L-1}^x\}_\nu, -s_{L}^x\rangle\right). 
$$
Note that the eigenstates do not show LRO in the bulk, and the connected correlator $C_{ij}$ is zero for any pair of bulk-spins as appropriate for a paramagnet. 

The Floquet eigenvalues for the even/odd boundary cats for a given $s_1^x s_L^x$ are $u(s^x, p) = p e^{-i \sum_{j=2}^{L-1} g_j s_j^x}$, independent of $s_1^x$ and $s_L^x$. Upon taking a log, the $\pm$ sign difference in the eigenvalues of the opposite parity boundary cat states translates into a quasienergy difference of $\pi$ \emph{i.e.} the two cat states are diametrically paired across the quasienergy circle.
Further, the quasienergies for the $\{|\rightarrow\rightarrow\rangle, |\leftarrow\leftarrow\rangle\}$ doublet are exactly degenerate with those of the $\{|\rightarrow\leftarrow\rangle, |\leftarrow\rightarrow\rangle\}$
doublet. Thus, each quartet of states represents two degenerate doublets (quasienergy splitting 0) with quasienergy splitting $\pi$ between the states forming the doublet. In the dual fermionic language, this spectral pairing structure comes from the presence of two Majorana modes, at quasienergies $0$ and $\pi$ respectively.

The combination of the bulk paramagnetic correlations and the quasienergy splittings in the quartet justify the name $0\pi$PM. Upon perturbing away from this line, these  properties 
continue to hold throughout the region labeled $0 \pi$PM, but the quasienergy splittings are more generically exponentially close to $0$ and $\pi$.

Strikingly, this phase also exhibits TTSB on the boundary---as such it is an example of a ``boundary time crystal''. This is apparent from our discussion above: starting from a product state in the $x$ basis, the action of the unitary flips the boundary spins so that they return to their original state only after two-periods, exhibiting period doubling on the boundary. The $\pi$ quasienergy splitting goes hand in hand with this period doubling. This can also be seen directly from the equation of motion of the boundary operators in the Heisenberg picture: $\sigma_{1,L}^x(nT)= (U_F^\dagger)^n \sigma_{1,L}^x (U_F^\dagger)^n = (-1)^n \sigma_{1,L}^x$ due to the anticommutation of the $\sigma^x$ and $\sigma^z$ operators. In contrast, the bulk $\sigma_j^x$ operators are constants of motion. 
For the most part we will not explore this aspect of Floquet phases in this review. Interested readers should peruse \cite{vonKeyserlingk2016b} instead.

Finally, we note that since the eigenstates in this phase are non-trivial only in the presence of a boundary, the phase is an
example of a Floquet SPT phase. The existence of a $\mathbb{Z}_2$ SPT is novel to the driven problem, since one needs a symmetry group at least $\mathbb{Z}_2\times \mathbb{Z}_2$ to realize a static SPT. The presence of the dTTS generated by $\mathbb{Z}$ compensates for the missing $\mathbb{Z}_2$.

\subsubsubsection{Phase Transitions} Before leaving the non-interacting model, we note that the critical behavior at the transitions between the different phases in Fig.~\ref{fig:DrivenIsing} can be completely understood. A set of symmetries dictate the particularly simple phase boundaries along $\overline{J}=\overline{g}$ and $\overline{J}=- \overline{g} + \frac{\pi}{2}$~\cite{FloquetcriticalityPNAS}. As mentioned in Sec.~\ref{sec:MBLorder}, these transitions represent \emph{eigenstate phase transitions} across which the behavior of individual highly-excited eigenstates changes in a singular way, even as thermodynamic functions look non-singular. These transitions proceed via infinite randomness critical points, and the critical behavior can be understood within the strong disorder renormalization group~\cite{Fisher95, lpqo, PekkerHilbertGlass, FloquetcriticalityPNAS}. Notably, the the critical lines are less localized than the phases, displaying a logarithmic scaling for entanglement entropy and stretched exponentially decaying correlations (as compared to area law entanglement and exponentially decaying correlations within the phases).

\subsubsection{Adding Ising symmetric interactions}
\mlabel{sec:interactingIsingDrive}
Thus far, we have primarily discussed the phase diagram of the non-interacting driven Ising chain, albeit in a many-body spin language. Without interactions, any amount of disorder is sufficient for Anderson localization of the Floquet unitary. For the interacting problem, the system is expected to remain Floquet MBL provided (i) the interactions are weak (ii) the  disorder is strong (iii) the driving frequencies are large compared to the local energy scales in the problem. Ref.~\mcite{KhemaniPRL} considered the driven Ising chain in the presence of Ising symmetric interactions, and showed that all four phases discussed above persist even with interactions. We focus on two aspects below: (i) the persistence of MBL and (ii) the peristence of all four phases. 

\begin{figure}
    \centering
    \includegraphics[width=0.3\columnwidth]{DrivenIsing.pdf}
    \includegraphics[width=0.39\columnwidth]{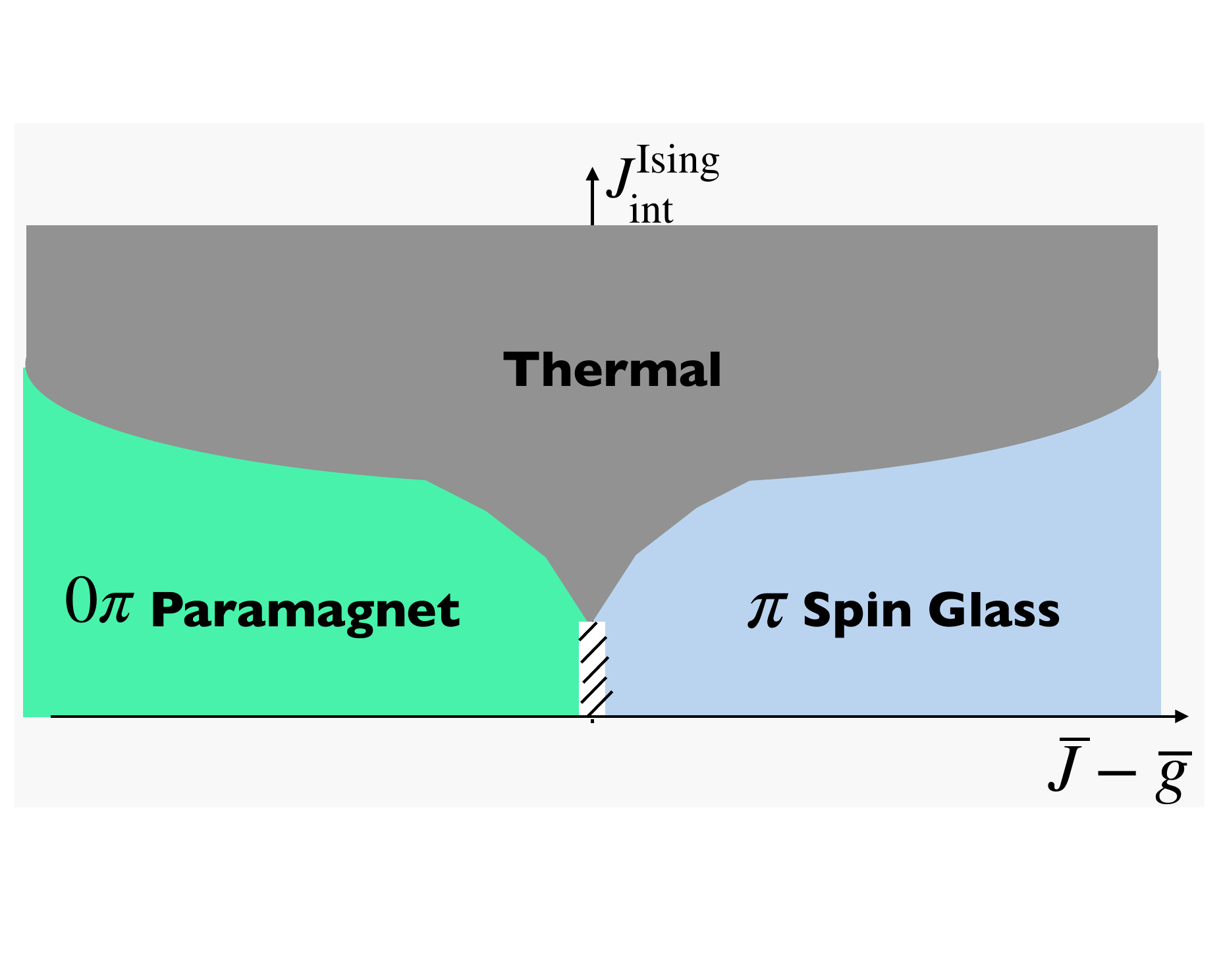}
    \includegraphics[width=0.3\columnwidth]{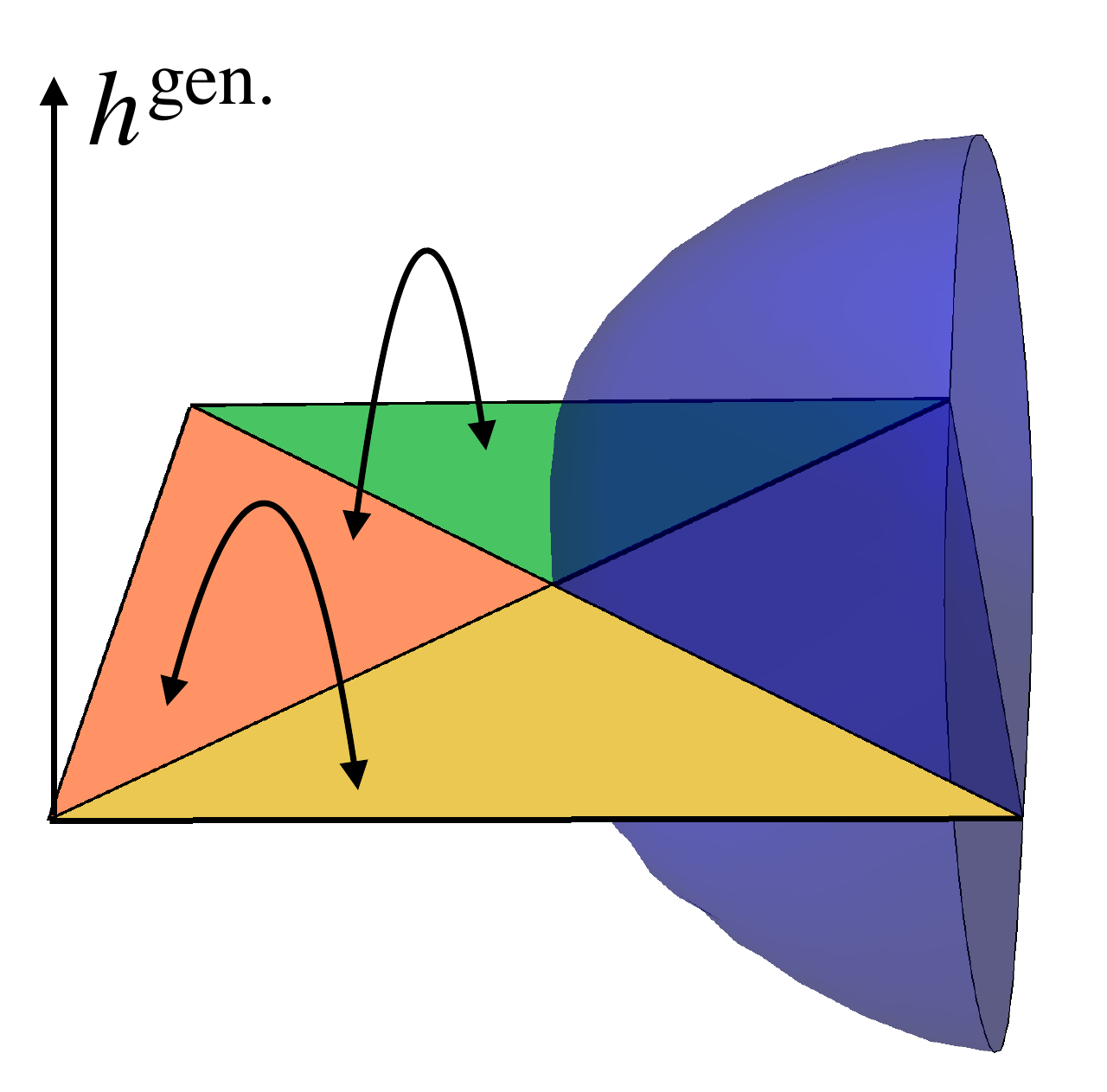}
    \caption{Perturbing the non-interacting Floquet Ising chain whose phase diagram if depicted in the left panel, and Fig.~\ref{fig:DrivenIsing}. With Ising symmetric perturbations, as in Eq.~\ref{eq:drivenIsing}, the general structure (middle) is analogous to  the interacting static case depicted in Fig.~\ref{fig:ising_interacting}, although the self-dual line corresponds to a different trace through the phase diagram, with the $\pi$SG giving way to the $0\pi$PM. (left) Unlike the static case, an entirely new feature appears when perturbing with interactions that break the Ising symmetry~\cite{ElsePRL, CVS}: the $\pi$SG is {\it absolutely} stable, i.e.\ stable to {\it all} such perturbations, while the remaining three phases only display purely paramagnetic correlations in absence of an explicit Ising symmetry. }
    \label{fig:perturbedDrivenIsing}
\end{figure}

\subsubsubsection{Persistence of MBL}
To start, we note that the strength of the transverse fields in the $\pi$SG ($g \simeq \pi/2$) is comparable to the driving frequency, as is the strength of the Ising couplings in the $0\pi$PM ($J\simeq \pi/2$). Thus, both non-trivial Floquet phases do \emph{not} immediately satisfy the large frequency requirement for Floquet MBL discussed in Section~\ref{sec:floqMBL}
Indeed, drives with couplings comparable to the frequency are essential for obtaining phases that are \emph{novel} to the driven setting. One of the central challenges addressed in Ref.~\mcite{KhemaniPRL} was establishing that these novel phases can, nevertheless, still be many-body localized in the presence of interactions. 

Intuitively, one can understand why the strong couplings in the $\pi$SG/$0\pi$PM phases are relatively benign (for the purposes of heating) by considering the Floquet unitary over \emph{two} driving periods. We will see that the effect of the strong terms can be eliminated over two periods. Explicitly for the drive in Eq.~\ref{eq:drivenIsing} in the $\pi$SG phase with $g=\pi/2-\epsilon/2$: 
 \begin{align}
    U(2T) &= \left(P^x_{\pi-\epsilon} \exp\left[{-i H^x_{\rm int} }\right] \exp\left[-i H_z  \right]\right) \left(P^x_{\pi-\epsilon} \exp\left[{-i H^x_{\rm int} }\right] \exp\left[-i H_z  \right]\right) \nonumber\\
     &= \left(P^x_{-\epsilon} \exp\left[{-i H^x_{\rm int} }\right] \exp\left[-i H_z  \right]\right)\left(P^x_{-\epsilon} \exp\left[{-i H^x_{\rm int} }\right] \exp\left[-i H_z  \right]\right)\nonumber\\
     &\simeq e^{-i 2H_{\rm MBL}}
     \mlabel{eq:2T}
 \end{align}
where $P^x_{\rm \theta} = \exp\left[{-i \frac{\theta}{2}\sum_i \sigma_i^x }\right]$ represents a rotation by angle $\theta$ about the $x$ axis, and $P^x_\pi\equiv P_x$.  We have eliminated $P^x_\pi$ in going from the first line to the second by using the fact that $[H^x_{\rm int},P^x_\pi] = [H_z, P^x_\pi]=0$, $(P^x_\pi)^2 = 1$ and $P^x_{\pi - \epsilon} = P^x_{\pi} P^x_{-\epsilon}$  The remaining terms in $U(2T)$ are now small compared to the frequency, and the usual conditions for Floquet MBL are satisfied. Since $U(2T)$ generates a stroboscopic time series probed at even (or odd) times, the system cannot heat to infinite temperature if $U(2T)$ is Floquet MBL. Equivalently, one can arrive at the same result by unitarily transforming to a ``toggling" frame in which the spins are flipped every period by acting with $P^x_\pi$. In this frame, the drive lacks the problematic $P^x_\pi$ piece and can be many-body localized. Of course, cranking up the strength of interactions should eventually destroy MBL and lead to thermalization to an infinite temperature Floquet ergodic phase. 

The preceding discussion points to a natural generalization of the criterion for Floquet MBL: we require all couplings in the driven system to be smaller than the drive frequency,  either for the original drive \emph{or} for the drive in a local unitarily transformed frame. When this is true, the system (possibly in a rotating frame) is described by a time-independent (quasi)-local MBL Floquet Hamiltonian, $H_{\rm MBL}$, as in the third line of Eq.~\eqref{eq:2T} above. This phenomenon was recently given the name ``crypto-equilibrium" in Ref.~\mcite{YaoPhysToday}.

\subsubsubsection{Persistence of Phase Structure} So far we have discussed the stability of \emph{localization} to the addition of interactions for the full range of parameters in the phase diagram in Fig.~\ref{fig:DrivenIsing}--- including cases for which the couplings are comparable to $\omega$ such as in the $\pi$SG/$0\pi$PM. We now show that the \emph{phase structure} in Fig.~\ref{fig:DrivenIsing} is also stable, so that one still obtains four MBL Floquet phases even in the presence of Ising symmetric interactions.

The connection between the stability of the four phases and the stability of localization may not be obvious \emph {a priori}. Indeed, when the system is probed stroboscopically in multiples of two periods (or examined in an appropriate frame), the dynamics is well described by a time-independent MBL Floquet Hamiltonian, $U(2T) \simeq e^{-i 2 H_{\rm MBL}}$. 
Since this evolution is effectively governed by a \emph{static} Hamiltonian, it does not immediately follow that phases that are novel to the Floquet setting could be obtained. Nevertheless, it turns out that upon transforming back to the original frame of reference (or looking stroboscopically every period),  observables can nevertheless show distinct behavior novel to the driven setting. 

There are two key ideas that rely on the l-bit formalism discussed in Sec.~\ref{sec:mbl}:
\begin{itemize}
    \item The stability of localization to perturbations about the four exactly solvable lines (with Floquet unitaries $U_0^{a}$) implies the existence of a local, finite-depth unitary transformation, $\mathcal{V}$, which relates the eigenstates of $U_0^a$, to those of the perturbed unitaries, denoted $U_\lambda^a$, upto exponentially small in $L$ corrections. Here $a$ denotes which of the four phases we are considering, as in Eq.~ \eqref{eq:U0pm}, \eqref{eq:U0sg}, \eqref{eq:U0pm2}, \eqref{eq:U0sg2}. For Ising symmetric perturbations of $U_0^a$, the unitary $\mathcal{V}$ is also Ising symmetric~\cite{CVS}.  Thus, there exist dressed or ``smeared out" operators, $\tau_i^\gamma = \mathcal{V}^\dagger \sigma_i^\gamma \mathcal{V}$ so that $\tau_i^\gamma$ has the same commutation/anticommutation relation with $P_x$ as $\sigma_i^\gamma$. Here $\gamma = x/y/z$. To wit, $\{\tau_i^x\}$ operators are Ising even, while $\{\tau_i^{y/z}\}$ are Ising odd.

    For the unperturbed models, the l-bits are the $\{\sigma_i^x\}$/$\{D_i\}$ operators in the paramagnetic/spin-glass phases respectively. Likewise, the l-bits for the perturbed models can be written in terms of the $\tau^\gamma$ operators. Thus every eigenstate of $U_\lambda^a$ can be characterized by its eigenvalues under the dressed l-bit operators and $P_x$. This, in turn, implies that \emph{$U_\lambda$ can be written entirely in terms of $P_x$ and the dressed l-bit operators.}  
    
    \item  Next, one can show from locality and continuity that the commutation/anticommutation relations between dressed $\tau_i^{x/z}$ operators and $U_\lambda^{p/s}$ are the same as those between $\sigma_i^{x/z}$ and $U_0^{p/s}$, again upto exponentially small in $L$ corrections~\mcite{vonKeyserlingk2016a, vonKeyserlingk2016b, CVS}. Here $p/s$ refer to either of the two paramagnets/spin-glasses respectively. 
    
    In the two paramagnets, we have that $ (U_0^a)^\dagger \sigma_i^x U_0^a = s_{ia} \sigma_i^x + O(e^{-c L})$, where $s_{ia} = \pm1$ depending of whether $i$ is in the bulk/boundary and whether $a$ denotes the PM or the $0\pi$PM. Likewise, in the two spin-glasses, we have $ (U_0^a)^\dagger \sigma_i^z U_0^a = s_{ia} \sigma_i^z + O(e^{-c L})$, where $s_{ia} = +1/-1$ in the $0$SG/$\pi$SG phases respectively. \emph{Because the transformation of these operators under the action of the unitary is quantitized to pick up a phase of either $+1$ or $-1$, this represents a ``topological" property that cannot be continuously deformed~\mcite{vonKeyserlingk2016a, vonKeyserlingk2016b, CVS}}. 
\end{itemize}

Putting it together, the two points above imply that the perturbed unitaries $U_\lambda$ generalize the expressions for $U_0^a$ in Eqs. \eqref{eq:U0pm}, \eqref{eq:U0sg}, \eqref{eq:U0pm2}, \eqref{eq:U0sg2} and have the form:
\begin{align}
U_\lambda^{PM} &= \exp\left[ -i \mathcal{H}_x(\{\tau_i^x\})  \right] \mlabel{eq:pmperturbed}\\
U_\lambda^{0\pi PM} &= \tau_1^z \tau_L^z  \exp\left[ -i \mathcal{H}_x(\{\tau_i^x\})  \right] \mlabel{eq:pm2perturbed}\\
U_\lambda^{0SG} &=  \exp\left[ -i \mathcal{H}_z(\{\tilde{D}_i^x\})  \right] \mlabel{eq:sgperturbed}\\
U_\lambda^{\pi SG} &=  P_x\exp\left[ -i \mathcal{H}_z(\{\tilde{D}_i^x\})  \right]\mlabel{eq:sg2perturbed}. 
\end{align}
The l-bits in both paramagnets are the $\{\tau_i^x\}$ operators (at least in the bulk), while those in the spin-glass phases are the dressed domain wall operators $\{\tilde{D}_i = \tau_i^z \tau_{i+1}^z\}$. The functions $\mathcal{H}$ represent quasi-local l-bit MBL Hamiltonians that take a form similar to Eq.~\eqref{eq:lbit}: 
\begin{align}
\mathcal{H}_x(\{\tau_i^x\}) &= \sum_j \tilde{g}_j \tau_j^x +  \sum_{jk} \tilde{\mathcal{J}}_{jk} \tau_j^x \tau_k^x 
+ \sum_{jkl} \tilde{\mathcal{J}}_{jkl} \tau_j^x \tau_k^x \tau_l^x  + \cdots \mlabel{eq:lbitx} \\
\mathcal{H}_z(\{\tilde{D}_i^x\}) &= \sum_{jk} \tilde{\mathscr{J}}_{jk} \tau_j^z \tau_k^z  + \sum_{jklm} \tilde{\mathscr{J}}_{jkml} \tau_j^z \tau_k^z \tau_l^z\tau_m^z + \cdots. 
\mlabel{eq:lbitz}
\end{align}
with exponentially decaying couplings $\mathcal{J}, \mathscr{J}$.
Note that each term in $\mathcal{H}_z(\{\tilde{D}^x_i\})$ contains an even number of $\tau^z$s since $\mathcal{H}_z$ is an Ising even functional of the dressed domain wall operators. From the functional forms above, it is clear that all our previous statements about the exactly solvable unitaries $U_0$ can be transposed to $U_\lambda$, but now in terms of the dressed l-bit operators (and hence with exponentially small in $L$ corrections). In particular, the characteristic spectral pairing of the eigenstates and the equations of motion for the bulk and boundary order parameter operators for all four phases persist to the perturbed case.  

We emphasize again that while the stability of MBL implies that $U_\lambda^{0SG}$ and $U_\lambda^{\pi SG}$ can be written in terms of the $\{\tilde{D}_i\}$ and $P_x$ operators, the exact functional forms above are determined by the requirement that $\tau_i^z$ satisfy the right commutation/anticommutation relations:
$(U^{0SG}_\lambda)^\dagger \tau_i^z U_\lambda^{0SG}=\tau_i^z$ and $(U^{\pi SG}_\lambda)^\dagger \tau_i^z U_\lambda^{\pi SG}=-\tau_i^z.$ This naturally leads to an extra factor of $P_x$ in $U^{\pi SG}_\lambda$, which means that $U_\lambda^{\pi SG}$ \emph{cannot} be written as the exponential of a static local Hamiltonian --- reflecting the novel and non-trivial dynamical properties of this phase~\footnote{Put differently, the transformation $g_i \rightarrow g_i + \pi/2$ takes the system from the $0$SG to the $\pi$SG, and leads to $U_\lambda^{\pi SG} = P_x U_\lambda^{0SG}$ as in Eq.~\eqref{eq:sgperturbed} and \eqref{eq:sg2perturbed} above. We can also see this from Eq.~\ref{eq:2T}, where $H_{\rm MBL}$ in that expression is the same as what we call $\mathcal{H}_z$ here, and the unitary in the second line of Eq.~\ref{eq:2T} (after absorbing the $P_\pi^x$) is simply two actions of the perturbed unitary in the $0$SG phase.}.  We also emphasize that the unitary $\mathcal{V}$ only relates the \emph{eigenstates} of $U_0$ and $U_\lambda$ and not the unitaries themselves. This is important because $U_0$ and $U_\lambda$ have different eigenspectra, and the presence of interactions in $U_\lambda$ is apparent from the long-range terms in $\mathcal{H}_{x/z}$ in Eq.~\eqref{eq:lbitx}, \eqref{eq:lbitz} which are absent in $H_0$\footnote{This is also true for the conventional description of MBL Hamiltonians via l-bits. A local unitary $\mathcal{V}$ relates the MBL eigenstates $|e_n\rangle$ to computational basis states in the $\sigma^z$ basis, $|p_n\rangle$ so that $|e_n\rangle = \mathcal{V}^\dagger|p_n\rangle$ (the unitary $\mathcal{V}$ diagonalizes $H_{MBL}$).  Since the $\{|p_n\rangle\}$ are eigenstates of the p-bits $\sigma_i^z$, it follows that the $\{|e_n\rangle\}$ are eigenstates of $\tau_i^z = \mathcal{V}^\dagger \sigma_i^z V$. However, the p-bit Hamiltonian that $H_{MBL}$ should be unitarily related to is not apparent \emph{a priori} and is only obtained once we know the form of the $H_{MBL}$ in the l-bit basis.}.

Finally, we note that while all four phases persist in the presence of weak interactions near the edges of the non-interacting phase boundary, the fate of the localized-to-localized infinite randomness transition lines is at present still unresolved. It is believed, though not conclusively established, that many-body resonances destabilize the (weakly) localized infinite randomness non-interacting critical points, so that the transitions between different MBL phases (either static or driven) proceeds via an intervening sliver of thermal phase. If we work with Ising self dual interactions $J_{\rm int}^x = J_{\rm int}^z$ and choose parameters so that $g_i \sim J_i $ (on average), then various symmetries dictate a phase diagram comprising the $0\pi$PM, $\pi$SG and thermalizing phases which qualitatively looks like Fig.~\ref{fig:perturbedDrivenIsing}.

\subsubsection{Dynamical Signatures}
Thus far our discussion has focused on the eigenstate and eigenspectrum characteristics of the four driven Ising phases. While these features are not directly measurable, they do have crisp measurable consequences that are amenable to experimental detection in the dynamics of the system. 

The consequences for dynamics are particularly transparent when considering the time evolution of the l-bit operators under the effective l-bit drives $U_\lambda^a$, Eqs.~\eqref{eq:pmperturbed}, ~\eqref{eq:pm2perturbed}, ~\eqref{eq:sgperturbed}, ~\eqref{eq:sg2perturbed}. These, in turn, have sharp consequences for the dynamics of physically measurable operators written in terms of the physical spins. 
We will rely on known wisdom on quenches in MBL systems, and the ensuing dephasing dynamics. 
In all cases below, we have in mind dynamics from easily preparable low-entanglement product states that are a superposition of the eigenstates of $U_F$. 

\begin{figure}
    \centering
    \includegraphics[width=\columnwidth]{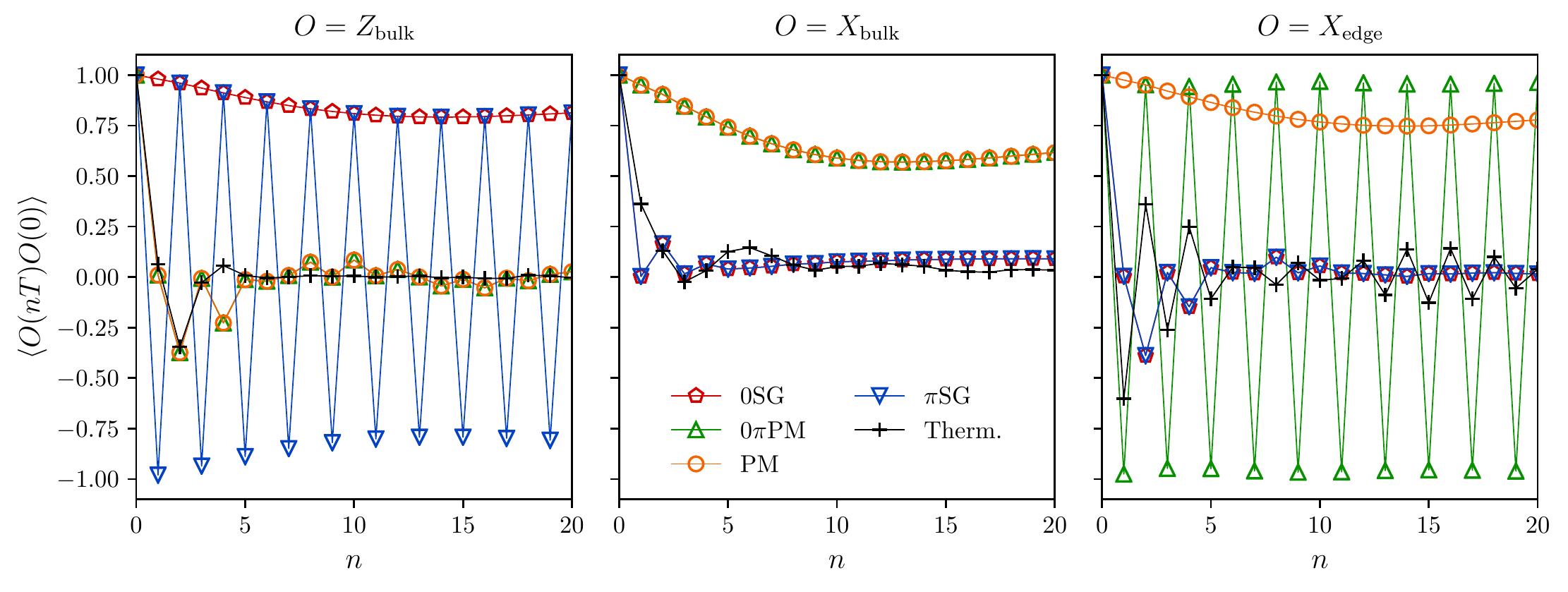}
    \caption{Dynamical signatures of Floquet MBL phases: telling apart all the four different phases occurring in the phase diagram, Fig.~\ref{fig:perturbedDrivenIsing}, as well as these from the thermal phase, is possible by considering measurements of spin autocorrelators, as follows. Left: a nonvanishing long-time autocorrelator of the $z-$component of spin in the bulk, $Z_\mathrm{bulk}$ distinguishes the two spin glasses, 0SG and $\pi$SG, from the other three phases (the two paramagnets and the thermal phase) in which it decays to zero. The two SGs in turn are distinguished by the (non)synchronised nature of the autocorrelator, i.e.\ by period doubling with a non-zero amplitude in the $\pi$SG, compared to synchronization to a non-zero value in the $0$SG. 
    Middle: the two paramagnets, PM and $0\pi$PM, are distinguished from the thermal phase by their nonvanishing of the $X_\mathrm{bulk}$-autocorrelator; while they can be distinguished from one another by the presence/absence of the period doubling in the non-vanishing edge autocorrelator of $X_\mathrm{edge}$ (Right). The data reflect simulations of the model in Eq.~\ref{eq:drivenIsing} with open boundary conditions. In the PM and $0\pi$PM phases, we set $J_i = 0.1$ and $\frac{\pi}{2}-0.1$ respectively, and sample $h_i$ from a uniform distribution of mean $\pi/4$ and full width $\pi/3$. In the 0SG and $\pi$SG we use the same distributions but swap the roles of $h$ and $J$. In all four non-thermal phases the interaction is $J_{\rm int}=0.05$. The thermal curve has $J_{\rm int}=1$ and random fields $J_i$, $h_i$ sampled within the PM slice of the $\overline{h}-\overline{J}$ phase diagram.}
    \label{fig:dynamical_signatures}
\end{figure}

\subsubsubsection{Paramagnet} This dynamics in this phase looks  synchronized and paramagnetic.  

The stroboscopic evolution within the PM phase is described by the effective static l-bit Hamiltonian $\mathcal{H}_x$, Eq.~\eqref{eq:lbitx}. Thus, the expectation value of each of the $\tau_i^x$ operators is conserved, while $\langle \tau_i^z(t) \rangle$ decays to zero as a power law in time due to the slow dephasing dynamics inherent to interacting MBL systems, as discussed in Section~\ref{sec:mbldynamics}. 

The description of $U_F$ by an effective static Hamiltonian  (for which local observables tend to steady-state values) means that all observables are \emph{synchronized} with the drive and look periodic with period $T$. 
    
If we work with physical spin operators $\sigma_i^\alpha$ rather than l-bits (which are not directly experimentally accessible), then Ising even operators such as $\sigma_i^x$, denoted $O_e$,  look synchronized and saturate to a (possibly non-zero) value which depends on (i) the overlap of $O_e$ with the conserved l-bits $\tau_i^x$ and (ii) the expectation value of $\{\tau_i^x\}$ in the initial state. In contrast, Ising odd operators such as $\sigma_i^z$, denoted $O_o$, have no overlap with the l-bits and decay to zero as a power law in time for all initial states. Note that local Ising odd operators are order parameters for Ising symmetry breaking. Thus, the decay of these operators behooves a \emph{paramagnet} for which the late-time expectation values of Ising order parameters vanish. 
    
\subsubsubsection{$0\pi$PM} This dynamics in this phase looks  synchronized and paramagnetic, except at the boundaries where one can get period doubling. 

In this case, all $\tau_i^x$ operators in the bulk are constants of motion, while the equation of motion for boundary operators shows period doubled dynamics: $\tau_{1,L}(nT) = (U_\lambda^\dagger)^n \tau_{l,L}^x (U_\lambda)^n = (-1)^n \tau_{1,L}^x$, 
In terms of the physical spins, Ising even operators near the edges of the system show period doubling, while those in the bulk look synchronized (up to an exponentially small piece showing period doubling, which comes from the overlap of the operator with the boundary $\tau^x$). Again, all Ising odd operators vanish as a power law in time for all initial states, as is appropriate for a paramagnet.

\subsubsubsection{$0$SG} This phase shown synchronized dynamics with symmetry breaking. 

The stroboscopic evolution within the $0$SG phase is described by the effective static l-bit  Hamiltonian $\mathcal{H}_z$, Eq.~\eqref{eq:lbitz}. Thus, the expectation value of each of the $\tau_i^z$ operators (and $\tilde{D}_i$ operators) is conserved, while $\langle \tau_i^x(t) \rangle$ decays to zero as a power law in time. 

If we start from Ising symmetry-broken initial states with a non-zero expectation value of the Ising order parameter $\langle \tau_i^z\rangle$, then this expectation value persists in time. On the other hand, Ising symmetric initial states have $\langle \tau_i^z\rangle = 0$ because $\tau_i^z$ is Ising odd. For such states, we need to examine two point functions of $\tau_i^z$ to diagnose symmetry breaking. 

Generalizing to physical spin operators, it follows that for initial states that break Ising symmetry, one-point expectation values of Ising odd operators $O_o$ are synchronized and non-zero. By contrast, ymmetric initial states require one to examine Ising even operators, such as two point functions of $\sigma_i^z$, or $\sigma_i^x$. The former have large overlaps with l-bits $\tilde{D}_i$ and retain a large portion of their initial value, while the latter have small overlap.  

\subsubsubsection{$\pi$SG} The dynamics within the $\pi$SG shows period doubling (\emph{i.e.} dTTSB) along with Ising symmetry breaking. 

In this case, the Floquet unitary $U_\lambda^{\pi SG}$ does \emph{not} look like the exponential of a static quasi-local Hamiltonian, and the system can depict non-synchronized dynamics. Indeed, within the l-bit formalism, we see that the equation of motion for Ising order parameter satisfies $\tau_i^z(nT) = (-1)^n\tau_i^z$. Thus Ising symmetry broken initial states with a non-zero expectation value $\langle \tau_i^z\rangle$ show period two oscillations, with stroboscopic snapshots that look like Fig.~\ref{fig:perpetuum_mobile}(d). The expectation value of $\tau_i^x$ decays algebraically to zero as before. 

The behavior generalizes to physical spin operators, so that all local Ising odd operators (such as the Ising order parameter $\sigma_i^z$) show period doubling upon starting from symmetry broken initial states. In general, examining the time-trace in Fourier space also shows responses at a slew of incommensurate frequencies (as we elaborate in 
Sec.~\ref{ss:dynamicsSRE} below), in addition to a Fourier peak at $\omega/2$ corresponding to the period doubling. However, the weight on these incommensurate frequencies decays as a power law in time due to slow dephasing.  

Instead, if we start with symmetric initial states and examine two point equal time correlation functions of Ising odd operators, then these look synchronized and non-zero.

Finally, all expectation values decay to zero in the Floquet ergodic phase, as is appropriate for thermalization to infinite temperature. 

To summarize, we have discussed above how all four non-trivial Floquet phases, together with the thermal phase, can be distinguished from each other by measuring local experimentally accessible operators starting from experimentally accessible initial states. These dynamical diagnostics are depicted in Fig.~\ref{fig:dynamical_signatures}.

A caveat: we have largely ignored the exponentially small corrections to the l-bit commutation relations in our discussion so far. In reality, $\tau^z(nT) = (-1)^n \tau^z + O(e^{-L}).$ These mean that the period doubling response eventually decays at the longest times, of $O(e^L)$ in a finite sized system. Note also that in our theoretical discussions, we can refer to dynamical signatures such as synchronization or period doubling, that last forever or up to times $O(e^L)$. However, any real experiment will have a finite lifetime that is typically orders of magnitude shorter than $O(e^L)$. Thus, a central challenge for current state-of-the-art experiments entails drawing conclusions about \emph{asymptotic} dynamical phases from finite-time data. These issues will be discussed over the course of Sections~\ref{sec:prethermalTC} and \ref{sec:expt_real}. 

\begin{figure}
\begin{tabular}{|c|c|c|c|}
\hline
    PM & 0SG & $\pi$SG & $0\pi$PM \\
    \hline \hline
 \includegraphics[width=2cm]{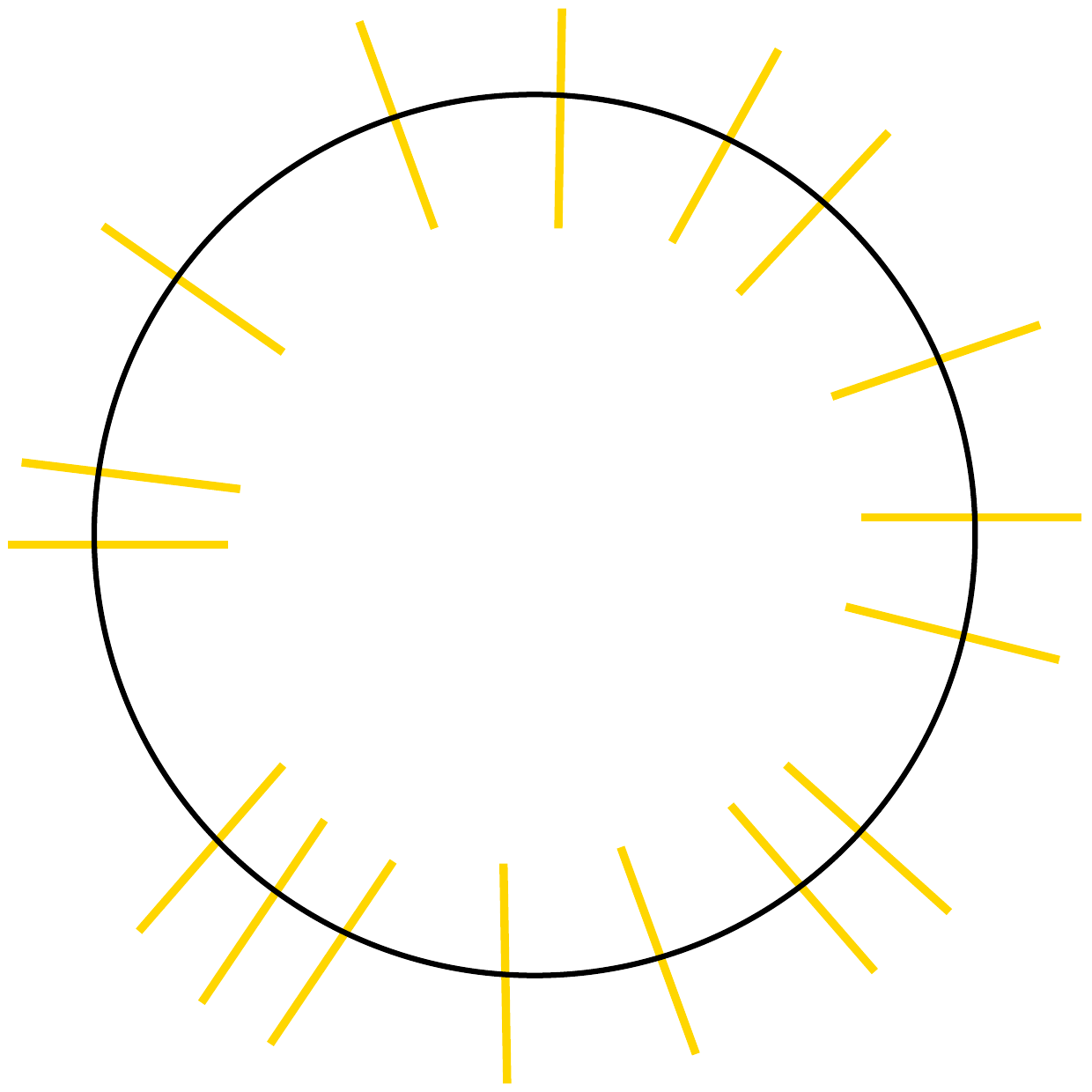}& 
 \includegraphics[width=2cm]{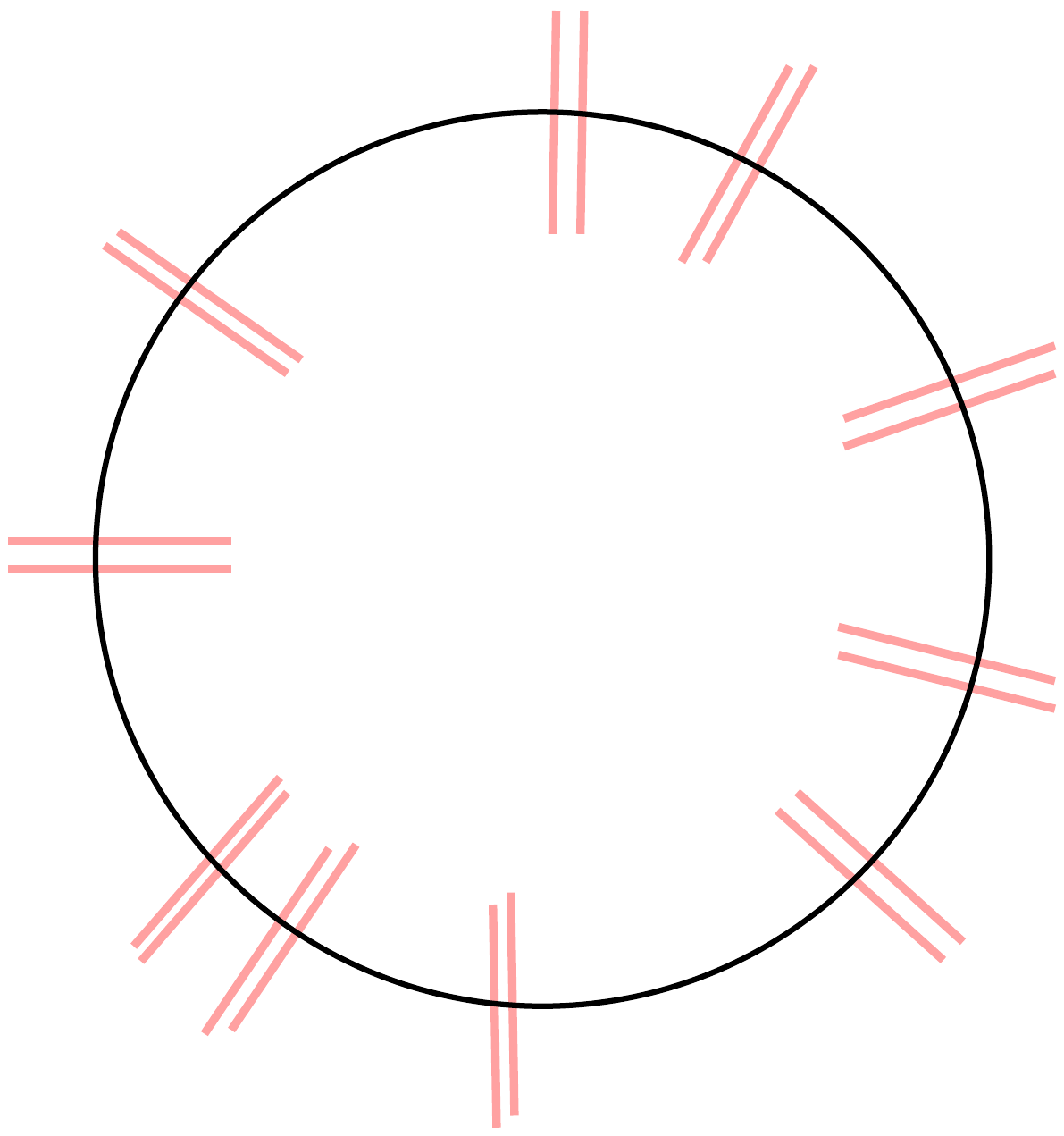}&
 \includegraphics[width=2cm]{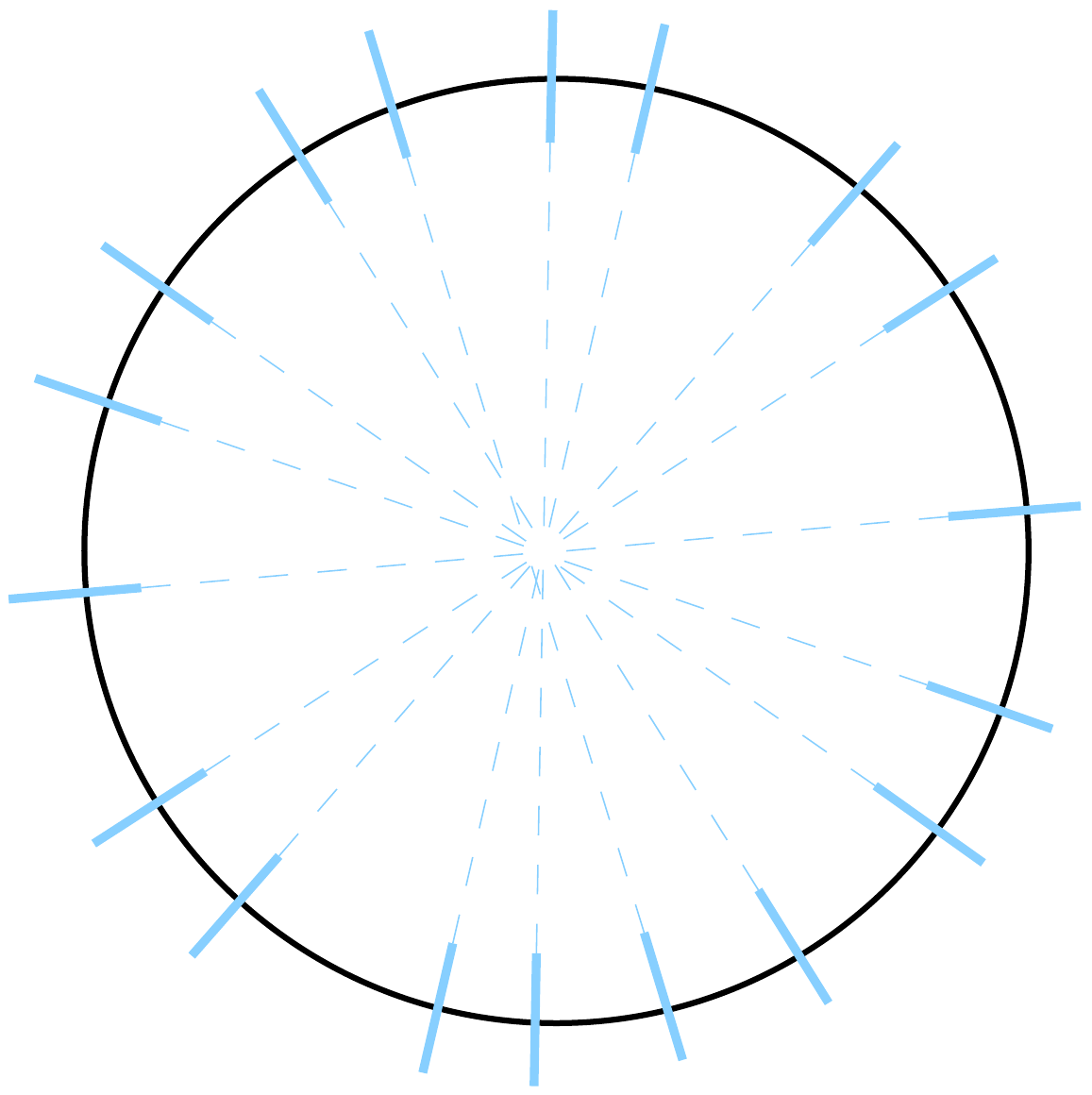}&
 \includegraphics[width=2cm]{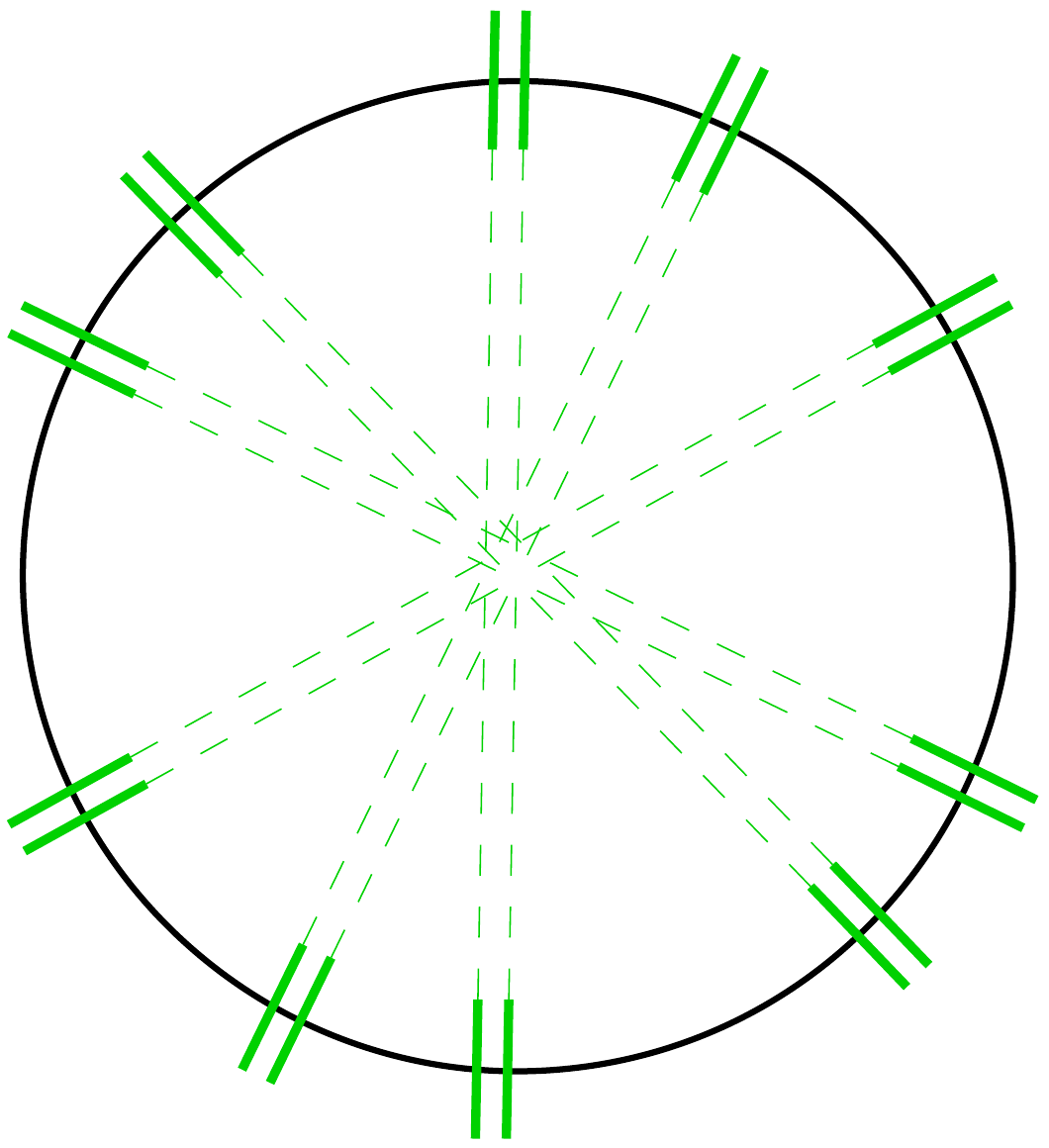}\\   
 \hline
 \includegraphics[width=2cm]{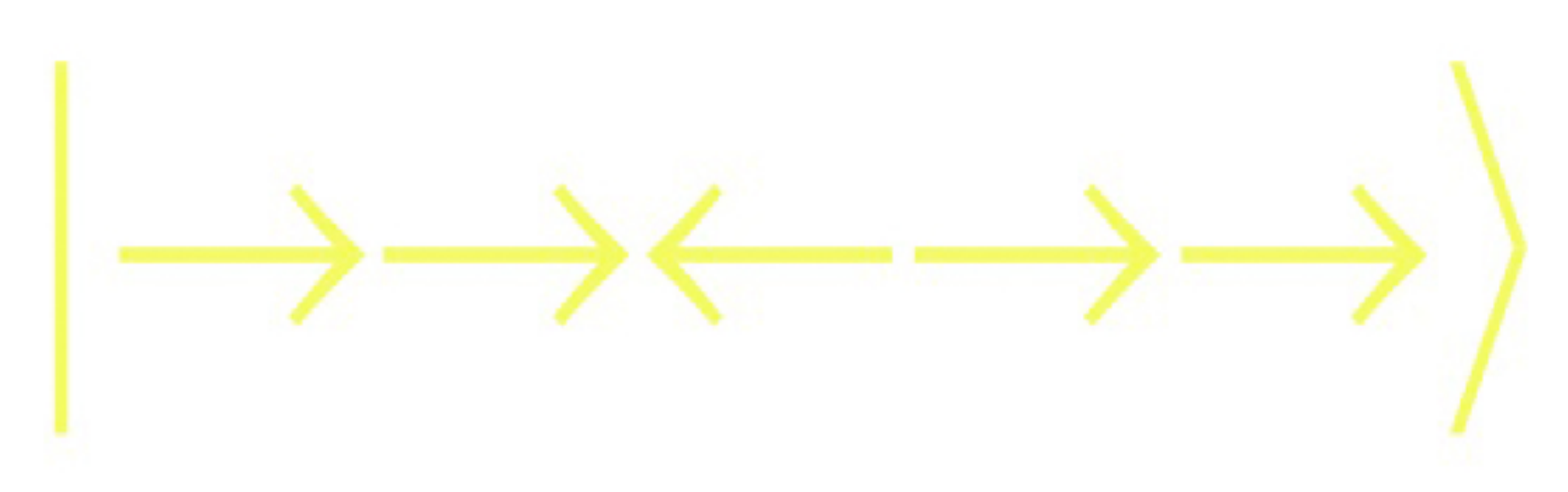}& 
 \includegraphics[width=1.4cm]{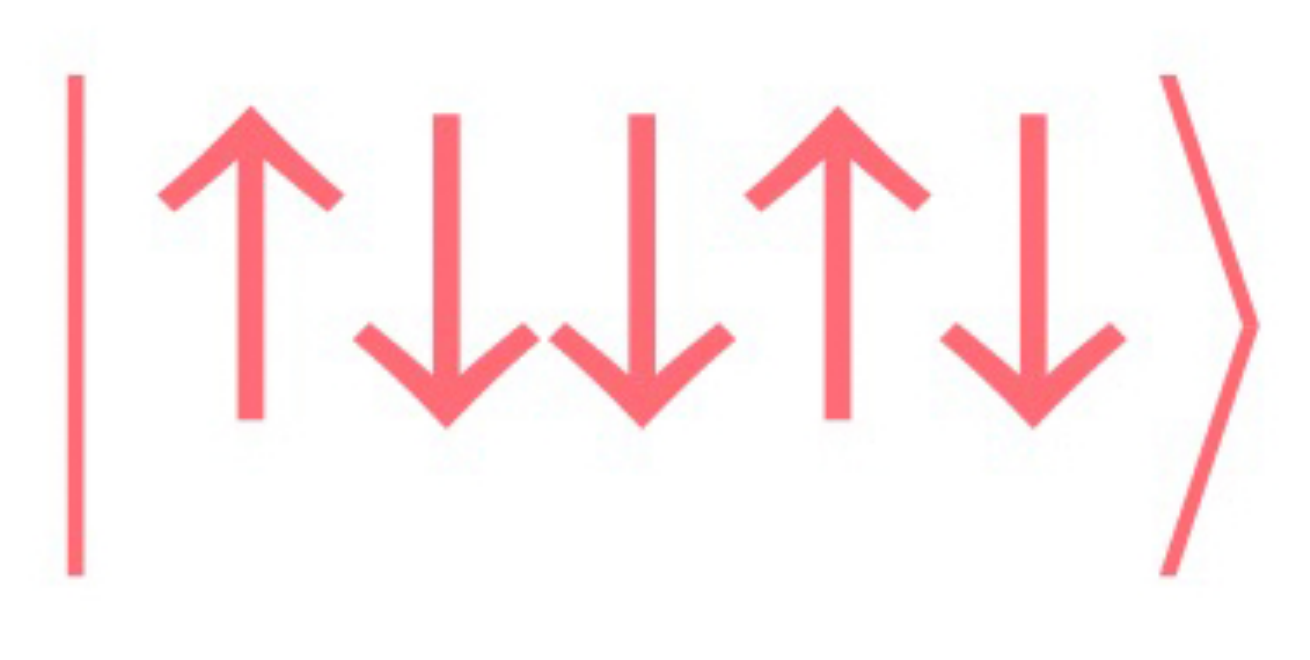}  \includegraphics[width=1.4cm]{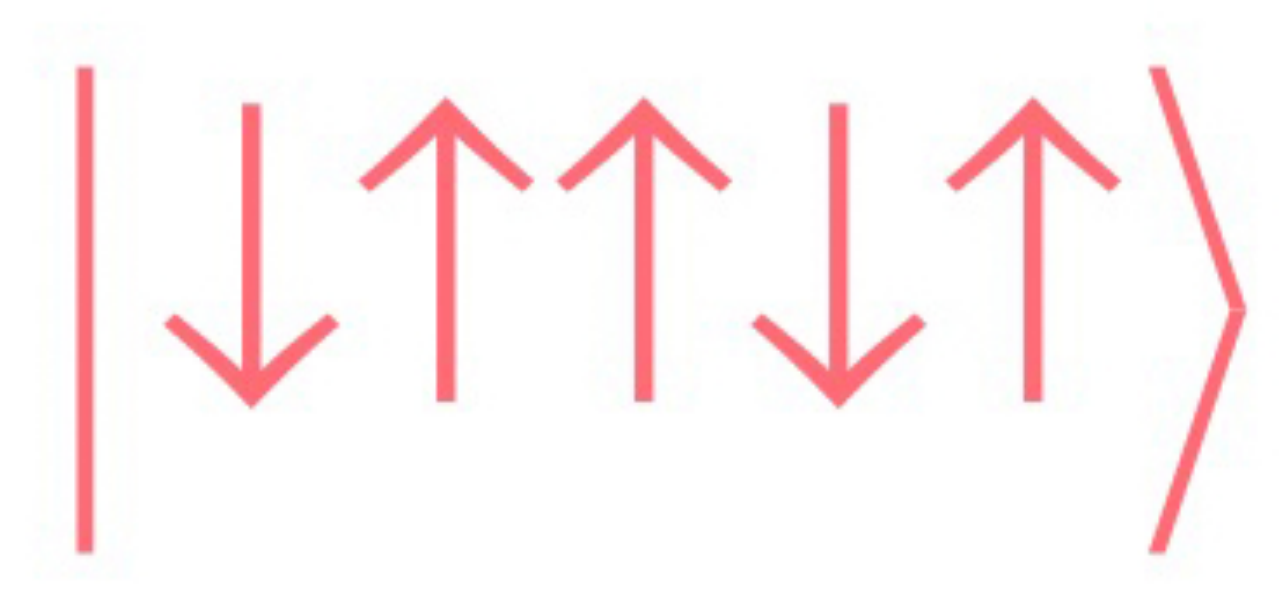}& 
 \includegraphics[width=1.7cm]{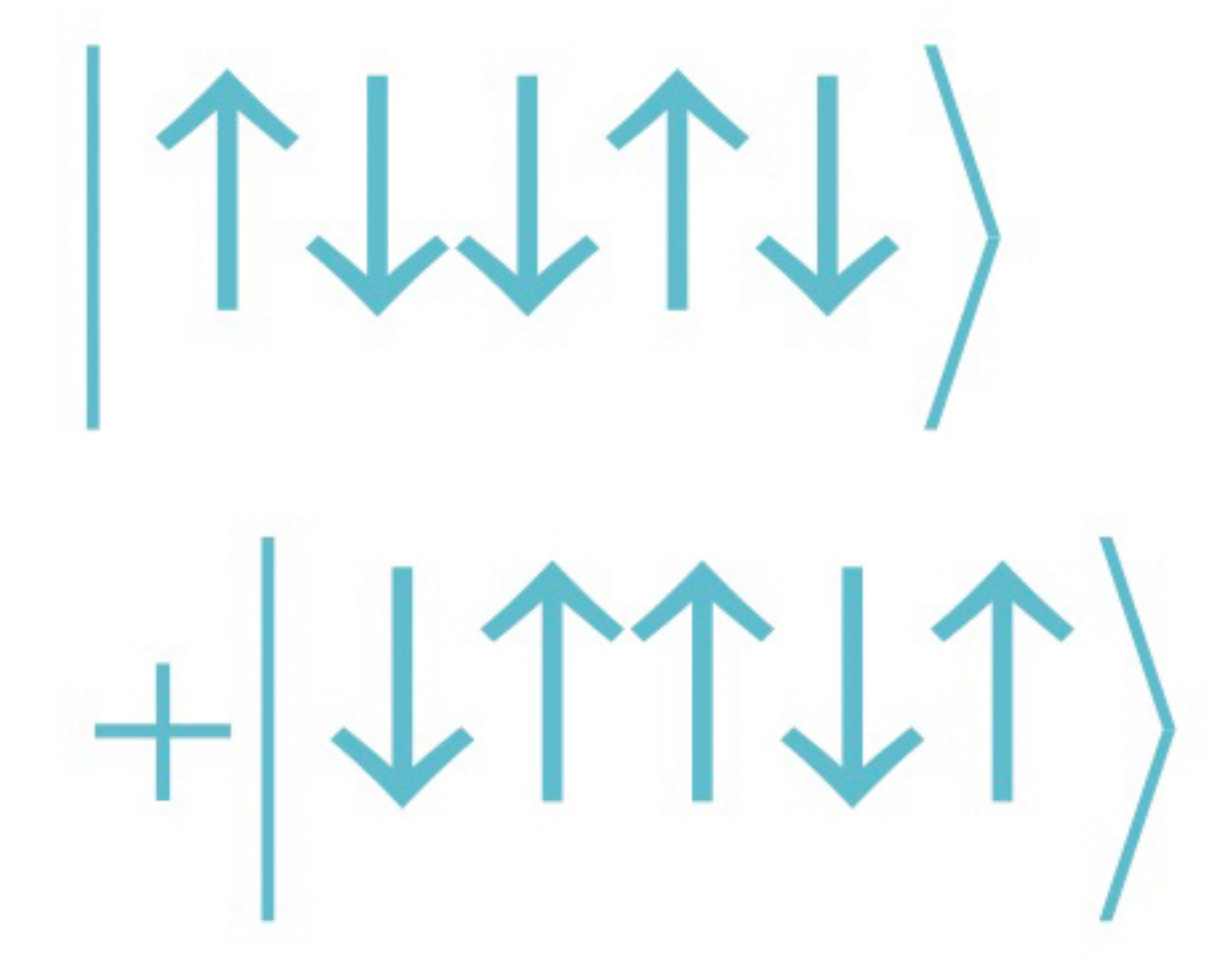}  \includegraphics[width=1.7cm]{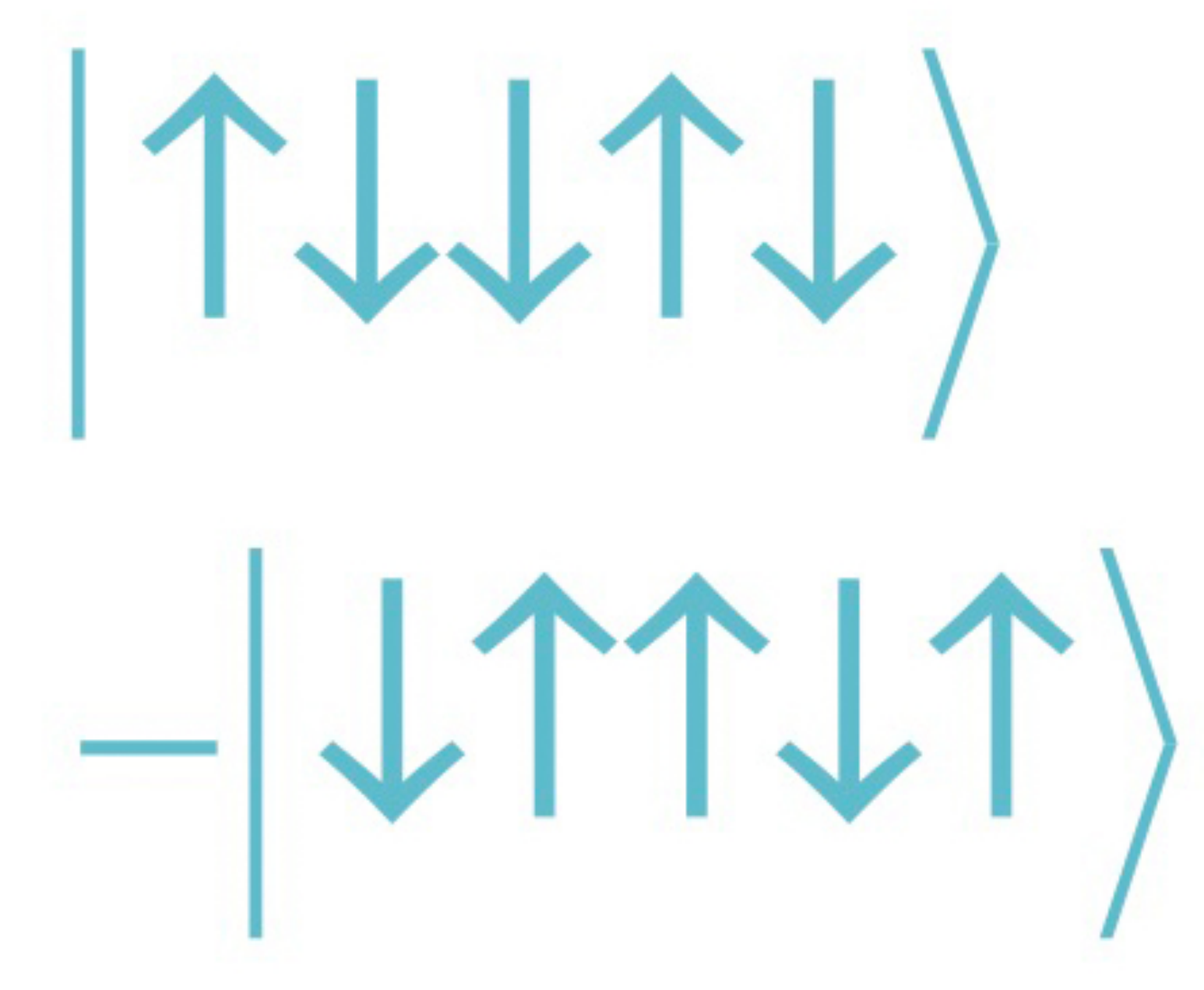}& 
\includegraphics[width=2.3cm]{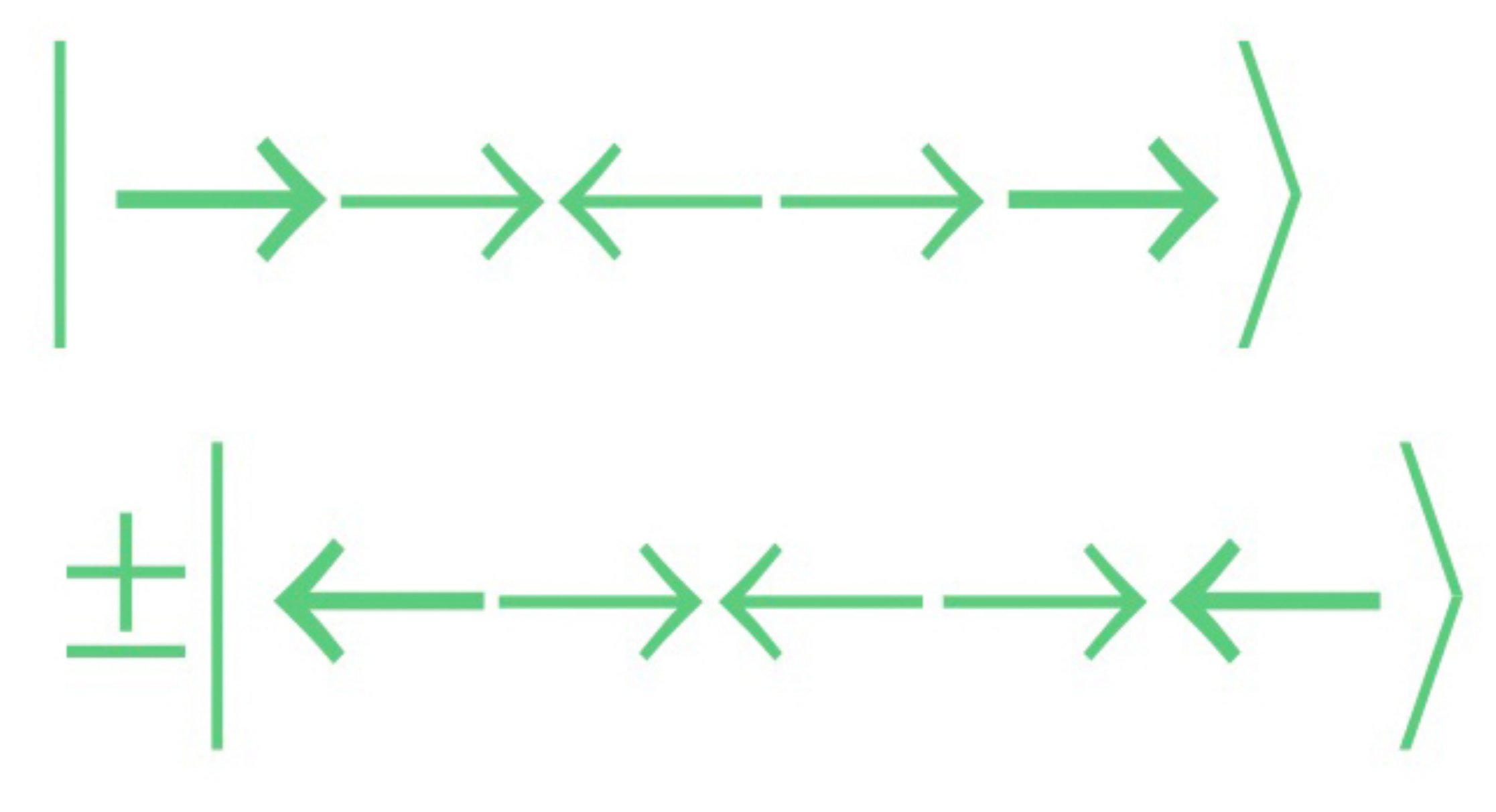}  \includegraphics[width=2.3cm]{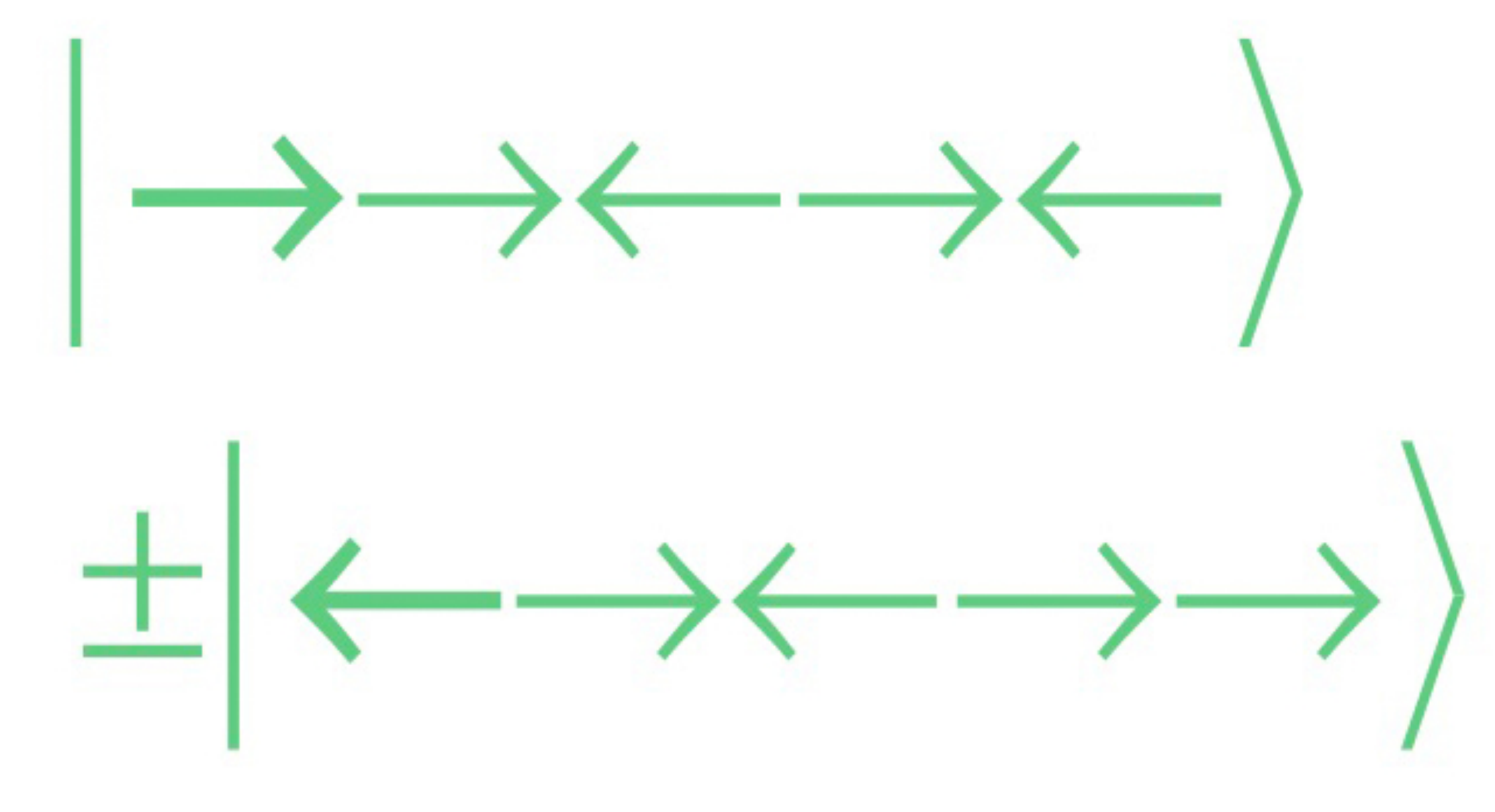}
\\
     \hline
$\langle Z\rangle=0, \langle X\rangle\neq0$& $\langle Z\rangle\neq0$& subharmonic $\langle Z\rangle\neq0$&$\langle Z\rangle=0, \langle X\rangle\neq0$, subharmonic $\langle X_{\mathrm{edge}}\rangle\neq 0$\\
\hline
\end{tabular}
\caption{Properties and diagnostics of the differnt phases of the driven disordered Ising chain. Top row: Eigenspectrum order of the localised phases. In turn, the thermal phase is distinguished from these by its Wigner-Dyson eigenvalue statistics. Middle row: eigenstate multiplets. The paramagnet has unique states with a Poissonian level distribution. For the 0SG and $\pi$SG, the states come in pairs, with an quasidegeneracy with energy difference $0$ and $\pi$ giving the names to the respective phases. The $0\pi$PM has a quadruplet of states, with pairs shifted by quasienergy $\pi$; these are distinguished by the parity of the edge spins. Vertical arrows represent the $z-$direction, horizontal ones the $x$ direction. Bottom row: Autocorrelator  diagnostics. The 0SG and $\pi$SG have non-vanishing $z$-component correlators, with the $\pi$SG showing TTSB in the form of the subharmonic response. The PM and $0\pi$PM lack these, but have non-zero $x$-component signal, with the symmetry-protected topological order of the $0\pi$PM showing TTSB in a subharmonic edge response.}
\end{figure}

\subsection{ The $\pi$SG: Absolute stability and emergent symmetries}\mlabel{s:ases}

Thus far we have discussed the phases of the driven non-interacting Ising chain in detail, and shown that these  persist in the presence of weak Ising symmetric interactions. 
In addition, it is easy to check that the $U_0^{\pi SG}$ also displays a set of ${\mathbb{Z}_2}$ anti-unitary time-reversal symmetries: $\mathcal{T_\alpha} U_0^{\pi SG} \mathcal{T_\alpha}^{-1} = (U_0^{\pi SG})^{-1}$, where $\mathcal{T_\alpha} = K P_\alpha$ and $K$ denotes complex conjugation --- making this a highly symmetric point in parameter space~\footnote{$\mathcal{T}_{y,z}$ are symmetries only for even length chains.}.

Remarkably, the $\pi$SG is not only robust to weak Ising symmetric perturbations, but also to \emph{all} weak $T$ periodic perturbations---including those that break the underlying unitary and anti-unitary Ising symmetries! The perturbed system continues to display Schrodinger cat eigenstates with $\pi$ spectral pairing and, correspondingly, long-range spatiotemporal order. The $\pi$ pairing of the eigenstates is essential for this stability --- in sharp contrast to the $0$SG phase where the addition of an Ising symmetry breaking field immediately lifts the degeneracy between the cat states and gives paramagnetic correlations.

We proceed much as our discussion in Sec.~\ref{sec:interactingIsingDrive} which discussed the stability of the phases in the presence of Ising symmetric interactions. 

First, the stability of MBL for weak enough interactions and strong enough disorder is, by itself, not a consequence of symmetries\footnote{Although symmetries will dictate which types of disorder lead to MBL and which do not.}. As before, this implies the presence of a local finite-depth unitary transformation relating the eigenstates of the perturbed unitary $U_\lambda$, denoted $|\nu\rangle_\lambda$ to those of $U_0$ so that  $|\nu\rangle_\lambda = \mathcal{V}^\dagger_\lambda |\nu\rangle_0$. In the $\pi$SG phase, the eigenstates $|\nu\rangle_0 = |\{d_i\}_\nu,p_\nu\rangle$ are eigenstates of the domain wall operators $D_i$ and the Ising parity $P_x$. Thus, the new eigenstates are eigenstates of dressed domain wall and dressed Ising symmetry operators,  $\tilde{D}_i = \mathcal{V}^\dagger D_i \mathcal{V}$ and $\tilde{P}_x = \mathcal{V}^\dagger P_x \mathcal{V} = \prod_i\tau_i^x$ and can be written more suggestively as $|\{\tilde{d}_i\}_\nu, \tilde{p}_\nu=\pm1\rangle$. This implies that the Floquet l-bits, $\{D_i\}$, and the dressed Ising symmetry $\tilde{P}_x$ must also commute with $U_\lambda$. \emph{Note that because the perturbation explicitly breaks Ising symmetry, it is no longer the case that $\mathcal{V}$ commutes with $P_x$ so that $\tilde{P}_x$ is generically different from $P_x.$ }

Next, we again appeal to the fact that $\tau_i^z$ must anticommute with $U_\lambda$ due to locality and continuity~\cite{vonKeyserlingk2016a, CVS}. Together with the commutation properties of $\tilde{P}$ and $\tilde{D}$ with $U_{\lambda}$, it follows that
\be\mlabel{eq:Ufcanonical}
U_{\lambda}^{\pi SG}=\tilde{P}_x e^{-i\mathcal{H}_z(\left\{  D^\lambda_{i} \right\} )}\punc{,}
\ee
where $\mathcal{H}_z$ is a local functional of $\tilde{D}$ (or, equivalently, an even functional of the $\tau^z_r$'s) as in Eq.~\eqref{eq:lbitz}. 

Written this way, the Floquet unitary Eq.~\eqref{eq:Ufcanonical} clearly has a $\mathbb{Z}_{2}$ symmetry $\tilde{P}_x$ --- although we say it is \emph{emergent} because $ \tilde{P}_x$, in general, depends on the details of the underlying Hamiltonian. $U_{f\lambda}$ similarly has an emergent antiunitary symmetry $\tilde{\mathcal{T}} \equiv  \tilde{P_{x}} \tilde{K}$ where $\tilde{K}$ is complex conjugation defined with respect to the $\tau^\alpha$. 

Thus Eq.~\eqref{eq:Ufcanonical} takes much the same functional form as the model unitary Eq.~\eqref{eq:U0sg2}, and correspondingly its eigenstates exhibit long-range order in the dressed order parameter $\tau^z_r$ (associated with spontaneous breaking of $\tilde{P}_x$). The statements about $\pi$ spectral pairing and the temporal dependence of observables (in particular  $\tau^z(nT)=(-1)^n \tau^z(0)$) also follow directly, establishing the robustness of spatiotemporal order in this phase.

So far our discussion has exactly paralleled the one in Sec.~\ref{sec:interactingIsingDrive}, which discussed the stability of all \emph{four} phases to Ising symmetric interactions. We now explain why the arguments of this section do \emph{not} generalize to the $0$SG and $0\pi$PM phases, so that these destabilize to a short-range correlated paramagnet with no special spectral pairing features upon adding Ising symmetry breaking interactions.  
In other words, once we explicitly break the $\mathbb{Z}_2$ Ising symmetry the distinction between the paramagnet, the SG {\it and} the $0\pi$ PM disappears but the $\pi$SG continues to have a sharp existence. This is pictorially depicted in Fig.~\ref{fig:perturbedDrivenIsing}.

We discuss the $0$SG phase here, and the argument for $0\pi$PM proceeeds analogously. 
The key distinction between the $0$SG and the $\pi$SG lies in the nature of the local unitary $\mathcal{V}_\lambda$ relating the unperturbed and perturbed eigenstates. While the existence of this unitary is implied by MBL (for both phases), $\mathcal{V}_\lambda$ is only local for the perturbed $0$SG phase if one starts with the \emph{symmetry broken} eigenstates of $U_0^{0SG}$, which are product states of $\sigma_i^z$, $|\nu\rangle_0 = |\{s_i^z\}_\nu\rangle$. Note that these are \emph{not} eigenstates of $P_x$. 
Thus, the perturbed eigenstates $|\nu\rangle_\lambda = \mathcal{V}^\dagger|\nu\rangle_0$ are product states of the l-bits $\tau_i^z$ (rather than Ising even domain wall operators). Thus, $U_\lambda$ commutes with the $\{\tau_i^z\}$ operators and generically has the form: 
\begin{equation}
    U_\lambda^{0SG} = e^{- \mathcal{H}_z^o(\{\tau_i^z\})}
\end{equation}
where $\mathcal{H}_z^o$ is a local Ising odd functional of $\tau_i^z:  \mathcal{H}_z^o = \sum_i \tilde{h}_i \tau_i^z + \tilde{J}_{ij}\tau_i^z\tau_j^z + \cdots$ which does \emph{not} have an emergent Ising symmetry. Thus, the degeneracy of the Ising symmetric doublets is lifted by the addition of a $\mathbb{Z}_2$ breaking field, and the perturbed eigenstates are short-range correlated product states. 

The discussion above illustrates why the $\pi$ pairing of the eigenstates is essential for the stability of the $\pi$SG phase to \emph{all} perturbations --- in the absence of any degeneracies, the Ising symmetric cat states remain cat states in the presence of perturbations and continue to the display $\pi$ spectral pairing.  In contrast, in the $0$SG (or $0\pi$PM), the basis choice stemming from the spectral degeneracies implies that the perturbed eigenstates can either look like long-range ordered cats or short-range correlated product states, depending on the nature of the perturbation. 

The stability of the $\pi$SG to all $T$ periodic perturbations was termed {\it absolute stability} in Ref.~\mcite{CVS}. This remarkable aspect of the $\pi$SG phase is akin to the absolute stability of topologically ordered phases like the $\mathbb{Z}_2$ ordered phase of the toric code. 
Indeed, much as the perturbed toric code displays an \emph{emergent} $\mathbb{Z}_2$ symmetry, the generically perturbed $\pi$SG phase has an \emph{emergent} Ising symmetry which then gets broken along with dTTS, in a manner identical to the Ising symmetric $\pi$SG~\footnote{We note that the $\pi$SG is not stable to perturbations that break the $t\rightarrow t+T$ periodicity of the drive. However, demanding that perturbations respect the temporal properties of the unperturbed models is quite natural. Indeed, if one perturbs the static toric code with generic time-varying perturbations then the system heats up and topological order is destroyed.}. This connection between emergent symmetries and stability is quite general. It applies also to MBL which is absolutely stable as a dynamical phenomenon, in contrast with integrability, because of the emergence of local conservation laws in the form of the lbits. 

The stability of the $\pi$SG is one of the central aspects of the phenomenology of the DTC phase, with favorable consequences for experiments. As such, it is worth illustrating this idea  with a concrete example. Consider the perturbed unitary~\cite{ElsePRL, CVS}:
\begin{equation}
    U_{h^z}^{\pi SG} = P_x \exp\left[ -i \sum_j J_i \sigma_j^z \sigma_{j+1}^z + h_j^z \sigma_j^z\right],
\mlabel{eq:longitudinalfieldsonly}
\end{equation}
where the longitudinal fields $h_i^z$ both break the Ising symmetry and make $U_0^{\pi SG}$ interacting. However, for this particular model, the effect of these fields on the properties of the $\pi$SG is benign. One can easily verify that the local unitary $\mathcal{V} = e^{i \sum_j \frac{1}{2} h_j^z \sigma_j^z}$ satisfies $U_{h^z}^{\pi SG} = \mathcal{V}^\dagger U_0^{\pi SG} \mathcal{V}$, so that both the perturbed and unperturbed models have identical spectra and exact $\pi$ spectral pairing\footnote{More generally, $\mathcal{V}$ only relates the perturbed and unperturbed eigenstates and not the unitaries themselves, so that the exact $\pi$ spectral pairing of the $U_0^{\pi SG}$ gets exponentially small corrections.}. The eigenstates $\mathcal{V}^\dagger|\{d_i\},p\rangle$ are still cat states which are eigenstates of the emergent parity operators $\tilde{P}_x =\mathcal{V}^\dagger P_x \mathcal{V}$. The ineffectiveness of the symmetry breaking fields is also indicated by looking at the unitary over two periods, $(U_{h^z}^{\pi SG})^2 = \exp\left[ -2i \sum_j J_i \sigma_j^z \sigma_{j+1}^z\right]$ in which case the fields get ``echoed out" by the action of $P_x$. In contrast, if one similarly adds longitudinal fields to $U_0^{0SG}$ so that $U_{h^x}^{0SG} = \exp\left[ -i \sum_j J_i \sigma_j^z \sigma_{j+1}^z + h_j^z \sigma_j^z\right]$, then it is apparent that the Ising symmetry and `cat-ness' of the eigenstates is destroyed, along with the degeneracy structure of the eigenspectrum. 

While the model in Eq.~\eqref{eq:longitudinalfieldsonly} is particularly transparent, it still has $\mathbb{Z}_2$ anti-unitary symmetries~\cite{CVS}. A family of unitaries that break \emph{all} symmetries of $U_0^{\pi SG}$ is the one presented at the beginning of this section in Eq.~\eqref{eq:TCgeneral} with $h^{x,y,z}\neq 0$~\cite{CVS}.

\subsection{Diagnostics of TTSB in the $\pi$SG}\mlabel{s:dynamics}

We now turn to thinking more explicitly in the time domain and evaluating the merits of identifying the $\pi$SG as a DTC. We have already discussed above 
that at general points in the absolutely stable $\pi$SG phase the emergent order parameter operators,
$\tau^z_i$, change sign every period. Prima facie, this implies the spatiotemporal order sketched in Fig.~1b: spin glass order in space and antiferromagnetic order in time. We now explain what this means from the viewpoint of the three diagnostics of TCs, see Sec.~\ref{sec:DefineTTSB}. 
\noindent

\subsubsection{Temporal and Spatiotemporal order in eigenstates }
\label{sec:pisg_correlators}

As discussed in Sec.~\ref{sec:DefineTTSB}, all single time stroboscopic operator expectation values $\langle O(nT) \rangle$ evaluated in Floquet eigenstates are strictly periodic with period $T$. Hence the the temporal component of the order is invisible to such operators and we must turn to unequal time correlators of operators $O_{r/s}$ localized near sites $r,s$, as discussed in Section~\ref{sec:TTSBdef_spatiotemporal}:
\begin{align}
\langle \nu |f(nT; r,s)| \nu \rangle &\equiv \langle \nu | O_r(nT) O_s |\nu  \rangle \nonumber \\
& = \sum_\mu  e^{-inT (E_{\mu} - E_{\nu})}\langle \nu| O_r|\mu \rangle \langle \mu| O_s |\nu \rangle.
\mlabel{eq:f_pisg}
\end{align}
These are evaluated in the Floquet eigenstates $|\nu \pm \rangle = |\{\tilde{d}\}_\nu, \tilde{p}= \pm 1\rangle$. Time dependence in these correlators (in the infinite volume limit) signals TTSB, Eq.~\eqref{eq:f-spatiotemp}.

Note that if the operators $O_r$ coincide with $\tau_r^z$, then it follows from the equation of motion $\tau_r^z(nT) = (-1)^n \tau_r^z$ that $\langle f(nT,r,s)\rangle  = (-1)^n \prod_{j=r}^{s-1} \tilde{d}_j$. Thus, $f(nT,r,s)$ is $2T$ periodic and shows the time-dependence required of a DTC. However, it is also illuminating (and physically pertinent) to consider generic local operators $O_r$ in the basis of physical spins. In this case, one expands $O_r$ in the basis of l-bits. 

The operator expansion of $O_{r/s}$ in the $\tau^\alpha$ basis generically contains terms that are odd combinations of $\tau^z$s. In the $\pi$SG phase, these have off-digonal matrix elements between $|\nu+ \rangle$ and its parity flipped partner $|\nu - \rangle$, which are separated in quasienergy by $\pi/T$. Then, irrespective of the spatial separation between $r,s$, Ising odd terms in the expansion of $O_r$ connect $|\nu+ \rangle$ to $|\nu- \rangle$, and those in $O_s$ will ``annihilate" this action and connect $|\nu- \rangle$ back to $|\nu+ \rangle$, thereby generically giving a $\pi/T$ frequency component to $f(nT)$ in Eq.~\eqref{eq:f_pisg}, corresponding to period doubling.  

In addition, the off-diagonal terms in the operator expansion for $O_{r,s}$ that involve $\tau^{\{x,y\}}_{r,s}$ locally flip spins in the l-bit basis, thereby making local domain wall excitations near sites $r/s$. Now a crucial point related to our prior discussion of MBL time-glasses: if $r,s$ are held a fixed distance apart in the infinite volume limit, then $f(nT; r,s)$ breaks TTS for {\it any} MBL-Floquet system. The reason is that flipping a spin in the l-bit basis near position $r$ connects eigenstate $|\nu\rangle$ with a different state $|\mu\rangle$ with a quasienergy difference set by $h_{\rm eff}$ a local ``effective field" that captures the cost of flipping an l-bit and depends on the configuration of all other l-bits in state $|\nu\rangle$. If $r,s$ are a finite distance apart, then the expansion of $O_s$ generically also has finite weight on $\tau^x_r$ and can annihilate the action of $O_r$ to reconnect $|\mu\rangle$ to $|\nu\rangle$. The net action due to the particular pair of states $|\nu\rangle, |\mu\rangle$ is a response at a frequency $h_{\rm eff}$ that is incommensurate with the driving frequency~\footnote{Note that the dephasing effects discussed in prior sections are not significant here because a local operator $O_r$ dominantly connects $|\nu\rangle$ to finitely many states $|\mu\rangle$, so the sum in Eq.~\eqref{eq:f_pisg} only involves finitely many frequencies and exhibits a quasiperiodic temporal dependence. The dephasing would result if $f$, in addition, involved a superposition over states $\nu$. }. More heuristically, one can crudely view a Floquet MBL system as a set of weakly interacting localized modes (the effective domain wall operators in this case) each with their own local spectra.  As in the simplest case of 2-level systems whose physics is that of Rabi oscillations, these local subsystems (which are excited by $\tau^{x/y}$) exhibit response at frequencies incommensurate with the driving frequency. The presence of these incommensurate frequencies means $f(nT)$ in {\it all MBL-Floquet systems always look glassy}, although for the $\pi$SG there is generically also a response that is rigidly pinned at $\pi/T$. 

The short distance temporal glassiness, however, goes away when we examine long distances in space by placing the operators arbitrarily far apart in an infinite system, {\it i.e.}, by taking $\lim_{L \rightarrow \infty}$ before examining the limit $|r-s| \rightarrow\infty$.  Since the operator expansions of $O_{r/s}$ are exponentially localized near sites $r/s$,  the off-diagonal terms in the expansion of $O_r$ which create domain-wall excitations near site $r$ {\it cannot be annihilated} by the action of $O_s$ in the limit $|r-s| \rightarrow\infty$ under the assumption of locality. Thus, the only terms that contribute to 
$f(nT; r,s)$ in this limit are diagonal in $\tau^{z}$s. Terms odd in 
$\tau^{z}$ give a response at $\pi/T$ while the even terms give a  response at frequency $0$. Thus we can write
$$\lim_{V\rightarrow \infty} f(nT; r,s) \xrightarrow{\mbox{large } |r-s|} c_0(r;\nu) c_0(s;\nu)  + (-1)^n c_1(r;\nu) c_1(s;\nu) $$
where the second piece reflects the spatiotemporal order of the odd $\tau^z$ terms, as well as  the connected part of the correlation function. The dependence of the coefficients on $r$, $s$ and $\nu$ has been made explicit to emphasize the glassy nature of the order in space. This establishes a connection between the long range spatial order in the eigenstates and the period $2T$ temporal order. The expectation values of the space-time correlator $f(nT;r,s)$ in various limits: (i) equal time, large $|r-s|$; (ii) unequal time, finite $|r-s|$; (iii) unequal time, large $|r-s|$ is qualitatively depicted in Fig.~\ref{fig:Rabi}(a) showing spin-glass, time-glass and spatiotemporal time-crystalline order respectively.   

\begin{figure}
\centering
\includegraphics[width=.4\columnwidth]{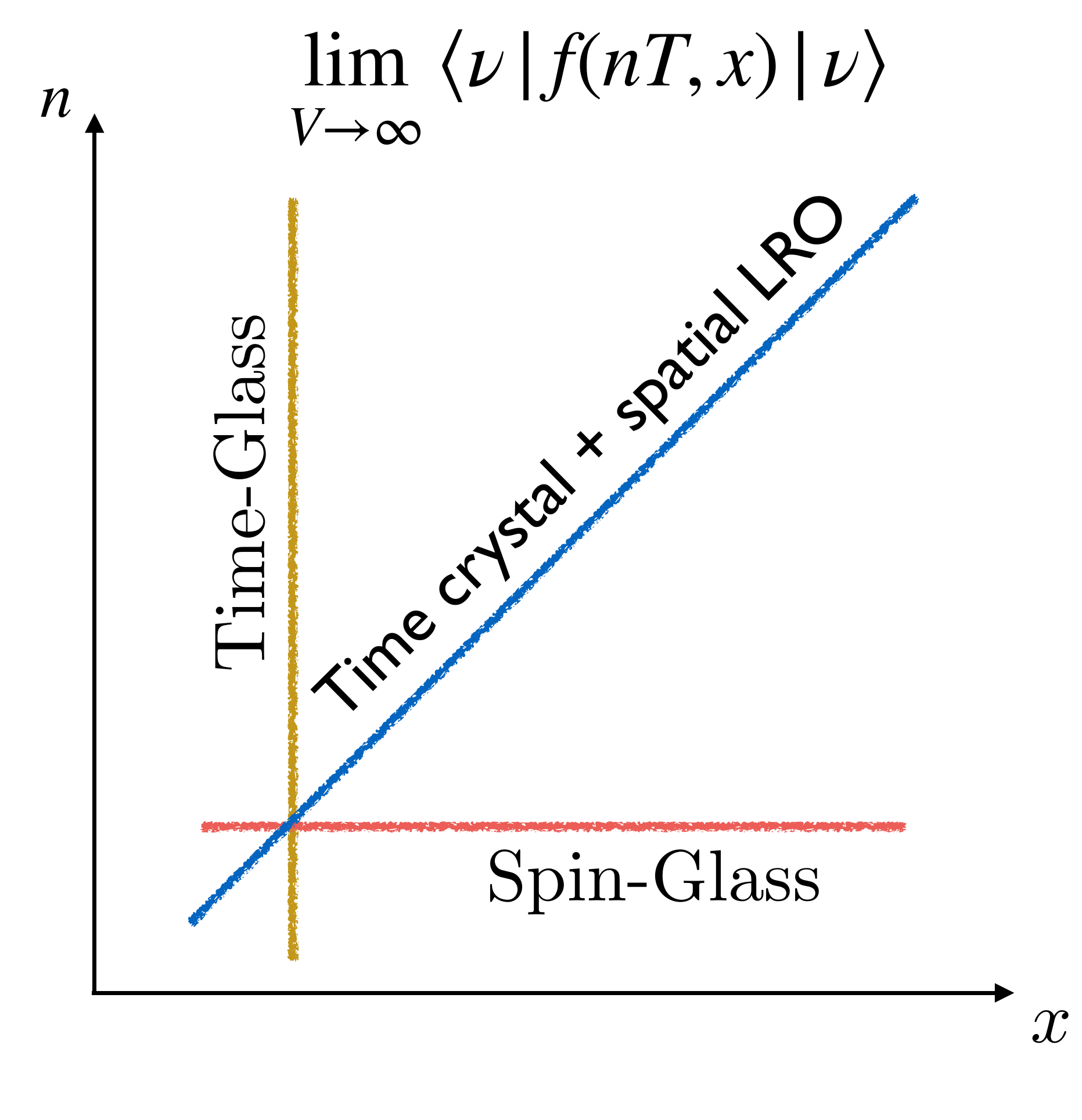}\hspace{15pt}
\includegraphics[width=.45\columnwidth]{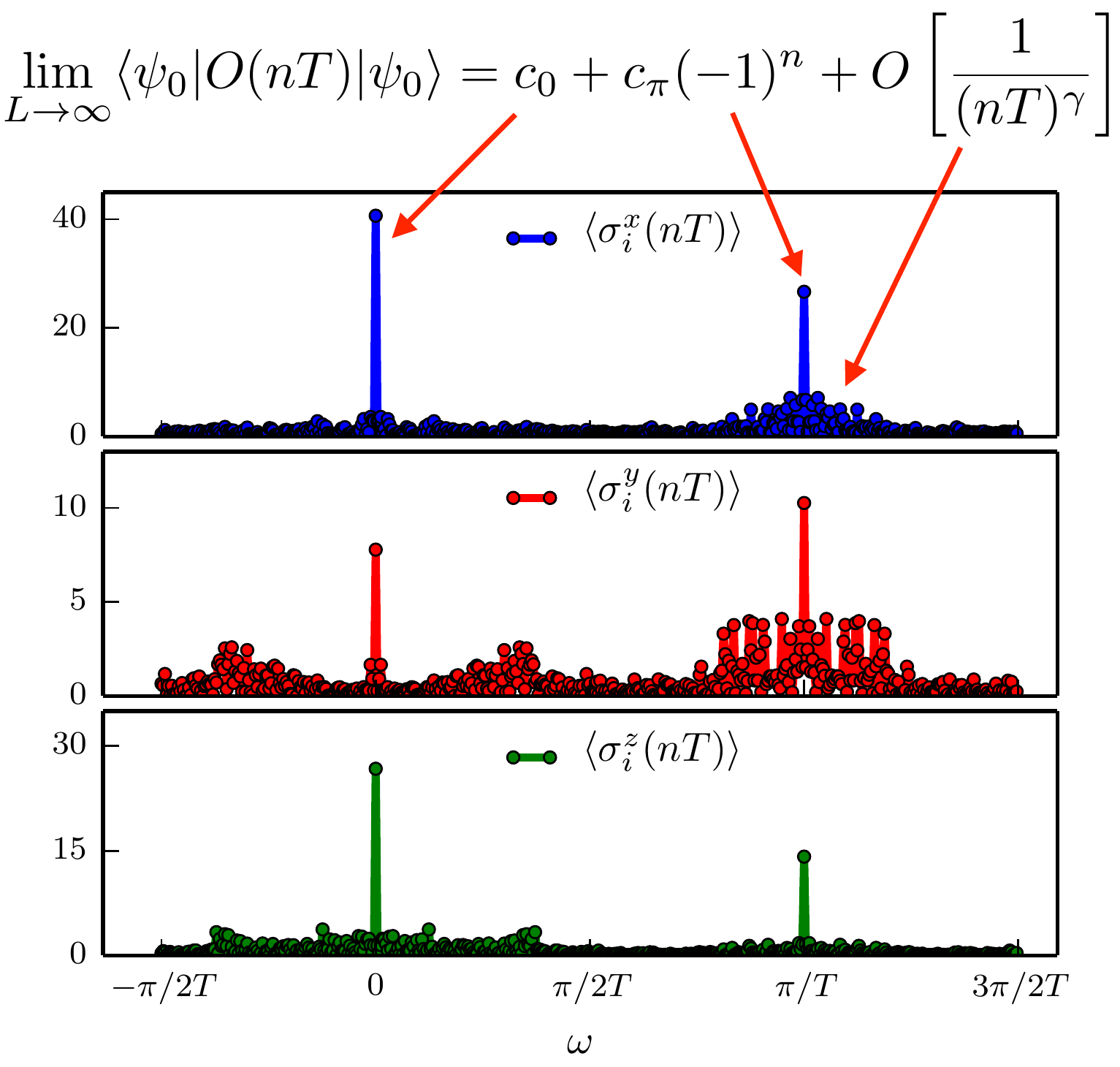}
\caption{(left): Eigenstate correlations and dynamics from generic initial states in the $\pi$SG phase. Unequal space-time correlation functions of local operators evaluated in Floquet eigenstates. The system looks like a spin-glass for finite $t$ and large $x$; it looks like a time-glass for finite $x$ and large $t$; and it looks like a spatiotemporally ordered time-crystal for large $x$ and $t$. See Section~\ref{sec:pisg_correlators} for details. 
(right): Fourier transform over time window $\Delta t=500 T$ of one point time-dependent expectation values $\langle \psi_0|\sigma^{\{x,y,z\}}(nT)|\psi_0\rangle$ in the ``generically'' perturbed model Eq.~\eqref{eq:TCgeneral}. The initial state $|\psi_0\rangle$ is a product state with physical spins $\sigma^\alpha$ randomly pointing on the Bloch sphere and uncorrelated from site to site. As discussed in the text, the response looks ``glassy'' with several incommensurate Fourier peaks in the addition to the peak at $\pi/T$, although we expect these to decay as a power law in the $L\rightarrow \infty,  t \rightarrow \infty$ limit. Data is shown for a single disorder realization in a system of length $L=10$.   }
\mlabel{fig:Rabi}
\end{figure}

Finally, we note that if  we perturb $U_{0}^{\pi SG}$ in a manner that respects an explicit symmetry like $P_x$ or $\mathcal{T}_x$, the resulting models reside in a special submanifold of the absolutely stable phase. The presence of the exact symmetries constrains the form of the dressed $\tau^\alpha$ operators and leads to concrete predictions about the order in and temporal dependence of different operators. Readers interested in how this works can consult Ref.~\mcite{CVS}.

\subsubsection{Susceptibility }
The above analysis can be complemented by considering the second diagnostic for TTSB discussed in Sec.~\ref{sec:TTSBdef_susceptibility}, and adding to $H(t)$ a ``staggered field'' in time of the form
$\epsilon \sum_n (-1)^n V \delta(t -nT)$, where $V$ is odd and diagonal in $\tau^z$. Note that $H(t)$ in the presence of the perturbation has a period of $2T$ (or a frequency of $\omega/2$), and the goal is to see a diverging response to this $2T$ periodic perturbation. 

To this end, consider the time-dependent expectation values of generic local operators $O_r$ (which have a projection on odd $\tau^z$ terms) in the \emph{perturbed} Floquet eigenstates $|\nu\rangle_\epsilon$ for the new period $2T$ unitary. The new unitary can be reshuffled to the form
$U_{\epsilon}(2T)=e^{- i 2\epsilon V} (U^{\pi SG}_{\lambda})^2$. This problem looks like the classic Ising symmetry breaking problem. At $\epsilon=0$,
$U_{\epsilon}(2T)=(U^{\pi SG}_{\lambda})^2$ has two degenerate states in the infinite volume limit (the $\pi$ pairing of cats in $U^{\pi SG}_\lambda$ translates to a degeneracy of the cats in the squared unitary). If $V$ breaks the symmetry between two members of the  doublet, then the perturbed period $2T$ eigenstates $|\nu\rangle_\epsilon$ just look like product states of $\tau^z$ in this limit, and are thus superpositions of the opposite parity eigenstates of $U_{f\lambda}$. 
Ising odd terms in the expansion of $O_r$ has a non-zero expectation value in these perturbed states. Coupled with the equation of motion for Ising odd operators, we get:
 $$  \lim_{ \epsilon \rightarrow 0} \lim_{ L \rightarrow \infty} {}_\epsilon\langle \nu |O_r(nT)| \nu \rangle_\epsilon = b_0(r;\nu) + b_1(r;\nu) (-1)^n $$
On the other hand, the opposite order of limits gives $\lim_{ L \rightarrow \infty}   \lim_{ \epsilon \rightarrow 0} {}_\epsilon\langle \nu |O_r(nT)| \nu \rangle_\epsilon =  b_0(r;\nu)$. 

Thus, an infinitesimal perturbation which breaks the dTTS symmetry corresponding to time translations by $T$ down to time translations by $2T$ is sufficient to induce period two oscillations (in the right order of limits), analogous to usual notions of SSB wherein infinitesimal Ising symmetry breaking perturbations can induce a net magnetization. Note, however, that a non-trivial response to a period two perturbation requires $V$ to also have a component that is odd with respect to the emergent Ising symmetry \emph{i.e.} the perturbation $V$ must break \emph{both} the emergent Ising symmetry and dTTS --- yet another restatement of spatiotemporal order.

\subsubsection{Dynamics from generic initial states}\mlabel{ss:dynamicsSRE}
Finally, we turn to the third diagnostic of TTSB involving evolution of generic operators starting from  general initial states rather than eigenstates, as was discussed in Sec.~\ref{sec:TTSBdef_initstate}. This measure is the most relevant for experiments that do not have access to abstract many-body eigenstates or l-bits. 

For concreteness, consider starting from a generic short-range correlated state like a product state of the physical spins which is a superposition of Floquet eigenstates. We measure one-point expectation values of generic local operators from these states: $\langle \psi_0| O_r(nT)|\psi_0 \rangle$. As the expectation value is not taken in an eigenstate of $U_F$, one can obtain aperiodic behavior signaling dTTSB. However, as discussed in Sec.~\ref{sec:TTSBdef_initstate}, such expectation values are expected to decay to a constant at late times in generic MB systems with no special eigenspectrum correlations. 
Instead, the special $\pi$ spectral pairing in the $\pi$SG leads to non-trivial temporal dependence in these expectation values.

If we consider the perturbed $\pi$SG with all symmetries broken, then the expansion of a local operator $O_r$ in the l-bit basis generically has Ising even and odd combinations of diagonal $\tau^z$ operators, along with off-diagonal terms like $\tau^{x/y}$. The even/odd diagonal operators commute/anticommute with $U_\lambda^{\pi SG}$, corresponding to synchronized/period doubled dynamics with Fourier peaks at $0$ and $\pi/T$ respectively\footnote{We are assuming here that the initial state exhibits a non-zero expectation value for the emergent order parameter, i.e. $\langle \psi_0|\tau^z_i|\psi_0 \rangle \ne 0$, which is generically true for short range correlated states in the physical basis of spins.}. 

In addition, off-diagonal spin operators like $\tau_r^x$ precess in an effective field that depends on the configuration of l-bits, as discussed in Section~\ref{sec:mbldynamics}. These give rise to \emph{glassy} time dynamics, with additional incommensurate Fourier peaks in the power-spectrum corresponding to the frequencies of precession.  However, because the initial state is a superposition of exponentially many l-bit eigenstates, and as the effective frequencies of precession are different for the different l-bit states, these dephase against each other. The standard logarithmic in time dephasing dynamics characteristic of MBL systems then predicts that the aperiodic behavior stemming from the additional Fourier peaks decays with a power law envelope $\sim t^{-\gamma}$, where $\gamma>0$ depends on the localization length\mcite{SerbynQuenches}. 

{\it Thus, at intermediate times, the $\pi$SG phase looks like a time-glass, with an additional quantized ``crystalline" response at $\omega = \pi/T$; however, upon waiting longer and longer, the strength of the incommensurate peaks decay, isolating the crystalline response.} This is depicted in Fig.~\ref{fig:Rabi}(b)

One subtlety deserves further exposition. In a finite sized system, $\tau^z$ only anticommutes with the Floquet unitary up to exponentially small in $L$  corrections, which in turn introduce corrections to the equation of motion: $ \tau^z(nT) = (-1)^n\tau^z(0) + O(e^{-L})$. This leads to exponentially small shifts in the spectral pairing at $\pi/T$ which vary randomly between pairs of eigenstates. Thus, if one waits a time $t \sim e^{L}$ that is long enough to (i) resolve the exponentially small many-body level spacings and (ii) to resolve the shifts in the spectral pairing away from $\pi/T,$ both the peak at $\pi/T$ and the extra incommensurate peaks almost entirely decay away due to usual dephasing mechanisms leaving behind aperiodic oscillations with a magnitude of $O(e^{-L})$. The precise details of the time dependence reflect the choice of initial state and disorder realization.

We can formalize the above in two non-commuting limits: (a) $\lim_{ t\rightarrow \infty} \lim_{ L\rightarrow \infty}$ and (b) $\lim_{L\rightarrow \infty} \lim_{ t\rightarrow \infty}$. While (a) characterizes the ``intrinsic'' quench dynamics of this phase, experiments as best only have access to limit (b). In (b) the late time aperiodic oscillations with envelope $O(e^{-L})$ discussed above also go away, and the one-point functions are constants. In (a), we never reach times of $O(e^{L})$ and instead observe persistent oscillations with period $2T$ out to $t \rightarrow \infty$ with all additional incommensurate oscillations decaying away as a power of time. 

Thus, the intrinsic dynamical response of this phase is characterized by a single rigid Fourier peak at $\omega = \pi/T$ which goes along with formally exact spectral pairing at $\pi/T$ and LRO in $\tau^z$. In this limit, the late time state exhibits a precisely doubled period for every single realization of disorder and combined space-time measurements would lead precisely to the kind of snapshot sketched in Fig.~\ref{fig:perpetuum_mobile}(d). 

\subsection{Generalizations}\mlabel{s:gen}

The $\pi$SG involves an emergent broken $\mathbb{Z}_2$ symmetry. There are analogous phases with other symmetries.
Ref.~\mcite{vonKeyserlingk2016b}  presents a family of models with an explicit global symmetry group $G$ which exhibits eigenstate long-range order,  protected spectral pairing and temporal crystallinity.  Many of these models are absolutely stable to local perturbations. Meanwhile bosonic SPT Floquet drives\mcite{vonKeyserlingk2016a,Else16,Potter16,Harper16} which generalize the $0 \pi$ PM are not stable to the inclusion of symmetry breaking perturbations, but in the presence of the protecting symmetry they exhibit time crystallinity at their edges. 

\subsubsection{$\mathbb{Z}_{n}$ and non-abelian models }
As a specific example, consider models with a global $\mathbb{Z}_n$ symmetry\mcite{vonKeyserlingk2016b, MoessnerParafermion, RussomannoClock}. There are $n$ possible phases with completely spontaneously broken symmetry\mcite{vonKeyserlingk2016b}, labelled by $k=0,1,\ldots n-1$. The eigenvectors of the corresponding unitary are the $\mathbb{Z}_{n}$ equivalents
of cat states i.e., macroscopic superpositions of $n$ spin configurations. In cases with $k\neq0$, and in the presence of $\mathbb{Z}_{n}$ symmetry, the spectrum consists of multiplets of $n$ cat states appearing in $n/g$ distinct groups each with degeneracy $g\equiv\text{gcd}\left(n,k\right)$. The $n/g$ distinct groups are split by quasienergy multiples of $2\pi g/nT$. As for the $\pi$SG, some
of these statements survive even when $\mathbb{Z}_{n}$
symmetry is explicitly broken. In particular, while the $g$-fold degeneracy for each group of cat states can readily be broken,  it remains the case that each eigenstate is paired in a multiplet of $n/g$ related cat states, separated by quasienergy $2\pi g/nT$. A similar statement holds for the non-abelian models in Ref.~\mcite{vonKeyserlingk2016b}. These more general drives have an explicit unitary non-abelian symmetry $G$,
and are classified by an element of the center of the group $z\in Z(G)$. Let $q$ denote the order
of $z$. The spectrum consists of $q$ groups of $G/q$ degenerate cat-like states, and  the $q$ groups are separated by quasi-energies which are multiples of  $2\pi/qT$. The $|G|/q$ degeneracy at each quasienergy can once again be lifted using symmetry breaking perturbations, but each eigenstate is still paired with $q$ cat state partners, split by quasienergy multiples of $2 \pi / qT$.

\section{Prethermal time-crystals}
\mlabel{sec:prethermalTC}

The previous sections laid out the necessary ingredients --- MBL and periodic driving --- for obtaining infinitely long lived many-body time-crystals, in infinitely large systems. These conceptual limits are essential for formally defining a dynamical \emph{phase} of matter in an asymptotic sense, but are difficult to achieve in practice in an experimental setting.  
Nevertheless, in many systems of interest, it may be possible to engineer an extremely long lifetime for the driven system by creating a suitably large separation of scales in the system parameters. 

As discussed previously, the primary issue when considering a driven many-body system is that of heating to a trivial infinite temperature state with no long-range order. However, one can intuitively imagine that if the drive frequency $\omega$ is much larger than the local energy scales in the system, $\sim h$, then the absorption of a single ``energy quantum" requires many local rearrangements of the degrees of freedom under the action of the drive. This is a high order process that can take a very long time when the drive acts as a sum of local terms --- even in a clean system with no disorder and no MBL 

The dynamics of the system in the interim can be usefully thought of as energy preserving or, more concretely, as being generated by an effective time-independent ``prethermal Hamiltonian", $H_{\rm eff}$, which to leading order in $1/\omega$ is just the time-average of $H(t)$. This intuition was formalized in a set of recent papers which showed that the time scale for heating $t_*$ could be \emph{exponential} in the separation of scales: $t_* \sim \exp[{ \omega/h}]$~\cite{DimaPrethermal_linearresponse, DimaPrethermal_Heff, DimaPrethermal_rigorous, MoriPrethermal1, MoriPrethermal2}. Within this long time window, the system can display non-trivial dynamics upon starting from certain low-temperature initial states, including realizing a ``prethermal Floquet time crystal"~\cite{ElsePrethermal, LuitzInfT}. 

The theory of prethermalization also has important implications for symmetry-protected TCs in time-independent Hamiltonians. As discussed previously, cTTSB can only be realized in the presence of an additional symmetry which is broken along with continuous time-translation, as in a ferromagnet showing precession in an external field. However,  demanding the presence of a symmetry requires fine tuning, and any realistic system will inevitably be plagued with stray symmetry-breaking interactions. Nevertheless, once can again engineer a regime where a symmetry (distinct from energy conservation) is present for exponentially long times~\cite{DimaPrethermal_rigorous}, thereby enabling cTTSB for a long time~\cite{ElsePrethermal}. 

Finally, it was recently pointed out that systems can also show prethermalization \emph{without} temperature~\cite{LuitzInfT}. This apparent oxymoron is resolved if one engineers a drive to display a long-lived symmetry, which may or may not also be accompanied by the presence of a prethermal Hamiltonian and energy conservation. This mechanism can show long-lived oscillations in global observables even upon starting from high temperature initial states, and is of relevance to understanding recent experiments on NMR systems~\cite{RovnyPRL, RovnyPRB}, as discussed in Section~\ref{sec:nmrexp}. 

We next discuss these various developments in more detail. 

\subsection{Slow heating within linear response theory} 
Consider a periodic Hamiltonian $H(t+T)=H(t)$ with period $T = 2\pi/\omega$. For the purposes of this section, we have in mind translationally invariant (or weakly disordered) Hamiltonians that are not many-body localized. 
It is convenient to split $H(t)$ as 
\begin{equation}
    H(t) = H_0 + V(t), 
\end{equation}
where $H_0 \equiv \frac{1}{T} \int_0^T dt\; H(t)$ is the time-averaged Hamiltonian, and $V(t) = H(t)-H_0$ is the drive. When $||V(t) ||\ll ||H_0||$, the drive is ``weak" and the heating can be analysed using familiar tools from linear response theory. By $||O||$ we mean the Hilbert-Schmidt operator norm of $O.$

We consider Hamiltonians and drives that are global but act as a sum of local terms, 
so that $H_0 = \sum_i h_i$ and $V(t) = \sum_i v_i(t)$, where $h_i$ and $v_i(t)$ are local operators with bounded range near site $i$ on a lattice. 
We denote by $h$ the maximum local operator norm, $h = \max_i ||h_i||$, and this represents the maximum energy that can be gained or lost by the system as a result of local rearrangements of any state. 

We assume the system is initially in thermal equilibrium at inverse temperature $\beta$ with respect to $H_0$. It is conceptually simplest to think of a monochromatic drive so that $V(t) = A \cos(\omega t) \sum_i v_i$.
If the drive is weak enough, $A ||v_i|| \ll h$, then the energy absorbed can be described within linear response theory (LRT) as $dE/dt = A^2 \omega \sigma(\omega)$, where $\sigma(\omega)$ is the dissipative part of the linear response function:
\begin{equation}
    \sigma(\omega) = \sum_{ij} \sigma_{ij}(\omega) = \sum_{ij} \int_{\infty}^{\infty} dt e^{i\omega t} \langle [v_i(t), v_j]\rangle_\beta. 
\end{equation}
It was shown in Ref.~\mcite{DimaPrethermal_linearresponse} that $\sigma(\omega)$ (and hence the heating rate per site) is bounded to be exponentially small in $\omega$: $$\frac{|\sigma(\omega)|}{N} \leq C \exp(-\kappa \omega/h),$$ where $N$ is the number of degrees of freedom, and $\kappa >0$ and $C$ are constants. This formalizes the intuition for slow heating expressed previously: the energy exchange between the system and the periodic drive is quantized in units of $\omega$, while the energy change from local rearrangements is bounded by $h$; when $\omega \gg h$, absorbing a single quantum $\omega$ requires a high-order cooperative process involving $n\sim \omega/h$ local changes, corresponding to a heating rate exponentially suppressed in $n$. 

\subsection{Effective prethermal Hamiltonians}
\mlabel{sec:prethermalTC-effH}
In fact it is possible to go beyond linear response theory and demonstrate not only slow heating, but also the existence of an effective \emph{prethermal Hamiltonian} that is approximately conserved during the period of slow heating~\cite{DimaPrethermal_Heff, DimaPrethermal_rigorous, MoriPrethermal1, MoriPrethermal2}. This can also be done in the regime of strong driving beyond LRT, say with $||v_i(t)||\sim h$, as long as $\omega \gg ||v_i(t)||, h.$

The central idea is to try to construct a (time-independent) effective Hamiltonian, $H_{\rm eff}$, that generates the time-evolution  over one period:
\begin{equation}
    U(T) = \mathcal{T}e^{-i\int_0^T dt H(t)} \equiv e^{-i H_{\rm eff} T}.
\end{equation}
Thus, the stroboscopic properties of the driven system can be understood via the eigenspectrum properties of $H_{\rm eff}$. 
While such a Hamiltonian can always be formally defined, we are interested in cases where $\Heff$ is a (quasi) \emph{local} Hamiltonian.

In the limit of high-frequencies, there are various approaches for systematically constructing $\Heff$ order by order in a high-frequency expansion, formally known as a Magnus expansion~\cite{PolkovnikovFloquetReview}: 
\begin{equation}
    \Heff = \sum_{m=0}^\infty \frac{1}{\omega^m}\mathcal{H}_{m}. 
    \mlabel{eq:magnus}
\end{equation}
The leading term in the expansion is simply the time-averaged Hamiltonian,  $\mathcal{H}_0 = H_0$. The Magnus expansion can be convergent in non-interacting systems when the driving frequency is large compared to the bandwidth of single-particle excitations. However, in an interacting system, the many-body bandwidth is extensive in the system size while the frequency $\omega$ is $O(1)$; hence the expansion is generically expected to have a shrinking radius of convergence with increasing system size~\cite{PolkovnikovFloquetReview}. 
This also follows from Floquet ETH --- if \emph{all} eigenstates of $U(T)$ look like maximally random infinite temperature states, then these cannot also be eigenstates of a local time-independent Hamiltonian.

While it is not thermodynamically sensible to drive the system at frequencies comparable with the (extensive) MB bandwidth, one can nevertheless consider cases where $\omega$ is much bigger than the \emph{local} energy scales in the problem.  In this case, it has been shown that the Magnus expansion looks convergent up to an order $n_o\sim(\omega/g)$, but starts to diverge at higher orders due to the eventual heating~\cite{DimaPrethermal_rigorous, DimaPrethermal_Heff, MoriPrethermal1, MoriPrethermal2}. Here $g$ is the maximum local energy scale set by the combination of $H_0$ and $V(t)$, since neither is assumed to be smaller than the other: $g \leq ||h_i + v_i(t)||$. Again, the scaling of $n_o$ intuitively follows from the need to make $O(\omega/g)$ local rearrangements to absorb a quantum of energy, a process that only takes place at order $n_o \sim~(\omega/g)$ for a drive that acts as a sum of local terms. 

The convergence of the Magnus expansion up to some order $n_0$ implies that the best approximation to a local $H_{\rm eff}$ is obtained by truncating the Magnus expansion at this order, $$\Heff^{(n_0)} = \sum_{m=0}^{n_0} \frac{1}{\omega^m} \mathcal{H}_m.$$ Moreover, since each higher order term in the expansion contains longer and longer ranged spatial rearrangements, the effect of truncating the higher order terms should only be felt at later times (due to locality and the Lieb-Robinson theorem).  Thus, each truncation order $n$ defines a time-scale $t_n$, so that the difference between the exact time-evolution and time evolution under the truncated $\Heff^{(n)}$ is only sizable at times longer than $t_n$. 

More precisely, Refs.~\cite{DimaPrethermal_rigorous, DimaPrethermal_Heff, MoriPrethermal1, MoriPrethermal2} showed that the $n_o$th order effective Hamiltonian  $\Heff^{(n_0)}$ is approximately conserved up to a time exponentially large in $n_0$, meaning that:
\begin{equation}
    \frac{1}{N} || U^\dagger(t) \Heff^{(n_0)}U(t) - \Heff^{(n_0)}|| \leq C 2^{-{n_0}}t, \;\;\;\;\,\, \quad \qquad n_o \sim \omega/g
    \mlabel{eq:pretherm_constime}
\end{equation}
where $C$ is a constant. The RHS is only sizeable, \emph{i.e.} $O(1)$, when $t_* \sim t_{n_0} \sim \exp(n_0) \sim \exp(\omega/g)$.  This defines the exponentially long time scale with approximate energy conservation. 

Note also from Eq.~\eqref{eq:magnus} that the difference between $\Heff^{(0)} = H_0$ and $\Heff^{(n_0)}$ is $O(1/\omega)$. Thus, if we simply work with the leading order time-averaged Hamiltonian $H_0$ -- often easiest to do in practice ---  the energy density looks approximately constant for an exponentially long time, up to a correction of $O(1/\omega)$ .

It may be useful to mention here that while Refs~\mcite{DimaPrethermal_rigorous} and ~\mcite{MoriPrethermal1} ultimately obtain the same results regarding the existence of a prethermal Hamiltonian for exponentially long time-scales, the method of analysis used in the two works is quite different. Our discussion above follows that in \mcite{MoriPrethermal1}. Instead, \mcite{DimaPrethermal_rigorous} takes the approach of constructing a time-dependent unitary transformation $Y(t)$, so that the Hamiltonian in the `rotating frame' is 
\begin{equation}
    H_{\rm rot}(t) = Y(t) H(t) Y^\dagger(t) -i Y(t) \partial_tY^\dagger(t). 
    \mlabel{eq:dimarotate}
\end{equation}
The transformation $Y(t)$ is constructed order by order in a high-frequency expansion $Y(t) = \prod_{m=1}^{\infty} Y_m(t)$, with $Y_0 = \mathbb{I}$. The goal of each successive rotation is to reduce the time-dependence in $H_{\rm rot}(t)$ by a factor of $1/\omega$. The effective (time-independent) Hamiltonian at $n$th order is obtained as the time-average: $\Heff^{(n)} = \frac{1}{T}\int_0^T dt H_{\rm rot}^{(n)}(t)$, where $H_{\rm rot}^{(n)}(t)$ is the Hamiltonian $H(t)$ transformed by a truncated $Y^{(n)}(t) = \prod_{m=1}^{n} Y_m(t)$ according to Eq.~\eqref{eq:dimarotate}. In fact, the $H_{\rm eff}$ constructed in this manner agrees -- at each order -- with that obtained from a Magnus expansion as in Eq.~\eqref{eq:magnus}. Once again, there is an optimal level of truncation for $Y(t)$, which gives a prethermal $\Heff^{(n_0)}$ valid for exponentially long times in the frequency.

\subsection{Prethermal discrete time crystals from symmetry breaking}
\label{sec:prethermalSSB}
The previous subsection described how a system under a high frequency drive can be approximately described by an effective time-independent Hamiltonian --- at least up to a prethermal time $t_*$ that is exponentially long in the frequency. Thus, if $\Heff$ is a Hamiltonian with non-trivial phase structure and dynamics, the driven system will inherit these until time $t_*$. Indeed, this type of ``Floquet engineering" is often used to generate desired Hamiltonians in a laboratory setting where $t_*$ could, in practice, be longer than the experimental coherence times~\cite{FloqEngineeringReview, EckardtEngineering}. 

However, phases realized in the manner above are so-called Type I Floquet phases, since they exist already for a static system described by $\Heff$. The natural question then is whether one can also realize Type II prethermal Floquet phases in clean, many-body systems -- namely, prethermal versions of phases that are genuinely new to the driven setting such as the $\pi$-SG/DTC. It was shown in Ref.~\mcite{ElsePrethermal} that this is indeed possible via a suitable generalization of the technology in Refs.~\mcite{DimaPrethermal_rigorous, MoriPrethermal1}.  The prescription in \mcite{ElsePrethermal} relies on (i) the existence of a long-lived prethermal Hamiltonian $H_{\rm eff}$ for large enough driving frequencies, \emph{and} (ii) the presence of an emergent Ising symmetry in $H_{\rm eff}$, with a symmetry breaking transition at a temperature $\beta_c$ defined with respect to the energy density of $H_{\rm eff}$. With these ingredients, symmetry broken initial states at low temperatures with respect to $\Heff$ show TC order for a long prethermal time scale. 

The basic picture of Ref~\mcite{ElsePrethermal} is similar in spirit to the construction of the MBL $\pi$-SG phase, up to prethermal time scales. The $\pi$-SG phase requires the local energy scale of some terms in $H(t)$ (those enacting the approximate $\pi$ flip) to be comparable to the drive frequency --- even though the existence of Floquet MBL nominally requires frequencies to be larger than local bandwidths. Likewise, we will again need some local energy scales to be comparable to $\omega$ to obtain prethermal TCs, even though prethermalization generally requires frequencies to be larger than local bandwidths. 

The redeeming feature enabling MBL TCs is that the ostensibly problematic ``flip" part of the drive can be eliminated when looking over two periods (or, equivalently, by transforming to a ``toggling frame"), so that all terms in $U(2T)$ are small compared to (half) the frequency and thus the usual theory of Floquet MBL can be applied. Likewise, one can split the putative prethermal drive as $H(t) = H_{L}(t) + H_{S}(t)$, where the local energy scales of terms in $H_{L}(t)$ are large, \emph{i.e.} comparable to the frequency $\omega$, while the local scales in $H_{S}(t)$ are small compared to $\omega$ and thus within the purview of the theory of prethermalization. Importantly, $H_L(t)$ is chosen so that there is an integer $N$ for which $X^N =\mathbb{1}$, where $X = \mathcal{T}e^{-i\int_0^T dt H_L(t)}.$ For the Ising DTC, $N=2$ and $X= P_x$. 

Thus, if one looks over two periods, then $U(2T)$ is well described by an effective prethermal Hamiltonian $\Heff$ for times exponentially large in $\omega/2$. However, to obtain a prethermal time-crystal, we also need something more interesting to happen over one period. In fact, this is possible since it turns out that, 
\begin{equation}
        \mathcal{V} U(T) \mathcal{V}^{\dagger} \approx X e^{iDT},
     \mlabel{eq:prethermalU}
\end{equation}
where $\mathcal{V}$ is a local (time-independent) unitary transformation, $D$ is a time-independent Hamiltonian, and $[D, X] = 0$~\mcite{ElsePrethermal}. This is approximately true up to times $t_*\sim \exp(\omega/g)$. 
This construction implies that the system has an \emph{emergent} symmetry $\tilde{X} = \mathcal{V}^\dagger X \mathcal{V}$ which commutes with $U(T)$, similar to our discussion on absolute stability and an emergent Ising symmetry in Sec.~\ref{s:ases}. Likewise, from this construction, it follows that $\mathcal{V} U(2T) \mathcal{V}^\dagger \approx e^{2iDT}$, so that the effective prethermal Hamiltonian describing the evolution of $U(2T)$ is $\tilde{D} = \mathcal{V}^\dagger D \mathcal{V}$, which also has the same emergent symmetry $\tilde{X}$.  

In general, in a clean many-body system, $\tilde{D}$ will be a non-integrable quasi-local Hamiltonian that thermalizes. Thus, if the system is observed at times that are multiples of $2T$, local observables (rotated by $\mathcal{V}$) equilibrate to a thermal state of $\tilde{D}$ on some timescale $t_{\rm th}$, that can typically be much shorter than $t_*$. The inverse temperature $\beta$ of the initial state is set with respect to $\tilde{D}$ according to $\langle \psi_0|\tilde{D}|\psi_0\rangle = \frac{1}{Z} \mbox{Tr} e^{-\beta \tilde{D}} \tilde{D}$.
In particular, if $\tilde{D}$ has a symmetry breaking transition, so that $\tilde{X}$ is spontaneously broken above inverse temperature $\beta_c$, a local order parameter such as a magnetization $M_i = \mathcal{V}^\dagger D \sigma_i^z \mathcal{V}$ thermalizes to a non-zero $O(1)$ value on starting from a symmetry broken initial with $\beta > \beta_c$\footnote{This discussion does not consider slow phase ordering~\cite{Bray} or coarsening dynamics of a symmetry broken system \emph{en route} equilibrium}.  By contrast, the order parameter evaluated starting from initial states at high temperatures thermalizes to zero:
\begin{equation}
    \lim_{V\rightarrow \infty} \langle \psi_0|M_i(2nT)|\psi_0\rangle \xrightarrow{t_{\rm th}\ll 2nT  \ll t_* } \langle M_i\rangle_\beta\;\; 
\begin{cases}
= 0 &  \beta < \beta_c\\
\neq 0 &  \beta > \beta_c
\end{cases}      
\end{equation}

By definition, the order parameter transforms non-trivially under the symmetry which, in turn, enables the realization of a prethermal time-crystal. For times $ t \gg t_{\rm th}$, the system toggles between positive and negative magnetizations, $\pm \langle M_i\rangle_\beta$, at even and odd times. Thus the amplitude of oscillations in the prethermal TC is set by thermal expectation value of the magnetization in the initial state, which in turn is set by the energy density of the initial state with respect to $D$. For example, in the Ising case,  $X^\dagger M_i X = -M_i$ with $X = P^x_\pi.$ Thus,
\begin{align} 
\lim_{V\rightarrow \infty} \langle \psi_0| M_i([2n+1]T)  |\psi_0 \rangle &= \lim_{V\rightarrow \infty} \langle \psi_0| e^{-i\tilde{D}[2n+1]T} \tilde{X}^\dagger  M_i \tilde{X} e^{i\tilde{D}[2n+1]T}|\psi_0 \rangle \nonumber\\
&= -  \lim_{V\rightarrow \infty}\langle\psi_0| e^{-i\tilde{D}[2n+1])T}   \tilde{M}_i  e^{i\tilde{D}[2n+1])T}|\psi_0 \rangle  \nonumber \\
&\xrightarrow{t_{\rm th} \ll 2nT \ll t_*} - \langle M_i \rangle_{\beta} = - \langle M_i[2nT]  \rangle,
\end{align}
thereby showing period two oscillations in time with amplitude $\langle M_i\rangle_\beta$. 

To summarize, the main ingredients for realizing prethermal Floquet DTCs \`a la Ref.~\mcite{ElsePrethermal} are: 
\begin{itemize}
    \item The existence of an effective prethermal Hamiltonian $\tilde{D}$ that is thermalizing. This approximately governs the evolution of $U(2T)$ for times exponentially large in system parameters 
    
    \item The existence of an emergent symmetry $\tilde{X}$ in the effective Hamiltonian, accompanied by a symmetry breaking transition at a critical temperature. Note that equilibrium constraints on symmetry breaking from Peierls-Mermin-Wagner type theorems will constrain the classes of systems for which prethermal TCs may be obtained using this prescription. For example, short range systems do not show SSB in one spatial dimension, and hence cannot realize a prethermal TC in this framework. By contrast, the existence of MBL is on most solid theoretical ground in short-ranged 1D systems. 
    
    \item Initial states that are symmetry broken and at energy densities below the critical temperature of $D$, so that order parameters evaluated in such such states thermalize to a non-zero value $\langle O\rangle_\beta$.

\end{itemize}
The last point furnishes one signature that can be used, in principle, to distinguish between prethermal and MBL TCs. The former only shows oscillations for low temperature symmetry-broken initial states, while MBL TCs do so for initial states regardless of their energy density. In Sec.~\ref{sec:prethermalTC_infT} below, we discuss a complementary prescription for obtaining prethermal TCs where the initial state can be at infinite temperature and SSB is not needed, thereby relaxing some of the conditions above. This prescription relies on the emergence of an additional conserved quantity, and does not rely on energy conservation or low temperature. In Sec.~\ref{sec:distinguishprethermal}, we put these all together, and discuss ways of distinguishing between all possible cases of MBL or prethermal TC order. 

It may also be the case that the system defined by $\tilde{D}$ is slowly thermalizing so that $t_{\rm th} \gtrsim t_*$, in which case the pictures above need modification. The amplitude of oscillations in this case is modulated by the decay of the memory of the initial state out to time $t_*$, instead of being set by the thermal equilibrium value. We will encounter one example of such a slow thermalizing system while discussing experiments on NV centers in diamond in Section~\ref{sec:NVexpt}.

Finally, one can consider longer ranged models with power-law interactions decaying as $J(r)\sim r^{-\alpha}$. These \emph{can} show a symmetry breaking transition in lower dimensions, say for $1 <\alpha < 2$ in 1D, and thus aid with some of the conditions required for observing prethermal TCs. These are also of interest from the point of view of experiments which often realize long-range dipolar and/or Coulomb interactions~\cite{MishaTCExp, MonroeTCExp}.  While rigorous results on prethermalization for long-range interactions are less established, such systems have still been shown, both analytically and numerically, to exhibit slow heating~\cite{DimaPrethermal_longrange, ElsePrethermal_longrange, machado_exponentially_2017} although the sharpest results are for $\alpha > 2d$ which are not long-ranged enough to evade Peierls. 
In general, more theoretical studies are needed to understand the full scope of prethermalization in long-range systems.

\subsection{Prethermal (symmetry protected) continuous time crystals}
We now briefly discuss how these ideas of prethermalization can also help with realizing symmetry protected continuous time-crystals. The central idea is that there are static analogs of the statements on effective Hamiltonians presented in Sec.~\ref{sec:prethermalTC-effH}, for certain classes of symmetries~\cite{DimaPrethermal_rigorous}. 

As an example, consider a static spin 1/2 Hamiltonian 
$$
H = H_c + H_{nc} + h^z S^z_{\rm tot}
$$
where $H_c$ denotes all terms that commute with the $U(1)$ symmetry generated by $S^z_{\rm tot}$, so that $[H_c, S^z] = 0$. This could include, for example, hopping terms of the form $\sum_{ij} J_{ij}(S^+_iS^-_j + h.c.)$, or interactions of the form $\sum_{ij} J_{ij}^z S^z_iS^z_j$. By contrast, terms in $H_{nc}$ do \emph{not} commute with $S^z_{\rm tot}$ and could include spin flip terms of the form $\sum_i h_i^x S^x_i$. We again consider Hamiltonians that are written as a sum of local terms, $H_{c/nc} = \sum_i h_i^{c/nc}$ respectively, and let us denote  $g > || h_i^c + h_i^{nc}|| \;\; \forall i$ as the maximum change in the energy of the system as a result of local rearrangements. 

Then, it was shown in Ref.~\mcite{DimaPrethermal_rigorous}, that if $h^z \gg g$, one can construct a (quasi) local effective Hamiltonian $\Heff$ order by order in an expansion in $1/h^z$,  so that $\Heff$ conserves a dressed version of $S^z_{\rm tot}$ to exponentially long times, $t_* \sim \exp(h^z/g)$. In other words, the effect of experimental imperfections that may lead to violations of $S^z_{\rm tot}$ can be \emph{exponentially} suppressed by adding a strong applied field $h^z$. The large field plays a similar role to the large frequency in the driven problem. 

The intuition for this is similar to the driven case: $H_c$ has sectors labeled by their values under $S^z_{\rm tot}$. These sectors are extensive and overlap with each other in a MB system. Adding a global field of strength $h^z$ displaces the energies of states in these sectors relative to each other, but there cannot be an exact conservation of $S^z_{\rm tot}$  in the presence on non-conserving terms $H_{nc}$, if $h^z$ is only of $O(1)$ strength. However, because the eigenvalues of $S^z_{tot}$ are integers, changes in the energy due to changing $ S^z_{\rm tot}$ in any state are quantized in units of $h^z$. Again, when $h^z$ is much bigger than the local scales in the Hamiltonian, $g$, one must make a high order rearrangement $\sim h^z/g$ in order to change $S^z_{\rm tot}$, leading to an approximate long-lived conservation law. 

One key requirement for the proof in Ref.~\mcite{DimaPrethermal_rigorous} is that the spectrum of the symmetry one is trying to protect must be discrete -- in analogy with a periodic drive that allows for energy absorption in packets of $\omega$. Thus, of the various SPTCs discussed in Sec.~\ref{sec:history}, the ones protected by symmetries with discrete spectra such as $U(1)$ are afforded an extra degree of (prethermal) stability.

\subsection{Prethermalization without temperature}
\mlabel{sec:prethermalTC_infT}
Finally, the ideas above lend themselves to yet another intriguing possibility -- that of prethermalization  \emph{without} temperature, recently presented in Ref~\cite{LuitzInfT}. We consider two scenarios for this. 

In the first, the system is driven at a high frequency $\omega$ compared to the local bandwidths, and thus the system has an effective prethermal Hamiltonian for a long time $t_*$. However, one can engineer things so that $\Heff$, in turn, has an additional (say $U(1)$) conservation law so that $M_{\rm tot} = S^z_{\rm tot}$ is (approximately) conserved in $\Heff$. In this case, one can measure an autocorrelator for $M_{\rm tot}$ even in an \emph{infinite} temperature state: $\mbox{Tr}[M_{\rm tot}(t)M_{\rm tot}]$, and this would look conserved for a long period of time. 

We can additionally also engineer for a large $\pi$- flip scale in the drive as in the case of prethermal TCs, Eq.~\eqref{eq:prethermalU}.  However, instead of needing $\Heff$ to show SSB, it only needs to show a global long-lived conservation of $M_{\rm tot}$, and then $\langle M_{\rm tot}(t)\rangle$ can be made to oscillate between positive and negative values. 

Engineering a global conservation law is not constrained by equilibrium considerations of SSB like the Peierls Mermin-Wagner theorems, and hence this can be achieved even in 1D systems with short-ranged interactions and in various systems for which the prethermal mechanism of Ref.~\cite{ElsePrethermal} may not apply. We note, however, that engineering an effective conservation law in $\Heff$ is not entirely trivial because the Floquet driving restricts various parameters to only be defined mod $2\pi$, and simply arranging for a large global field in $\Heff$, as the last subsection might suggest, is challenging~\cite{LuitzInfT}.  

A mechanism similar to this turns out to be important for understanding the physics of a recent TC experiment on an NMR system~\cite{RovnyPRL, RovnyPRB}, discussed in Section~\ref{sec:nmrexp} 

One can go a step further, and consider drives that are instead at \emph{low} frequencies so that there is not even a long-lived $\Heff$, but there can still be a long-lived conservation law~\cite{LuitzInfT, AsmiStrongDriving}. In this case, the concept of temperature is not well defined, but one can nevertheless measure the magnetization of different classes of initial states, and observe non-trivial long-lived dynamics. 

\subsection{Distinguishing between prethermal and MBL TCs}
\label{sec:distinguishprethermal}
For realistic experiments with a lifetime limited by extrinsic factors, it may often be the case that prethermal time window is longer than the experimental lifetime. Thus, the question naturally arises on how to distinguish a prethermal DTC from a bona fide (MBL-localised) infinitely long-lived one, and also how to distinguish between prethermal $U(1)$ DTCs of the previous section, and prethermal DTCs relying on SSB (Ref.~\cite{ElsePrethermal}). 

To achieve this goal, we avail ourselves of the fundamentally distinct origin of the respective longevities. While the emergence of locally conserved quantities -- the l-bits -- underpin MBL, the prethermal $U(1)$ DTC only offers a global conservation law which is not in conflict with local spin diffusion. The prethermal SSB DTC relies on yet a distinct mechanism which requires low-temperature initial states.

The crispest way to distinguish between these mechanisms is by considering a variety of different initial states and measuring \emph{local} spin autocorrelators in the $z$ basis. If we start with a random infinite temperature product state of $z$ spins in the $\Sztot=0$ sector, only an MBL DTC displays oscillations in local autocorrelators $\langle \sigma_i^z(nT) \sigma_i^z\rangle$. These states are too high in temperature for prethermal SSB DTCs, and they have $\Sztot = 0$, leading to zero net magnetization density for the $U(1)$ DTC. 

Further, to distinguish between prethermal $U(1)$ TCs and prethermal SSB TCs, one can start with initial states with $\Sztot = 0$, but still at a low temperature with respect to $\Heff$ (say product states with a single domain wall in the center of the chain). Local autocorrelators in the $U(1)$ TC will thermalize within the $\Sztot=0$ sector, showing no net magnetization and zero amplitude of oscillations. On the other hand, the prethermal SSB DTC will show oscillations in local correlators starting from such states, with occasional ``phase slips" at late times due to slow coarsening dynamics of domain walls. 

By contrast, if one starts from a polarized initial product state, then all three categories give virtually indistinguishable signatures. This is a drawback of existing TC experiments on disordered systems, discussed next,  which only consider a very limited class of initial states. In addition, often experiments do not have single-site resolution and can only measure \emph{global} sample magnetizations. This could again very similar signatures between MBL TC and $U(1)$ TCs depending on the initial state probed.

These considerations further emphasize the remarkable robustness of MBL TCs, and their ability to show a clear oscillatory signal from almost any typical (short range correlated) initial state, and for almost every observable. By contrast prethermal TCs -- both the symmetry breaking and U(1) kinds --- are more fine tuned with respect to the choices of initial states and observables that show a clear signal. 

\section{Experimental Realizations}
\mlabel{sec:expt_real}

The previous sections established the remarkable stability of the TC phase to \emph{all} weak perturbations respecting the periodicity of the drive. We also showed that there are \emph{measurable} signatures of time-crystalline order in dynamics from generic initial states --- even though the crispest diagnostics of TCs rely on abstract eigenstate measures. These twin features of absolute stability and measurable signatures make the TC phase especially well suited to experimental detection. While any realistic experiment is typically far from the idealized limit of an infinitely long-lived MBL TC, signatures of time-crystalline order have nevertheless been observed across a variety of different platforms. 

Much excitement was generated by a first pair of experiments, one on trapped ions~\cite{MonroeTCExp} and the other on nitrogen vacancy (NV) centers in diamond~\cite{MishaTCExp}. Since then, NMR experiments have added to the mix of systems under consideration~\cite{RovnyPRL, RovnyPRB, SreejithStars}. Rydberg atom arrays and superconducting qubits represent other promising platforms, and proposals for observing TCs are actively being developed and refined across all these different setups.

In any such experiment, the system is subject to a periodic drive similar in form to the idealized models described in Sec.~\ref{sec:FMBLTC}.
These experiments have in common that period doubling (\emph{i.e.} a subharmonic response) is observed over a broad window in time set by the lifetime of the experiment, typically on the order of 100 driving periods (Fig.~\ref{fig:experimentaloverview}).  Crucially, the observed period doubling is robust across a range of parameters (such as a systematic offset of a spin-flip angle away from perfect inversion, 
as in Eq.~\eqref{eq:TCgeneral}. By further tuning experimental parameters, crossovers to regimes without such robustness or without period doubling can be induced. 

Experiments of this kind are a relatively recent possibility, representing years of sustained effort towards creating well-isolated, controllable quantum systems that remain quantum coherent for long enough times to observe interesting out-of-equilibrium dynamics. Between them, the various platforms realize a broad matrix of experimentally tunable parameters such as: 
\begin{itemize}
  \setlength\itemsep{0em}
    \item type (and number) of degrees of freedom
     \item microscopic controllability
    \item spatial dimensionality
    \item coherence time
    \item range, type  of interactions
    \item nature of disorder
    \item types of preparable initial states
    \item nature of observables
    
\end{itemize}
Each platform represents a particular optimization of trade-offs across various parameters. For example, a high degree of addressability and control over individual constituents almost always comes at the cost of limiting the size and number of degrees of freedom in the system. 

\begin{figure}[t]
    \centering
     \includegraphics[width=\columnwidth]{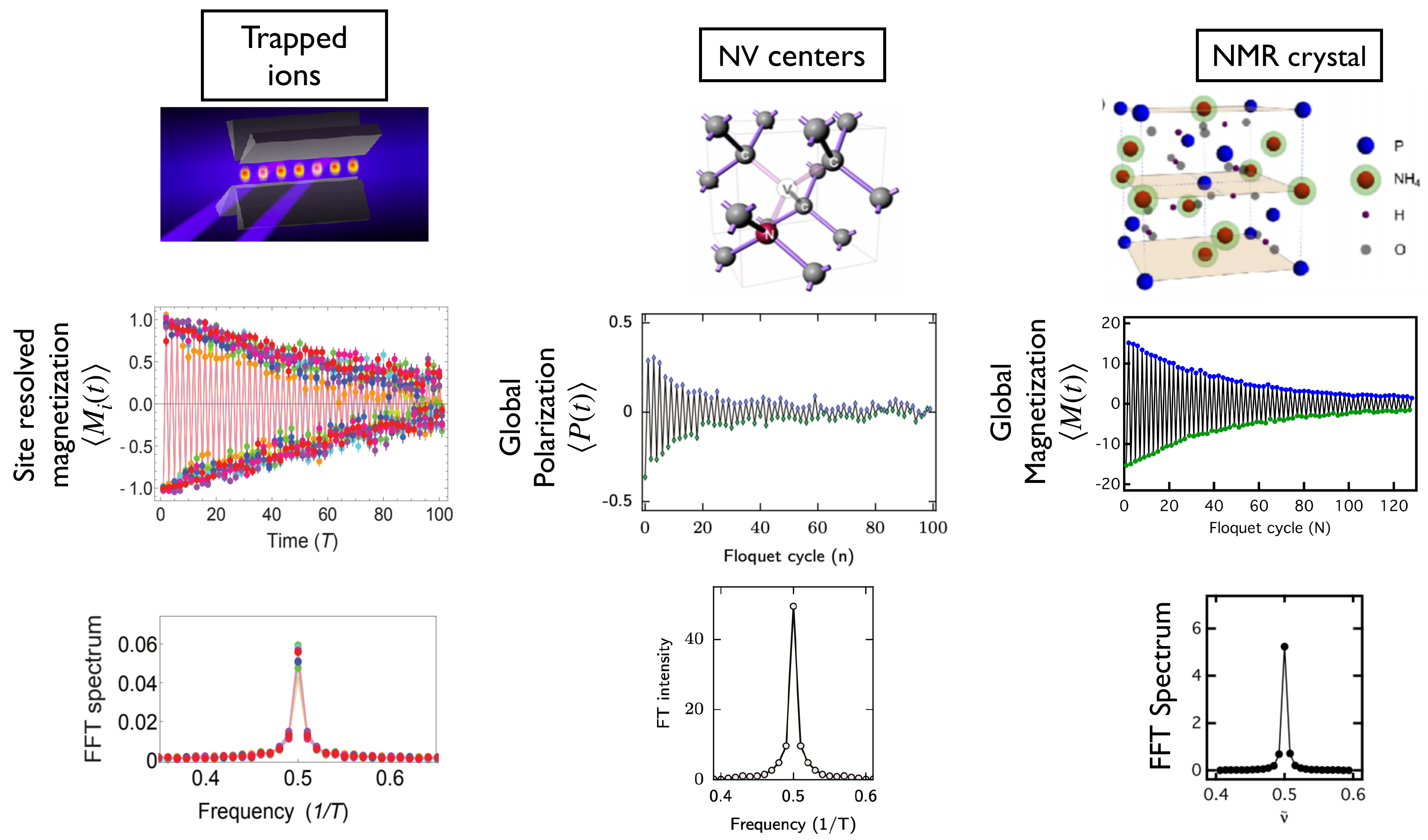}
     \caption{Overview and similarity of results from trapped ion, NV centre and NMR experiments. Top row indicates geometric arrangements of the set-up: a short chain of ions, randomly located nitrogen vacancies in diamond, and nuclear spins in a clean three-dimensional crystal. Middle row: the period doubled response observed stroboscopically persists for a duration of the order of 100 periods. It is robustly locked to period doubling, at least for the lifetime of the experiment, as evidenced by the peak at $2/T$ in the Fourier spectrum (bottom).}
    \mlabel{fig:experimentaloverview}
\end{figure}

Our primary goal in this section is to provide a discussion of each of the existing experiments on time-crystals, in turn. Despite the apparent similarities in observed signatures, the origin of period doubling is quite distinct across the different platforms. None turn out to realize (or even approximate) a true MBL time-crystal. The trapped ion and NMR setups look to be prethermal (although in conceptually distinct ways), while the diamond experiment is thought to realize a so-called `critical' TC with power-law slow thermalization, at least in a certain regime. 

Nevertheless, while there is still much work to be done in convincingly demonstrating the presence of long-range spatiotemporal order -- in an asymptotic sense -- each of the existing experiments represents a valuable step on the way forward. By establishing feasabilities and phenomenologies which, in turn, have led to a deeper understanding, the existing experiments help spotlight promising variations and extensions and point towards tools needed for future advances.

\subsection{Trapped ions}
\label{sec:trappedexp}
\begin{figure}
    \centering
    \includegraphics[width=0.8\columnwidth]{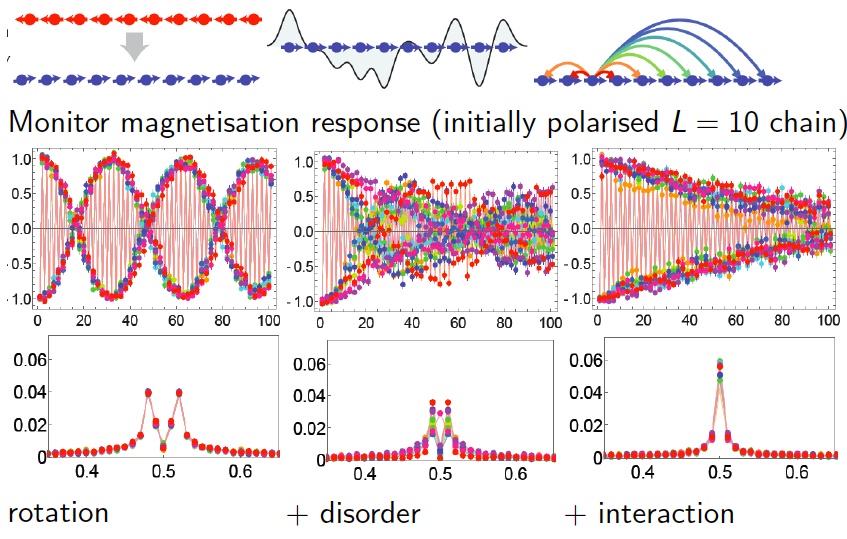}
    \caption{Basic structure of the trapped ion experiment. Top: ternary drive consists of imperfect $\pi$ flip, on-site disorder, and interactions (see Eq.~\eqref{eq:MonroeDTC}). These components can be switched on individually. Left column: the imperfect $\pi$ flip in the absence of disorder and interactions yields regular oscillations with beats whose frequency is set by the imperfection (middle), visible as peaks displaced from $2/T$ in the Fourier spectrum (bottom). Middle column: adding disorder makes individual ions (different colours of points in middle row) oscillate out of sync, degrading the signal in Fourier space. Right column: further adding interactions locks the response at period doubling, with a slowly decaying envelope, with the peak in the Fourier spectrum shifted to $2/T$, as required for a time-crystalline signal, its width reflecting the finite lifetime.}
    \mlabel{fig:monroe}
\end{figure}

The experiment in Ref.~\mcite{MonroeTCExp} is based on trapped ions arranged in a linear, one-dimensional array. Ion traps are a highly tunable platform that afford a high degree of control over each individual degree of freedom,  making for a conceptually crisp setup. The ions can be cooled and trapped at fixed lattice positions using external electromagnetic fields which generate trapping forces that balance the Coulomb repulsion between ions. Effective spin degrees of freedom can be realized using the internal states of the ions; specifically, two hyperfine states of $^{171}$Yb$^+$ are used to isolate an effective spin 1/2 degree of freedom on each site in the experiment in Ref.~\mcite{MonroeTCExp}. The internal state of each ion can be initialized, manipulated and measured with near perfect efficiency using lasers.

The experimental protocol consists of subdividing the drive period into three pieces which act in alternation to realize a model drive similar to 
Eq.~\eqref{eq:TCgeneral}. These involve a spin flip operation, the application of a disordered onsite field, and the addition of Ising interactions (Fig.~\ref{fig:monroe}):
\begin{equation}
H(t) = \begin{cases}
H_{\rm flip} = g(1-\epsilon) \sum_i \sigma_i^x &  0\leq t \leq t_1\\
H_{\rm int} =  \sum_{ij} J_{ij} \sigma_i^z \sigma_j^z &  t_1 <  t \leq t_1 + t_2\\
H_{\rm onsite}= \sum_i h_i \sigma_i^z &  t_1 + t_2 < t \leq t_1 + t_2 + t_3,
\end{cases}    
\mlabel{eq:MonroeDTC}
\end{equation}
with $t_1 + t_2 + t_3 = T$. The spin-flip realizes a deliberately imperfect $\pi$ pulse by setting $g t_1 = \pi/2$ with a small deviation $\epsilon$: under one application of this unitary, a fully polarized initial state in the $z$ basis is mapped to an oppositely  polarised state, but with an offset  angle $\epsilon \pi$. The onsite fields $h_i$ are disordered and picked uniformly with $h_it_3 \in [0, \pi]$, and the interactions are long-ranged and well approximated as $J_{ij} \sim J_0/r_{ij}^\alpha$ ,where $r_{ij}$ is the distance between ions $i$ and $j$
and $\alpha = 1.5$. 
The experiment prepares a pure state $|\psi_0\rangle$ in the $z$ basis. An attractive feature of this system is that it permits single-site addressability,  and the experiment is thus able to measure the autocorrelator $\langle \psi_0 | \sigma_i^z(nT) \sigma_i^z(0)|\psi_0\rangle$ for each spin individually. This is done for about 100 periods. 

In the absence of interactions ($J_{ij}=0$), each spin oscillates independently and is sensitive to the deviation $\epsilon$ in the $\pi$ flip, thereby showing period-doubled oscillation subject to beats. This corresponds to two split peaks in Fourier space at frequencies $\omega = \pi/T \pm \epsilon/T$ (Fig.~\ref{fig:monroe}). The onsite fields applied during the last segment of the drive can be made site-dependent and random (in a completely controlled way), so that the behavior of different spins ceases to be identical. Each spin then independently precesses about a different direction in spin space at a different rate set by the local random field. There is no long-range spatial order, and no robust period doubling.

\begin{figure}[t]
    \centering
    \includegraphics[width=\columnwidth]{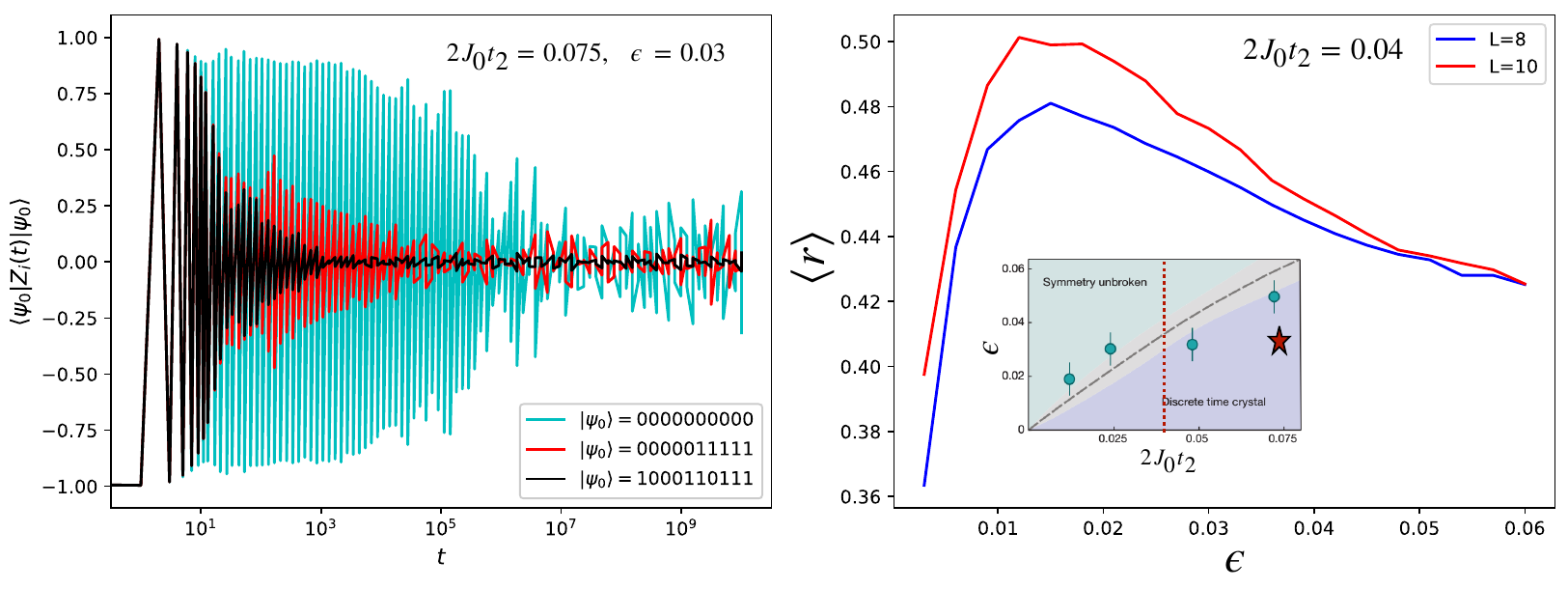}
    \caption{(Left): A simulation of the trapped ion DTC experiment, using the \emph{actual} experimental matrix of inhomogeneous Ising couplings.  Autocorrelators measured from initial states at different energy densities show a strong initial state dependence, unlike what is expected for an MBL TC, but consistent with predictions for prethermal TCs. The parameters are chosen to lie within the putative DTC phase using the phase diagram reported in Ref.~\cite{MonroeTCExp} (see inset, parameter choices denoted by star). The dominant source of disorder in this setup is in the onsite fields rather than the Ising couplings, which is \emph{not} sufficient to stabilize a DTC. 
    (Right): Averaged level spacings ratio \cite{OganesyanHuse} for $2J_0t_2=0.04$ and varying $\epsilon$ (dashed line in inset), again simulated using the actual matrix of experimental couplings and averaged over onsite field disorder realizations and eigenstates. This ratio flows towards the Poisson value of 0.38 with increasing system size in an MBL system, while it flows towards the GOE value of 0.54 in a thermalizing system. The data shows a flow towards a thermal phase for the \emph{entire} range of $\epsilon$, showing no evidence of an MBL phase.  
    Simulations using a uniform power law approximation to the Ising couplings look qualitatively similar in showing the absence of an MBL DTC.
    }
    \mlabel{fig:monroeED}
\end{figure}

However, as discussed in Sec.~\ref{sec:interactingIsingDrive}, the presence of Ising interactions can lead to correlated long-range order and a locking of the subharmonic oscillations at period doubling, removing the offset generated by the nonzero $\epsilon$ (Fig.~\ref{fig:monroe}). In the trapped ion setup, the interactions between the spins are necessarily long-ranged, stemming from the underlying Coulomb interaction between the ions. By modulating the interactions with external lasers, it is possible to realize a variety of different spin-spin interactions (Ising/XY/XYZ) with a variety of different power law decays: $J_{ij} \sim r_{ij}^{-\alpha}$, where $\alpha$ can be tuned from $\alpha =0$ to $\alpha = 3$. The explicit form of the interaction depends upon the details of the lasers used and the vibrational normal modes of the trapped ions in the cavity. These can be calculated and measured with great accuracy, and they approximately take the form $J_{ij} \sim J_0/r_{ij}^{\alpha}$ in the experiment in Ref~\mcite{MonroeTCExp}, with $\alpha = 1.5$. 

The great controllability of this platform is purchased at the expense of a limit in system size. The majority of results in Ref.~\mcite{MonroeTCExp} are presented for a system of 10 ions, starting with a globally polarized intial state $|\psi_0\rangle = |\uparrow\uparrow\cdots \uparrow\rangle$. As the interaction strength, set by $J_0$, is tuned relative to $\epsilon$, the strength of the subharmonic peak at $\pi/T$ varies --- eventually splitting into two peaks for large enough $\epsilon$ -- and this is used to extract an approximate phase-diagram delineating the boundaries of the DTC phase (inset, Fig.~\ref{fig:monroeED}). Of course, since the experiment is only run for a finite amount of time and is also subject to dephasing from the external environment, there is a decay envelope in the period doubled time-trace which is reflected in a broadening of the peak at $\pi$ in the Fourier transformed signal.

Thus far, the observed experimental signatures are entirely consistent with the general theoretical expectations for TCs discussed in 
Sec.~\ref{s:dynamics}. While one cannot extrapolate and make statements about asymptotically large sizes and times from the experimental data alone, the fact that the system does display such clear signatures of a crossover represents an experimental \emph{tour de force} in the quest for programmable quantum simulators -- a feat that would have been unthinkable just a few years ago.

Despite the inevitable experimental limitations on accessible system sizes and times, the various parameters in the experiment represent a choice for the best optimization towards the ideal limit of an MBL TC --- \emph{i.e} the expectation that an (idealized) perfectly isolated experiment run to infinitely long times would realise  the pristine DTC phase. The disordered onsite fields were introduced to engineer many-body localization, and the range of interactions, $\alpha = 1.5$, was chosen to be short-ranged enough so as to be compatible with MBL~\cite{Burin} (ignoring avalanche type non-perturbative instabilities that may delocalize the system at the largest sizes and times~\cite{deroeckAvalanche}). 

\emph{Nevertheless, we find that the dynamics in this experiment actually shows a strong initial state dependence in line with the theory of prethermalization discussed in
Sec.~\ref{sec:prethermalTC} (see Fig.~\ref{fig:monroeED}), rather than the predictions of MBL TCs.} The reason is that the dominant disorder source in this problem comes from disorder in the online $z$ fields rather than the Ising interactions. 

Due to a subtle effect discussed here, disorder in the onsite $z$ fields alone is \emph{not} sufficient to achieve an MBL TC phase, not even in an idealized limit. 
To understand why, it is instructive to consider the drive over two Floquet periods. The onsite longitudinal fields are odd under the global $\pi$ flip and thus change sign from one period to the next. This leads to the disordered field being effectively ``echoed out" over two periods, leaving a spatially uniform system to leading order in a high-frequency expansion.
That is, the leading order time-averaged Hamiltonian over two periods, defined via $U(2T) = e^{-i 2 H_{\rm eff}T}$,
\begin{equation}
    H_{\rm eff}^{(0)} \propto  t_2 \sum_{ij} J_{ij} \sigma_i^z \sigma_j^z +  \epsilon\frac{\pi}{2} \sum_i \sigma_i^x
    \mlabel{eq:Heff_Monroe}
\end{equation}
is disorder free and hence not localized (higher order terms give weak corrections in the regimes under study, where the small values of $\epsilon$ and $J_0 t_2$ correspond to a high-frequency drive of a small, finite system\footnote{At the longest times for large enough systems, the system should heat to infinite temperature and $H_{\rm eff}$ ceases to be a good description.}). 
This is the main reason it was crucial that the model drives considered in Sec.~\ref{sec:FMBLTC} had disorder in the Ising couplings, which are symmetric under the $\pi$ flip and therefore do not average out. By contrast, a Floquet MBL paramagnet far from the limit of near-perfect $\pi$ flips (\emph{i.e.} with $gt_1 \approx 0$) can be realized with disorder in the onsite fields alone, since the drive no longer has a ``flipping" or echoing out action. This is a subtle --- and \emph{a priori} non-obvious --- difference in the nature of disorder necessary to stabilize different Floquet MBL phases, one that was only appreciated and sharpened as a result of a careful consideration of the trapped ion experiment.

We note that in Ref.~\mcite{YaoDTC}, Yao et. al. numerically study the drive in Eq.~\ref{eq:MonroeDTC}, both with nearest-neighbor and long-range interactions, with a view to modeling the trapped ion experiment. As they note in a subsequent Erratum, they too found it necessary to add disorder in the Ising couplings to stabilize the DTC phase. They justify this as a reasonable model of the experiment, stating that the actual experimental interactions are  not described by a perfectly uniform power law due to inhomogeneities in the spacings between the ions. However, the source of the experimental inhomogeneity is \emph{not} uncorrelated randomness in the ion positions, but rather \emph{deterministic} effects coming from the interplay of the long-range Coulomb interactions with the edges of the sample from the confining trap --- in particular, the couplings are  perfectly reflection symmetric about the center of the chain. Thus the rationale for modeling the setup using uncorrelated disorder in the interactions is not entirely clear to us. 

Now, one may argue that there are cases where deterministic inhomogeneities are sufficient to stabilize MBL, the case of quasiperiodic onsite fields being a good example~\cite{IyerQP, KhemaniCPQP}. So one can ask whether the (deterministic) inhomogeneities in the Ising interactions in the ion trap is sufficient to stabilize MBL.  
Fortunately, due to the high degree of control in this setup, the exact matrix of experimental couplings $J_{ij}$ can be explicitly calculated and independently measured. With experiments run on systems containing $10-14$ spins, one can perform full exact diagonalization using the \emph{actual} experimental couplings. This represents a simulation of an ideal, perfectly isolated version of the experiment, in particular providing access to arbitrarily late times. Fig.~\ref{fig:monroeED} shows such a simulation for a system of 10 ions with parameters chosen to be in the putative TC phase. The figure displays the autocorrelator $\langle \sigma^z_i(nT)\sigma^z_i(0)\rangle$ as in the experiment. We run the simulation for much longer times and with different initial states. The data is averaged over 100 statistically independent samples of the disordered onsite fields.

The first visually striking result is a strong initial state dependence in the lifetime of the TC, contrary to what is expected for an MBL TC. Second, the signal eventually decays to zero for all initial states at this system size. 
The longest lifetime is achieved for the perfectly polarized initial state used in the experiment, which is also a very low (negative) temperature state with respect to the effective Hamiltonian in Eq.~\eqref{eq:Heff_Monroe}. Due to the long-range interactions, $-H_{\rm eff}$ can display an Ising symmetry breaking transition at some inverse temperature $\beta_c$, and the polarized initial state lies above $\beta_c$ (and hence in the ordered phase) for small enough $\epsilon$. Such a state can then show long-lived oscillations for times that can be much larger than 100 periods, as discussed in Sec.~\ref{sec:prethermalTC}, and shown in Fig.~\ref{fig:monroeED} (blue line).
By contrast, a randomly picked half-filled initial state is at a high temperature with respect to $H_{\rm eff}$ and has a much shorter lifetime, as shown in Fig.~\ref{fig:monroeED} (black line). Note that the Supplemental Material in Ref.~\mcite{MonroeTCExp} did consider the effect of different initial states, and also presented experimental data starting from a non-polarized initial state.  However, the chosen alternate initial state had only a single domain wall in the middle of the chain $|\psi_0\rangle = |\uparrow \uparrow \uparrow \uparrow \uparrow \downarrow\downarrow\downarrow\downarrow\downarrow\rangle$, which is still a relatively low temperature state with repect to $-H_{\rm eff}$. Indeed, the lifetime of this state is intermediate between the polarized state and a randomly chosen half-filled one (Fig.~\ref{fig:monroeED}, red line). 
Again, if one only looks over the experimental lifetime of 100 periods, the state with one domain wall also shows a subharmonic response with a Fourier peak at $\pi$. But simulating to longer times clearly shows the varied initial state dependence and late time decay.

While the correlation between the energies and lifetimes of the initial state are consistent with expectations for a prethermal Floquet TC, more careful studies on the scaling of these lifetimes with different parameters and system sizes will be needed to conclusively demonstrate prethermalization\footnote{For example, if we consider dynamics solely under the time-dependent Hamiltonian $H_{\rm eff}$, then we expect results very similar to those plotted in Fig.~\ref{fig:monroeED}. In this case, symmetry broken initial states in the ordered phase ($\beta > \beta_c$) thermalize to a non-zero value of the magnetization until a time $O(\exp(L))$, beyond which dephasing between the Ising even and odd paired states destroys order in a finite system. By contrast, high temperature initial states thermalize to zero magnetization at much shorter time scales set by the energy density of the state. For a large enough Floquet system in the prethermal regime, $H_{\rm eff}$ is a good description of the dynamics (at even times, $t=2n$), but only until an $O(1)$ prethermalization time $t_* \sim \exp(\omega/J) \ll O(\exp(L))$. Thus, the lifetime of the period doubled signal will be cut-off by $t_*$. Of course, if the system size is small enough, then the system may never thermalize to infinite temperature for large enough frequencies, so the time-scale $t_*$ may only be visible for simulations of larger systems. }. It is also interesting that the observed experimental phase boundary looks roughly linear in the $J_0-\epsilon$ plane, which is consistent with the expected zero temperature phase boundary for symmetry breaking in $-H_{\rm eff}$. In other words, it is our hypothesis that the experimentally observed transition out of the DTC phase actually represents a conventional Ising symmetry breaking transition in $H_{\rm eff}$, which is reflected in the dynamics of a low-temperature polarized state. 

In any case, while a careful analysis of prethermalization warrants further study, it is clear from the initial state dependence in Fig.~\ref{fig:monroeED} that the system is not an MBL TC {\it sensu stricto}, as this would statistically show the same behavior across all initial states in the $z$ basis. To test this further, we also look at the level statistics of the eigenvalues of the Floquet unitary across a fixed cut in the phase diagram of Ref.~\mcite{MonroeTCExp}, again using the exact matrix of experimental couplings. The data is averaged over all Floquet eigenvalues, and several different disorder realizations. We show in Fig.~\ref{fig:monroeED} that the level statistics shows a clear flow towards thermalization with increasing system size, with no hint of localization for \emph{any} value of $\epsilon$, include for parameters nominally supposed to be in the TC phase. 

This brings us to a point worth emphasizing: the lack of DTC order in the trapped ion setup does \emph{not} require us to consider \emph{asymptotic} limits of an infinitely large system, that may never be experimentally accessible in any case. For instance, we are not relying on theoretical predictions of non-perturbative ``avalanche" instabilities induced by rare thermal bubbles that may destroy MBL for the largest sizes and times in a system with power-law interactions. Instead, the lack of DTC order in such a system is already visible using straightforward exact diagonalization studies on the exact experimental matrix of couplings on systems of only 8-12 sites --- as is clearly shown in both the time-dynamics and level statistics data in Fig.~\ref{fig:monroeED}.

To summarize: The trapped ion setup arguably realizes the conceptually cleanest experimental platform in which time-crystals have been probed thus far. The high degree of control in this setup enables us to understand various features in great detail, helping identify several \emph{a priori} non-obvious ingredients that future experimental efforts should incorporate. 

While the trapped ion setup does not realize an MBL DTC, it shows signatures consistent with a prethermal DTC. In particular, the lifetime of the period doubling signal shows a strong initial state dependence. Examining this systematically for larger systems and a range of frequencies is an interesting future direction in its own right. In addition, two primary lessons for future efforts towards realizing MBL TCs may be: (a) it is desirable to engineer drives with disorder in the Ising interactions; disordered static onsite fields are generally not enough and (b) one should analyze dynamics from a wide range of initial states using a  range of observables, to ensure that the signal is not inadvertently due to fine-tuned/special initial conditions. In principle, the trapped ion platform --- with additional tuning knobs --- also permits one to control each individual $J_{ij}$, and hence to engineer disorder in these. It would be most instructive for a future experiment to systematically examine the effect of this.

\subsection{Nitrogen Vacancy Centers}
\label{sec:NVexpt}
A complementary platform for studying time-crystals is furnished by nitrogen vacancy (NV) centers in diamond. This platform comprises a dense and disordered many-body ensemble of $\sim10^6$ NV centers in three dimensional diamond, and thus probes a completely different parameter regime from the trapped ion experiment. Each NV center represents an $S=1$ electronic spin with three levels, from which two levels can be isolated by applying an external magnetic field to form an effective spin 1/2 degree of freedom.  These spins can then be initialized, manipulated and measured via microwave radiation. The dense ensemble of spins strongly interact with each other via magnetic dipolar interactions decaying as $1/r^3$ in 3D. There are multiple sources of disorder in the problem including lattice strain, the presence of paramagnetic impurities, and random spatial positioning of the NV centers. This leads to randomness both in the interaction between any pair of spins, and in onsite fields. However, despite the strong randomness, the long-range dipolar interactions prevent localization in three dimensions, even at a perturbative level~\cite{Burin}. 

The system is controlled via strong microwave fields. A strong ``spin-locking" field in the $z$ direction can be used to realize a Floquet drive with disordered Ising interactions~\cite{MishaTCExp}: 
\begin{equation}
    U(T) = P^x_{\theta} \exp\left[-i  \tau_1 \sum_{ij}\frac{J_{ij}}{r_{ij}^3} \sigma_i^z \sigma_j^z\right].
    \mlabel{eq:NV}
\end{equation}
As before, $P^x_\theta$ represents a rotation by angle $\theta$ about the $x$ axis and we work with imperfect $\pi$ pulses to probe the DTC phase, $\theta = \pi(1-\epsilon)$. The imperfect $\pi$ pulse is realized by acting with a strong microwave field in the $x$ direction for a short period of time $\tau_2 \ll \tau_1$. 

The experimental protocol is very similar to the trapped ion case. The system is prepared in a polarized initial state $|\uparrow\uparrow \cdots \uparrow\rangle$ and the global magnetization of the sample $M = \sum_i \sigma_i^z$ is measured after every period, for 100 periods or so. Note that due to the large density of spins, one can only measure a global magnetization rather than site-resolved autocorrelators as in the trapped ion setup. The relative strength of the interactions is encoded in $\tau_1$, the time for which the interaction piece of the drive acts. As before, for small $\tau_1$, the offset $\epsilon$ is visible in beats in the stroboscopic magnetization corresponding to split Fourier peaks at $\pi/T \pm \epsilon$. As $\tau_1$ is increased, the beats disappear and the response robustly locks into period doubling indicating a crossover into a TC-like phase. This system is large enough that the limiting factor for probing asymptotic dynamics is finite time rather than finite system size. Nevertheless, as in the trapped ion case, the finite time signal can be used to extract an approximate ``phase-diagram" in the $\tau_1 -\epsilon$ parameter space, delineating the regimes with robust period doubling together with crossovers to regimes without. In this space, it was shown that the lifetime of the TC (extracted from the width of the Fourier peak at $\pi$) is only limited by the extrinsic decoherence time due to environmental couplings, rather than intrinsic thermalization times~\cite{MishaTCExp}.

A follow up experiment on the same platform exploited the intrinsic spin-1 degree of freedom of the NV centers to also realize a $\mathbb{Z}_3$ time crystal with periodicity $3T$~\cite{Misha3T}. The basic setup is a simple generalization of the period doubled case. Instead of toggling between the `up' and `down' states of spin 1/2 degrees of freedom, the drive periodically cycles through the three $S=1$ states. Once again, there are systematic deviations in cycling from one state to the next, but strong enough interactions can nevertheless give a robust signal locked to period tripling. 

A few comments are in order. 

First, despite the strong disorder in the system, long-range dipolar interactions in 3D are incompatible with many-body localization~\cite{Burin}. Thus, this system is expected to eventually thermalize to infinite temperature, corresponding to a decay of the period doubled signal. Evidently, the disorder is still sufficient to impede the rate of thermalization, so that time scale for this decay is longer than the experimentally accessible lifetime of 100 periods. 

Second, one can again wonder whether the slow decay of magnetization is a result of prethermalization due to a special polarized initial state. However, if we again look at the leading order effective Hamiltonian over two periods,
\begin{equation}
    H_{\rm eff}^{(0)} \propto  \tau_1 \sum_{ij} \frac{J_{ij}}{r_{ij}^3} \sigma_i^z \sigma_j^z +  \epsilon\frac{\pi}{2} \sum_i \sigma_i^x,
    \mlabel{eq:Heff_NV}
\end{equation}
then it is clear that initially polarized state is actually at a very high temperature with respect to $H_{\rm eff}$ due to the disorder in the Ising couplings $J_{ij}$ (which are not echoed out). As a result, this state does not lie within the usual prethermalization framework discussed in Sec.~\ref{sec:prethermalTC}, which requires one to start with a low temperature initial state in a symmetry broken phase\footnote{While it is impossible for us to numerically simulate the 3D system with millions of NV centers, a toy simulation (not shown) of 14 sites in a one dimensional system interacting via disordered $1/r$ interactions confirms that there is no strong dependence of the thermalization time on the initial state chosen, unlike the trapped ion case.}. 

Instead, it has been proposed that behavior of this model can be understood as a result of a power-law slow approach to thermal equilibrium in $H_{\rm eff}$ due to the disordered dipolar couplings --- realizing a so-called ``critical time-crystal"~\cite{CriticalTCPRL}. The power-law approach to thermalization follows from some straightforward power-counting arguments that apply to systems with disordered long-range interactions, $J \sim r^{-d}$,  in $d$ spatial dimensions. Such a theory would predict an algebraic decay of the envelope of oscillations, within the regime in which the system is described by $H_{\rm eff}$. This is the expected dynamics for all typical initial states, and even for single-site auto-correlation functions (although these are experimentally inaccessible). 
While this theory does qualitatively capture the broad crossovers in the ``phase diagram" as a function of $\tau_1$ and $\epsilon$, it should be noted that the experiment did not actually observe a power-law slow relaxation. Neither is such a relaxation seen in numerical simulations of analogous models with $1/r$ interactions in one dimension (not shown). Finally, we note that $H_{\rm eff}$ is only a good approximation to the dynamics within a high-frequency regime for small enough $\tau_1$ (compared to the fields and interactions) --- and then too, only for some finite  prethermal time-scales. The behavior of the system for the largest $\tau_1$ values and at longer times times is at present not well understood\footnote{For the 3T periodic TC where the interaction terms are more general and not just Ising like, the system enters a universal Markovian dephasing regime for the longest $\tau_1$, which gives an exponential decay rate $\Gamma \sim 1/\epsilon^2$,  corresponding to a decay in magnetization $\sim \cos(\epsilon)^N$ over $N$ periods. This was, however, not found to be a good description for the case of a drive with purely Ising interactions~\cite{Misha3T}}. 

To summarize, the NV center experiment realizes a dense, macroscopic 3D system that is strongly in the many-body regime. Despite the messy and complex nature of this system, robust period doubling (and tripling) has remarkably been observed over an extended time scale for a range of parameters. While the time-crystalline signatures are not expected to survive to infinite times in this system due a lack of MBL, the thermalization is nevertheless slow enough to obtain a broad regime in time with period doubling. While this slowness has been attributed to ``critical thermalization" --- at least for a subset of parameters ---  a more careful characterization of mechanisms for slow thermalization in models like Eq.~\eqref{eq:Heff_NV} is an important direction for future work.

\subsection{NMR Platforms}
\label{sec:nmrexp}
The next experimental platform we consider executes a very similar drive protocol as the trapped ion and diamond experiments. However, it is distinct in being a nominally disorder free system in which the the degrees of freedom -- the nuclear spins of $^{31}P$ in ammonium dihydrogen phosphate (ADP, NH$_4$H$_2$PO$_4$) -- form an ordered and atomically dense three-dimensional crystal. As in the diamond experiment, the interactions between the spins are long-ranged and dipolar, scaling as $1/r^3$ in three dimensions. 
This system comprises the largest number of individual degrees of freedom of all the candidates yet. 

The system is initialised in a high-temperature weakly polarised mixed state
and driven under a sequence that alternates between an interaction term and a near perfect $\pi$-pulse, as before:
\begin{equation}
    U(T) = P^x_{\theta} e^{iT H_{NMR}}
\end{equation}
with $\theta = \pi(1-\epsilon)$. The presence of the large magnetic field means that the time evolution is best considered in the rotating frame, where the effective
interaction Hamiltonian between the $^{31}$P nuclear spins takes a slightly different form from the normal three-dimensional dipole-dipole interaction. 
In the secular approximation, it takes the form:
\begin{align}
    H_{NMR} = 
    {\gamma_P^2\hbar^2}\frac{\mu_0}{4\pi}
    \sum_{i>j}
    \frac{1-3\cos^2\theta_{ij}}{2 r_{ij}^3}
    \left(3I_i^z I_j^z-\vec{I_i}\cdot\vec{I_j}
    \right)\ ,
\end{align}
where $I$ are the nuclear spin operators, $\gamma_P$ the gyromagnetic ratio, and the interacting pair of nuclear spins is separated by the vector $\vec{r}_{ij}$, which makes an angle $\theta_{ij}$ with the axis along which the field is applied. 

The experiment varied the relative duration of the interaction and spin-flip terms, along with $\epsilon$, and measured the global magnetization $M_{\rm tot}(t)$. Once again, robust period doubling was observed in a range of parameters, and a ``phase diagram'' similar to the diamond experiment was observed. Again, the limiting factor was finite time rather than finite system size. 

The similarities in experimental signatures between the diamond and NMR experiments is striking, particularly since there is no disorder or mechanism for slow thermalization in the latter. The high temperature of the initial state also puts it squarely outside the prethermal TC framework that relies on SSB at low temperatures, discussed in Ref.~\cite{ElsePrethermal} and Sec.~\ref{sec:prethermalSSB}.
Instead, as discussed recently in \cite{LuitzInfT}, the NMR experiment is best understood as an example of the phenomenon of $U(1)$ prethermalization 
\emph{without} temperature, discussed in Sec.~\ref{sec:prethermalTC_infT}. Note that when $\epsilon = 0$, the global magnetization is conserved by $H_{NMR}$ and this toggles between positive and negative values. On increasing $\epsilon$ from 0, the violation in the global conservation law is weak, and one can still observe a long-lived signal. In fact, by changing experimental parameters to add an additional effective $z$ field in $H_{NMR}$ (which is not large), the lifetime of the TC can be enhanced by orders of magnitude~\cite{LuitzInfT}.

As discussed in Section~\ref{sec:distinguishprethermal}, in principle one should be able to distinguish between the NMR and diamond systems if one had access to \emph{local} autocorrelators -- since these would decay rapidly in the NMR setup but not in the diamond one. However, measuring these is currently beyond the capabilities of both experiments.

Finally, we note that since this system is probed and manipulated by standard nuclear magnetic resonance techniques, a large body of knowledge is available on how to detect, and potentially enhance, coherence --- in particular making use of novel spin echo sequences such as the one presented in Ref.~\mcite{RovnyPRL}. Going forward, NMR crystals therefore provide a promising platform for improvising drive sequences that could enhance the lifetime of the DTC signal.

The final experiment we survey is again an NMR one: the objects of study are star-shaped molecules~\cite{SreejithStars}. This limits the total number of degrees of freedom to be studied to a few dozen, a respectable few-body system well beyond the capacity of today's computers for full diagonalization. 

In these molecules, a central spin interacts with a group of satellite spins, which otherwise only interact weakly within the group. There are several advantages of this platform. One is the availability of a series of molecules containing 4, 10 and 37 spins. Another is the possibility of choosing the central spin to be momentless, which leaves the satellite spins as an effectively non-interacting control system. 

The experimental protocol again consists of an imperfect spin flip imposed on an interacting system. A robust period doubled response is again observed, with the non-interating control system exhibiting beats with a period set by the value of $\epsilon$. This star-shaped set-up is simple enough to permit a detailed analysis in the idealized representation of a disorder-free system. 
Such an analysis, presented in Sec.~\ref{sec:zoology} below, shows that the ``star-shaped" interactions in this system make it an effectively two-body or ``mean-field" TC in disguise, instead of exploiting all the MB degrees of freedom. We discuss such classes of systems in the next section, but the reader should be reminded of our previous discussion of a pendulum viewed as a one-body vs. a many-body object.

\subsection{Outlook}

In conclusion, there is much reason to be optimistic about the realization of Floquet time crystals. First, the abovementioned experiments have taken great strides towards this goal, exhibiting a number of desirable signatures in nascent form, only a short time after the original theoretical proposals. These experiments have helped considerably sharpen the ingredients needed for future demonstrations of asymptotic MBL TCs with true long-range spatiotemporal order.
Importantly, the missing ingredients appear not to be truly fundamental stumbling blocks. Of course, more systematic experimental studies of prethermalization and prethermal TCs would be interesting in their own right, and perhaps more approachable than MBL TCs in the near term. 

Some lessons that can be drawn from the experimental efforts thus far are: 
\begin{itemize}
    \item It is important to consider a wide range of initial states at varying energy densities (with respect to some effective Hamiltonian). MBL TCs should be relatively insensitive to the change in initial states, but the lifetime of prethermal TCs will be strongly dependent on the initial state.
    
    \item It is desirable to measure site resolved observables to demonstrate long-range spatiotemporal order. In some cases, globally averaged observables can give a long-lived oscillatory signal if the system has an approximate global conservation law, even without the presence of long-range order. 
    
    \item It is important to engineer disorder in the Ising couplings --- disorder in onsite fields alone is generally not sufficient
    
    \item From general MBL considerations, short ranged interactions are more suited for localization
    
\end{itemize}

With these ingredients in mind, a complementary promising platform for obsering an MBL time crystal is provided by Rydberg atom arrays, which have seen much progress of late~\cite{BlochDressedRydberg, MishaRydberg}. They have relatively short-range interactions of van-der-Waals type, scaling as $r^{-6}$. Importantly, they can allow for the introduction of bond disorder in the Ising couplings by varying the relative locations of atoms, while also maintaining full spatiotemporal control using the usual cold atom microscopy techniques. Their primary current limitation is a relatively short many-body coherence time, which limits the available observation window for pinning down spatiotemporal long-range order.

\newpage
\section{A zoo of time crystals}
\mlabel{sec:zoology}
The term time crystal has been widely used of late.
In this last section, we broaden our scope considerably to examine a variety of phenomena in classical, open, and few-body systems that ostensibly share some dynamical signatures of the MB TCs considered until now. 
 As outlined in the previous sections, one can identify a number of ingredients which should (not) be present for a pristine incarnation of this concept. Ultimately, it is an issue of personal choice and convenience where to draw the line. 
However this may be, the rate at which proposals appear is sufficiently high that it is illusory to list all of them even in the context of a review such as this. Rather than attempting to decide what proposal `deserves' the label time crystal, it is considerably more satisfying to appreciate the broad range of settings and physical underpinnings which have been proposed in this context, as these range widely, from condensed matter physics via quantum optics to nonlinear and open system physics, and beyond into other disciplines. 

In this spirit, this section is devoted to listing -- without any pretence at completeness -- a number of proposals, and discussing unifying and distinguishing features across these. Two important ingredients turn out to be the `effective size' of the systems, i.e.\ whether one is dealing with genuine many-body physics; as well as the role of energy conservation, in particular existence and nature of any source or sink of energy and entropy --- in other words, the notions of \emph{macroscopic} and \emph{conservative}, emphasized earlier. 
Naturally, one challenge faced by all proposals is the actual demonstration of temporal symmetry breaking; or, put differently, the question of the timescale for which the subharmonic reponse persists, and what, if anything, limits it.

\subsection{Interdisciplinary chaos, and complexity}

First, to note the importance of `time-crystallinity' in settings far removed from physics, we mention that period doubling is by no means an uncommon phenomenon in classical systems, where it has played a conceptually immensely important role. One setting is provided by nonlinear classical systems, where it is one of the characteristic fingerprints of a system {\it en route}  to classical chaos~\cite{Feigenbaum1978}. This has been a hugely influential observation,  radiating out in many directions. In particular, an equation known as the logistic map appeared in the genesis of the field of theoretical ecology and mathematical biology in general, and our understanding of predator-prey/food-web systems in particular \cite{May645}. The logistic map generates a discrete time series (sequence) $\{x_n \}$ according to
the rule 
\begin{equation}
    x_{n+1}=r x_n (1-x_n)
\end{equation} 
for $x_n\in [0,1]$ and $r\in[0,4]$. The idea is that for small population size $x_n$, reproduction is proportional to the number of individuals present, while environmental restrictions (e.g.\ scarce  resources such as food or space) limit the maximal population to what is known as the carrying capacity. 

For $r<3$, this time series approaches a time-independent steady state -- a fixed point. For $3<r<1+\sqrt{6}$, the series almost always  oscillates between two values at long times. For larger $r$, a period-doubling cascade (period 4, 8, 16, $\ldots$) presages the onset of chaos at $r\approx 3.56995$, with the resulting aperiodic behaviour of the time series depending sensitively on the initial condition. 

One of the lessons of this analysis is that regular underlying laws can, in real life, lead to irregular observables: even in the absence of fluctuations in external conditions, population sizes can behave non-trivially from one year to another, in particular oscillating regularly between two distinct values. This biological insight is remarkable even without attaching the label time crystal to it.  

Indeed, this is an instance of a distinction which is sometimes colloquially made between a system which is complicated, and one which is complex. While no one would have been surprised that population size fluctuations are hard to predict, this intuition would have most naturally been based on the observation of how {\it complicated} its description must be: in a real-life ecosystem, there may be many species of animals interacting with each other; a large number of uncontrollable parameters, such as the weather; as well as considerable variations spatially as well as between individuals. Nonetheless, even in a maximaly simple model with a single parameter, chaotic behaviour -- {\it complexity} -- emerges. In this sense, several models we will encounter in the following are effectively few-body classical ones, even when they are used to describe many-body systems.

\subsubsection{Towards higher dimension: limit cycles and strange attractors}
The logistic map with its discrete time series may be thought of as a stroboscopically observed (non-Hamiltonian) Floquet problem. Its period doubling is superficially reminiscent of the discrete time-crystallinity, and one may thence ask whether similar phenomena appear in models where the  time variable is continuous.

The first exhibit in this direction is the limit cycle for a system with two degrees of freedom, the study of which dates back to work of Poincare~\cite{PoincareLimit}. A stable limit cycle is a one-dimensional counterpart of the stable fixed point: a system initialised near such an object is attracted to it. While the fixed-point corresponds to a time-independent configuration, the limit cycle implies a nontrivial time evolution which repeats periodically. 

Many models exhibiting such behaviour  have been studied. For dynamical systems, a single degree of freedom gives rise to two-dimensional phase space, and one important model is the van der Pol oscillator~\cite{PolOscillator} described by the equation of motion
\begin{equation}
   \frac{d^2x}{dt^2} - \mu(1-x^2)\frac{dx}{dt}+x=0 \ .
\end{equation}
For $\mu=0$, this is just the familiar harmonic oscillator, while for $\mu>0$, a limit cycle arises. Trajectories for different initial conditions for the harmonic oscillator ($\mu=0$) can be parametrised by their energy, and energy conservation ensures that they do not intersect, nor approach, each other. By contrast, for $\mu>0$, the `friction' term allows trajectories with different initial conditions to approach a common limit cycle. 

This analysis was undertaken in the context of electrical circuits but related problems have appeared in many branches of science. One prominent early example of a nonlinear two-component system is the Lotka-Volterra model, arising for oscillatory chemical reactions, a fascinating subject in its own right \footnote{The Nobel lecture of Prigogine\cite{prigogine} provides a readable overview from the perspective of non-equilibrium (thermo)dynamics.}. This also includes models in theoretical ecology akin to the logistical map above. The latter was triggered by an observation of an unexpected relative increase in the population of predator fishes in the Adriatic Sea as a result of decreased human fishing activities during the first World War~\footnote{A visiting physicist surveying marine wildlife from the balcony of the Adriatico Guest House of ICTP in Trieste may today note a somewhat analogous population increase of fish-eating birds in response to changed human fishing practises following enlargement of the European Union. We thank Fabio Perco for pointing this historical note out to us.}.

From a many-body physics perspective, the existence of a limit cycle is not much more suprising than the periodic motion of the harmonic oscillator, and the natural question is thus how these cycles fare as the number of degrees of freedom is increased. 

This question takes us further into the realm of deterministic chaos, and touches upon two concepts in particular, namely that of the strange attractor, and that of stability of periodic orbits, which is the subject of the Kolmogorov-Arnold-Moser theorem. 

The strange attractor is a generalisation of the zero-dimensional fixed point and the one-dimensional limit cycle to higher-dimensions. This is not entirely straightforward: a limit cycle encodes a unique trajectory in phase space, as both the cycle and the trajectory are one-dimensional. While the limit cycle turned out to be more complex than the fixed point in that it allowed for time-dependence, the strange attractor in turn offers more complex behaviour: it can exhibit positive Lyapunov exponents, i.e.\  trajectories approaching each other arbitrarily closely can diverge exponentially with time at short times, with the size of the attractor in phase space limiting the divergence at long times.     

Finally, the fate of all of these low-dimensional considerations in the thermodynamic limit are not settled to this day. Perhaps for this reason, there is a certain disconnect between recent developments in quantum many-body physics, for which the $N\rightarrow\infty$ limit was the sociologically natural starting point, and the non-linear dynamics/chaos community which has tended to adopt an intrinsically more few-body perspective. 
 
A survey of these fields with the aim of establishing connections and vistas for future work is obviously highly desirable but also well beyond the scope of this review. Suffice it to say that one way of phrasing a typical question regards the fate of the KAM theorem for many degrees of freedom.  

This theorem deals with the question how an integrable system turns chaotic as its degree of nonlinearity increases. Here,
integrable is used to mean that the Hamiltonian evolution can be described by angle-action variables, i.e.\ effectively by independent harmonic oscillator coordinates. This corresponds to the dynamical evolution in phase space taking place on high-dimensional tori, with each pair of angle-action variables contributing one non-contractible loop. 

For few-body systems, the KAM theorem essentially states that, as nonlinearities are increased to turn an integrable system into a chaotic one, this does not happen in one abrupt step. Rather, phase space continues to host  pieces  exhibiting periodic orbits, while an increasing fraction of it is filled with chaotic dynamics. 

A quantum many-body version of this phenomenon, and its possible relation with ideas such as quantum scars, is high on the list of desired items, but alas not yet available. In the following, we concentrate on a survey, embedded in the conceptual context developed in the review so far, of well-known phenomena in the classical and quantum dynamics.

\subsection{Classical few-body, and effectively few-body, systems}

In this section, we focus on a number of models which between them realise a range of remarkable phenomena related to subharmonic responses in the broad sense. 

We have emphasized time and again the important role of the many-body nature of the spatio-temporal ordering phenomena underpinning time crystallinity. The converse of this is that simplified treatments of a many-body system, in particular via  mean-field theories or systems with infinite-range interactions, are not in themselves reliable guides to the behaviour of the many-body system. For instance, if a mean-field theory exhibits period-doubling, this by no means stops the driven many-body system it purports to describe from heating to infinite temperature instead. Absent a demonstration that such heating does not happen, it may not be prudent to jump to conclusions about the long-time stability of time-crystalline phenomena thus obtained.

\subsubsection{Period doubling in the parametric oscillator}
We first consider a variant of the ubiquitous harmonic oscillator: 
the classical {\it parametric} oscillator  \mcite{kuehn}
presents another simple set-up in which robust period doubling arises. Here, it may be the coefficients of the friction and restoring terms, $\beta_f(t)$ and $\omega(t)$, which are weakly, and periodically, modulated. The concomitant equation of motion for the displacement $x$
\begin{equation}
    \frac{d^2 x}{dt^2}+\beta_f(t) \frac{dx}{dt}+\omega(t)^2 x=0
\end{equation}
exhibits a period doubled behaviour of $x$ provided the modulation takes place at around twice the natural frequency of the oscillator. 

Note that in the above we have slipped in dissipation in the form of friction. Absent this, the period doubling is in fact not stable in the sense that the amplitude of the stroboscobic response is not constant. This phenomenon is thus not realisable in a closed system with Hamiltonian evolution, which does not permit a sink of energy. We return to this point in the context of quantum systems in the section on open systems below.

\subsubsection{Synchronisation}
A closely related phenomenon with a venerable history of its own harks back to Huygens' experiments in the 17$^{th}$ century\mcite{huygens}, who formulated what is now known as synchronisation~\cite{Scholarpedia_synchronization}. Here it is an external forcing, rather than an internal parameter, of a driven system that is modulated in time, leading the system to respond at a frequency related to that of the driving's. In particular, forcing by an external periodic stimulus can stabilise a response at a frequency a rational multiple $p/q$ of the forcing's. The range of resulting entrainment can be plotted in a planar parameter space, one dimension of which records the frequency offset between the driving and the natural frequency of the driven system;  and the other the forcing amplitude. This yields  a triangular motif, known as Arnold tongue, at weak forcing, with a devil's staircase describing the relative stabilities of the various rational lockings. 

Recent work \cite{chitratc} has taken up this set of ideas in the context of the present discussion of time crystals. The analysis  considers the period doubling bifurcation in a mean-field `all-to-all' coupled set of coupled nonlinear oscillators, with accompanying illustrative experimental work on a pair of modes provided by the vibrations of two macroscopic strings. 

Altogether, these ideas not only underline the beauty of nonlinear dynamics and chaos precursors, but also is of tremendous importance for applications, not least in biosystems such as ourselves, e.g.\ in setting the periodicity of our internal clock(s), or in cardiac pacemakers. 

Also, several instances from everyday life can be subsumed under this heading, although the distinction
between what constitutes an internal (parametric oscillator) and external parameter (synchronisation) is not always clear-cut. 
Examples include the motion of a child  on a swing who increases the amplitude of the oscillatory motion by varying his/her position, and hence the swing's moment of inertia, periodically; or an analogous action on the censer in medieval cathedrals, as in the botafumiero in Santiago de Compostela \mcite{schlichting}.

\subsubsection{Faraday waves}
Moving on from the above parametric oscillator, a similar phenomenon has been known to occur in hydrodynamics for considerably longer than the present interest in time crystals, in an observation named after Faraday \mcite{faradaywaves,goldstein}. The eponymous waves occur as an instability of a many-body system -- the surface of a liquid. When this surface is subjected to a periodic drive
perpendicular to the surface (which may be induced by a loudspeaker at the bottom of the container of the liquid), various normal modes of the surface may be excited. 

The time dependence of the normal mode amplitudes of the surface waves is given by Mathieu's equation. 
For an appropriate choice of parameter values,  an instability occurs in which the \emph{one} mode on the surface of the liquid 
oscillates with \textit{twice the period} of the drive. 

As in the case of the parametric oscillator, this phenomenon is an instability in the sense that the amplitude of the mode in fact grows without bound, i.e., the system absorbs energy from the drive indefinitely. In practise, nonlinear effects will modify the behaviour at large amplitudes, and presumably shift energy into other modes, causing the liquid to heat. To reach
a steady state, it is thus again necessary to deposit this energy somewhere -- in this
case, the bulk of the liquid thus can act, literally, as a bath for the surface mode. Thus, this is an effectively one-body system --- with the many-body nature of the liquid manifesting itself in that the system explicitly contains the degrees of freedom into which the energy from the unstable mode can be transferred.

\subsubsection{Kuramoto model and phase locking}
The Kuramoto model\mcite{Kuramoto1975} considers particles on a circle, parametrised by angles $\{\theta_i\}$, which individually execute a regular circular motion with frequency $\{\theta_i\}$. These are coupled, and the model can be thought of as capturing a variant of synchronisation, where it is the interaction with other particles, rather than an external drive, which leads to a joint response.

Their coupling aims to lock the phases at a joint value:
\begin{equation}
    \frac{d\theta_i}{dt}=\omega_i+\frac{K}{N}\sum_{j=1}^N\sin(\theta_i-\theta_j) \ .
    \mlabel{eq:kura_eom}
\end{equation}
This is another mean-field--type model, in that all $N$ particles interact pairwise with each other, with a strength $K/N$
vanishing in the thermodynamic limit. 
For sufficiently strong $K$, the particles phase lock completely, that is to say, they move on the circle with fixed angular frequency, and fixed relative phases. 

There are plenty of variants of the Kuramoto model \mcite{KuramotoRMP}; for example, rather than a fully connected set-up, there also exist studies of models on sparser graphs, with time-delayed interactions, or involving differing distributions of oscillator frequencies. 

On this level, the Kuramoto model does exhibit a long-lived stable response. In the context of the above discussion on many-body versus few-body physics, the equation of motions Eq.~\eqref{eq:kura_eom} involve only `angle' variables. In a Hamiltonian picture, the corresponding  amplitude (`action') variables are not dynamical, and as such, the model can be considered to be fine-tuned, with the possibility of a perturbed version of such a model exhiiting heating, like in the above examples.

\subsubsection{Towards quantum systems: semiclassics}
The above examples were all classical, and the implications of a classical subharmonic response in a quantum setting was first studied by  Holthaus and Flatte \mcite{holthaus} for the case of a particle in a driven triangular well, consisting of a hard wall at $x=0$ and a time-dependent potential  for $x>0$, with driving strength $\lambda<1$, of $$V(x,t)=x\ [1+\lambda \sin(\omega t)]\ .$$

This pioneering work presaged a number of developements: it constructed the Floquet solutions of the quantum problem corresponding to the period doubled regime in classical phase space, and noted (i)  the appearance of a pair of eigenfunctions with $\pi$ quasienergy difference, (ii)  the role of their linear combinations in the genesis of the subharmonic response, as well as (iii) the finite lifetime of the subharmonic response.

\subsection{Effectively few-body quantum systems}
Given the long history of period doubling/locking in nonlinear systems, which we are not in a position to do justice beyond the few snippets presented here, it is perhaps not too surprising that systems for which the analysis reduces to an equation in this class, can in turn appear to exhibit robust time translational symmetry breaking. However, 
for the purposes of the present review, we have emphasized the 
importance of the stability of such a phenomenon in a many-body setting. 
Therefore, and without wanting to 
diminish the  importance and beauty of the few-body realisations, we note that 
 in mapping a many-body problem onto a (classical) few-body problem, the properties of the latter may only apparently be inherited by the former. 
However, it is only upon treatment of the problem as a full many-body problem -- like in the case of the grandfather clock -- that it can be settled whether the subharmonic oscillations are a featureless transient, a prethermal phenomenon, or a fully robust time crystal in the above sense. Several many-body examples are, in this sense, few-body in disguise. 

A case in point is the pioneering work by Sacha \mcite{2015PhRvA..91c3617S}, which identified the possibility of, and interest in,  realising  {\it discrete} time crystals, perhaps more easily than the fully fledged time-translational symmetry breaking. 

Its basic set-up is that of a cloud of interacting bosons bouncing on an oscillating horizontal mirror, subject to a gravitational potential, akin to the triangular well studied by Holthaus and Flatte \cite{holthaus}. The theoretical analysis focuses on the one-dimensional vertical motion, and by considering only a single mode in a Gross-Pitaevskii framework, reduces to an effective single-particle problem. The nonlinearity arising from the interparticle interactions then is responsible for the appearance of a response subharmonic relative to the oscillation frequency of the mirror.

\subsubsection{NMR experiments on stars}

A recent NMR experiment on `star-shaped' clusters \mcite{2017arXiv170808443P} appears to be a rather crisp realisation of an intrinsically few-body phenomenon, in the sense that the geometric arrangement of $N$ spins surrounding a central one quite simply does not allow, for geometric reasons, for an $N\rightarrow \infty$ limit when embedded in finite dimension. 

One particularly attractive feature of these
experiments, indeed, is the simplicity of the concomitant model involving a
set of $N+1$ spins-1/2, the analysis of which in a disorder-free setting is already
very revealing. Denoting the central spin by $S_0$ and the surrounding ones by $S_j$, $j=1\ldots N$, the Floquet unitary reads:
\begin{equation}
 U\left(J,\theta;T\right)=\exp\left[-\imath\theta{\textstyle \sum_{j=0}^{N}}S_{j}^{x}\right]\exp\left[-\frac{\imath JT}{\hbar}S_{0}^{z}{\textstyle \sum_{j=1}^{N}}S_{j}^{z}\right] .
\end{equation}
Here, $J,\theta,T$ denote the interaction strength between the outer and the central spin, the flip angle (chosen to be close to $\pi$) and the Floquet period.

A simple Fourier transform of the outer spins,
\begin{equation}
  \tilde{S}(k)=\frac{1}{\sqrt{N}}\sum_{j=1}^{N}S_j\exp(i k j)  
\end{equation}
shows that only the 
uniform $k=0$ Fourier mode is coupled to the central spin:
\begin{equation}
 U\left(J,\theta;T\right)=\exp\left[-\imath\theta\sqrt{N}  \tilde{S}^{x}(0)\right]\exp\left[-\frac{\imath JT}{\hbar}S_{0}^{z}  \sqrt{N}\tilde{S}^{z}(0)\right] .
\end{equation}

The model hence
becomes effectively a two-body one, in which -- crucially -- one body, $\tilde{S}(0)$
comes with a prefactor $\sqrt{N}$.  The effective energy scale thus grows
with system size, as the interaction term reads $\sqrt{N} S_0^z \tilde{S}(0)$. 
Hence there is a `mismatch' ratio
of scales between the rotation imperfection $\epsilon$ and the interaction term, 
which vanishes as $N$ grows, and accounts for the exponential growth
of the subharmonic oscillation lifetime with system size.

\subsubsection{Other `mean-field' models}

The preceding model of course, mathematically, admits the $N\rightarrow\infty$ limit, but the effective model remains a two-body one all along--the $N$ outer spins together provide an effective mean field.  Closely related to this  is the Lipkin-Meshkov-Glick model, an infinite-range (i.e., effectively zero-dimensional) model in which all $N$ spins interact equally with all others. In the absence of disorder, its simple Hamiltonian only involves the total spin $S$ 
\begin{equation}
   - H_{LMG}=\frac{\gamma}{N}
   \left(S_x^2+S_y^2\right)+h S_z \ .
\end{equation}
This in various guises/deformations has been analysed extensively, in particular with view to the relation of few- and many-particle physics, see \mcite{2008PhRvE..78b1106R,2014arXiv1407.1217G}. 

In the present context, it was found that in the presence of an approximate $\pi$-flip, the system manages to exhibit robust, infinitely long-lived subharmonic oscillations even in the absence of disorder. 
Underpinning this is again  the dimensional reduction of the many-body Hilbert space \mcite{PhysRevB.95.214307}.

A precursor of some of the ideas presented here is a study of a driven $O(N)$ model \mcite{2016PhRvB..93q4305C}. Here, it was found that the heating (leading to Floquet-ETH) generically present in an interacting many-body system can be terminated in the limit of large-$N$, 
leading to a state with
non-trivial correlations and period doubling.

The much-studied phenomenon of the pseudogap in the cuprates\mcite{PhysRevLett.63.1700} 
has also received attention in the context of time crystal studies. A mean-field treatment of a static 
two-band model 
incorporating a combination of repulsive and attractive interactions was claimed to lead to a free-energy minimum
with a ground-state wavefunction oscillating in time \mcite{2019arXiv190207520E}.

\subsubsection{Two-level systems}
We include in this mini-overview the perhaps effectively simplest quantum mechanical system exhibiting a non-stationary response are two-level systems. These exist in various familiar incarnations such as, e.g., in Rabi oscillations which describe a two-level atom coupled to an optical cavity mode.  

Most simply, consider an isolated system with an energy
spacing $\delta E$ between its two levels. For any initial state in the form of a superposition of the two levels, there will be observables which forever oscillate with frequency $\hbar\delta E$. 
Without wanting to detract from the importance of two-level systems in any way, this example perhaps  
serves to underline the necessity of considering actual {\it symmetry-breaking} in a many-body setting as a central ingredient for time crystalllinity.

\subsubsection{The laser}
The question how the laser -- paragon of twentieth century technological innovation -- fits into the time crystalline story arises frequently: it does, after all, yield a practically stably oscillating light field. 

Its simplest description as an ensemble of (otherwise non-interacting) two level atoms\footnote{For practical implementations, three or four level systems offer great advantages.}, the gain medium, interacting with a joint light field. This yields a pair of `mean-field' rate equations, one for the number of excited atoms, the other for the number of photons in the light field.

Like some of the above examples, the laser does not represent a stable phase of matter of a closed system: either lasers produce a sequence of pulses, or they are sustained in a steady state by balancing energy input and loss.  

A distinctive feature of the laser is the role of the pumping action which provides this energy input: the gain medium needs to be driven out of equilibrium to reach a critical level of population inversion (the excess occupancy of the excited level with respect to the lower level), at which lasing action onsets like an avalanche.

\subsection{Open many-body quantum systems}
\mlabel{sec:open}

Let us turn to open {\it many-body} systems next.
Beyond the classical realm of most of the above instances, the addition of a bath in 
quantum mechanics takes us outside the paradigm of unitary evolution:
we need to turn our attention to systems the dynamics of which is not
purely Hamiltonian. This, of course, increases the number of systems and settings enormously and presents the challenge of defining a notion of a time-crystal in this setting which is not so broad that it encompasses separately well-known and well-understood items such as battery-operated clocks. 

It will not have escaped the attention of the reader that 'being quantum mechanical' is not an entirely self-explanatory concept. A given experimental material or system might in general very well be under the impression that it is governed by the laws of quantum mechanics; from this perspective, behaviour described by purely classical physics would then be an emergent phenomenon. This poses the natural question to what extent time-crystallinity can occur in a classical setting, discussed in some detail in \mcite{2018arXiv180102628Y}. This work also emphasized the special nature of a thermal bath, to which we will return shortly, which does not just extract energy (`friction') but also is a source of (thermal) fluctuations.

In quantum mechanics, unlike in classical mechanics, it is not so easy to include a friction term. 
One way to formulate the new situation is to consider 
an `open' system,  obtained when coupling the
degrees of freedom under observation (`system') to an external set of
degrees of freedom (`environment'). This increases the space of models
under consideration immensely.

The consideration of Floquet systems is a rather restricitive special case of this -- the drive needs to be supplied somehow, be it in the form of an optical  field of a laser, or the mechanical displacement of a mirror or loudspeaker mentioned above. The choice underlying the Floquet formalism is that the laser and the mirror themselves have no further intrinsic dynamics, and in particular do not get entangled with the system under consideration. The evolution of the system thus still remains fully unitary,
and this is one of the distinguishing features of the Floquet time crystal.  

Going beyond this, by relaxing the constraints of unitarity, allows the 
addition of a wide variety of phenomena. One
is the loss of conservation of energy even in an undriven setting with
full time-translational invariance, as energy may be exchanged with
the environment. Most simply, this may effectively amount to including friction  which drives the system to a preferred steady state.  

If
the environment has particular structure, and is e.g.\ described by a
temperature, thermal fluctuations are introduced to the system. This point is quite important: a system in contact with a thermal bath necessarily is subject to fluctuations, i.e.\ it will not exclusively experience friction, i.e.\ lose energy but will necessarily -- possibly very rarely -- also absorb energy.    

There
is then  no limit, as a matter of principle, to the complexity of the
environment, which may itself have considerable internal structure and a
non-trivial quantum many-body dynamics of its own.
Already the wide variety of settings for time crystallinity discussed
above is therefore but a subset of the present collection of
models, and identifying systematic structure in this collection is not
only challenging for technical reasons -- open, in particular quantum,
systems are not particularly easy to treat -- but also because of the
`intuitive' decisions one needs to take for what type of bath and
system-bath coupling is still sufficiently `simple' and `natural' to
be included herein.

One can formulate a general setting in which to observe discrete time crystals
in the following way. Consider subdividing the Hilbert space of a quantum 
many-body system into $M$ sectors. Next, consider a Floquet 
dynamics with period $T$ which, loosely speaking, approximately
cycles the sectors into each other, $m\rightarrow m+1 \mod{M}$. This promises to yield a discrete time crystal with period $M T$, 
and several proposals follow this strategy. 

However, it is clear that 
any imperfection in the transfer of weight from one sector to its 
neighbour risks spreading out any state initially concentrated in one sector,
resulting in a decay of the period-$M$ signal at long times.   
The result is
a density matrix in a mixed state spread equally over all sectors. Thence, if thermal fluctuations degrade the transfer between the $M$ states, they may limit the lifetime of the time crystal. 

The way in which the $M$ states may arise can be very varied. Specifically, for the case of open quantum systems, one original proposal is motivated by a study of Rydberg gases~\mcite{PhysRevLett.122.015701}. It considers a system with not only a single steady state, to which the system will relax at long times, but in addition a long-lived metastable state ($M=2$): a system prepared near the metastable state will take a time to reach the true steady state which diverges in the thermodynamic limit. 

Formally, the Lindblad equation describing this system has an eigenvalue $\lambda_0=0$ for the steady state, but in addition a gap to $\lambda_1$ which vanishes with system size $N$.
One can then consider a dynamics which exchanges the neighbourhoods of the steady and the metastable state in order to produce a period-doubled time crystal. 

Another idea involves an open version of the Dicke model from cavity/circuit QED \mcite{PhysRevLett.122.015701}, which describes the coupling of $N$ two-level atoms to a light field: as an atom absorbs or emits a photon, it changes its internal state.  For sufficiently strong light-atom coupling, there is a phase with a well-defined parity involving the sum of the number of photons in the cavity and the number of atoms having transitioned into, say, the lower energy state. This parity then defines the $M=2$ sectors.

In these proposals, the transfer between sectors is explicitly arranged (assumed) to be perfect and hence
the time-crystalline behaviour long-lived despite the openness.

A natural question is thus how to repair any imperfection, which is in general unavoidable,  in the weight transfer
between sectors. One option is to consider 
systems with long-range interactions
and the resulting effectively simplified structure of Hilbert space, as described above. 
Another is to arrange for the system to create the 
requisite inertia cooperatively \mcite{2019arXiv190404820L}. The simplest avenue is provided by 
symmetry-breaking, where the ordered moment of a system becomes 
effectively macroscopic, and thus long-lived. For concreteness, consider an Ising system, in $d\geq2$, where there is 
a finite-temperature phase transition in equilibrium. 

Any not too large 
imperfection, be it through 
noise or via a spin flip process detuned from an exact $\pi$-rotation, will leave the majority of the spins pointing in a preferred direction which alternates between up and down. The defects can thus be identified and removed by cooling the system, i.e.\ by making the small fraction of
spins antialigned
with the majority direction fall in line, thus pushing the system back towards the fully ordered
ground state.

The $\pi$-spin glass is distinct from the above phenomena in that it
has a time-crystalline phase in the narrower sense outlined above, of not
requiring energy input or extraction from the bath. We
next enquire about its stability when put in communication with an
environment. It can in fact be stable in a rather straightforward
way, or rather by a sleight of hand, basically by cutting a $\pi$-spin glass into two, and labelling one half the bath and the other half the system.

To achieve this, 
consider two essentially identical systems which are $\pi$-spin
glasses in isolation, and couple them weakly. This may be done, e.g.\
with two of the disordered spin chains introduced above, arranged as a
ladder, with the additional coupling in the form of an Ising exchange along the rungs. If
this rung coupling is weak enough not to cause a transition out of the
$\pi$-spin glass phase, the combined system will `continue' to be a
$\pi$-spin glass, and labelling one rung the system and the other the
environment achieves our aim.

The question then is how much structure one allows the environment to
have. This is to some degree a question of choice, a point which
we illustrate next after noting that for a sufficiently featureless,
`generic', environment, time-crystalline order does not survive.

The demonstration of the disappearance of the
DTC signal for the case of a Markovian thermal
bath, in particular one without any further potentially useful fine-tuned structure,
is perhaps not entirely
surprising: thermal fluctuations are known to destroy many-body
localisation, and it would thence seem natural that they should
destroy a feature which has many-body localisation as one of its core
ingredients. The mechanism outlined here shows how the eigenstate
order underpinning the $\pi$ spin glass is in itself fragile to 
contact with the Markovian bath. 

The basic ingredient to the demonstration is the `block-diagonal ensemble' 
structure of the density matrix of the $\pi$ spin glass. This is a related to
the diagonal ensemble in conventional systems, which states that at long
times, expectation values of an operator ${O}$ for a system starting in a well-defined pure state $\psi_0$
depend only on the size of the coefficients, $c_n$, of that state when expressed in a
basis of eigenstates $\left\{|\nu\rangle\right\}$, as the off-diagonal terms average to zero on account
of their oscillation with a `random' energy difference 
$E_\nu-E_\mu$ between two states $\mu$ and $\nu$, see Eq.~\eqref{eq:Ot-diag}.

In the presence of eigenspectrum order, we group the eigenstates into multiplets, $\nu,\zeta_\nu$. Here, the internal index $\zeta_\nu=1\ldots M$ labels the $M$ different states with fixed relative eigenenergies. For the case of the Floquet $\pi$-spin glass, $M=2$ and $E_{\nu,1}-E_{\nu,2}=\pi$ for all $\nu$. 

Eigenspectrum order thus eliminates the  `randomness' 
in the quasienergy differences allowing a restriction to the diagonal ensemble. 
Indeed, this underpins the subharmonic
oscillations described in detail in Sec.~\ref{sec:TTSB}. 

\begin{eqnarray}
    \langle\psi_0(t)|{O}|\psi_0(t)\rangle
    &=& \sum_{\nu,\zeta_\nu,\mu,\zeta_\mu}
c_{\nu,\zeta_\nu}^* c_{\mu,\zeta_\mu} 
\langle \nu,\zeta_\nu|{O}|\mu,\zeta_\mu\rangle
\exp[i t(E_{\nu,\zeta_\nu}-E_{\mu,\zeta_\mu})]\\
&\longrightarrow\atop{t\rightarrow\infty}&
\sum_{\nu,\zeta_\nu,\zeta^\prime_\nu}
c_{\nu,\zeta_\nu}^* c_{\nu,\zeta^\prime_\nu} 
\langle \nu,\zeta_\nu|{O}|\nu,\zeta^\prime_\nu\rangle\exp[i t(E_{\nu,\zeta_\nu}-E_{n,\zeta^\prime_\nu})] \ ,
\label{eq:bda}
\end{eqnarray}
as the energy differences within the block of multiplet $n$ $E_{\nu,\zeta_\nu}-E_{\nu,\zeta^\prime_\nu}$,  in Eq.~\ref{eq:bda} are those determined by the eigenstate order. For the $\pi$-SG, these are either 0 (for $\zeta=\zeta^\prime$) or $\pi$ (for $\zeta\neq \zeta^\prime$).  

Now, the action of a generic bath will lead to transfer of weight between 
different multiplets $\nu\neq \mu$. However, between these, and unlike in the
case of the ordered Ising magnet mentioned above, there is in 
general no robust relationship between observables in the glassy correlators
between different multiplets. As the weight gets distributed over more and more
multiplets due to contact with the bath, the subharmonic signal washes out and eventually
gets lost entirely. 

An ingenious proposal 
for survival of a DTC despite a specific form of 
communication with the environment
-- much less artificial than the two-leg ladder 
constructed above -- 
\mcite{2019arXiv190204986D} 
was made in the context of a kicked random Ising chain
subject to uncontrolled 
radiative decay. Using information on the photons
likely to be emitted, one can place a mirror to reflect these 
at a judiciously chosen distance so that their subsequent reabsorption by the 
system occurs with a phase shift conducive to stabilising the DTC. 
The resulting lengthening of the DTC lifetime can be considerable, as
well as robust to deviations from the optimal location of the mirror. 

If one is interested in observing DTCs in practise, on finite laboratory
length- and time-scales, this kind of stabilisation engineering may very
well be called for, and an autonomously
interesting exercise in understanding the underlying bath-system
physics.

\section{Concluding remarks and outlook}
\mlabel{sec:outlook}

We have tried to present a self-contained overview over the physics of time crystals, from its underpinnings in 
statistical mechanics to its realisations in condensed matter and cold atom physics, as well as its connections to other fields.  In this concluding section, we provide a few comments on perhaps some
broader lessons one can take from this, and we also try to indicate some promising avenues for future
developments.

The general disdain in which perpetuum mobile proposals are held these days might be taken a 
powerful disincentive to studying time crystals. Further down the path, the relentless tendency of 
systems to approach thermal equilibrium, along with a no-go theorem, would also
seem to present  a formidable obstacle on the path to the destination time crystal.

However, as is so often the case for a fundamental physics 
question, a cooperative phenomenon in condensed matter
physics can provide an entirely new angle on this problem, 
as is famously familiar from the case of fractionally charged
quasiparticles \mcite{2001cond.mat..3366R}.
Assumptions too obvious or common sense even to be stated
explicitly may turn out not to be so indispensible after all:
the physics of cooperative phenomena which is the subject of
condensed matter physics takes relatively simple ingredients
to create complex cooperative behaviour. 

The search for time crystals is a case in point, where a number
of separate developments combined to evade the various obstacles
discussed above, in particular many-body localisation and eigenstate order, albeit
at the price of sacrificing continuous time-translation symmetry at the outset. 

The resulting product -- the $\pi$ spin glass or Floquet time crystal --  then was studied in more detail, which threw up a host of further surprises. Perhaps the most spectacular one was the notion of absolute stability: what looked like an Ising symmetry-breaking phase actually does not require an exact Ising symmetry to start with, but only an emergent one, which is  stable to small generic terms breaking the Ising symmetry.

The resulting deeper understanding then allows to relax various demands, which has in turn
led to the identification of a rich set of prethermal time crystalline phenomena which, while unstable in 
the long-time limit, nonetheless persist for extended regimes. These insights in turn were related to experimental developments, as time crystalline signals were detected in settings which did not fulfil all the requirements identified for the Floquet time crystals.

Indeed, this highlights another highly attractive feature  of condensed matter physics, namely the quick turn-around time between theory and
experiment. On a scale of many months to a few years, different experimental platforms and realisations
were identified and investigated. 

The breakneck speed of such developments is awesome to behold,
and if one were to search for a fly in the ointment, it may be that the time budget for ensuring that all 
dead ends, alternative explanations and limiting aspects are explored is also curtailed, 
putting a premium on primacy over  longevity. 

From the incomplete yet extensive lists of developments in this review, it is already clear that in the near term, 
there are plenty of interesting avenues to explore; be it to identify and classify exhaustively the various 
time-crystalline phenomena; to pin down the simplest models for their theoretical study, and 
 the optimal regimes for their experimental observation.

From a broader perspective, recent studies of the real-time dynamics of many-body quantum systems have yielded rather spectacular progress, providing a new view on the physics of thermalisation and equilibration, and the alternatives to it. Floquet time crystals are a particular highlight, as they have broadened our notions of spatiotemporal order in a fundamental way. Especially in the fields of driven, and open systems, it seems highly likely that many other interesting phenomena are still awaiting discovery. We hope that this review has given sufficient guidance to, and will  motivate, some readers from outside the field to join in this research effort.

\section{Acknowledgements}
\mlabel{sec:ack}
We are especially grateful to Arnab Das, Curt von Keyserlingk and Achilleas Lazarides for key collaborations on our work in this area and for generously offering comments on the manuscript, and to Matteo Ippoliti for extensive help with preparing figures. VK thanks Joonhee Choi, Soonwon Choi, Renate Landig and Mikhail Lukin for experimental collaboration. We are grateful to Sean Barrett, Immanuel Bloch, Robert Blum, Christian Gross,  Jared Rovny, and Johanness Zeiher for many discussions of their experimental platforms.  
We also thank Patrick Bruno, Anushya Chandran, Bert Halperin, Wen Wei Ho, David Huse, Ray Goldstein and Lucas Wetzel for insightful discussions. This work was supported with funding from the Defense Advanced Research Projects Agency (DARPA) via the DRINQS program. The views, opinions and/or findings expressed are
those of the authors and should not be interpreted as representing the official views or policies of the Department
of Defense or the U.S. Government. VK was partially supported by the Harvard Society of Fellows and William F. Milton Fund. This research was supported in part by the National Science Foundation under Grant No. NSF PHY-1748958, and by  the  Deutsche  Forschungsgemeinschaft through  ct.qmat, the Cluster of Excellence EXC 2147 (project-id 39085490).

\appendix
\renewcommand*{\thesection}{\Alph{section}}

\section{Watanabe and Oshikawa's Proof}
\label{sec:OW}

In ``Absence of Quantum Time Crystals", Watanabe and Oshikawa offer two proofs for the absence of spatio-temporal order in static systems. The first proof holds strictly in ground states while the second is claimed to hold in Gibbs states at $T>0$. We will now show that the second proof has an implicit assumption without which it fails.

WO consider two hermitian operators:
\begin{eqnarray}
    A = \sum_{\bf x} a({\bf x}) \nonumber \\
    B = \sum_{\bf x} b({\bf x})
\end{eqnarray}
where $a({\bf x})$, $b({\bf x})$ are local and the sum is over sites in a finite lattice of $N$ sites. They define
\begin{eqnarray}
    g_{AB}(t) = \frac{1}{N^2}\langle [A(t),B(0)] \rangle \\
    f_{AB}(t) = \frac{1}{N^2}\langle A(t)B(0) \rangle
\end{eqnarray}
with the expectation values being understood as those in the Gibbs state with inverse temperature $\beta$. They invoke the fluctuation dissipation theorem in a finite volume to relate the Fourier transforms of these two functions as
\begin{equation}
    g_{AB} (\omega) = (1 - e^{-\beta \omega}) f_{AB} (\omega) \ .
\end{equation}

Next, they note that the Lieb-Robinson bounds on the commutator in (A.2) implies that 
\begin{equation}
    \lim_{N \rightarrow \infty} g_{AB}(t) = 0
\end{equation}
at any fixed $t$. This follows upon examining the double sum in
\begin{equation}
   g_{AB}(t) = \frac{1}{N^2} \sum_{\bf x} \sum_{\bf y} \langle [ a({\bf x},t),b({\bf y},0)] \rangle 
\end{equation}
where (essentially) only the terms for which $|{\bf x - y}| < v_{LR} t$ are nonzero and there are a finite number of these for each $\mathbf{x}$ at a fixed $t$ in the limit $N\rightarrow \infty$.

Next, WO argue that the vanishing of $g_{AB}(t)$ implies that 
\begin{equation}
    \lim_{N \rightarrow \infty} g_{AB}(\omega) = 0
\end{equation}
as well, from which they conclude that $f_{AB}(\omega) \propto \delta(\omega)$.

But does (A.5) imply (A.7)? This would be true if
\begin{equation}
    \lim_{N \rightarrow \infty} g_{AB}(\omega;N) = \lim_{N \rightarrow \infty} \int dt e^{i \omega t} g_{AB}(t;N) \stackrel{?}{=}  \int dt e^{i \omega t} \, \lim_{N \rightarrow \infty} g_{AB}(t,N)
\end{equation}
where we have made the $N$ dependence explicit. But we cannot interchange the integration over $t$ and the $\lim_{N \rightarrow \infty}$ unless $g_{AB}(t,N)$ decays sufficiently rapidly in $t$ at every fixed $N$ --- in other words, one must assume a lack of temporal order in order to prove a lack of spatiotemporal order, rendering the proof almost tautological.

For example, we may wish to invoke the dominated convergence theorem which requires that $|g_{AB}(t,N)| < h(t)$ for every $N$ where $h(t)$ is integrable. WO do not offer such a bound. The Lieb-Robinson bound itself does not imply the desired long time asymptotics. It is perfectly consistent with a function of the form $g_{AB}(t) \sim O(1)$ for {\it all} $t > N^{1/d}/v_{LR}$.
Hence the proof does not go through.

\newpage
\section*{References}
\bibliography{tcrev}
\bibliographystyle{iopart-num}

\end{document}